# Recent advances in DNA origami-engineered nanomaterials and applications

Pengfei Zhan, Andreas Peil, Qiao Jiang, Dongfang Wang, Shikufa Mousavi, Qiancheng Xiong, Qi Shen, Yingxu Shang, Baoquan Ding*, Chenxiang Lin*, Yonggang Ke*, Na Liu*

**Abstract**

DNA nanotechnology is a unique field, where physics, chemistry, biology, mathematics, engineering, and materials science can elegantly converge. Since the original proposal of Nadrian Seeman, significant advances have been achieved in the past four decades. During this glory time, the DNA origami technique developed by Paul Rothemund further pushed the field forward with a vigorous momentum, fostering a plethora of concepts, models, methodologies, and applications that were not thought of before. This review focuses on the recent progress in DNA origami-engineered nanomaterials in the past five years, outlining the exciting achievements as well as the unexplored research avenues. We believe that the spirits and asset that Seeman left for scientists will continue to bring inter-disciplinary innovations and useful applications to this field in the next decade.

## 1. Introduction

Molecular self-assembly plays a fundamental role in the structural complexity and functionality of biological systems. Nature evolves sophisticated ways to self-assemble information-carrying materials into well-organized cellular architectures from the nanoscale to the microscale. In particular, the cell can be viewed as a biological factory containing various molecular machines that work in concert. The ultimate goal of synthetic biology is to build biological mimics that can complete individual tasks as well as support artificial signaling and communication in a fully controllable manner. In comparison to proteins or other biomolecules, DNA is ideally suited for constructing biological mimics or, more broadly speaking, functional structures and materials by molecular self-assembly due to many reasons. First, DNA has a well-defined B-form structure: a right-handed double helix formed by two complementary single strands with 10.5 bases per helical turn and about 2 nm in diameter. Second, DNA strands undergo predictable interactions. Single-stranded DNA (ssDNA) hybridizes into a double helix following the Watson-Crick base pairing, where A pairs with T and G pairs with C. Third, facile synthesis and chemical stability of DNA render many practical applications possible. Fourth, DNA can be readily modified and functionalized with a variety of nanoscale entities that possess interesting biological, chemical, magnetic, electrical, or optical properties. Fifth, DNA self-assembly is a highly parallel bottom-up fabrication method with spatial accuracy and resolution at the nanoscale. Sixth, dynamic DNA structures exhibit excellent spatiotemporal responses to a multitude of external stimuli with the inherent sequence specificity, programmability, and addressability of DNA.

The idea of using nucleic acids, the primary genetic material in cells, as building blocks for the construction of functional structures and materials, was conceived by Nadrian Seeman in 1982.[1] In the past four decades, the field of DNA nanotechnology has made significant advances, from assemblies of small DNA motifs to giant DNA superstructures of gigadalton molecular weight, from static to complex dynamic structures in response to environmental factors, from unmodified DNA/RNA nanostructures to functional constructs for a wide range of applications in biomedical engineering, drug delivery and therapy, nanophotonics and electronics, energy harvesting and transfer, biochemical sensing, super-resolution imaging, nanomachinery, biomimetics, and synthetic cells. A plethora of nucleic acid self-assembly approaches have been developed, enabling the creation of complex DNA assemblies and novel hybrid nanomaterials from one dimension (1D) to two and three dimensions (2D and 3D). Meanwhile, the nucleic acid self-assembly mechanisms have also been better elucidated, facilitating the design, prediction, simulation, and optimization of diverse DNA-based systems.

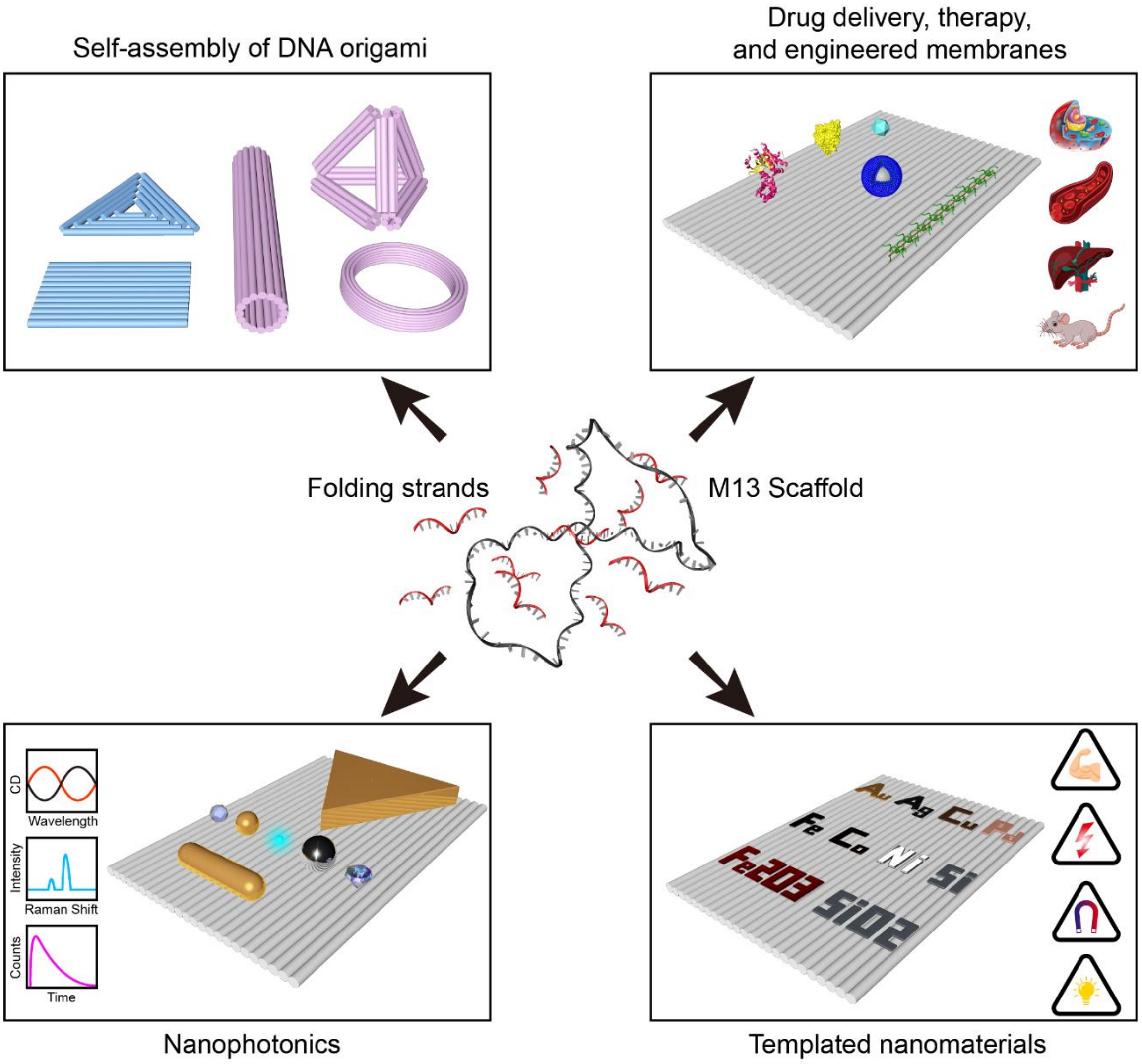

**Figure 1.** Schematic of DNA origami-engineered nanomaterials and applications.

This review focuses on the most recent advances in DNA-origami-based nanomaterials, including DNA assemblies, superstructures, nanodevices, functional hybrid systems, and many others in the recent five years. Previous important achievements have been discussed in the review by Hong *et al*.[2] and others.[3-4] Our review is structured as follows (see Figure 1). We start with a brief overview of the milestones in DNA self-assembly, followed by a comprehensive presentation of the recent breakthroughs in self-assembly methodologies and advanced DNA architectures. Subsequently, we recapitulate the working mechanisms and experimental realizations of DNA origami-templated functional nanomaterials, ranging from inorganic nanoclusters to biologically relevant molecules. Furthermore, we summarize the recent progress in bridging DNA nanotechnology with other research fields, such as drug delivery and nanomedicine, membrane biology, and nanophotonics. Finally, we finish this review with conclusions and an outlook to highlight the future avenues and promising directions in this vigorous, multidisciplinary field.

## 2. DNA self-assembly

### 2.1 Brief overview of the milestones

In 1982, Seeman realized that the branched DNA junctions could be connected into a 3D crystalline scaffold through DNA hybridization of single-stranded overhangs for protein crystallographic studies (Figure 2A). This conception laid the foundation of DNA nanotechnology. Nevertheless, naturally occurring branched DNA junctions often suffer from conformational instability due to branch migration induced by internal sequence symmetry. To circumvent this issue, in 1983 Seeman showed that the branch migration could be eliminated by using unique sequences on the DNA junctions to yield immobile Holliday junctions.[5] The immobile DNA branched junctions make DNA an ideal construction material to build DNA nanoobjects through single-stranded sticky ends. Following this strategy, many DNA nanostructures, such as 3D DNA cubes,[6] truncated octahedrons,[7] and Borromean rings[8] were created by connecting the branched DNA junctions. However, these branched junctions were relatively flexible, and the resulting DNA structures generally had very low rigidity. Later, Seeman and co-workers developed double-crossover molecules (DX),[9-10] which comprised two double helices connected through crossovers. The DX molecules are also called DX tiles, which are stiffer than a single DNA helix. The DX tiles were successfully constructed into periodic 2D lattices and tubes (Figure 2B).[11-12] Since then, several other DNA tiles were also demonstrated, such as triple-crossover tiles,[13] four-helix,[14] eight-helix, twelve-helix tiles,[15] parallelogram tiles,[16] cross-shaped DNA tiles,[17] triangular,[18-20] and three-point star motifs.[21] These DNA motifs significantly enriched the toolbox to design DNA nanostructures. On the other hand, the self-assembly of DNA tiles usually leads to DNA structures with uncontrolled sizes because DNA tiles repeatedly participate in the assembly process. The resulting DNA structures have thus low addressability, limiting site-specific modifications of DNA structures for many applications. It is also possible to assemble DNA tiles with unique sequences into finite structures with addressability. For instance, a 10-tile array and a 16-tile array were created using cross-shaped DNA tiles.[22-23] However, general methods to designing finite and addressable structures of arbitrary shapes were highly desirable.

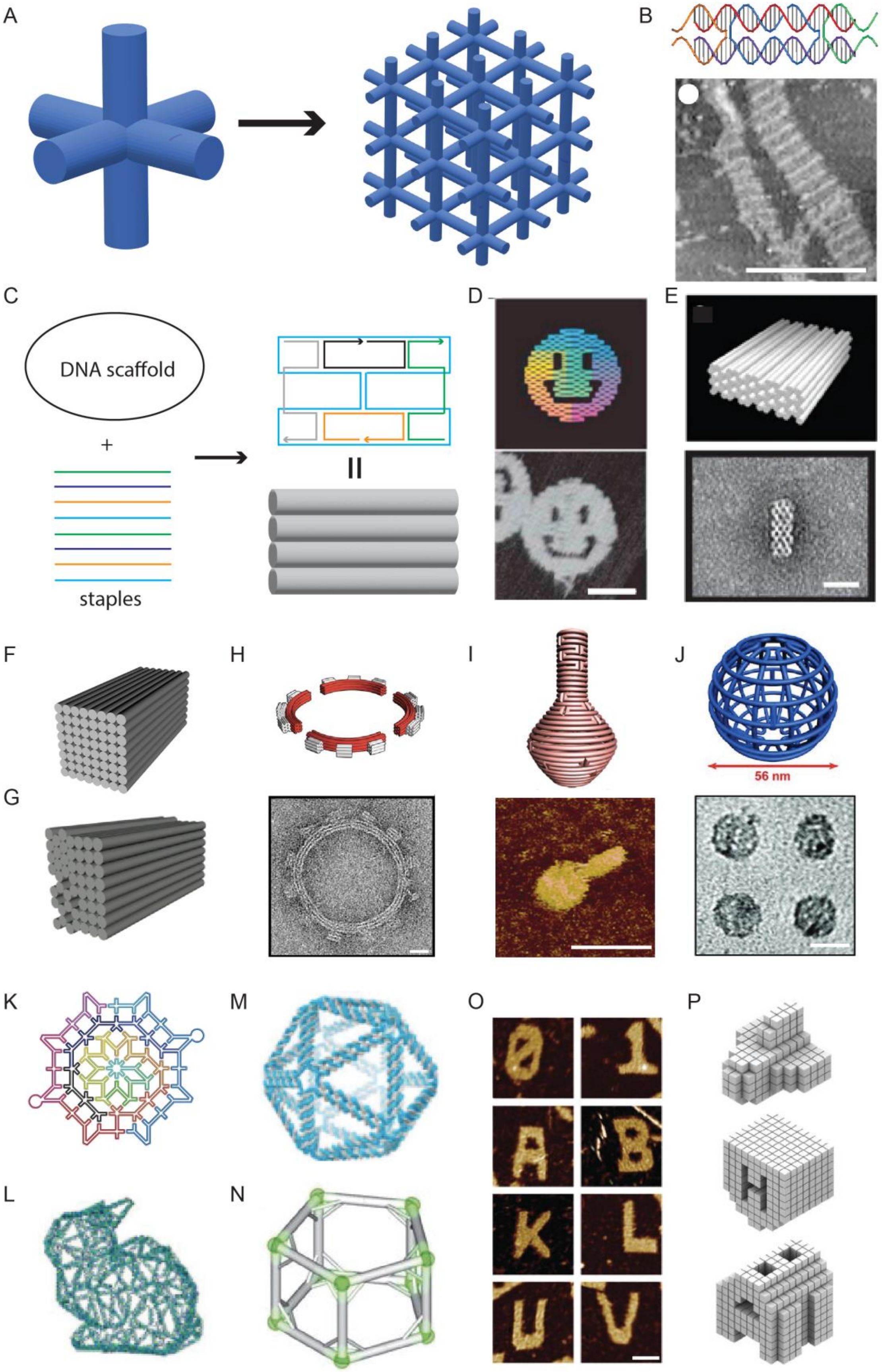

**Figure 2.** Overview of fundamental achievements in structural DNA nanotechnology. (A) A proposed 3D DNA crystal that is self-assembled from branched DNA motif for protein organization. (B) Double crossover DNA motif and the assembled 2D lattice imaged by AFM. Scale bar: 300 nm. Reproduced with permission from ref 11. Copyright 1998 Macmillan Publishing Ltd. (C) DNA origami assembly. A single-stranded scaffold of DNA is folded into a target shape by multiple short synthetic strands. Every short strand binds different regions of the scaffold and connects them together through crossovers. (D) A representative DNA origami structure with a smiley face shape and its AFM image. Scale bar: 50 nm. Reproduced with permission from ref 26. Copyright 2006 Macmillan Publishing Ltd. (E) A multiple layer DNA origami constructed from honeycomb DNA lattice. Scale bar: 20 nm. Reproduced with permission from ref 27. Copyright 2009 Macmillan Publishers Ltd. (F) A eight-layer DNA nanostructure packed from a square lattice. Reproduced with permission from ref 29. Copyright 2009 American Chemical Society. (G) A 3D nanostructure with 52 HB packed on honeycomb/square/hexagonal lattice. Reproduced with permission from ref 30. Copyright 2009 American Chemical Society. (H) Curved DNA structures formed the ring structure by hierarchy assembly. Reproduced with permission from ref 31. Copyright 2009 Macmillan Publishers Ltd. (I) A nanoflask constructed by concentric rings of DNA helices. Scale bar: 75nm. Reproduced with permission from ref 32. Copyright 2011 AAAS. (J) A sphere gridiron structure based on DNA four-arm junctions. Scale bar: 50 nm. Reproduced with permission from ref 33. Copyright 2013 AAAS. (K) A complex wireframe flower-and-bird structure. Reproduced with permission from ref 34. Copyright 2015 Macmillan Publishing Ltd. (L) A wireframe rabbit. Reproduced with permission from ref 35. Copyright 2015 Macmillan Publishing Ltd. (M) A wireframe icosahedron. Reproduced with permission from ref 36. Copyright 2016 AAAS. (N) A 3D DNA polyhedron made of origami tripods. Reproduced with permission from ref 37. Copyright 2014 AAAS. (O) Modular assembly of 2D DNA structures using single-stranded tiles. Scale bar: 50 nm. Reproduced with permission from ref 39. Copyright 2012 Macmillan Publishers Ltd. (P) 3D shapes assembled from a master set of DNA bricks. Reproduced with permissions from ref 40. Copyright 2012 AAAS.

To this end, two important strategies were developed. The first one was DNA origami realized by Paul W.K. Rothemund in 2006.[24] As shown in Figure 2C, long ssDNA as a scaffold is folded by hundreds of short synthetic DNA strands, called staples, into prescribed shapes. It is worth mentioning that before the era of DNA origami, early works also explored the use of long ssDNA for assembling DNA structures. For example, Yan *et al*. assembled DNA nanostructures by connecting multiple DNA tiles with a ligated scaffold.[25] Shih *et al*. used 1669-nt long ssDNA to self-assemble into an octahedron wireframe nanostructure with five other short DNA strands.[26] These works provided inspirations for the development of DNA origami, which is a milestone in DNA nanotechnology and has significantly advanced the field. In Rothemund's original paper, various 2D origami shapes were created, including square, rectangle, star, smiley face (Figure 2D), and triangle. Later, Shih and co-workers introduced 3D DNA origami by packing multilayers of DNA helices into a honeycomb lattice (Figure 2E).[27] The caDNAno software was also developed,[28] making the design process much more convenient. The multilayer DNA origami was later extended to form square (Figure 2F),[29] hexagonal, and hybrid lattices (Figure 2G).[30] By controlling the curvature and the DNA helix length, twisted and curved DNA origami was realized (Figure 2H).[31] Using a different strategy, Han *et al*. achieved 3D curvatures by bending double DNA helices to follow the rounded contours of the target object and then adjusted the position and pattern of crossovers to follow the natural twist density (Figure 2I).[32] Furthermore, DNA origami with crossed DNA helices using four-way junctions was reported and it greatly facilitated the design of DNA wireframe structures (Figure 2J).[33] Zhang *et al*. extended this strategy using multi-arm junctions, creating more complex wireframe structures (Figure 2K).[34] Nevertheless, this method required complex routing of the scaffold. Two computational algorithms for wireframe polyhedrons were developed to solve this problem (Figures 2L and 2M).[35-36] A 3D target geometry can be rendered by routing the DNA scaffold on the surface, followed by an appropriate arrangement of DNA staples. To construct large structures, Yin and co-workers used DNA origami tripods as the building blocks to build wireframe polyhedrons with edges over 100 nanometers in length (Figure 2N).[37]

The second strategy for constructing addressable DNA structures is the modular assembly of DNA structures by ssDNA tiles (SSTs). Unlike the tiles containing a structure body and sticky ends on both ends, SSTs only comprise four concatenated sticky ends, and different SSTs can bind together through the floppy sticky ends. Yin *et al*. initially used SSTs to assemble DNA nanotubes with different circumferences.[38] The authors later extended the SSTs for assembling different 2D and 3D DNA nanostructures.[39-41] Due to the modularity of this method, DNA nanoobjects of various shapes can be assembled from a complete set of master strands (Figures 2O and 2P).

## 2.2 Scaffolded self-assembly

### 2.2.1 Design principles

DNA origami enables the self-assembly of prescribed DNA nanoobjects from the bottom up. The design principle of DNA origami is that multiple short synthetic DNA strands bind to different regions of the DNA scaffold and fold it into a prescribed shape. This method is also called scaffolded self-assembly. The procedures for designing an arbitrary single-layer DNA origami are as follows. The target shape of a DNA nanostructure is converted into cylinders, which represent the DNA double helices, and the length of the cylinder is an integral multiple of the full helical turn. Next, these cylinders are connected by crossovers with an interval of 1.5 helical turns. The DNA scaffold continuously routes through each helix. Its complementary strand is divided into short DNA strands as staples, typically in the range of 15-60 nucleotides (nt). For designing multi-layered DNA origami, a 3D structure can be viewed as folding single-layered origami into multi-layers. The cross-section of the DNA helices can be packed into a honeycomb, square, or hexagonal lattice.

Both the DNA helices from the same layer and between the two layers are connected by crossovers at the permitted positions. Depending on the type of the cross-section pattern of the helices, the angles between adjacent crossovers are varied according to the parameter of the DNA helix (34.3° twist per base pair, or 21 base pairs every two turns). The angle is 120° for the honeycomb lattice, 60° for the hexagonal lattice, and 0° or 90° (0° is for the helices within the same plane and 90° is for the helices between two planes) for the square lattice. The staples on both helical edges are usually extended to avoid nonspecific multimer aggregations or act as connectors to form multimers. For an arbitrary shape, there could be multiple routing pathways, which lead to different assembly yields. Different long ssDNA can be used as scaffolds, one of which is the M13 virus genome (7249-nt long circular ssDNA). It was chosen and used by Rothemund due to its well-identified sequence and commercial availability. After the design process, the sequences of the staples are generated based on the scaffold sequences, following the Watson-Crick base pairing rule. The target DNA nanostructures are subsequently folded by mixing the DNA scaffolds with an excess amount of the staple strands, typically in a 1×TAE buffer with 12.5 mM magnesium chloride. The samples are slowly annealed from a specifically optimized temperature, for instance, 90° to room temperature. During the annealing process, the staple strands find unique binding positions on the DNA scaffold, and the target structure is eventually formed. Before the invention of DNA origami, it was commonly believed that eliminating secondary structures, purified DNA strands and precise stoichiometry were indispensable to correctly fold DNA structures. However, the success of the DNA origami technique proved that the assembly process is of high error tolerance.

### 2.2.2 Mechanism of the DNA origami assembly

Although remarkable success has been achieved, the assembly process of DNA origami still needs further exploration and optimization. In particular, a deep understanding of the folding mechanisms and principles will greatly help to improve the design and assembly yield of large and complex structures. Recent progress has revealed some details of the origami assembly, such as its cooperative behaviors[42-43] in a narrow temperature range and under magnesium or magnesium-free conditions.[44-46] A number of factors, including annealing temperatures, staple concentration, scaffold routing, *etc*., can affect the assembly yield, shape, and kinetics of the origami. A structure with different designs and sequences of strands may result in very different assembly yields.[27] Even for structures of similar shapes, different routings of the scaffolds may change the overall kinetics and thermodynamics of the folding processes. It was shown that the nick point distribution of staples strongly influences the assembly.[46] In addition, the nucleation of a 2D rectangular origami was found more likely to occur on both edges, probably resulting from the favorable thermodynamics.[47] However, increasing the concentration of staples in the middle changed the nucleation position to the center region, suggesting that competitive nucleation existed among the staples. The competitive assembly was further confirmed in an origami design with two sets of staples.[48] Increasing the concentration of one set of staples promoted the corresponding structure in the final products. Furthermore, Bath and colleagues studied the folding pathway of DNA origami.[43] They designed a DNA origami system that could fold into multiple distinct shapes and found that the scaffold's early formation of long-range connection by the staples could strengthen a specific folding pathway. Their finding suggested that controlling the initial binding of staples could guide the assembly pathway.

To further understand the folding pathway, the Dietz group investigated the folding behavior of every single staple in a multilayer DNA origami structure.[49] The folding kinetics of every staple was measured using both intra-strand assay (fluorescent dye and quencher on two termini of the same staple) and inter-strand assay (fluorescent dyes on the two termini of two proximal strands). The authors observed that the time required for the complete folding of the DNA staples varied from 20 to 200 min. These strands followed a defined sequence of events during the folding. The segment truncation experiments further verified this observation. When deleting the 20 fastest incorporation terminal segments, the intermediate structure no longer formed, but the full structure could still form after several hours. Nevertheless, deleting the 20 slowest incorporating terminal segments did not prevent the formation of the intermediate structure, but the fully folded structure was delayed. Further analysis of the folding curves revealed that each folding reaction presented two-phase kinetics. The first phase took place between 2 to 50 min and the second phase ranged between 100 to 350 min. The inter-strand assay showed that a single strand could display independent binding kinetics on both termini. This result indicated that the origami folding occurred at the individual binding domain level rather than the whole strand level. In addition, they found that the incorporation time of the strands did not correlate with the free energy of hybridization, in contrast to the previous hypothesis, which suggested that the strand sequence did not determine the incorporation sequence during folding.[44] To further interpret this observation, the authors performed permutation experiments using another set of strand sequences with the same design pattern. It was found that the two folding pathways were strongly correlated. This finding confirmed that the folding pathway was determined by the cooperative effects rather than the local sequences of the strands. Additionally, they mapped the mean folding time data with scaffold and strand crossovers onto a graph. The results revealed that once crossovers connected certain distant regions on the scaffold, the folding began cooperatively. Nevertheless, it is still challenging to conclude what factors direct the sequence of the folding events using simple design rules.

### 2.2.3 Higher-order DNA origami assembly

Since the invention of DNA origami, building large DNA structures has been a pursuit in DNA nanotechnology. In general, DNA structures can be made larger by increasing the length of the DNA scaffold or the hierarchical assembly of DNA origami monomers. A straightforward solution is to use a longer DNA scaffold, which can be obtained by polymerase chain reaction (PCR), rolling circle amplification, or extracting from λ/M13 hybrid virus.[50-51] However, there are several issues with a longer DNA scaffold to make large DNA structures. First, the increased length of the DNA scaffold can give rise to more internal secondary structures, which reduce the folding yield of DNA origami. Second, the cost of constructing large structures increases with the length of the scaffold.

In contrast, the hierarchical assembly uses origami monomers with programmed connectors to assemble into periodic or finite superstructures. This method can significantly reduce costs. Hybridization and base stacking are the two basic driving forces for hierarchical assembly. DNA origami units with extending sticky ends hybridize based on the sequence-specific Watson Crick base pairing to form polyhedral origami structures,[37] 1D ribbons,[52-54] 2D lattices, and crystals.[55-57] Base stacking is another route for stabilizing DNA double helices. It is enabled by intramolecular forces between aromatic rings of adjacent base pairs in a DNA helix. It was first observed by Yan *et al*. and studied for the assembly mechanism of DNA nanostructures.[25, 58] Blunt-ended DNA helices protruding from DNA origami edges can be programmed to act as chemical bonds to achieve specific binding. DNA origami with shape complementarity can assemble into higher-order structures. For example, Dietz and co-workers developed dynamic 3D multilayered origami,[59-60] while Qian and co-workers designed square-shaped DNA origami with 4-fold symmetry as tiles to assemble into DNA origami arrays.[61] Sugiyama and co-workers combined shape complementarity and hybridization for higher-order assembly.[62-64] In addition, the hierarchical assembly was achieved by surface assistance.[65-75]

Different methods of hierarchical assembly were reported successively. The Dietz group combined base stacking with self-limiting structures for hierarchical assembly with sizes up to micrometers.[76] The advantage of the self-limiting structures is that higher-order structures can be assembled from a few monomers. Like many homo-oligomeric proteins, which contain only one type of subunit, DNA origami monomers can be assembled into higher-order structures according to the designed binding rules. The authors used a V-shaped DNA origami structure (V brick) as the building block for the self-limiting assembly (Figure 3A). The shape complementarity mediated the oligomerization of the V bricks from the two asymmetric, interlocked self-complementary surfaces. The opening angle of the V bricks could be adjusted by changing the length of double-helical spacers between two surfaces. This angle determined the number of building blocks in oligomers as well as the diameter of the ring-shaped structure. By adding another set of self-complementary docking sites into the V bricks, the oligomerization perpendicularly to the plane of the ring was enabled, forming long tubes with thick walls. Further complex structures in 3D were also explored using this method, including tetrahedrons, hexahedrons, or dodecahedrons.

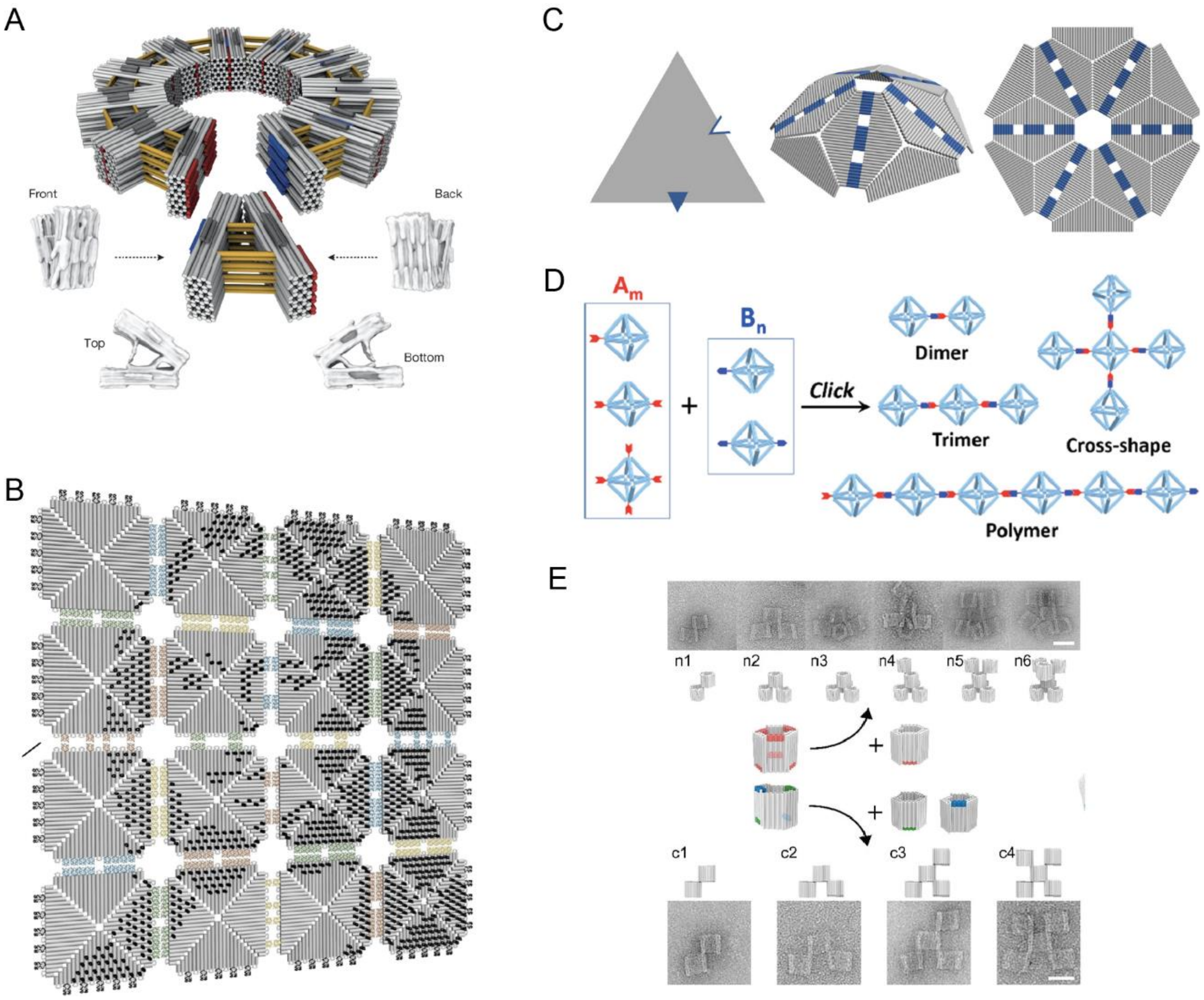

**Figure 3.** Scaffolded assembly for higher-order structures. (A) A ring DNA structure constructed by self-limiting V-shape multi-layer DNA origami. Reproduced with permission from ref 76. Copyright 2017 Macmillan Publishers Ltd. (B) A 4 × 4 DNA origami array with an example pattern of Mona Lisa constructed from the fractal assembly. Reproduced with permission from ref 77. Copyright 2017 Macmillan Publishers Ltd. (C) A 3D DNA structure constructed by triangular origami tiles. Reproduced with permission from ref 78. Copyright 2018 American Chemical Society. (D) Octahedral DNA frames were constructed into higher-order nanoscale objects through controlled bonds at the vertex and linkage by chemical reactions. Reproduced with permission from ref 79. Copyright 2019 American Chemical Society. (E) A hexagonal prism DNA origami was used as the building block for higher-order assembly through the orthogonal and directional bonds at both ends. Scale bars: 40 nm. Reproduced with permission from ref 81. Copyright 2022 Wiley.

Almost simultaneously, the Qian group developed another hierarchical method called fractal assembly.[77] The authors used square DNA origami as the building block to create 2D DNA origami arrays. The square DNA origami could bind each other through the connectors at the interfaces. They used one set of edge staples as connectors, and each DNA origami only had a subset of the connectors. A large DNA origami array was created using the hierarchical assembly of DNA origami tiles according to their recursively local assembly rules. In addition, each DNA origami monomer could be decorated with surface patterns, so that the DNA origami array revealed the entire pattern image (Figure 3B). Notably, the authors developed the FracTile Compiler software to help non-experts design DNA sequences and experimental procedures for making large DNA patterns. Several patterned images were created on the DNA origami arrays, including the Mona Lisa, a rooster, and a chess-game pattern with the help of this software. As a complementary approach to square origami tiling, they used triangle origami tiling to enrich the design space for the fractal assembly.[78] Each triangle edge could be programmed to be complementary to itself or to the other two edges. Surprisingly, the triangle tiles could assemble into 3D instead of 2D structures in a 20-tile design (Figure 3C), probably because the tiles bent at a small angle along the seams between the three isosceles triangles. Other different 2D and 3D arrays were further constructed by programming the binding specificity of the edges. The programmability of the edges in the DNA origami tiles makes them ideal building blocks for the realization of complex higher-order DNA structures.

DNA origami can also be programmed to mimic structures of molecules formed by atoms depending on their valences and anisotropic binding modes. In recent years, significant progress has been made in developing valence-controlled origami monomers for higher-order assembly. The Gang group reported an approach for the valence control of DNA origami polyhedrons using click chemistry.[79] DNA self-assembly usually uses non-covalent bonds (hydrogen bond or base stacking) to connect different species. The use of click chemistry offers permanent and site-specific ligation between DNA nanoobjects. In their work, the authors modified the vertex of an origami octahedron with azide or dibenzocyclooctyl (DBCO). Different DNA origami octahedrons carrying these groups could be ligated together by copper-free azide-alkyne click reaction. By controlling the valence of the functional groups on the origami octahedron, the DNA objects were connected into different architectures, including dimers, trimers, crosses, and polymers (Figure 3D). In another work, the same group designed a DNA nanochamber, whose bonds were programmed on the three orthogonal axes.[80] The bonds were fully addressable and differentiated, which allowed the formation of 1D, 2D, and 3D organized arrays. Using different binding modes, the authors demonstrated that heteropolymers, helical polymers, 2D lattices, and mesoscale 3D nanostructures could be encoded by the sequence of the bonds. The Wang group realized finite hierarchical assembly through orthogonal and directional bonding on the hexagonal prism DNA origami (HDO).[81] The HDO unit contained six distinguishable bonds at each end for directional bonding. Each bond had two opposite orientations for binding, yielding multiple possible directional binding configurations between two HDOs. A library of finite HDO superstructures was constructed by combining HDOs with appropriate valences (Figure 3E). Also, other methods were reported for the hierarchical assembly of DNA origami. For example, the Bathe group realized supramolecular assemblies with wireframe DNA origami.[82] The Ding group used flexible and covalent-bound branched DNA structures to template the assembly of triangular DNA origami into super-DNA origami.[83] The Gu group took advantage of cholesterol-modified DNA to drive the DNA origami assemblies through the cholesterol-encoded hydrophobic interaction.[84]

### 2.3 Scaffold-free self-assembly

#### 2.3.1 DNA tiles

DNA tile assembly is a versatile method of creating DNA structures using small DNA motifs, with the DX tile being one of the most used DNA motifs. Individual DX tiles connect to each other through the complementary sticky ends. Various DNA structures such as 2D lattices,[11, 85-86] nanotubes,[87-89] automation pattern,[90] DNA Sierpinski triangles,[91] and algorithmic self-assembly systems,[92-94] have been constructed by DX tiles. There are several types of DX tiles, such as DAE-E, DAO-E, DAE-O, DAO-O. The names are chosen according to their structural configurations, such as the number of crossovers (double or single), the orientation of the strands through the crossover (parallel or antiparallel), the number of half-turns between intramolecular crossovers (even or odd), and the number of half-turns between intermolecular crossovers (even or odd). Depending on the number of half-turns between the tiles, the architecture of the DNA tile assembly can be classified into E-tiling and O-tiling, which contain an even number of half-turns and an odd number of half-turns, respectively. Usually, E-tiling produces DNA tubes, and O-tiling generates planar DNA ribbons. This phenomenon is due to the intrinsic curvature of DNA tiles.[17, 88] In E-tiling architecture, the tiles are aligned identically, which results in tube formation due to the intrinsic curvature of each tile, whereas O tiling generates planar ribbons because of the cancellation of cumulative curvatures. The Xiao group investigated the quantitative physical description of the chirality of DNA nanotubes induced by intrinsic tile curvature and with the arm twist (Figure 4A).[95] Three types of tile cores were designed using three- and four-way junctions. Tile cores were joined with arms having an odd and an even number of half turns, forming planar 2D lattices and tubes, respectively. By coupling the intrinsic curvature of tile cores with the arm twist, right-handed (RH) or left-handed (LH) chiral DNA tubes were obtained reliably. They also provided a method to measure the chirality of the DNA tubes by resolving the structure with atomic force microscopy (AFM) at the single-tile level.

Other than using DX tiles to form DNA tubes, SSTs were employed to assemble into DNA tubes with prescribed circumferences.[38] Given an even number of DNA double helices, they can roll up into a tube with zero offsets along the two long edges, generating a straight and unstressed tubular structure. Insertion or deletion of base pairs into a double helix causes expansion/compression and RH/LH torques. Expansion or compression in turn leads to bending, and torque results in twisting.

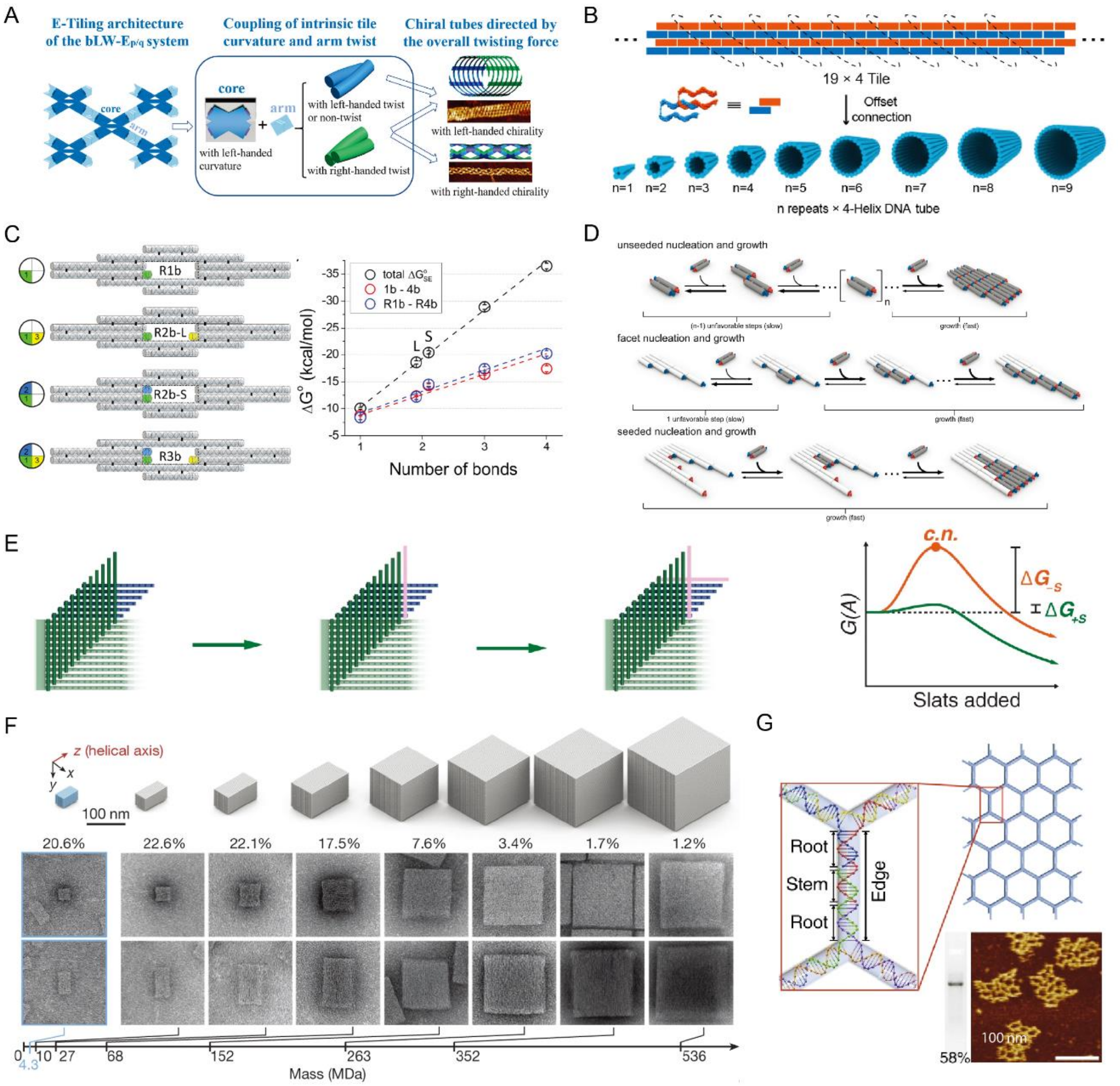

**Figure 4.** Scaffold-free assembly. (A) Regulation of tube chirality coupled with arm twist. Reproduced with permission from ref 95. Copyright 2022 American Chemical Society. (B) Programming the tube circumferences by offset connection. Reproduced with permission from ref 97. Copyright 2019 American Chemical Society. (C) Structural designs for different scenarios of tile attachments and observed energy penalty in tile attachment. Reproduced with permission from ref 99. Copyright 2017 American Chemical Society. (D) Regulation of DNA tile assembly with controlled nucleation. Reproduced with permission from ref 100. Copyright 2021 American Chemical Society. (E) Crisscross assembly with ssDNA slats. Reproduced with permission from ref 101. Copyright 2021 Macmillan Publishers Ltd. (F) DNA brick nanostructures with up to 500 MDa were built from 52-nt single-stranded tiles. Reproduced with permission from ref 102. Copyright 2017 Macmillan Publishers Ltd. (G) A 2D wireframe array of hexagonal tessellation pattern with 3-arm vertices. Scale bars: 100 nm. Reproduced with permission from ref 114. Copyright 2019 Macmillan Publishers Ltd.

The combined effects force the tube into a helical shape. For an even number of DNA double helices in a DNA tube, it must close the double helices with a discrete offset, which generates a supertwisted helix bundle. This supertwist could be tuned by reducing the closure offset through shifting a single tile, one nucleotide position at a time within the structure.[96] Interestingly, it was found that the Cy3 dye modification of the tube structure exerted a small amount of bending to the tube structure, similar to the base insertion effect. By combining the Cy3 modification with the fine-controlled supertwisting *via* tile shifting, the stepwise control of the microscale diameter of the DNA helical tubes could be achieved. The Ke group employed the "offset connection" to program the tube circumference (Figure 4B).[97] The boundary SSTs were designed to close the tube in an offset manner. The SST array grew wider to compensate for the offset and generated a wide tube with some degree of twist. Using this strategy, they assembled DNA tubes with various helices from the 19×4 to 19×14 tiles.

#### 2.3.1.1 Kinetics of DNA tiles

It has been shown that the kinetics of the DX tile homodimerization was 2-fold higher than that of ssDNA hybridization,[98] although DNA tiles are much bigger than ssDNA and less diffusive. This indicated that the nucleation efficiency, as well as the rate constant of association of DNA tiles, could be enhanced by introducing more sticky ends. It was confirmed that DNA tiles with two sticky ends had a higher association rate than the ones with a single sticky end. The higher rate was mainly attributed to the lower activation energy of dimerization. Abstract tile assembly model (aTAM) and kinetic tile assembly model (kTAM) were proposed to help with the design, simulation, and prediction of the tile assembly. In the kTAM, assumptions were made as follows. First, the attachment rates were constant and equal, regardless of the number of correct or incorrect bonds for a tile at the binding site. Second, the relation between the binding strength and the number of bonds was linear. For the attachment rate, although this assumption was verified on the mica surface, the kinetics of DNA tile assembly in solution might be different because tile diffusion in solution is not a rate-limiting step. Instead, nucleation is the key factor in determining assembly kinetics. Liu and co-workers experimentally tested these hypotheses.[99] The authors designed a multimer using a DAE-E tile and monitored the single-tile attachment with different numbers and orientations of bonds (Figure 4C). The free energy change of each individual pair of sticky ends ($\Delta G_{SE}^{\circ}$) and the free energy change of single tile attachment ($\Delta G^{\circ}$) were independently measured. They found that the binding free energy was linearly correlated with the number of bonds. However, the sum of $\Delta G_{SE}^{\circ}$ was large than $\Delta G^{\circ}$ (for n-bound attachment, n=1-4), and the differences were proportional to the number of bonds, which indicated the energy penalties after the tile attachment. The authors attributed the energy penalty to the formed loop by the constrained configuration after tile attachment. After considering the loop penalty, they proposed the independent loop model, in which the free energy changes of the tile attachment were described as the combination of sequence-dependent bond strength and sequence-independent loop penalty with the assumption of no interloop tension. In addition, the authors measured the kinetics of a single DAE-E tile attachment. The on-rate of kinetics was not equal for all binding scenarios. It depended on multiple factors, such as the number of sticky ends, the sequence of sticky ends, and the steric crowding effect (structural flexibility and the accessibility of the binding site). The different rates could result in different final products, such as tubes or long narrow ribbons. Furthermore, they confirmed that mismatched sticky ends did not contribute to free energy changes for the tile attachment.

The tile self-assembly starts with nucleation, which can be classified into three modes: unseeded nucleation, facet nucleation, and seeded nucleation. Unseeded nucleation is when monomer tiles bind together through one sticky end, forming a dimer, trimer, or other higher-order assemblies. It needs to reach a critical nucleus of n tiles before the seeded nucleation. After the nucleation barrier is overcome, tiles can continue to grow in the presence of a nucleation seed. There are two scenarios for the tiles growing directly onto the seed. A free tile attaches to the seed through two bonds, named seeded nucleation, or by a single bond, named facet nucleation. Liu and co-workers investigated the dynamics of three nucleation modes in nucleated self-assembly of DNA tiles (Figure 4D).[100] They designed a "frame-filling" model system using a rhombic DNA origami template and a self-complementary DX tile. By adjusting the activities of bonds on the template, the tile nucleation favored a specific nucleation mode. The tile assembly was $Mg^{2+}$ dependent, so increasing $Mg^{2+}$ over a critical concentration initiated the lattice formation. Therefore, the assembly of the free tiles could be triggered by rapidly tuning the concentration of $Mg^{2+}$. Under a constant $Mg^{2+}$ concentration, the assembly of tiles was temperature dependent, which suggested that three nucleation modes could be differentiated by adjusting the temperature range. The unseeded nucleation occurred at low temperatures (22 to 12 °C). The kinetic curves showed a lag phase at the beginning, followed by a fast phase, indicating a slower nucleation stage and a fast growth stage. Since unseeded growth involved 1-bond attachment, the binding was unstable at elevated temperatures. Unseeded growth could be inhibited at temperatures over 22°C. Facet nucleation could initiate in the presence of the origami frame with only a 1-bond sticky end with temperatures over 22°C, but was eliminated over 24°C. The seeded assembly was found dominant in the range of 24-26°C. As a result, by using the programmable origami frame design and temperature-controlled nucleation, the tile assembly could be guided to follow a specific pathway.

For DNA tiles self-assembled into periodic arrays, the rate-limiting nucleation often leads to products with varied sizes. Although seeds provide fast nucleation for homogeneous growth kinetics, spontaneous nucleation can still happen in a seed-independent fashion, resulting in a sub-population of different assemblies. To obtain only seed-dependent self-assembly, a large kinetic barrier has to be engineered to prevent spontaneous nucleation. In the meantime, the present seeds should bypass the barrier easily for fast growth. The large kinetic barrier can be achieved in two ways: by making the monomers inactive or by limiting incorporation of free monomers to cooperatively-attached intermediates. For the latter, a large coordination number could completely suppress the spontaneous nucleation while keeping rapid and irreversible growth on the seeds. The Shih group utilized crisscross slats to demonstrate seed-dependent self-assembly of massive DNA origami structures (Figure 4E).[101] The ssDNA slat monomers interacted only with the nearest neighbors on the origami seeds to achieve complete seed-dependent assembly and yielded ribbon products with average lengths over 4 µm. The number of seeds controlled the ribbon copy number precisely, and robust kinetic control of nucleation was realized in a wide range of divalent-cation concentrations, slat concentrations, and temperatures. Crisscross assembly could be further expanded to gigadalton 2D and 3D architectures for its all-or-nothing formation and nanoscale addressability.

### 2.3.2 DNA bricks

Despite the robustness of DNA origami for designing custom DNA structures, the dimension of DNA origami monomers and sequences of the DNA strands are limited by the DNA scaffold. Yin and co-workers invented the modular assembly strategy using SSTs and circumvented the use of DNA scaffolds. Hundreds of SSTs are assembled into finite and fully addressable DNA structures through local interactions. This method enables the realization of large, intricate 2D and 3D DNA nanostructures. In contrast to DNA origami, which uses half biological and half synthetic materials, the modular assembly utilizes complete synthetic strands. In other words, every strand from the DNA structures is addressable, and the sequences can be arbitrarily assigned, whereas the staple sequences in DNA origami have to follow the DNA scaffold according to the base pairing rule, and thus, the sequence diversity of DNA origami is limited. In addition, every SST can be chemically modified precisely, whereas the modification of the DNA scaffold is not routinely applied. The most important feature of the DNA brick method is its modularity. For the 2D or 3D DNA structures made of SSTs, they can serve as molecular canvases and each individual SST acts as a pixel. DNA structures of distinct shapes can be obtained by selectively removing certain SSTs from the molecular canvas. The Yin group demonstrated more than 100 2D and 3D DNA structures assembled from a master set of SSTs.[39-40]

Since DNA scaffolds are not involved in the self-assembly of DNA bricks, the size of the formed DNA structures can be easily scaled up. Recent progress showed DNA-brick structures could be scaled up to 0.1-1 gigadalton.[102] The original design used a 32-nt DNA brick with an 8-nt binding domain for the self-assembly of 3D structures. The first generation of the DNA-brick led to DNA structures on the megadalton-scale from hundreds of unique bricks. In an updated version of a 52-nt DNA-brick with a 13-nt binding domain, the yield and thermal stability of the DNA structures were enhanced. The improvement was due to two reasons. First, the 13-nt binding domain had much higher sequence diversity than the 8-nt domain, reducing mis-binding between two DNA bricks when a large number of bricks were used to form superstructures. Second, the melting temperature of the 13-nt binding domain was higher than that of the 8-nt domain, which resulted in a faster nucleation rate and higher thermostability after the assembly of the 52-nt DNA-brick. As a result, the second generation of the DNA-brick enabled the construction of 0.1-1-GDa structures from 10,000 bricks. The largest structure was a 532.4 MDa cuboid with more than 30,000 unique bricks (Figure 4F). Large multimer structures were also constructed through the hierarchical assembly, such as a 1GDa tetramer built from four 262.8 MDa monomers. Furthermore, the structures comprising a large number of components enabled programmable patterns. Custom cavities were created inside the structures, yielding intricated superstructures from a master set of DNA bricks.

#### 2.3.2.1 Mechanism of the DNA-brick assembly

DNA-brick self-assembly has been utilized to realize DNA structures of various shapes. Nevertheless, the assembly pathways for the DNA brick assembly are complex because nucleation can happen randomly at any location. Since most DNA bricks are identical in length, they have similar free energy. Thus, the DNA brick assembly is more likely stuck in kinetic traps. The assembly yield usually ranges from a few percent to ~30%, which is much lower than that from the DNA origami assembly, *i.e.*, over 90%. Also, it generally takes a long time to complete the assembly process. In a study by Ke *et al.*,[40] the authors merged the 16-nt half brick on the boundary helices with the preceding 32-nt full brick along the direction of its helix and formed a 48-nt DNA brick. These elongated boundary bricks (BB) improved the assembly yield by 1.4-fold. Despite the effectiveness of the elongated BB, the underlying mechanisms were not further explored in the original paper. It was hypothesized that two possible mechanisms could explain the increased assembly yield by the elongated BB. First, the elongated BB had three binding domains, whereas the regular bricks had two domains, so the former could serve as large seed strands to speed up the nucleation. Second, the final structures had more stability with the long bricks. The Reinhardt group used Monte Carlo simulations to investigate the effect of the elongated BB and the assembly process.[103] They used four sets of samples, namely no-BB structure, edge-BB structure, face-BB structure, and all-BB structure, and simulated the brick assembly process at a series of fixed temperatures. The authors found that the assembly products aggregated when the temperature went below 315 K, probably because low temperatures favored incorrect bonding and the assembly suffered from kinetic traps. In the range of 317 K and 318 K, the all-BB structure assembled the fastest, followed by the face-BB structure, edge-BB structure, and no-BB structure. At 319 K and above, the all-BB structure could still assemble up to a temperature of 326 K, but the edge-BB and no-BB structures took much longer to nucleate. Further simulation analysis showed that the BB structure could reduce the energy barrier for self-assembly and the mean first-passage time (MFPT) for faster nucleation. Nevertheless, this effect varied for different positions or the numbers of BB. The edge-BB structure had a slight reduction in the free-energy barrier compared to the no-BB structure, followed by the face-BB structure and the all-BB structure. Regarding the MFPT on the face-BB structure and the edge-BB structure, both structures had the same number of BB, but the face-BB structure had a lower MFPT than the edge-BB structure, likely because the face-BB structure had more interaction chances of interaction with other bricks than the edge-BB structure.

#### 2.3.2.2 Kinetics of DNA bricks

Since the elongated BB could increase the assembly yields and the nucleation rate, it is reasonable to design even longer strands to further tune the assembly yield and nucleation kinetics. The Ke group introduced a long DNA seed strand for the DNA brick assembly to program the nucleation process and assembly pathway.[104] The authors designed four DNA bricks, including a small triangle, a small rectangle, a large rectangle, and a large 3D rod. They also designed a seeded structure on the boundary position for each brick structure using a 425-nt strand. The unseeded and seeded structures were tested under the same isothermal assembly conditions to compare the assembly performance. Through experiments and computer simulations, it was demonstrated that the long seed strand could increase the DNA brick assembly kinetics by reducing the nucleation free-energy barrier. The authors observed increased assembly kinetics in both the initial and overall periods. The main reason was that the free brick molecules grew directly on the seed strand, and no lead-in time for the nucleation was required. For the unseeded structure, it needed thermal fluctuation to overcome the nucleation free-energy barrier. In addition, the seeded structures had a broader temperature range for successful assembly products and a higher overall temperature for the assembly to occur. A higher temperature could reduce the chance of incorrect folding and favor the correct folding pathway. Mis-binding between DNA bricks was less likely to happen at high temperatures, and the nucleation free-energy barrier was large enough to inhibit the nucleation of an unseeded structure. As a result, nucleation mostly occurred on the seeded structure. Furthermore, the authors investigated the thermal stability of the pre-assembled DNA bricks with and without the seed strand. They found that the seeded structure began to disassemble at a much higher temperature than the unseeded structure, which indicated that the seed strand could protect the structure under elevated temperatures. This protective effect might be due to the reduced nicking points on the seed strand, which weakened the base stacking interactions. The seeded assembly strategy not only helps the reduced assembly time, but also guides the assembly pathway for the complete structure.

## 2.4 Wireframe DNA structures

Wireframe DNA nanostructures have been previously demonstrated using both scaffold-free and scaffolded approaches.[34-36] Recent advances with the scaffolded approach combined with computations facilitate the design process and offer new features,[105-109] such as increased edge stability[105] or surface planarity,[108] irregular boundaries, and internal structures.[107] However, it remains challenging to scale up the size of an individual wireframe structure. The scaffold-free approach offers many advantages. First, the sophisticated routing of the scaffold is not needed, thus making the design process much simpler. Second, the structure size is not constrained by the DNA scaffold, so that the structure can be easily scaled up. Earlier scaffold-free methods for wireframe DNA polyhedrons were reported by several groups. For instance, Seeman and co-workers assembled a cube and a truncated octahedron from short synthetic DNA strands.[6-7] Turberfield and co-workers assembled chiral tetrahedrons.[110] The Sleiman group synthesized cyclic DNA concatemers by connecting organic molecules and short DNA strands, and subsequently assembled them into different DNA polyhedrons with other short DNA strands.[111] Shih *et al*. folded ~1.7k-nt ssDNA into an octahedron with five other synthetic DNA strands.[26] Moreover, the Mao group assembled tetrahedron, dodecahedron, buckyball structures from three-arm junction tiles, and an icosahedron from five-arm-junction tiles hierarchically.[112] The authors later demonstrated a series of DNA polyhedrons by connecting two types of multi-junction DNA tiles through sticky ends.[113] Nevertheless, these small DNA motif-based wireframe structures lacked structural complexity. The Wei group presented a general strategy for the self-assembly of complex wireframe structures using entirely short DNA strands.[114] Their design rule was such that first, a target structure was represented by a graph and further rendered into a node-edge network. Each node represented vertices of DNA helical arms, and each edge represented a DNA duplex. The edges were then split into complementary domains. Under this design framework, they assembled sophisticated wireframe structures, including 2D arrays (Figure 4G), tubes, polyhedrons, and multi-layered 3D arrays. Notably, the rigidity of the wireframe structures was not weakened compared to that of the DNA origami approach. The authors demonstrated that marshmallow-like polyhedrons after triangulation arrangements obtained substantial improvement in structural rigidity.

## 2.5 Single-stranded DNA and RNA origami

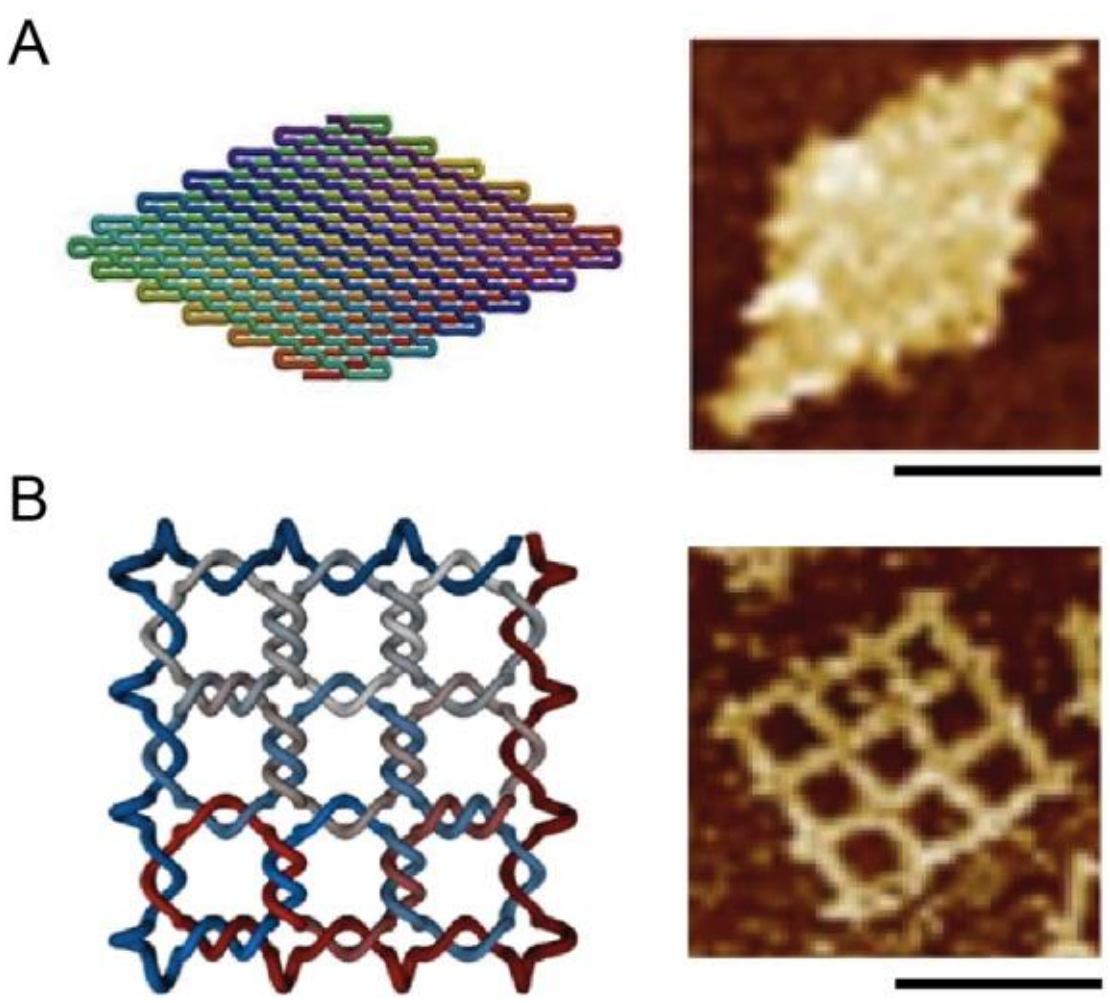

**Figure 5.** Single-stranded DNA and RNA structures. (A) A 12 × 12 ssOrigami containing 10,682 nt. Scale bar: 50 nm. Reproduced with permission from ref 120. Copyright 2017 AAAS. (B) A DNA 3 × 3 square lattice with a crossing number of 57. Scale bar: 100 nm. Reproduced with permission from ref 122. Copyright 2018 Macmillan Publishers Ltd.

Apart from the scaffolded DNA origami and DNA brick approaches, DNA or RNA structures can be assembled from single-stranded nucleic acids. ssDNA or ssRNA structures offer several advantages over multi-stranded structures. First, the single-stranded nucleic acid self-folding process is independent of the reactant concentrations, and thus the self-assembly yield is higher than that of multi-stranded nanostructures. Second, the rate of intramolecular folding is kinetically faster than intermolecular folding on multi-stranded structures. Third, the ssDNA or ssRNA can be synthesized *in vitro* by PCR or *in vivo* using a plasmid vector. The single-stranded nanostructures can be replicated and produced at a lower cost than those from the multi-strand self-assembly strategy. Early attempts along this direction included the folding of a 79-nt DNA strand into a four-arm DNA junction,[115] a 160-nt DNA strand into a multi-crossover DNA nanostructure,[116] a 286-nt

DNA strand into a tetrahedron structure,[117] and a 660-nt RNA strand into a six-helix rectangle tile.[118] These simple single-stranded structures could be replicated *in vitro* or in living cells.[118-119] Nevertheless, a general method for the self-assembly of a single-stranded nucleic acid into arbitrary shape and size with increased complexity was lacking. Han *et al*. introduced ssDNA and ssRNA origami (ssOrigami), in which DNA or RNA structures were self-assembled from ssDNA or ssRNA.[120] The idea of ssOrigami derived from the unimolecular folding of protein and RNA structures, in which minimal knotting complexity was achieved to avoid being kinetically misfolded during the folding process. The design strategy of ssOrigami is that ssDNA or ssRNA first self-assembles into partially complementary dsDNA and dsRNA and further coheres through parallel crossovers.[121] The main challenge of ssOrigami is to achieve structural complexity and programmability while avoiding the topological knots to ensure smooth folding. Several design strategies were employed to achieve minimal knotting complexity. First, the ssOrigami used parallel crossovers rather than commonly used antiparallel crossovers for interhelical cohesion. While DNA strands need to cross the central plane that contains all the parallel DNA helical axes at antiparallel crossovers, lacking a cross at the central plane for the parallel crossovers can reduce the knotting complexity of the structure. Second, secondary structures in the ssDNA present challenges for DNA synthesis. The helical domains are limited to 10 bp to reduce the local self-interaction of ssDNA. Third, the ssDNA strand is split into two approximately equal halves to separate all helical domains, reducing self-interaction significantly. Based on these rules, when an ssDNA strand is produced, it immediately folds into partially complementary ssDNA, which contains paired helical domains and unpaired single-stranded regions. These unpaired single-stranded regions can further form locking domains through base pair recognition. Every locking domain is between two adjacent parallel crossovers. The length of the locking domains is defined as 6 bp. The authors successfully created a variety of DNA ssOrigami (Figure 5A) using these design strategies. RNA ssOrigami could be readily created by changing the helical or locking domain lengths, due to the helical twist difference between the B-type DNA helix (10.5 bp/turn) and A-type RNA helix (11 bp/turn). The helical domain and locking domain for the DNA ssOrigami were 10 nt and 6 nt in length, respectively. In other words, 32-bp DNA contained two 10-nt helical domains and two 6-nt locking domains, noted as the 10-6-10-6 design. To adopt the scheme for RNA, the 10-6-10-6 design was modified to the 10-6-11-6 or 8-8-9-8 design, which corresponded to three full helical turns for an RNA A-type helix. The invention of ssOrigami provides a new blueprint for the design of user-specified nucleic acid nanostructures. The fact that nucleic acid nanostructures can be self-assembled *in vivo* enables the production of RNA nanomachines that can function inside cell environments.

Although the original ssDNA/ssRNA origami was designed with the knotting-free strategy, knotted DNA structures could also be constructed. Qi *et al*. reported complex molecular topology using single-stranded nucleic acids.[122] Like the knotting-free ssDNA/ssRNA origami, DNA parallel crossover motifs were used as modular building blocks. In contrast to the previous work, the dsDNA chain threaded itself multiple times in a wireframe network to yield a knotted structure (Figure 5B). Both the wireframe edge and paranemic crossover numbers were adjusted to program the knot crossing number so that knotted DNA wireframes could be designed with molecular topologies of varied complexity. The authors designed several 2D wireframe DNA structures and achieved molecular knots as high as 57, which was higher than the previously reported strategies based on the B-form/Z-form double-stranded DNA helices,[8, 123-125] paranemic crossovers,[126-127] and DNA four-way junctions.[128] This approach could be expanded to 3D structures to achieve more complex molecular topologies.

## 2.6 Meta DNA

DNA nanotechnology has mainly utilized DNA molecules based on DNA hybridization to build structures on the sub-micrometer scale. It is instructive to ask whether DNA nanostructures can act as an enlarged version of DNA molecules to promote structural constructions up to the micrometer range. Enlarged DNA structures retain the properties of individual DNA molecules, such as DNA hybridization, de-hybridization, and strand displacement. Importantly, they can act as building blocks to create superstructures based on higher-order binding rules. Zhang *et al*. initially proposed the idea of using DNA superstructures for exponential amplification.[129] Later, Chandran and co-workers systemically investigated meta-DNA as higher-order structures of natural DNA molecules.[130] Ideally, meta-DNA can keep both the structural and functional similarities of DNA molecules. The authors listed the properties of meta-DNA (mDNA) that resembled those of DNA molecules. (1) single-stranded mDNA comprised a linear polymer of meta-nucleotides. Each meta-nucleotide had the backbone to provide the rigidity and a base for the hybridization of double-stranded meta-DNA. (2) One meta-nucleotide from one strand could pair with one meta-nucleotide of another strand to allow two single-stranded mDNAs to form a double-stranded mDNA through the reverse complementary sequence (Figure 6A). (3) The meta-backbone was connected by a strong bond, while the meta-base was connected by a weak bond. (4) Double-stranded mDNA was separated by breaking the meta-base bond. (5) Single-stranded mDNA was more flexible and had a shorter persistence length than double-stranded mDNA. (6) mDNA underwent toehold-mediated strand displacement. During the process, the weak meta-base bonds were broken and re-formed. (7) mDNA could be synthesized by adding meta-nucleotides to the 3´ end of a primer strand with the help of a template mDNA strand. (8) mDNA restriction could be realized by cleaving the meta-backbone at the specific recognition site. The authors also suggested that the T-junction DNA motif could serve as the meta-nucleotide to achieve mDNA with connectivity and geometry.[131] The horizontal double helix of the T-junction served as the meta-backbone, and the vertical helix served as the meta-base. A linear chain of T-junctions formed a single-stranded mDNA with directionality. Two polymers of single-stranded mDNA could form double-stranded mDNA through meta-base binding. The T-junction should be protected before activation to avoid the spontaneous aggregation of the T-junctions into mDNA. The T-junction was sequentially activated and bound to the mDNA primer to synthesize mDNA. Also, the mDNA could mimic DNA-DNA and DNA-enzyme interactions, such as hybridization, denaturation, strand displacement, polymerization, and restriction cuts.

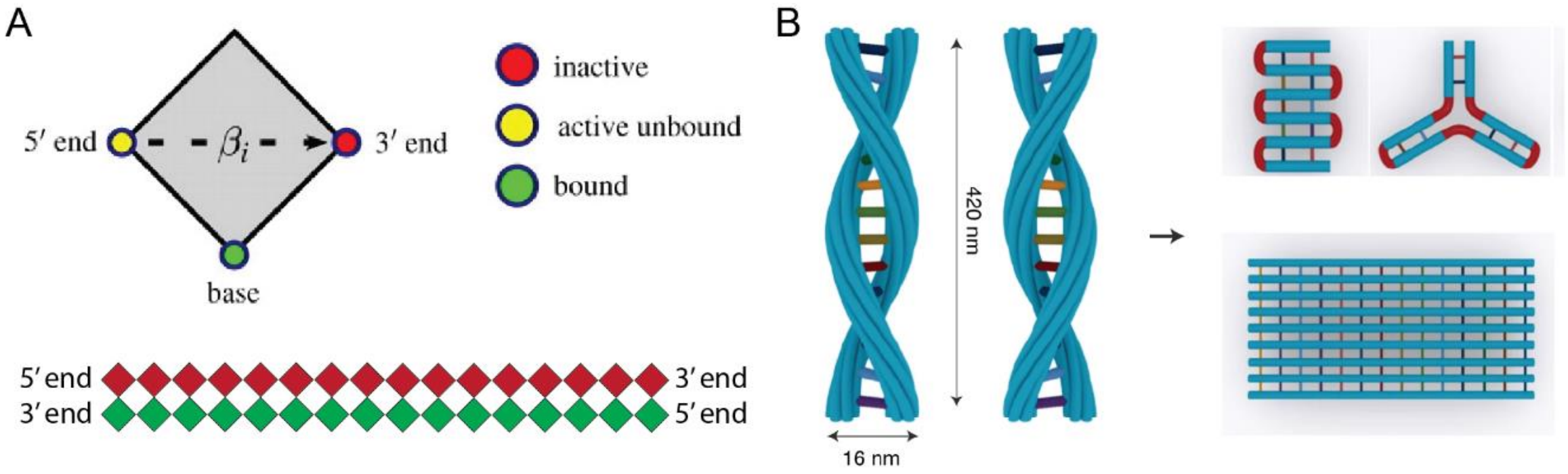

**Figure 6.** Meta-DNA structures. (A) Abstraction of a meta-nucleotide as an enlarged version of a nucleotide. Single-stranded meta-DNA made of meta-nucleotides can form double-stranded according to the complementary meta-bases. Reproduced with permission from ref 130. Copyright 2012 Royal Society. (B) Meta-DNA made of six-helix bundle DNA origami can assemble into higher-order of superstructures. Reproduced with permission from ref 132. Copyright 2020 Macmillan Publishers Ltd.

The experimental realization of mDNA was achieved by Yan and co-workers.[132] They used a six-helix DNA origami bundle as the mDNA unit to self-assemble into DNA structures in the sub-micrometer or micrometer scale. The mDNA carried ssDNA on the six-helix DNA origami bundle to work as a meta-base to form “bonds” with its complementary mDNA, while the six-helix DNA origami bundle acted as a meta-backbone. The single-stranded meta-base was 10-nt long, which in principle, could be programmed to have up to $4^{10}$ different types of bonds. The mDNA was 420 nm in length and 6 nm in diameter. After forming double-stranded mDNA, the width of the product was ~16 nm (Figure 6B), 8 times bigger than that of the dsDNA molecule. The meta-base spacing and placement could be programmed to adjust the directionality and chirality of the double-stranded mDNA. For example, if two mDNAs were arranged with spiral meta-bases on the helices, they self-assembled into double-helix structures, either in an RH or LH twist, resembling the natural double-helix DNA molecule. The twist angle of meta-bases could be adjusted to change the twisted double-stranded mDNA with different turn numbers. The authors further used the mDNA as the building block to construct a series of higher-order superstructures, such as meta-junctions, meta-double-crossover tiles, meta-polyhedron, and meta-lattices (Figure 6B), by changing the local flexibility of the mDNA and their interactions. These superstructures ranged from hundreds of nanometers to several micrometers. Interestingly, the mDNA could also undergo strand displacement reactions just as DNA molecules do, which enabled the design of dynamic DNA devices at the microscopic scale. Compared to the aforementioned methods, such as enlarging the DNA scaffold and connecting DNA origami nanostructures by sticky-ends or blunt-end stacking for building micrometer-sized DNA structures, the mDNA provides a conceptionally new idea for transforming DNA nanotechnology from the nanoscale to the microscale. It is anticipated that the mDNA can be used to construct more complicated static and dynamic systems that render many new applications possible. However, it is noteworthy that the mDNA proposed by Hao and co-workers mainly mimicked the properties of hybridization and strand displacement of DNA but still lacked other functionalities, such as the synthesis of mDNA from a primer and nicking reactions. The synthesis of mDNA is an important aspect, because it will allow for dynamic length control as well as the amplification of mDNA. The complete mimicking of DNA molecules is still challenging and certainly awaits more endeavors.

### 2.7 DNA crystals

Seeman's original vision was to use branched DNA junctions to construct rigid DNA crystal lattices that could host and resolve structures of biomolecules through X-ray crystallography. The X-ray diffraction technique was used to determine the structure of crystallized DNA molecules and led to the double helix model of DNA by Watson and Crick. The application of X-ray diffraction for the structural determination of proteins requires them to be crystallized and arranged into periodic patterns. The programmable properties of DNA structures could serve as a porous scaffold to orient and host guest proteins at specific positions. However, the assembly of 3D DNA crystals differs from conventional DNA nanostructures in several aspects. First, DNA structures created using DNA origami or DNA brick methods exhibit distinct sequences to form finite and stable geometries. Second, 3D DNA crystals should provide sufficient diffraction resolutions for resolving the details of the structures. Although several factors affect the resolution of DNA crystals, structural homogeneity is one crucial factor. Third, DNA crystals have to contain sufficient space for guest molecules. Therefore, the challenges of 3D DNA crystal self-assembly lie in the design of porous 3D DNA crystals with sufficient uniformity and large cavity size to host macromolecules.

After 27 years of Seeman's proposal, it was not until 2009 that the Seeman and Mao groups reported the first self-assembled 3D DNA crystal.[133] The DNA crystal was based on the tensegrity triangle motif, which consisted of three helices with different helix axis directions. Four-way junctions connected the three helices to form a three-fold symmetry and the crystal structure was formed by arrays of tensegrity triangles with periodic cavities. There were 21-nt between the triangles and 7-nt between crossovers within each triangle. The motifs could connect together into a higher-order periodic structure through the 2-nt sticky ends on the tails of the DNA helices. The 3D tensegrity DNA crystal contained a rhombohedral cavity size of ~103 $nm^3$ with a cross-section of ~23 $nm^2$ to accommodate guest molecules. The obtained diffraction resolution was 4 Å. Based on the tensegrity triangle crystal, other types of DNA crystals were also developed. For instance, DNA crystals containing two different triangles were constructed.[134] Each triangle had different sequences, and two triangles were assembled into an alternative pattern through sticky ends. The increased sequence complexity of the crystals allowed them to host heterogeneous guest molecules. In addition, the length of each helix on the triangle motif could be programmed to tune the cavity size of the DNA crystals. Both three- and four-helical turns for tensegrity triangle motifs were assembled into crystals.[135] Their cavity sizes were ~568 $nm^3$ and ~850-1100 $nm^3$, respectively. Nevertheless, these crystals resulted in decreased diffraction resolutions of 5.5 Å for the three-turn crystals and 10-14 Å for the four-turn crystals, respectively.

The original crystals used 2-nt sticky ends (GA:TC) for the tensegrity triangle motif cohesion. Seeman and co-workers found that the non-Watson-Crick sticky ends (AG:TC) yielded a hexagonal space group, where the DNA double helices bent at the crossovers (Figure 7A).[136] The hexagonal lattice had a cavity diameter of 11 nm and a cell volume of 470000 Å$^3$, larger than the rhombohedral crystal (103000 Å$^3$). The large cavity could potentially host large guest molecules for structural characterizations. In addition, it was shown that the sticky end length and sequences, as well as the 5´- and 3´- phosphate, could impact the crystal formation and resolution.[137] The triangle motifs with 1-,2-,3-, and 4-nt sticky ends and different sequence combinations of 2-nt sticky ends revealed differences in the diffraction resolution. The position of 5´-phosphate on the crossover, helical or central strands affected the resolution of the DNA crystals as well. It was found that 1-nt sticky ends and 5'-phosphate resulted in the best diffraction resolution, which was 2.62 Å. The 5´-phosphorylation was found to affect the crystallization process.[138] For the DNA crystal based on the tensegrity triangle motif, it was observed that the 5´-phosphorylation could substantially promote crystallization (Figure 7B). The crystallization took place at a 1/16 DNA concentration and a 1/8 of the lower buffer concentration compared to those in the original protocol. Furthermore, the crystallization kinetics was accelerated at both the nucleation and growth stages. This finding emphasized that phosphorylated DNA could have faster self-assembly kinetics than nonphosphorylated DNA by strengthening the cohesion of the sticky ends.

Another strategy by the Mao and Seeman groups used a hairpin with complementary sticky ends in the tensegrity triangle motif to modulate the crystallization kinetics.[139] The addition of the hairpin with the motif strands led to an improved resolution of the crystals (Figure 7C). It was hypothesized that the hairpin could affect the crystals in two ways: by reducing the number of nuclei and slowing down the crystal growth kinetics. The hairpin could competitively bind to the motif sticky ends, which reduced the chance of clusters from incorporating more motifs to grow bigger at the nucleation stage, giving rise to few but large crystals. The resulting crystals presented higher resolution than those without the hairpin addition. Interestingly, the hairpin also influenced the morphologies of the crystals. The hairpin strands specifically inhibited the growth direction of the corresponding DNA helices and ultimately changed the overall shape of the crystals.

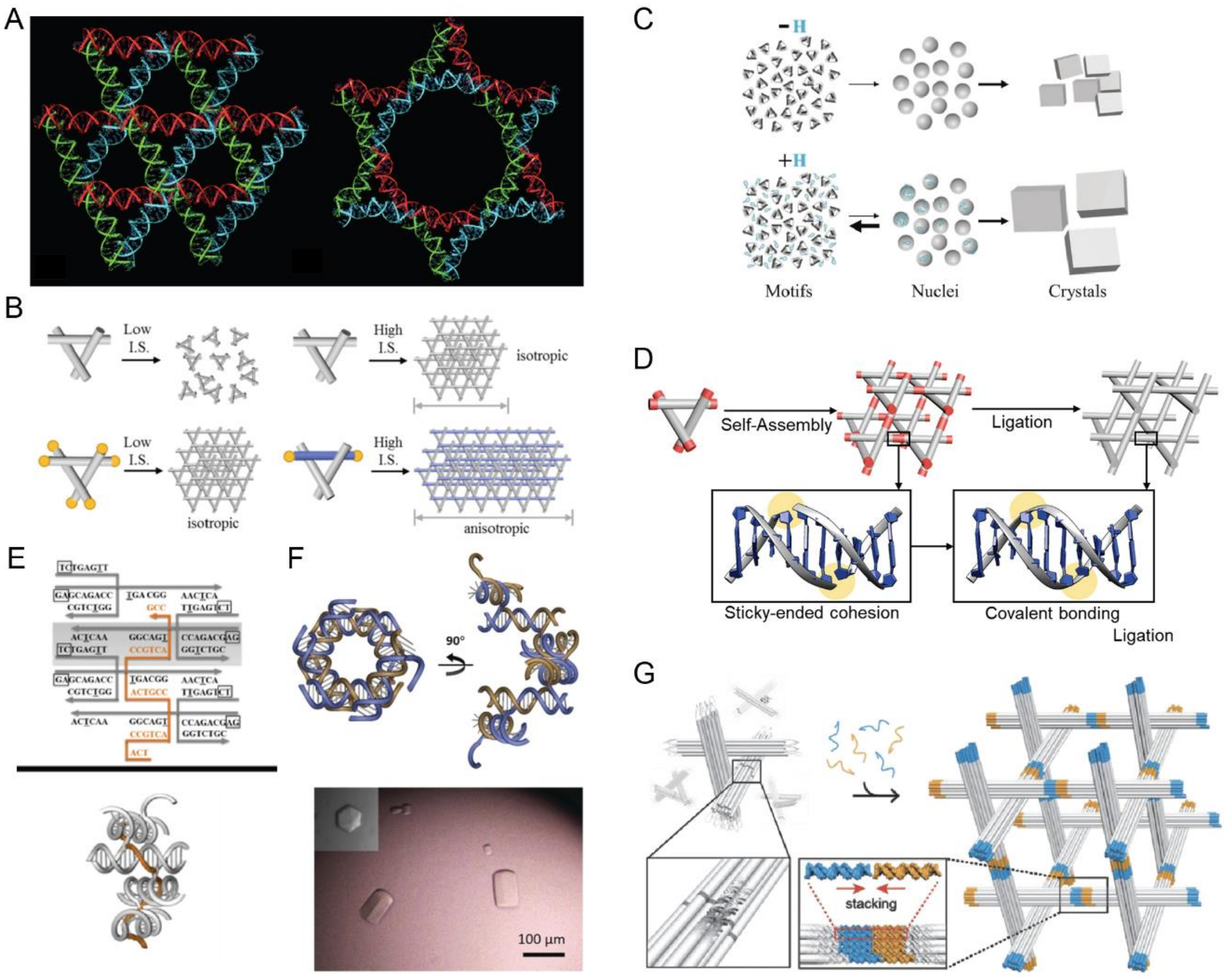

**Figure 7.** DNA crystals. (A) The tensegrity triangle motif with canonical sticky ends (GA:TC) can assemble into a rhombohedral DNA crystal (Left). While the noncanonical sticky ends (AG:TC) caused self-assembly into a crystal with a hexagonal space group (Right). Reproduced with permission from ref 136. Copyright 2021 American Chemical Society. (B) 5'-Phosphorylation strengthens sticky-end cohesions. Reproduced with permission from ref 138. Copyright 2021 American Chemical Society. (C) Modulating the DNA crystal with a hairpin. Reproduced with permission from ref 139. Copyright 2018 Wiley. (D) Making the DNA crystal robust by DNA ligation. Reproduced with permission from ref 143. Copyright 2019 American Chemical Society. (E) Tuning the cavity size and chirality of DNA crystals based on the tensegrity square motif. Reproduced with permission from ref 145. Copyright 207 American Chemical Society. (F) A 3D DNA crystal with six-fold symmetry. Reproduced with permission from ref 146. Copyright 2018 Wiley. (G) A 3D DNA crystal made of DNA origami tensegrity triangle. Reproduced with permission from ref 148. Copyright 2018 Wiley.

The interactions of motifs for the self-assembly of DNA crystals are intentionally designed to be weak to achieve DNA structures in micrometer sizes. However, this also leads to the instability of DNA crystals at low ionic concentrations or high temperatures. Several strategies were demonstrated to make the 3D DNA crystals more stable, including triplex bundling,[140] cross-linking,[141-142] and ligation.[143] In the case of the triplex bundling method, a triplex-forming oligonucleotide was added to bind to the inter-unit cohesion region. Nevertheless, this method required an acidic solution to form the triplex. In the case of the cross-linking method, some molecules formed inter-strand crosslinks within the DNA crystal under light irradiation. In the case of the ligation method, the DNA strands of the triangle motif were modified with 5′ phosphorylation.

The nick points at the sticky-ended cohesion were linked by T4 DNA ligase into covalent bonding (Figure 7D), yielding increased stability. It was experimentally demonstrated that after ligation, DNA crystals were still structurally intact under various harsh conditions, for instance, at 65˚C for 16 hours, dehydration for a month and rehydrating for 42 days in water, and being scratched 12 times. The increased stability of DNA crystals can promote their applications in many research fields and also foster new opportunities in DNA nanotechnology.

Inspired by the tensegrity triangle motif, other motifs were developed to create self-assembled DNA crystals. For example, the Yan group designed a tensegrity square with 5-nt per edge.[144] The square motif contained a central 20-nt strand with four sequence repeats of five nucleotides that formed four four-arm junctions at the corners of the central square motif. However, the 20-base central sequence could not make two full turns. As a result, this motif was crystallized into a series of duplexes tethered together by the central sequence. The crystals contained densely packed continuous helical layers, but lacked the uniform periodicity of void spaces. The authors redesigned the tensegrity square and changed the five nucleotides repeat sequence to six nucleotides (Figure 7E).[145] This modification reduced the structural strain caused by the single base deficit in the central strand. In addition, they synthesized the LH enantiomer of the motif using L-DNA. Both D-DNA and L-DNA motifs were successfully crystallized. The crystals revealed uniformed and periodic cavities with dimensions of ~ 4 × 2 × 5 nm$^3$. Importantly, the L-DNA crystals obtained nuclease resistance, making the crystals potentially applicable to cellular systems. Yan and co-workers further designed and crystallized a 3D DNA array with a layered hexagonal lattice (Figure 7F).[146]

Only two 21-nt strands were needed to form the array by hybridizing it into a repeating array of layered Holliday junctions. Two strands hybridized with 11 bp, and the other four sticky ends (two were 4 bases long, and two were 6 bases long) could associate into the Holliday junction separating the two layers. The junction also adopted a 120° angle between the two duplexes. The remaining two unpaired bases are further connected with each other along the helix in the same layer.

Branched DNA junctions serve as the central building block for 3D lattices. The preferences of DNA junctions can thus significantly affect the assembly of DNA crystals. Yan and co-workers probed the effect of all 36 immobile Holliday junction sequences on three separate assembled DNA crystals, the "4 × 5" and "4 × 6" designs, and a "scrambled" sequence variant of the "4 × 6" design.[147] It was demonstrated that most junctions yielded crystals, while few junction sequences were completely resistant to crystallization. Molecular dynamics simulations revealed that these fatal junctions lacked two ion binding sites crucial for crystal formation. The authors also found that the junction sequences could enhance resolution and influence crystal symmetry in both the "4 × 5" and "4 × 6" designs. Surprisingly, it was observed that the adjacent sequences to the junction had a large effect on the crystal assemblies. Changing downstream sequences adjacent to the junction could alter the salt buffer preferences and improve the crystal resolution modestly.

In addition to the aforementioned small DNA motifs, the DNA origami and DNA brick methods can also lead to the construction of DNA crystals. The DNA brick crystals could grow up to a micrometer in size in the lateral dimensions and with controlled depths up to 80 nm.[41] The DNA helices were designed to be parallel or perpendicular to the plane of the crystal. Sophisticated 3D cavities and channels were engineered in the DNA brick crystals. Alternatively, DNA origami offers high structural rigidity and large cavity space to host big guest particles. The Liedl group designed a DNA origami tensegrity triangle motif (Figure 7G), which could be considered as an enlarged version of the small DNA triangle motif. It consisted of three equal-shaped struts made of 14-helix bundles arranged in an over-under fashion.[148] The origami motif could polymerize into a rhombohedral lattice by keeping the same cross-section and axial orientation of the structs. The cavity size was 67 nm, much larger than that of the crystals made of small DNA motifs. The large cavity allowed the precise encapsulation of guest molecules, such as 10 nm and 20 nm gold nanoparticles (AuNPs). One could adjust the size of the DNA origami motif to tune the crystal cavity size, enabling the encapsulation of larger objects. In a recent work from the Ke group, the DNA gridiron design was exploited for DNA crystal formation.[149] The canonical DNA gridiron contained an even number of half-turns between adjacent junctions and resulted in a two-layer conformation, where two sets of parallel DNA helices were perpendicular and located in two planes. The new interweaving design assigned an odd number of half-turns between junctions, producing a weaving morphology between the vertical and horizontal helices. It was shown that although the interweaving design had a low self-assembly yield in the finite structure, it promoted crystal formation in the infinite structure. Moreover, tuning the interweaving thread number led to different crystal sizes. It was hypothesized that the weaving design introduced mechanical stress in the structure and the increased rigidity of DNA gridiron facilitated the crystal formation of the flat 2D lattices.

## 3. RNA self-assembly

### 3.1 Mechanism of the RNA assembly

RNA performs multiple functions in the cellular environment, such as the regulatory process. The primary role of RNA is being the genetic information processor for the expression of proteins and replication. Noncoding RNA has also been discovered, such as small nuclear RNA, microRNA, small interfering RNA, small nucleolar RNA, riboswitches, and catalytic RNA. The diverse functions of RNA originate from its structural complexity. RNA can spontaneously assemble into 3D conformations to perform specific tasks. Unlike DNA molecules, in which canonical Watson-Crick base pairings determine 3D conformations, RNA’s conformations are driven by both base pairings and noncanonical hydrogen-bonding interactions. It is challenging to assemble synthetic

RNA structures into prescribed shapes using the same rules used for DNA. However, RNA self-assembly has several advantages. For instance, RNA can be produced by transcription and potentially self-assemble in cells immediately post-transcription. In contrast, pre-assembled DNA structures have to be delivered into cells for cellular studies. RNA materials can also be obtained from bacteria for bulk production of RNA structures. Furthermore, RNA structures are compatible with cellular environments and have higher stability *in vivo*. In contrast, DNA structures usually suffer from low stability under physiological conditions, mainly caused by low magnesium concentration and enzymatic digestion.

In the last two decades, significant progress has been made in exploiting RNA self-assembly. Jaeger *et al*. designed tectoRNA, which self-assembled into RNA objects through tertiary interactions.[150-151] They found that hairpin tetraloops and their receptors could mediate the tectoRNA assembly. Jaeger and co-workers further designed an L-shaped tectoRNA containing two loops joined by a right-angle motif.[152] Four tectoRNA formed squared-shaped RNA (Tectosquares) through the loop-loop interactions, called kissing loop (KL). Tectosquares were further assembled into molecule jigsaw puzzles through specific sticky tail connectors. Other small RNA motifs were used to construct filaments,[153] square-shaped RNA particles,[154] RNA nanorings, and RNA polyhedrons.[155-156] These RNA motifs were demonstrated as building blocks for modular and hierarchical assembly of nanostructures.

### 3.2 Higher-order RNA assembly

Inspired by DNA tiles as building units to construct higher-order structures, the Andersen group designed RNA tiles to assemble into RNA origami with the help of molecular modeling.[118] They designed two RNA tiles containing four different strands using A-form RNA helices. RNA-AO tile had an odd number of half-turns between crossovers, and RNA-AE had an even number of half-turns. Because RNA helices had approximately 19° of inclination relative to the helical axis, the base pairs between crossovers in the two RNA tiles differed. In the RNA-AO tile, the base-pair inclinations of both helices were opposite, and the base pairs between crossovers defined a trapezoid. In the RNA-AE tile, the base-pair inclination was parallel, so that the base pairs between crossovers defined a parallelogram. The authors further converted the multistranded RNA tiles into ssRNA origami tiles. Four helix ends were connected to form hairpin loops. Subsequently, one of the inner crossover helical domains was replaced with a 180 ° KL interaction. One could further design RNA origami tiles with more than two helices with the introduction of 'dovetail seams' at the crossover. RNA origami tiles were assembled into lattices through the tile-tile interaction by RNA KL. The RNA KL interaction had varied angles, including ~180° or ~120°. The RNA KL angles defined the geometry of RNA lattices. For instance, a hexagonal lattice was formed using three different tiles with 120° KL interactions, while a rectilinear lattice was formed by two different tiles with 180° KL interactions. They also found that the RNA origami tiles could be assembled by cotranscription or on mica by annealing. The cotranscriptional folding of RNA origami tiles provided a low-cost method for producing RNA structures, which could potentially function in cells. Nevertheless, their method led to a low folding yield and a limited size (660 nt) of the folded structures. The Mao group constructed longer wireframe ssRNA structures with cotranscriptional folding.[157] The ssRNA followed a hierarchical pathway during the folding. ssRNA first formed a hairpin immediately after transcription. Then, unpaired domains further folded into complete structures through tertiary interactions. This hierarchical assembly of RNA nanostructures prevented kinetical traps during the folding and achieved a high assembly yield. Nevertheless, RNA nanostructures were usually stabilized by KL interactions and probably contained some pseudoknots. Later, the Andersen group developed RNA origami design tools to improve the cotranscriptional yield.[158] The software automatically selected structural modules from a library to build RNA origami. It could identify and bypass potential folding traps and optimize the sequences. The software allowed the rapid prototype of multiple RNA scaffolds and predicted the effect of different design parameters. The authors achieved RNA origami with up to 2360 nucleotides from 32 designs.

In addition to the self-assembly of RNA structures from ssRNA, multi-stranded RNA structures can be folded with identical RNA motifs. It resembles many homooligomeric proteins consisting of the same subunits. Weizmann and co-workers constructed RNA homooligomeric nanostructures using branched kissing loops (bKL).[159] The bKL, modified from the coaxially stacked kissing loop (KL), was a T shape motif formed by an L-shaped bulge and a hairpin loop by paranemic cohesion. bKL was used as a versatile construction module for its branched geometry and rigidity. They designed several RNA tiles containing bKLs to self-assemble into higher-ordered RNA nanostructures. For example, they constructed a Z-shape tile that contained two parallel double helical beams and one perpendicular struct. A bulge was designed at each beam-struct joint, and a complementary loop was designed at the end of the beam. The Z-tiles grew into ladder structures, and the beam length was tuned to cause twisting and bending, leading to circular or spiral RNA ladders with curvature. Furthermore, the strut length was adjusted to make a 360 ° turn, resulting in C-shaped tiles. Other RNA motifs were incorporated into the Z- or C tiles to increase the structural complexity. For example, the C-tile was modified into a 3- or 4- way junction, giving rise to RNA ladders with 3 or 4 rails. Adding a 5-nt bulge to each beam of the Z tiles made the structure an out-of-plane 90 ° turn, and the resulting tiles formed a 3D cage. The bKL-based tiles were demonstrated to be programmable to assemble into various RNA structures.

### 3.3 RNA topological structures

RNA can also be used to construct topological structures. Although various topological structures have been constructed using DNA in the past two decades, less attention has been paid to synthetic RNA topology. Synthetic RNA topological structures have not only helped to understand their physical and biological properties, but also assisted the discovery of RNA topoisomerase. Around 26 years ago, the Seeman group constructed the first synthetic ssRNA trefoil knot.[160] This knotted RNA structure led to the discovery of the first RNA topoisomerase Escherichia coli DNA Topo III.

Based on this study, other RNA Topo activities were discovered in Type IA DNA Topos.[161-162] The Weizmann group reported two methods to synthesize RNA topological nanostructures.[163] In the first one, two RNA strands and two DNA strands formed a four-way junction with the RNA strands as the helical strands and DNA as the crossover strands, respectively. The four-way junctions were connected to form knotted structures. The RNA scaffold was threaded into a knotted structure by the DNA staples. After linking the nicks and removing the DNA from the structure, a knotted ssRNA structure was generated. The second method used an ssDNA knotted structure as the template. RNA strands could bind to the ssDNA template. After ligating the nicks, the DNA was digested, and the ssRNA knot was produced. The synthetic ssRNA knot was used as a topological probe. The authors found that Escherichia coli DNA topoisomerase I had low RNA topoisomerase activity, whose unknotting activity for ssRNA could be inactivated by the R173A point mutation. They also discovered that RNA topology could inhibit reverse transcription, a similar phenomenon in DNA topology, which blocks the DNA polymerases.[128] The RNA topological structures have shown their significance in RNA biology and can help to solve many unexplored problems. For example, the discovered RNA Topoisomerase activity in some proteins could help to identify naturally topological RNA structures in cells. In addition, it is crucial to explore whether RNA-specific Topo exits as well as the relationship between RNA Topo and cellular functions of identified proteins.

## 4. Dynamic DNA nanotechnology

Structural DNA nanotechnology uses nucleic acids to construct DNA nanostructures with varying sizes and complexities from self-assembly. It explores DNA structures in equilibrium states. In contrast, dynamic DNA nanotechnology focuses on the active control over the geometry, motion, or growth of DNA structures at the nanoscale. The interest is mainly in the kinetics and non-equilibrium states of DNA structures.

In general, external fuels or environmental changes are needed to drive DNA structures in motion.

The first dynamic DNA nanodevice was based on the B-Z transition of DX molecules.[164] Two rigid DX molecules were connected by a central helix. By changing the condition for the transition between B and Z forms of DNA, the DX molecules achieved switchable motion. In the seminal work by Turberfield and co-workers, a molecular tweezer made of DNA could close and open upon the alternative addition of "fuel" DNA strands and anti-fuel DNA strands, respectively.[165] This process involved a fundamental mechanism called toehold-mediated DNA strand displacement reactions, in which long ssDNA displaces short ssDNA from a DNA duplex with the help of a few nucleotides as the toehold and forms a long new DNA duplex. Since then, the toehold-mediated DNA strand displacement has been a widely used strategy in dynamic DNA technology. Another mechanism is called hybridization chain reaction (HCR).[166] It involves cascaded reactions between two different hairpins initiated by a trigger strand. A variety of dynamic nanodevices have been developed based on strand displacement and HCR, such as DNA walkers,[167-171] switchable 3D DNA cages,[172] switches,[173-176] molecular gears,[177] nanorobots,[178] actuators,[179-180] DNA circuits, and other nanomachines.[181] A recent work from the Turberfield group utilized strand displacement for programmable printing on a 2D origami surface.[182] The DNA molecular printer was precisely positioned on the surface and wrote patterns by incorporating local ink strands through strand displacement reactions.

There are also many other strategies to induce dynamic structures changes, including molecular recognition, such as aptamer binding,[183] hybridization, and de-hybridization of nucleic acid duplexes,[184] environmental factors such as pH value,[185-192] temperature,[59, 193] light,[194-195] ionic concentration, and enzymatic reactions to DNA strands, such as strand degradation,[196] cleavage,[197] ligation,[168] extension.[198] Specifically designed DNA sequences can enable DNA structural changes regulated by metal ions or pH.[188, 192] Base stacking between blunt ends of DNA helices is sensitive to metal ions and temperature and has also been utilized to manipulate DNA structural change.[59]

**4.1 Reconfigurable DNA nanostructures**

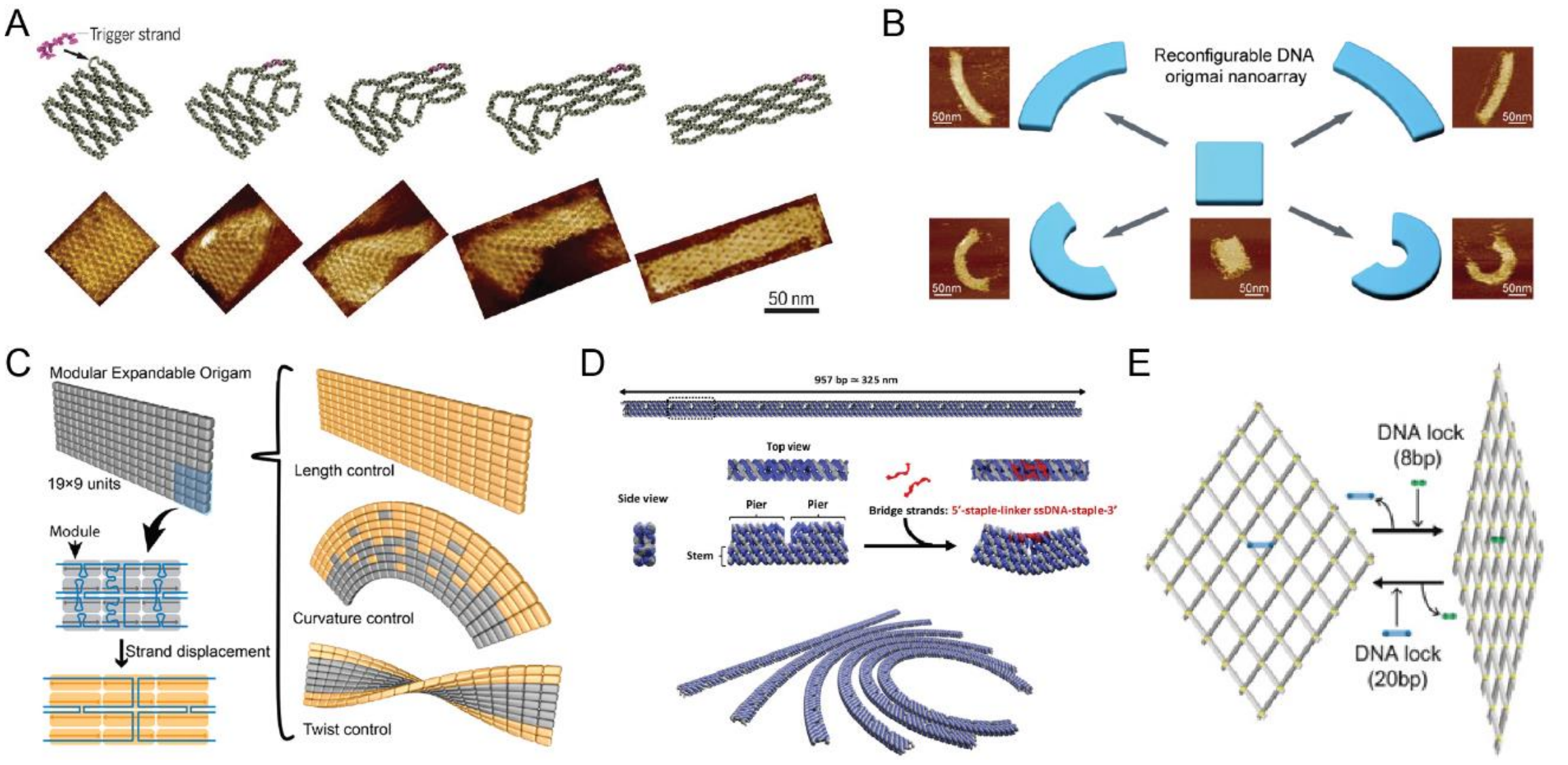

**Figure 8.** Reconfigurable 2D DNA nanostructures. (A) Cascaded transformation of DNA molecular array. Reproduced with permission from ref 184. Copyright 2017 AAAS. (B) Reconfigurable curved DNA origami. Reproduced with permission from ref 202. Copyright 2020 American Chemical Society. (C) Modular transformation of DNA origami with expandable DNA units. Reproduced with permission from ref 203. Copyright 2021 American Chemical Society. (D) Modular deformation of DNA origami with tension-adjustable modules. Reproduced with permission from ref 204. Copyright 2020 Wiley. (E) Reconfigurable DNA Accordion Rack. Reproduced with permission from ref 205. Copyright 2018 Wiley.

Reconfigurable DNA nanostructures refer to DNA nanostructures that change their conformations upon external stimuli. The fact that DNA nanostructures can interact with environmental factors suggests that the configuration of the assembled DNA nanostructures can be tuned for various applications, for instance, biosensing, drug delivery, nanorobots, and plasmonics. There has been intense research on reconfigurable DNA nanostructures over the last two decades. Early studies focused on relatively simple designs composed of several DNA strands, such as DNA tweezers or switches, whose dimensions were typically limited to a few nanometers. After the invention of DNA origami, arbitrary shapes of DNA nanostructures could be designed in the megadalton range. The dynamic range is extended to over 100 nm. Shape transformation can occur in simple structures, such as a DNA box, DNA barrel[183] or DNA nanoactuator,[180] DNA force spectrometer.[199] Reconfigurable structures may contain immobile bodies and dynamic flexible joints that can move in response to external stimuli.[200-201] Depending on the connectivity of DNA units in the DNA nanostructures, the transformation mechanisms can be classified into cascaded transformation, modular transformation, and global transformation. In the cascaded transformation, the structural transformation proceeds in a step-by-step fashion. The connections between units are made to be dynamically switchable to achieve this aim. Once a local DNA unit is transformed, it can trigger the transformation of the neighboring units *via* the connections. During this process, the previous connections are broken and new connections are subsequently formed. For example, the Ke group designed a synthetic DNA molecular machine, which transformed from one conformation to another *via* a cascading reconfiguration process (Figure 8A).[184] This molecular machine was constructed by interconnecting a small DNA unit called “antijunction”, which was switched between two conformations through an open conformation. The structure transformation was initiated by trigger strands, which provided external mechanical energy to transform the local DNA units. The transformation of one DNA unit could propagate to its neighboring unit and eventually led to the global conformation change of the DNA structure. This phenomenon was named “information relay” as an analog of “domino array”. During the transformation process, no hybridization or de-hybridization was involved. The conformation change was driven solely by the base stacking at the junction position between two antijunction units. The base stacking between two antijunctions experienced the breaking and re-forming process during the stepwise transformation. In another work, the authors demonstrated that the DNA origami domino array could switch between a noncurved conformation and a curved conformation.[202] In contrast to the original DNA origami domino array, where all the antijunctions were the same size, the revised DNA origami domino array comprised gradient-sized antijunctions along the DNA helices. Before the transformation, every helix had the same length, so that the DNA structure was noncurved. After transformation, because the orientation of the DNA helices was shifted, the DNA helices from one side were longer than those on the other side. This resulted in a curved conformation for the entire DNA structure. The curvatures could be programmed by adjusting the relative size of the DNA antijunctions (Figure 8B).

DNA nanostructures can also be transformed through independent DNA nanostructure modules. By selectively tuning specific modules, the overall shapes of the DNA nanostructures are changed. This type of transformation typically involves multiple stable conformations achieved by combining triggers to tune the individual modules. Suzuki and Ke groups reported two methods for the modular transformation of DNA nanostructures. In the work of the Ke group,[203] the authors designed a DNA origami structure that could transform *via* multiple modes through modular transformations. The DNA origami consisted of small modular dynamic units, having a double helix with a loop in the middle. Due to its modularity, each DNA unit could be selectively elongated by another DNA strand, thus changing the DNA origami's overall conformation. It was demonstrated that the DNA origami structure could transform in three modes (changing length, curvature, and twist) with the addition of different combinations of expansion strands (Figure 8C). In addition, the transformation from one conformation to another could be realized in multiple pathways. In the work of the Suzuki group,[204] a DNA origami structure consisting of multiple modules was designed. Each module contained a stem and a pier. Two neighboring piers had space intervals, but could be connected by bridge strands. After connections, the two modules were bent due to the tension caused by the bridge strands. By selectively connecting the modules in the DNA origami, the overall shape adopted a curved conformation due to the cumulative tension (Figure 8D). The curvature was programmable by using different numbers and locations of the bridge strands.

In the case of global transformation, the DNA units are connected by stiff joints, such as a double helix. All the units are strongly coupled and collectively transform from one conformation to another due to the strongly coupled connection. The Kwon group demonstrated a reconfigurable accordion rack made of a DNA beam lattice.[205] The DNA lattice could be viewed as arrays of cross-shape motifs connected by double helices. The cross-shaped motif comprised two interconnected dsDNA. Due to their flexible connection, the two double helices adopted different angles adjusted by the lock strands with different lengths. As a result, the whole structure transformed collectively to the same conformation due to stiff connections between the motifs (Figure 8E).

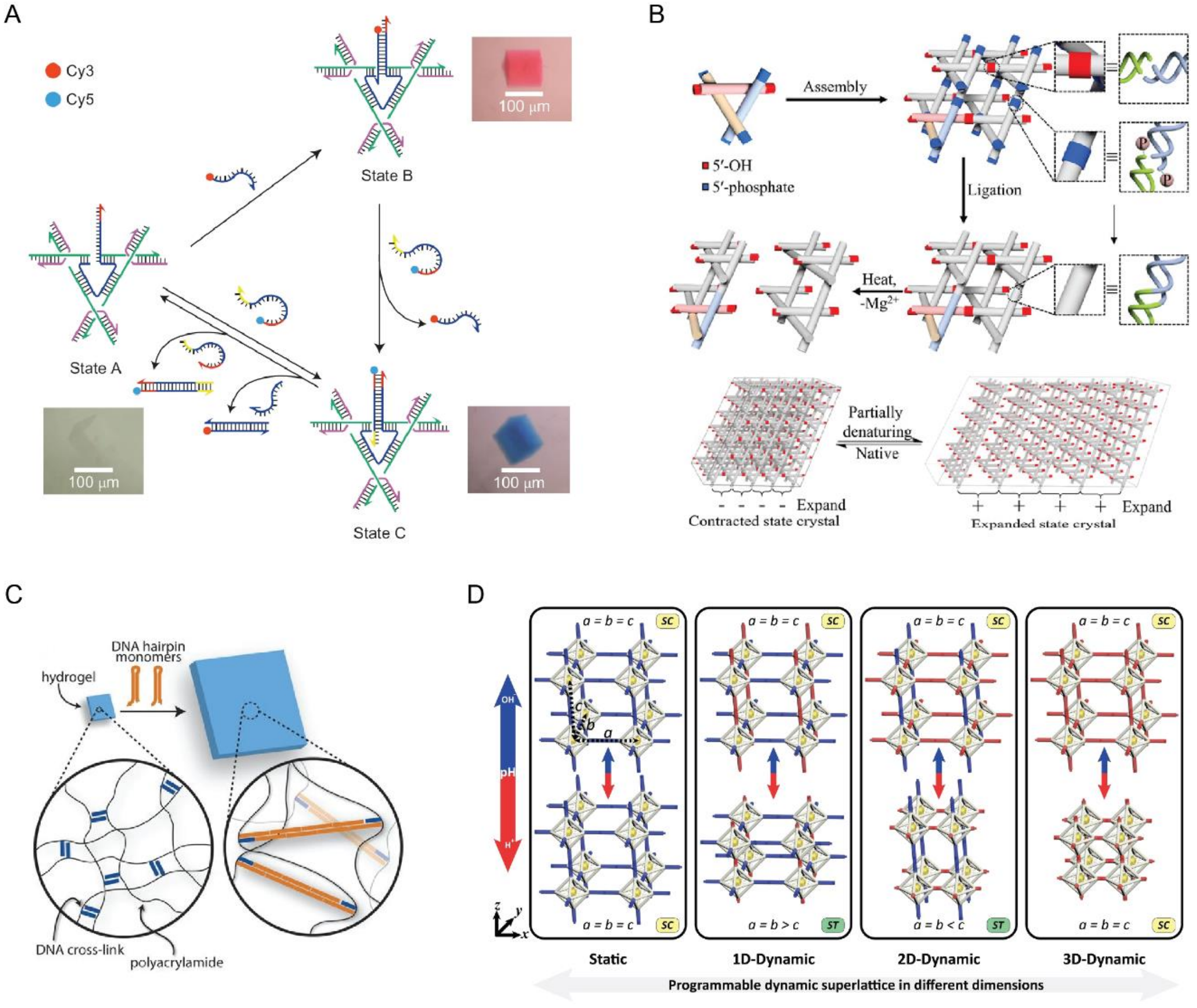

**Figure 9.** Reconfigurable 3D DNA structures. (A) Operation of the three-state color device using DNA crystals. Reproduced with permission from ref 208. Copyright 2017 Macmillan Publishers Ltd. (B) Dynamic motion in DNA crystals. Reproduced with permission from ref 209. Copyright 2022 Wiley. (C) DNA-directed expansion of DNA hydrogel. Reproduced with permission from ref 210. Copyright 2017 AAAS. (D) pH-induced symmetry conversion of DNA origami lattices. Reproduced with permission from ref 192. Copyright 2022 Wiley.

The reconfigurability of DNA structures can be further extended from 2D to 3D. In particular, complex reconfigurable structures, when integrated with other systems, can foster many new applications, such as with metallic NPs and/or single emitters for nanophotonic applications.[206-207] This aspect will be discussed in detail later. Moreover, target recognition-induced opening and closing of DNA structures are capable of cargo transportation and control of enzyme activity.[183, 200]

It is noteworthy that the reconfigurability of DNA structures can lead to evident macroscopic property changes. The Seeman group demonstrated that it was possible to manipulate a 3D DNA crystal by inserting or removing DNA strands (Figure 9A).[208] The 3D DNA crystal was assembled using a modified tensegrity triangle motif as the building unit, which comprised an extension of the central strand. The input of fluorescence dye-modified strands could bind to it and change the color of the crystal driven by strand displacement reactions. In a recent work from the Mao group,[209] the DNA tensegrity triangle motif was selectively modified with phosphate groups on two-component duplexes. Thus, the DNA was ligated only along two directions, and the third direction remained un-covalent. The sticky-end cohesion in the third direction could dissociate or re-associate under ionic concentration or temperature changes. The overall crystal displayed macroscopic expansion or shrinkage over 50 μm (Figure 9B). The Shulman group demonstrated a high-degree swelling of hydrogels by DNA-directed shape changes (Figure 9C).[210] The DNA-cross-linked polyacrylamide hydrogels experienced structural expansions triggered by HCR. In addition, the Tian group reported the dynamic conversion of DNA origami lattices induced by pH tuning (Figure 9D).[192] The DNA origami lattices assembled by octahedral origami frames were integrated with pH-reactive i-motif sequences in the connectors. Upon pH changes, the pH-responsive connectors switched between two states, leading to the DNA lattices' overall transition between simple cubic and simple tetragonal configurations.

### 4.2 Artificial DNA motors

Nature has evolved motor systems for transportation, cargo sorting, and assembly of biomaterials. Kinesin and dynein are two major motor proteins transporting intracellular cargo along microtubules powered by ATP hydrolysis. Inspired by such biological motors, a plethora of synthetic self-assembled DNA analogs have been developed, such as DNA walkers that can perform stepwise movements on nucleic acid substrates. A typical DNA walker system consists of a walker and a track for movement. The driving force for the movement is usually chemical fuels, which are converted into mechanical work. The external energy elevates the walker into a high-energy state. Relaxation from the high-energy to the low-energy state drives the walker for mechanical walking.

The Seeman and Pierce groups reported the earliest DNA walker systems.[211-212] They both designed biped walkers, which moved along well-defined tracks made of DNA. The walking was guided by external DNA strands through strand displacement reactions. When the linker between the walker and the track was removed by the external DNA strand, another DNA strand was added to form a new linker between the walker and the subsequent stator. The direction of the movement was controlled by the fuel strand, which linked the walker to a new stator. By repeating the dissociation and association of the DNA walker and the stators, the walker moved along the track. Several other types of DNA walkers driven by strand displacement reactions were reported.[170, 213-218] Some of them achieved autonomous motion.

Enzymatic reactions are another scheme to power DNA walkers. Yin *et al*. demonstrated a walker system that moved autonomously along a track through alternative cleaving and ligation of DNA strands.[168] Later, the Mao group developed a DNAzyme-based walker, which continuously digested the substrate and moved to the neighboring stator through strand displacement reactions.[219] Moreover, Bath *et al*. designed a DNA walker powered by a nicking enzyme,[197] where the walking mechanism relied on the nicking enzyme consuming the substrate followed by the walker moving to another stator through strand displacement reactions. This walker was further adapted on the DNA origami surface for long-range transportation and navigating the path under instruction.[220-221] Furthermore, exonuclease[196] and RNase H[222-223] were used to power motion. Other environmental factors include light,[224] pH,[169] and electric fields.[225-226]

In addition to walking, other types of motions were also developed, for instance, sliding,[227-229] rotation,[225-226, 229] and rolling.[222-223] Synthetic rotary systems are inspired by biological rotary motors, such as the bacterial flagellar motor and the $F_1F_o$-ATPase, which comprise rotor and stator components and perform rotary motion by consuming chemical energy. The Dietz group designed a self-assembled DNA origami rotary apparatus,[230], with a rotor unit and clamp elements. The rotary motor achieved Brownian rotation in both clockwise and counterclockwise movements. The authors later developed a DNA origami rotary ratchet motor that obtained directional motions (Figure 10A).[231] The new motor consisted of a rotor arm, a pedestal with a dock for the rotor unit, and a triangular platform. The triangular platform was installed with physical obstacles that protruded with an inclination of about 50° from the surface to achieve directional motion. These obstacles allowed the rotor arm to move only in one direction, as a movement in the opposite direction would trap the rotor in an energy barrier. Under an external electric field, the motor was driven to move with a preferred rotation direction. Liu and colleagues also developed rotary systems.[232-233] In a recent work, they built a gear system consisting of a small origami ring, a large origami ring, and AuNPs. These modular components were assembled to work cooperatively with bidirectional rotations mediated by the AuNPs.

Sliding is another essential mechanical motion in natural systems. For example, the kinesin-5 protein mediates the crosslink and sliding antiparallel movements of microtubules. The Liu group reported an artificial nanoscopic analog of the sliding system (Figure 10B)[227] using two antiparallel DNA origami filaments mediated by two AuNPs powered by DNA fuels. The sliding system could perform stepwise sliding and reversible movements. This two-layer sliding system was later extended into multilayer nanoarchitectures,[228] enabling complex dynamic motions on the nanoscale. Furthermore, multiple sliding components could be engineered to perform regulated and coordinated motion.[229] In the meantime, synthetic rolling motors were developed.[222-223] These rolling motors improved the low track affinity in the DNA walker system, while maintaining high motor velocity.

Biological systems have developed efficient cellular transportation systems. Motor proteins can move micrometer-long distances at the speed of μm/s. The early developed synthetic transportation system could hardly achieve comparable speeds. DNA walkers driven by strand displacement generally suffer from low speeds and require seconds to minutes to perform a single operation.[170, 220, 234] It was demonstrated that the strand displacement rate constant could reach up to $10^6$ $M^{-1}S^{-1}$.[235] Walter and co-workers improved the speed of strand displacement-based walkers with a cartwheeling design (Figure 10C).[171] The DNA walker comprised a central sequence and two toehold domains at both ends. The footholds were designed with sequences complementary to the central sequence and the toehold domains of the walker. The footholds with two different sequences were alternatively distributed on the DNA origami surface, so that the walker moved along the track by alternatively binding to two different footholds through the toehold. Since the toehold domain was always free at the distal termini, it eliminated the necessity of enzyme reactions to generate the free toehold before walking *via* strand displacement. Thus, the walker was driven only by strand displacement. The length of the toehold domain affected the strand displacement rate. The authors carefully optimized different toehold lengths and achieved the highest walking rate of ~300 nm/min with a step size of 7 nm, which was an order of magnitude higher than that of other DNA walker systems. New types of high-speed DNA motor systems were also reported. For example, an RNase H-powered rolling motor consumed the track by the enzyme and formed a new connection with the new track during the rolling (Figure 10D).[222-223] This type of motors could reach a translocation rate up to ~1 $\mu$m $min^{-1}$. Motor proteins were also utilized in constructing DNA motors. Furuta and co-workers realized the movement of motor protein dyneins along DNA tubes (Figure 10E).[236] The original track-binding domain of the dynein was replaced with a DNA binding domain, which enabled the motor to bind to and move along the DNA track. This hybrid motor retained the high speed (~220 nm/s) of the dynein on the reusable track by ATP hydrolysis. Using the programmable DNA track and track recognition of the hybrid motor, the motor system could be engineered to perform multiple functions of cargo molecules, such as disperser, aggregator, sorter, and integrator.

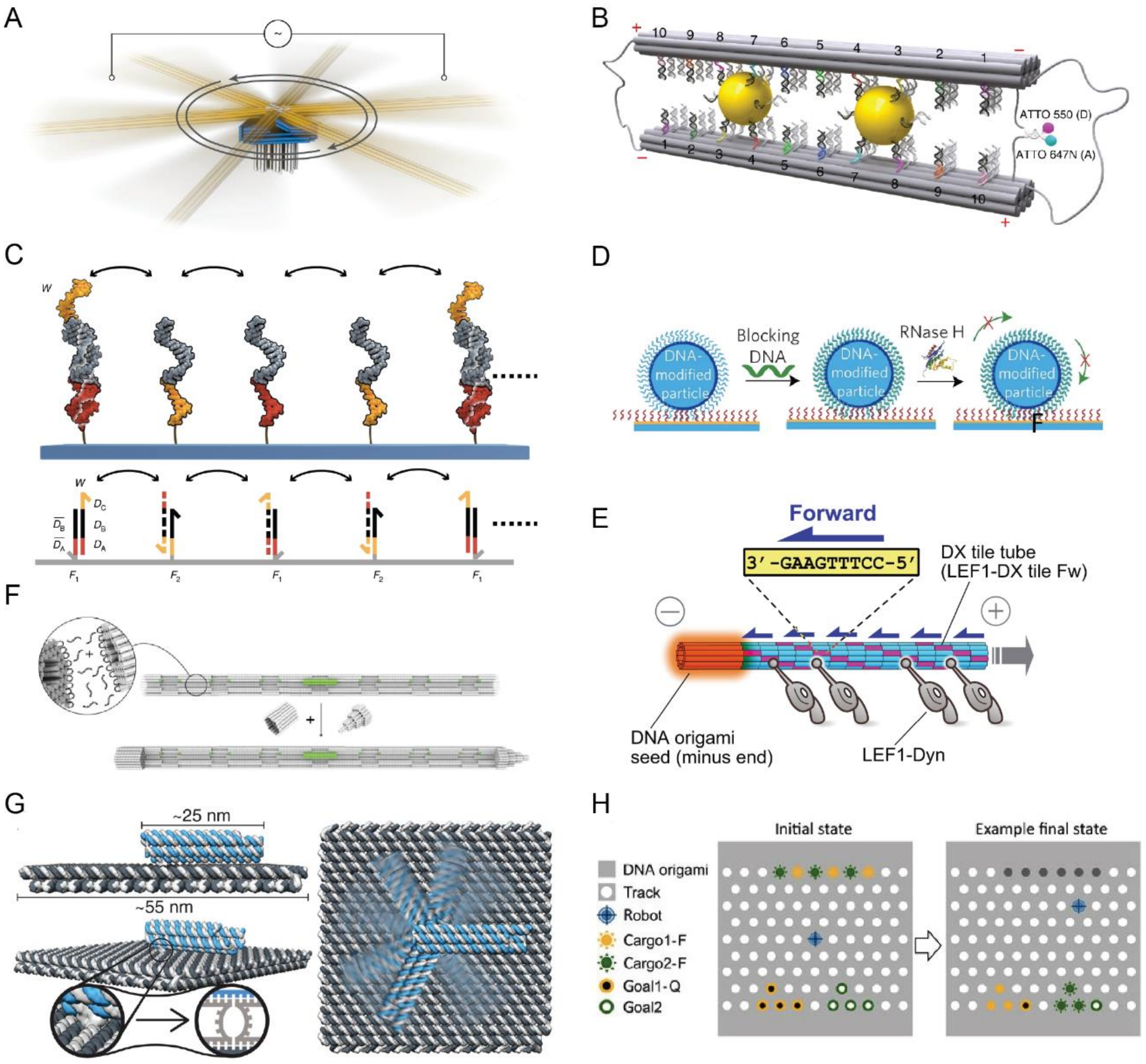

**Figure 10.** DNA motor systems. (A) A DNA origami rotary ratchet motor. Reproduced with permission from ref 231. Copyright 2022 Macmillan Publishers Ltd. (B) A sliding system made of doublet DNA origami filaments and gold nanocrystal. Reproduced with permission from ref 227. Copyright 2018 Macmillan Publishers Ltd. (C) A cartwheeling DNA acrobat based on DNA strand exchange. Reproduced with permission from ref 171. Copyright 2018 Macmillan Publishers Ltd. (D) A DNA rolling motor powered by RNase H. Reproduced with permission from ref 222. Copyright 2016 Macmillan Publishers Ltd. (E) A engineered protein motor with the DNA-binding module. Reproduced with permission from ref 236. Copyright 2022 AAAS. (F) A synthetic tubular molecular transport system. Reproduced with permission from ref 237. Copyright 2021 Macmillan Publishers Ltd. (G) A DNA self-assembled robotic arm powered by electric fields. Images were reproduced with permission from ref 225. Copyright 2018 AAAS. (H) A cargo-sorting nanorobot. Reproduced with permission from ref 214. Copyright 2017 AAAS.

Going further, novel motor systems can even eliminate the process of breaking and re-forming bonds with the track to further improve speed. The Dietz group developed a tubular molecular transformation system that traveled inside a tubular tunnel (Figure 10F).[237] This piston was made of DNA, and its transport in the barrel structure was driven by Brownian motion. There was no need to continuously form and break the bonds with the track, as the DNA walkers did. Thus, the speed of motion reached 0.3 $\mu m^2/s$ with total travel distances up to micrometers. The Simmel group developed a rotary DNA system powered by electric fields (Figure 10G).[225] A DNA robotic arm was positioned on a DNA origami platform. The rotation of the arm was actuated with external electric fields. By altering the electric field, this method offered precise control over the rotation direction and frequency. The achieved maximum rotation frequency was 25 Hz, much higher than the hybrid ATP-driven motor system.[238]

Early-developed DNA walkers could only perform simple functions, such as walking along a defined track.[239] Some walkers could perform more complicated work, such as product assembly,[170, 234] cargo transportation,[240] or deciding a path at a cross.[221] The Qian group explored the DNA walker-based nanorobot to perform complex cargo sorting tasks (Figure 10H).[214] The nanorobots had three building units. Each unit performed a specific sub-task: cargo pick-up, cargo sorting, or cargo drop-off. Two types of multiple cargos were initially anchored without order on the specific locations of a 2D DNA origami surface. The robot autonomously picked up cargos and delivered each type of cargos to the destination location until all cargos were correctly sorted. The cargo-sorting nanorobot followed an exquisite algorithm. Once the DNA walker picked up a cargo, it started random walking along the track. When it reached the destination, a cargo recognition procedure was performed, and only the correct cargo was released at the destination. By repeating the picking up, random walking, and releasing procedures, all cargos were sorted correctly. Another complex task was to solve mazes by DNA navigators.[215] The authors designed several crossed paths on a 2D DNA origami surface. One end was used as an entrance, and the other was defined as an exit. Among the possible paths for the DNA walker, only one path was defined as the correct one. With the introduction of a DNA walker, it could walk along the path and randomly decide a path when it met a cross. A molecular maze was realized by trial and error from a large number of walker molecules on the DNA origami with only a fraction of successful pathfinding.

**4.3 Dynamic self-assembly**

In cells, many cellular components frequently disassemble and reassemble by exchanging information and energy with their environment. A remarkable example is microtubule polymerization. The building units α and β-tubulins polymerize into microtubules of micrometers and depolymerize, when environmental conditions change. It is of great interest to mimic the dynamic assembly process of microtubules using DNA nanotechnology and ultimately control the dynamic assembly of DNA nanomachines. In contrast to conventional thermodynamic assembly, which explores the assembly of DNA structures with the lowest energy, dynamic assembly can yield products that form the fastest. This aspect was utilized by the Mao group to co-fold two complementary DNA strands into identical nanostructures simultaneously rather than a DNA duplex.[241] In addition, dynamic assembly allows for the dynamic control of DNA nanostructures using environmental factors, which are useful in many potential applications. To this end, a variety of DNA nanostructures were designed to exhibit disassembly and reassembly properties regulated by environmental triggers. For instance, the Mao group developed the ATP-triggered self-assembly of DNA nanostructures (Figure 11A),[241] where a DNA aptamer was incorporated into a DNA motif. The default conformations of the motif did not associate with each other to form large DNA nanostructures. In the presence of the ligand, the binding of the aptamer to the ligand caused the motif to change the conformation into a T-junction, which could assemble into 1D arrays or 2D lattices.

Fine control over the dynamic assembly process can be achieved by coupling with DNA circuits. The Winfree group demonstrated the regulation of the DNA tile self-assembly using a DNA strand displacement circuit.[89] The DNA tile that formed the DNA nanotubes was designed with two protected strands, which prevented their assembly into nanotubes. However, the protected strands could be removed by a deprotector, so that the active DNA tiles autonomously assembled into nanotubes. The release of the deprotector was coupled with a DNA circuit, which amplified the production of the deprotector in the solution with the catalyst. Based on this work, the Ricci group developed the pH-controlled assembly of DNA tiles.[242] Using a triplex-forming strand that switched between a double helix and a triplex at different pH values, the release of the deprotector was controlled by the pH-sensitive DNA circuit. Franco and co-workers further demonstrated the pH-driven reversible control of the self-assembly of DNA nanotubes using DNA tiles.[187] A regulator strand was designed to control the activity of the DNA tile monomers. The binding of the regulator with tile monomers prevented the monomers from being assembled into nanotubes. The regulator was further controlled by a pH-responsive sensor made of a triplex-forming structure. In acidic and neutral conditions, the sensor was locked and could not react with the regulator, so that inactive tiles were blocked by the regulator. Under basic conditions, the sensor was unlocked and could block the regulator. The active tile monomers activated the downstream self-assembly of the nanotubes. Furthermore, molecular oscillators can also control the assembly process. The Franco group demonstrated the dynamic control of the self-assembly of DNA tubular structures coupled with a molecular oscillator.[243] The authors used DNA tiles to assemble DNA nanotubes of micrometers in length. These DNA nanotubes were disassembled by RNA invaders that inhibited the DNA tiles and re-assembled by RNase H-digested RNA invaders isothermally. The dynamic assembly was further coupled with a synthetic molecular oscillator, which periodically regulated the concentration of the RNA invaders by transcription (Figure 11B). When the RNA invaders were at low concentrations, the RNA molecules on the DNA tiles were digested by the RNase H, so that the active DNA tiles assembled into DNA nanotubes. When the RNA invaders were present at high concentrations, which surpassed the effect of RNase H, the DNA tiles were inhibited and the DNA nanotubes thus were disassembled.

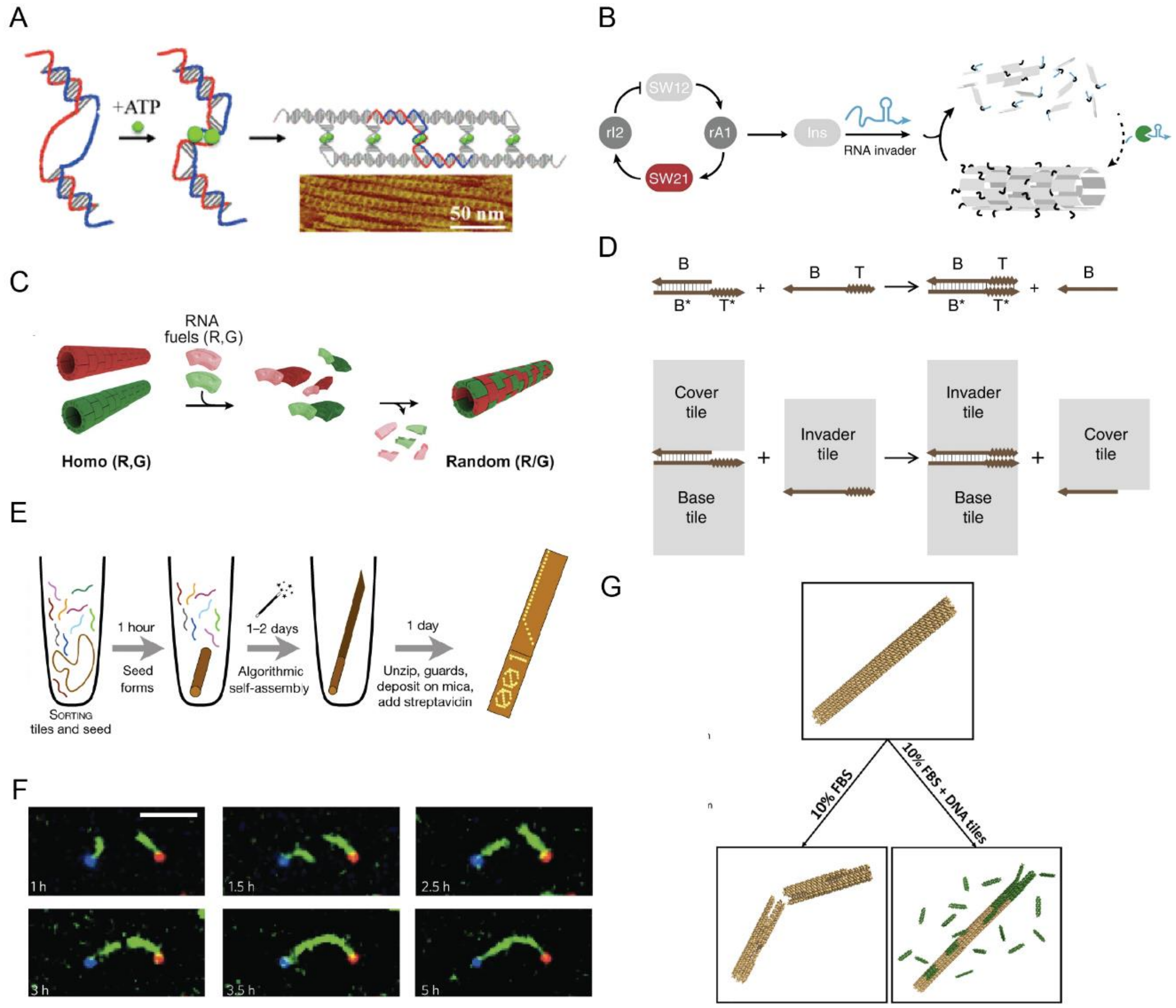

**Figure 11.** Dynamic DNA assembly systems. (A) A DNA motif was assembled into DNA arrays after binding with ATP. Reproduced with permission from ref 241. Copyright 2020 American Chemical Society. (B) A synthetic transcriptional oscillator-controlled dynamic assembly of DNA nanotubes. Reproduced with permission from ref 243. Copyright 2019 Macmillan Publishers Ltd. (C) RNA mediated DNA nanotube reorganization. Reproduced with permission from ref 245. Copyright 2021 American Chemical Society. (D) Reconfiguration of DNA origami tiles through tile displacement. Reproduced with permission from ref 246. Copyright 2018 Macmillan Publishers Ltd. (E) Complex DNA computing based on seeded growth. Reproduced with permission from ref 247. Copyright 2019 Macmillan Publishers Ltd. (F) Connecting molecular landmarks through seeded growth. Scale bar: 5 µm. Reproduced with permission from ref 250. Copyright 2017 Macmillan Publishers Ltd. (G) Self-healing of DNA nanostructures. Reproduced with permission from ref 251. Copyright 2019 American Chemical Society.

Biological systems generally use two strategies to sustain the stability and functional capacity of biomolecules: structural exchange and structural reorganization. Structural exchange refers to the exchange of subunits of biomolecular complexes with free units in the environment to adapt to environmental changes. This process involves the competitive dissociation and association between bound and free units. DNA nanostructures can act as templates to investigate such dynamic exchange reactions.[244] Structure reorganization involves the dynamic exchange of units between post-assembled structures. Recombination is the best example of such a process, where two DNA pieces recombine to produce new combinations of alleles. This recombination engenders genetic diversity at the molecular level and DNA nanotechnology can mimic such phenomena. The Ricci group demonstrated the reorganization of DNA nanotubes fueled by RNA (Figure 11C).[245] Two orthogonal addressable DX DNA tiles were designed with a sticky end at one binding domain, such that they could assemble into DNA nanotubes autonomously. When adding RNA fuels bound to one binding domain of the DNA tiles through the sticky ends, the DNA nanotubes disassembled into DNA tiles. However, introducing RNase H into the system could digest the RNA molecules and recover the activity of the DNA tiles. Two types of DNA tiles were randomly assembled into DNA nanotubes, which contained hybrid types of DNA tiles. The Qian group reported a DNA origami-level reorganization strategy.[246] The origami reorganization was achieved by the DNA tile displacement mechanism, in which an invader DNA origami tile displaced another tile from an array of tiles (Figure 11D). The invader tile had a longer binding domain on the tile edge acting as a toehold than the one that was displaced. This phenomenon was similar to DNA strand displacement reactions. This tile displacement had also properties similar to the DNA strand displacements, such as tunable displacement kinetics, competitive displacement, toehold sequestering, and cooperative hybridization. Based on these properties, complicated reorganizations of DNA origami tiles on the DNA array were achieved, demonstrating a simple-yet-powerful strategy for tuning the dynamic behaviors of DNA nanostructures.

Seeded growth is another type of dynamic assembly. It is a self-assembly model, in which DNA tile monomers autonomously add on the termini of the seeds. The seed provides the nucleation position and information to determine which structural form is grown as well as where and when. Unlike the self-assembly of DNA into thermodynamically lowest energy products, self-assembly by seeded growth provides kinetical control over the assembly process and thus may change the shape of products. Complex DNA nanostructures can be generated according to the seed information using the same building units. The Winfree group pioneered the seeded growth for the algorithmic self-assembly.[90, 94] It combined the properties of the periodic assembly of DNA tiles and uniquely addressable DNA origami. In the DNA tile assembly, a tile had up to four binding domains and could bind to the neighboring tiles through the complementary sticky ends. A few types of DNA tiles with different sequences could be programmed into periodic 2D or 3D lattices. Sticky ends protruding from the DNA origami acting as seed strands provided rules for the attachment of DNA tiles, yielding DNA crystals with different patterns. Also, the Winfree group extended the algorithmic assembly using seeded growth for complex DNA computing.[247] A variety of DNA circuits were executed by the algorithmic tile growth with the encoding information from the seed (Figure 11E), demonstrating the high programmability and information-richness of the seeds.

Beyond the applications in algorithmic growth, seeded growth can also guide the assembly of DNA nanostructures. Seeds provide the nucleation sites for DNA tiles and thus significantly accelerate tile growth. DNA tiles preferably grow on the seeds because of the lower free energy. The Schulman group used programmable seeds to direct the assembly of DNA nanotubes.[248] They found that the seeds increased the number of tube structures, suggesting that the seeds reduced the nucleation barrier. Also, the seeds acted as the template for the nanotube circumference. The nanotubes from the tile assembly had the same circumference as the seeds. Although the seeded growth provided precise control over the nucleation and growth, it lacked control over the termination of growth. The products usually yielded structures of uncontrolled lengths. The Schulman group found that the nanotube growth could be terminated by a cap structure that bound to the growing nanotubes and prevented the attachment of further tile monomers.[249] This method offered a programmable strategy to regulate the growth process of nanotubes. In addition, DNA nanotubes from seeded growth could grow to connect pairs of molecular landmarks with different separation distances and relative orientations.[250] DNA origami positioned on the surface provided nucleation for the self-assembly of DX tiles into DNA nanotubes. The free ends of the DNA nanotubes diffused in the solution and joined together to form stable connections (Figure 11F). This point-to-point assembly showed that the self-assembly process of the DNA nanostructures could be tuned by the assembling molecules, the existing nanostructures, and their locations to change the shape of the final products. As a variation of the seeded growth, free tile monomers in the solution could incorporate into the broken DNA nanotubes and repair the structure defects (Figure 11G).[251] The free tiles reached a dynamic equilibrium between the repair and degradation processes and thus allowed to keep the structures intact for a long time. It offered a convenient approach to reverse the degraded DNA structures in biological environments. Furthermore, the seeded growth could also guide the hierarchical assembly of DNA nanotubes.[252] Pre-assembled DNA origami seed junctions provided multiple nucleation positions for the tile growth. The multiple seed locations, relation orientation, and shapes encoded the overall morphology of the hierarchically assembled products.

### 4.4 Replication of DNA structures

Self-replication is an intrinsic feature of living systems. Replication in cells is a complex process with multiple enzymes involved. It is still unclear how exactly life evolved from a chemical system capable of self-replication. Developing synthetic systems that can self-replicate from a template has attracted much interest in the field. Artificial self-replication promises to achieve a deep understanding of the intricate processes in biology. Many synthetic self-replication systems have been developed, such as the modified autocatalytic RNA enzyme, which could perform self-sustained exponential amplification.[253] The Winfree group used DNA tile crystals to replicate combinatorial information.[93] However, few systems exhibited exponential growth, a prerequisite for Darwinian selection. The Seeman and Chaikin groups developed a DNA origami raft system that could exponentially replicate a seed pattern and presented environmental selection (Figure 12A).[254] The origami raft was decorated with the pattern “T” or “A” formed by DNA hairpins. On the opposite side of each raft, there were eight sticky ends by which T raft bound to A raft. On both edges of the daughter rafts, the second set of sticky ends was used to connect the same type of origami rafts into a dimer. The horizontal sticky ends contained a photocrosslinkable nucleotide, so that two sticky ends could covalently link under UV light. The seed “TT” was formed from two T rafts by hybridizing the sticky ends. The seed was mixed with a pool of T rafts and A rafts. The self-replication process of the dimer seed system was such that the seed rafts were recognized and hybridized with the daughter rafts by the vertical sticky ends. A new generation dimer was formed using horizontal bonds and the photocrosslinking reaction under UV light. Daughter rafts were separated from seed rafts by heating the system. The daughter dimer rafts then served as the seed for next-generation offspring. Therefore, the replication of origami rafts could exponentially continue under the UV and temperature cycles. The authors then subjected the replication system to an environmental condition that could regulate the replication. The vertical sticky ends used to bind the seed and the daughter rafts were coupled with pH-sensitive triplex-forming strands, which dynamically adjusted the parent-daughter templating under different pH values. It could program the pH-sensitive triplex DNA to enable or inhibit replication. In system I, the acidic condition enabled parent-daughter templating, while in system II the acidic condition inhibited the templating. The replication kinetics of system I was 1.3-1.4 times faster than that of system II in acidic conditions. The kinetics was reversed at the basic condition. The cases were similar, when two systems were mixed together. The environment-dependent replication rate allowed for the selective growth of species in multi-component systems. In this work, only a single copy of a parent template was made for every replication cycle. Increasing the offspring number per cycle can speed up the replication yield, which promotes the turnover of a species in a population. Chaikin and Seeman changed the rectangular origami rafts into DNA origami cross-tile motifs, which could produce multiple offspring per generation (Figure 12B).[255] A single-parent cross-tile dimer could template multiple daughter dimers on both sides, enabling ladder-structure formation at low temperatures. The ladder structure was then subjected to crosslinking by UV, and the offspring separated with the parent dimer by heating, yielding a series of daughter dimers.

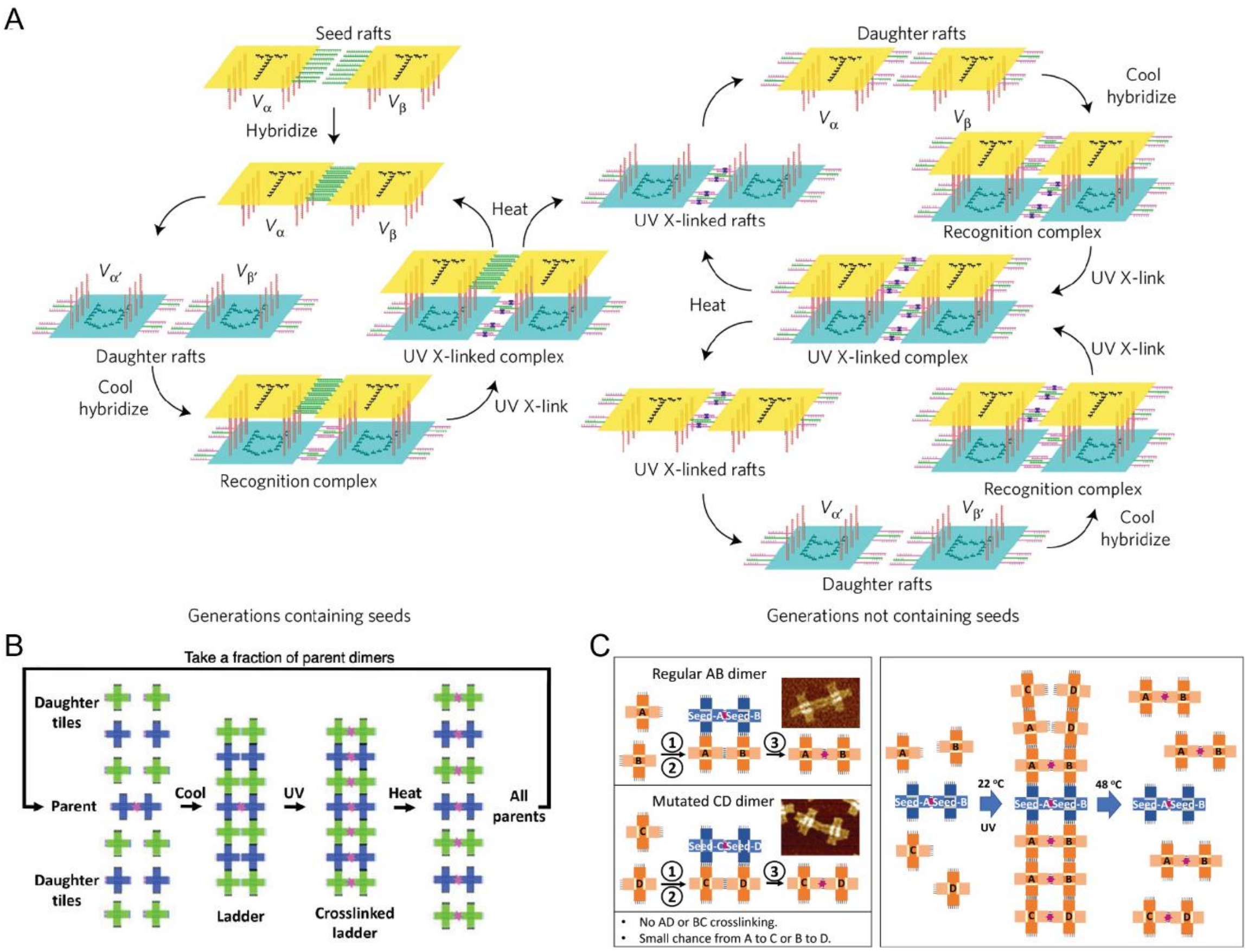

**Figure 12.** Self-replication of DNA nanostructures. (A) Self-replicating cycling using DNA origami raft system. Reproduced with permission from ref 254. Copyright 2017 Macmillan Publishers Ltd. (B) Self-replicating system on cross-tile DNA origami. Reproduced with permission from ref 255. Copyright 2019 National Academy of Sciences. (C) Mutations on the self-replicating system based on cross-tile origami. Reproduced with permission from ref 256. Copyright 2021 National Academy of Sciences.

Replication in biological systems is imperfect, but because the offspring is not always the same as the parent template, it leads to the Darwinian evolution of species. The development of error-prone replication systems enables directed evolution, which can be utilized to obtain offspring with desired properties. In order to mimic the evolution, the Chaikin group introduced mutations in the artificial self-replication tiles (Figure 12C).[256] In the replication system based on origami cross-tiles, the dimer AB seeds templated the tile assembly of monomer tiles A and B. For mutation experiments, a new set of tiles C and D was introduced into the system. C and D had slightly different complementary sticky ends compared with A and B, respectively. Therefore, A could bind to C, while B could bind to D with a small chance. In the replication system, starting with AB parent dimer, the AB templated ladder mostly contained the AB dimer and a small portion of the CD. After UV exposure and heating, the offspring contained mostly AB dimers and might be a small portion of "muted" CD dimers. The mutation rate could be regulated by the number of mismatches on the vertical recognition sticky ends. The mutation was the first step during evolution, and when mutated species obtained advantages like a faster growth rate, they soon dominated the population.

## 5. Nanomaterials templated by DNA origami

Given the high sequence specificity, addressability, and programmability of DNA, DNA origami structures are unique molecular pegboards to organize a variety of nanomaterials, such as silica, silica and metal composites, metallic NPs, metal oxide NPs, carbon nanotubes, magnetic nanoclusters, polymers, enzymes, proteins, biomolecules and many others with prescribed conformations and nanoscale accuracy. This advantage opens many opportunities to construct hybrid nanodevices, in which collective behavior, mutual interactions, and multi-functions can be readily designed and exploited.

### 5.1 Inorganic nanomaterials

#### 5.1.1 Silica and composites

In 2009, Jin *et al*. reported the construction of DNA-silica composites with *p4mm* symmetry.[257] Through the silane hydrolysis reaction, the authors created chiral DNA-silica structures,[258] hierarchical films with 2D mesostructured DNA-silica platelets[259], and chiral silica films.[260] In 2021, Auyeung *et al*. reported solid-state DNA-NP superlattices based on silica encapsulation on dsDNA.[261] 3D DNA superlattices consisting of AuNPs were fabricated in solution. Silica-generating precursors, N-trimethoxysilylpropyl-N,N,N-trimethylammonium chloride (TMAPS) molecules were added and absorbed onto DNA duplexes that connected the particles. Upon the subsequent addition of another precursor, triethoxysilane (TEOS) molecules, silica composites were generated to support the superlattices. The resulting silica-encapsulated lattices maintained good stability and integrity at high temperatures and in different solvents.

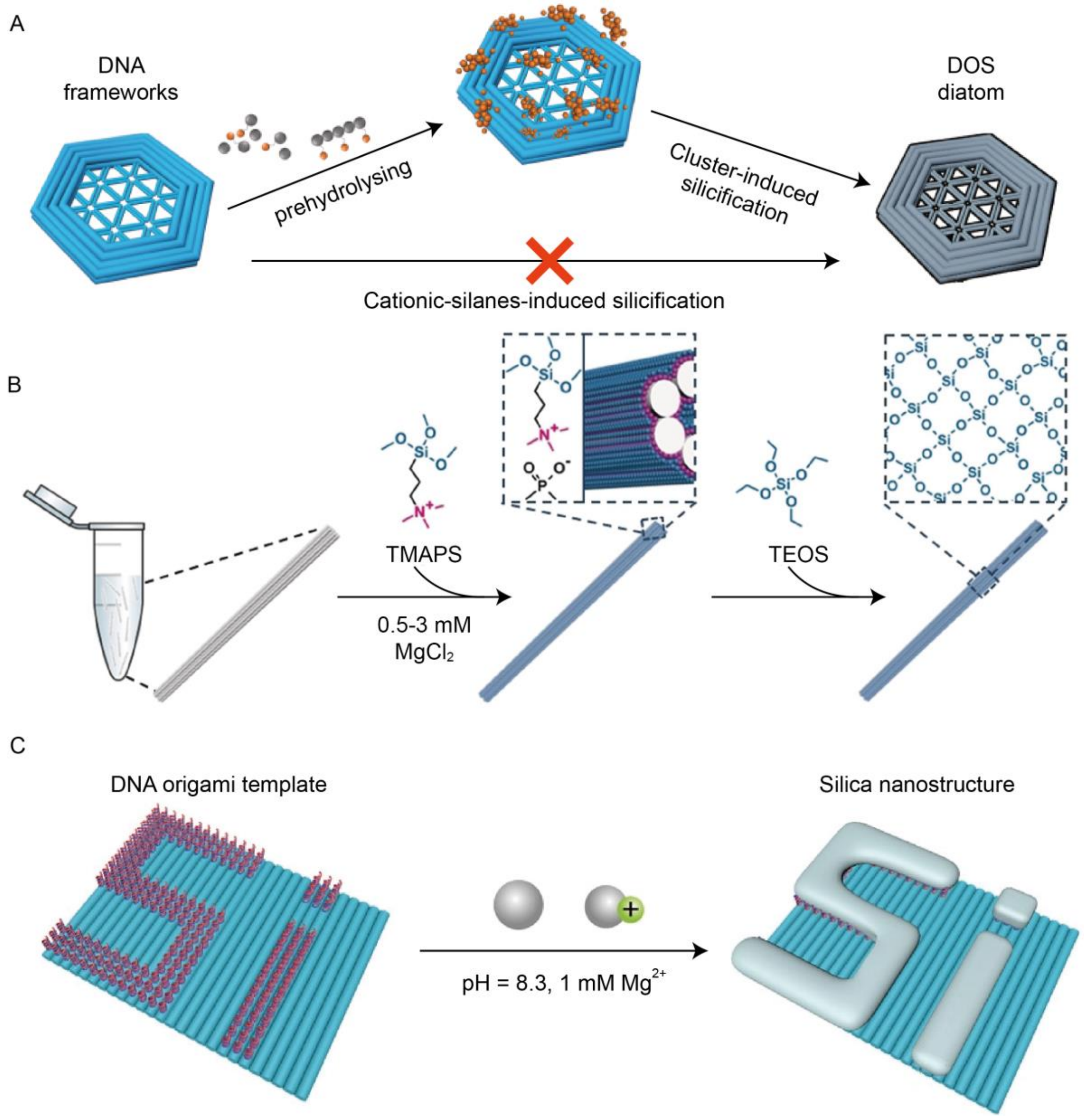

**Figure 13.** Silica composite templated on DNA origami. (A) Growth of silica composite guided by DNA origami templates. Reproduced with permission from ref 262. Copyright 2018 Macmillan Publishers Ltd. (B) DNA origami-enabled silica growth. Reproduced with permission from ref 263. Copyright 2018 Wiley. (C) Site-specific silica growth on customized DNA templates. Reproduced with permission from ref 265. Copyright 2020 Wiley.

Although DNA-assisted localized silane hydrolysis has been vastly investigated, DNA origami as customized templates to guide the growth of silica composites was first reported only in 2018 by Liu *et al*.[262] At a relatively low concentration of $Mg^{2+}$, the silicification of the DNA origami was achieved by a surface-assisted two-step mechanism. First, the pre-hydrolysis step with TMAPS and tetraethyl orthosilicate (TEOS) molecules formed silica clusters. Second, the positively charged pre-hydrolyzed silica precursor clusters containing multiple TMAPS molecules were then attached to the surface of the DNA origami, enabling a silica shell (~3 nm) covered on the user-defined DNA origami (Figure 13A). Nguyen *et al*. employed a similar approach *via* sol-gel chemistry to achieve origami-templated silica growth.[263] The silica shell covered and protected the encapsulated DNA origami, maintaining the structural integrity even in the dry state (Figure 13B). Nguyen *et al*. also utilized a similar silica-coating strategy to produce ultrathin silica shell-origami hybrid structures, which were stable in aqueous solutions and preserved their structures in polar organic solvents.[264] The coated structures also showed enhanced nuclease resistance, which is very useful in nanofabrication and biomedical applications.

Taking the unique addressability of DNA origami, Shang *et al*. reported the site-specific synthesis of silica-DNA hybrid materials.[265] As the guiding template for the silica growth, DNA origami was designed and constructed with extended dsDNA at selected sites, where the protrusions showed stronger electrostatic interactions with positively charged silica precursors than the bare origami surface. In the silicification process, the silica clusters formed by TEOS and TMAPS hydrolysis and absorbed preferentially onto dsDNA protrusions, leading to the silica composite growth (Figure 13C). With the *in situ* small-angle X-ray scattering (SAXS) technique, Ober *et al*. investigated the structural changes of DNA origami during silicification.[266] During the early reaction, substantial condensation of the DNA origami framework was induced by adding silica precursors. The silica deposition led to an increase in the overall size of the silicified DNA structures. The silica growth was also observed inside the DNA origami structures, revealing a strong condensation by the silicification process and displaying thermal stability up to 60°C.

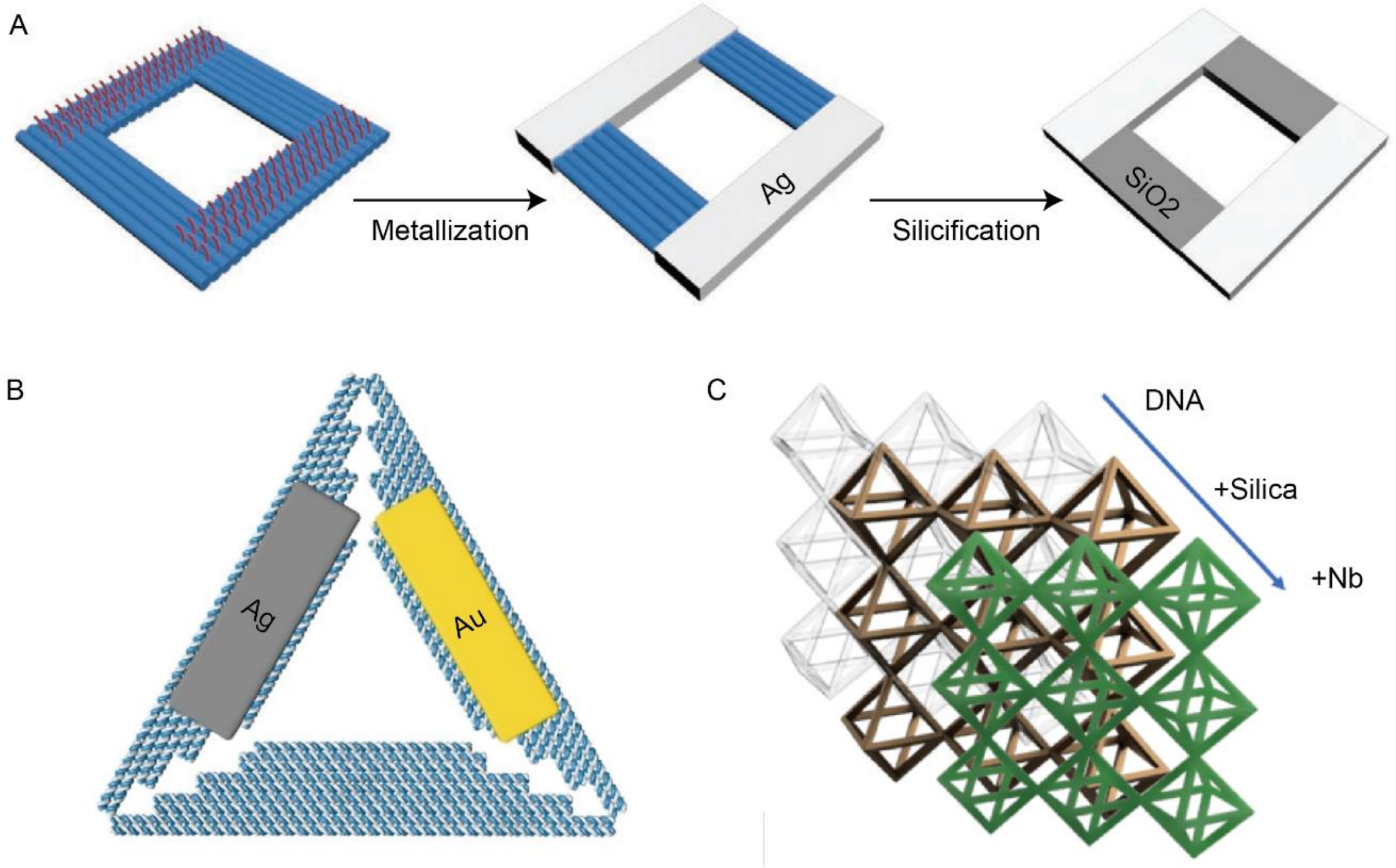

**Figure 14.** DNA origami engineered silica/metal heterostructures. (A) Site-selective growth of silica/silver on DNA origami. Reproduced with permission from ref 267. Copyright 2021 Wiley. (B) Synthesis of silica composite and gold cluster on desired positions of DNA origami templates. Reproduced with permission from ref 268. Copyright 2021 Wiley. (C) A superconductive superlattice based on silica and niobium (Nb) coating. Reproduced with permission from ref 269. Copyright 2020 Macmillan Publishers Ltd.

In addition to pure silica patterns, Zhao *et al*. developed a programmable deposition method for constructing silver (Ag)-silica-coating DNA origami.[267] Thiolated DNA strands were extended from tubular origami to guide Ag anion absorption as seeds and subsequent *in-situ* reduction, resulting in site-specific metallization of Ag. Cysteamine molecules were utilized for specific surface masking of Ag-coating areas with positively charged amino groups, leaving the negatively charged origami surface for addressable silica growth through similar sol-gel chemistry. The authors demonstrated 2D silica/Ag hybrid structures with arbitrary patterns on the DNA origami (Figure 14A). Dai *et al*. reported an alternative strategy for constructing anisotropic silica/metal heterostructures.[268] Triangular DNA origami with two types of ssDNA protrusions were utilized. After hybridization with complementary ssDNA-1, dsDNA-1 protrusions were created on the origami surface, enabling the binding of silica precursors and site-specific silicification. Subsequently, complementary ssDNA-2 was hybridized to create dsDNA-2 on the DNA origami, inducing the Au cluster attachment and growth (Figure 14B). Interestingly, Shani *et al*. realized 3D superconducting nanoarchitectures based on DNA superlattices.[269] Octahedral DNA origami frames were incorporated with 10 nm AuNPs in the centers of the units. Through connections at the vertices, cubic superlattices were formed. Silica layers were generated on the DNA bundles of the superlattices using sol-gel chemistry, converting the soft, DNA-based materials into solid structures. Subsequently, the silica-coated superlattices were dispersed on silicon chips for niobium (Nb) deposition. After coating ~10 nm thick Nb, the superconducting superlattices were created (Figure 14C). This work outlined a novel concept to build 3D superconducting nanodevices with arbitrary geometries.

### 5.1.2 Calcium phosphate and carbon nanotubes

Calcium phosphate (CaP) is an important inorganic material with excellent biocompatibility. In 2020, Liu *et al*. applied DNA frameworks to guide the mineralization of CaP. In order to avoid the inhibition of $Mg^{2+}$ on CaP crystallization, $Ca^{2+}$ was utilized in the assembly process to stabilize the DNA structures. In the solution that contained excessive phosphate, amorphous calcium phosphate (ACP) clusters (~ 1 nm) were formed on $Ca^{2+}$-stabilizing DNA structures. After introducing freshly prepared calcium chloride ($CaCl_2$), further crystallization of local ACP proceeded, resulting in DNA-templated mineralization of CaP nanocrystals. By carefully controlling the reaction conditions, both DNA tetrahedrons and DNA origami structures were uniformly coated with CaP nanocrystals. The CaP shell greatly enhanced the cellular uptake efficiency as well as the stability of DNA origami in the physiological environment.[270] In 2021, Wu *et al*. reported an alternative method to maintaining the site-specific decoration ability of DNA origami after the biomineralization process.[271] Streptavidin molecules were introduced to the predefined sites on the DNA origami *via* biotin-streptavidin interactions, preventing CaP from being absorbed onto these regions. After biomineralization, biotin-modified transferrin could still be allocated to the desired sites on the DNA origami.

Although carbon nanotubes possess interesting electronic properties, the arrangement of carbon nanotubes into well-defined complex geometries has been a central challenge in nanotechnology. Maune *et al*. first applied DNA origami to template ssDNA-labelled single-walled carbon nanotubes (SWNTs) into 2D geometries, such as lines and cross-junctions. DNA hybridization between the ssDNA tagged on SWNTs and DNA origami guided the SWNT assembly with nanoscale resolution.[272] Taking advantage of the simple streptavidin-biotin interactions, Eskelinen *et al*. presented an alternative method to constructing DNA origami-SWNT architectures.[273] Zhao *et al*. demonstrated DNA origami-templated SWNTs of discrete lengths obtained by high-performance liquid chromatography (HPLC) purification into intricate geometries.[274] To increase the assembly yield of DNA origami-SWNT structures, Zhang *et al*. introduced an approach mediated by AuNP-based spherical nucleic acids (SNAs).[275] More specially, they prepared ssDNA containing TAT-repeated sequence wrapped SWNTs *via* hydrophobic interactions. The ssDNA-tagged AuNPs (SNAs) were then conjugated on 2D DNA origami through DNA hybridization with ssDNA extensions. Serving as the bridges, the SNAs guided the ssDNA-labelled SWNTs to the prescribed positions on the DNA origami. Due to the higher ssDNA densities for hybridization (20 strands/100 $nm^2$ on SNAs vs 4 strands/100 $nm^2$ on DNA origami), the cooperative assemblies using SNAs resulted in an approximately five-fold yield of the DNA origami-SWNT architectures, which provided a more convenient and versatile way for the fabrication of electronic devices. In a recent work of Zhao *et al*., the authors integrated DNA-guided SWNT assembly with nanometer precision and applied them in building large arrays of high-performance solid-state electronics.[276] ssDNA sequences were modified at the channel interface of the templates to precisely organize SWNTs with intertube pitches as small as 10.4 nm on the field-effect transistor (FET) substrate. The FETs exhibited more than ten-fold improvement in key transport performance metrics compared to previously reported biotemplated FETs. This biomolecular assembly approach offered a new solution for the scalable production of biotemplated electronic devices.

### 5.1.3 Metal, semiconductor, and magnetic nanoclusters

Metal and metal oxide NPs are important components in nanoelectronics and nanophotonics. Li *et al*. demonstrated the growth of metal and metal oxide nanoclusters (MMONs) on DNA origami.[277] Thiol-tagged ssDNA strands were positioned at the designated sites on the DNA origami surface, which possessed a strong affinity for multiple metal ions, including palladium (Pd), iron (Fe), cobalt (Co), nickel (Ni), silver (Ag), and gold (Au). In the redox reactions, the metallization proceeded and MMONs were generated on the DNA origami (Figure 15A). Zhang *et al*. reported the realization of Ag spiral structures on DNA origami.[278] 1D and 2D DNA origami structures were created with protruding ssDNA strands, which induced the local enrichment of Ag precursors, enabling site-specific metallization on the DNA origami. Ag spiral patterns with well-defined chirality were obtained (Figure 15B). Furthermore, Jia *et al*. reported that condensation and metallization of Cu and Ag on DNA origami could lead to arbitrary patterns with 10 nm precision.[279] DNA origami with ssDNA protrusions was first constructed and served as the guiding templates for efficient and site-specific condensation of metal ions. Subsequently, localized metal plating proceeded by redox reactions, giving rise to metallization on the origami (Figure 15C).

Hybrid nanoarchitectures of different metal/metal oxides can also be assembled using DNA origami. Aryal *et al*. reported the assembly of gold/tellurium nanorods (AuNRs/TeNRs) on DNA origami.[280] Bar-like DNA origami was fabricated with ssDNA strands at selective positions as binding sites for the attachment of complementary ssDNA-coated AuNRs. Using an Au plating solution, the redundant ssDNA on the bound AuNRs was removed. TeNRs were then positioned in the gaps of the tethered AuNRs on the bar template. Polybenzimidazole (PBI) coating of the AuNR/TeNR attached DNA origami was carried out. Subsequently, the annealing and polymer removal procedures were used to establish the tight junction between the AuNRs and TeNR, resulting in a continuous Au/Te/Au wire-like nanostructure. In another work of the same group, electroless Au plating was used to fill the gaps between the AuNRs and TeNR to create Au-Te-Au junctions.[281] The electronic property of the metal-semiconductor junctions was examined by two-point electrical characterizations.

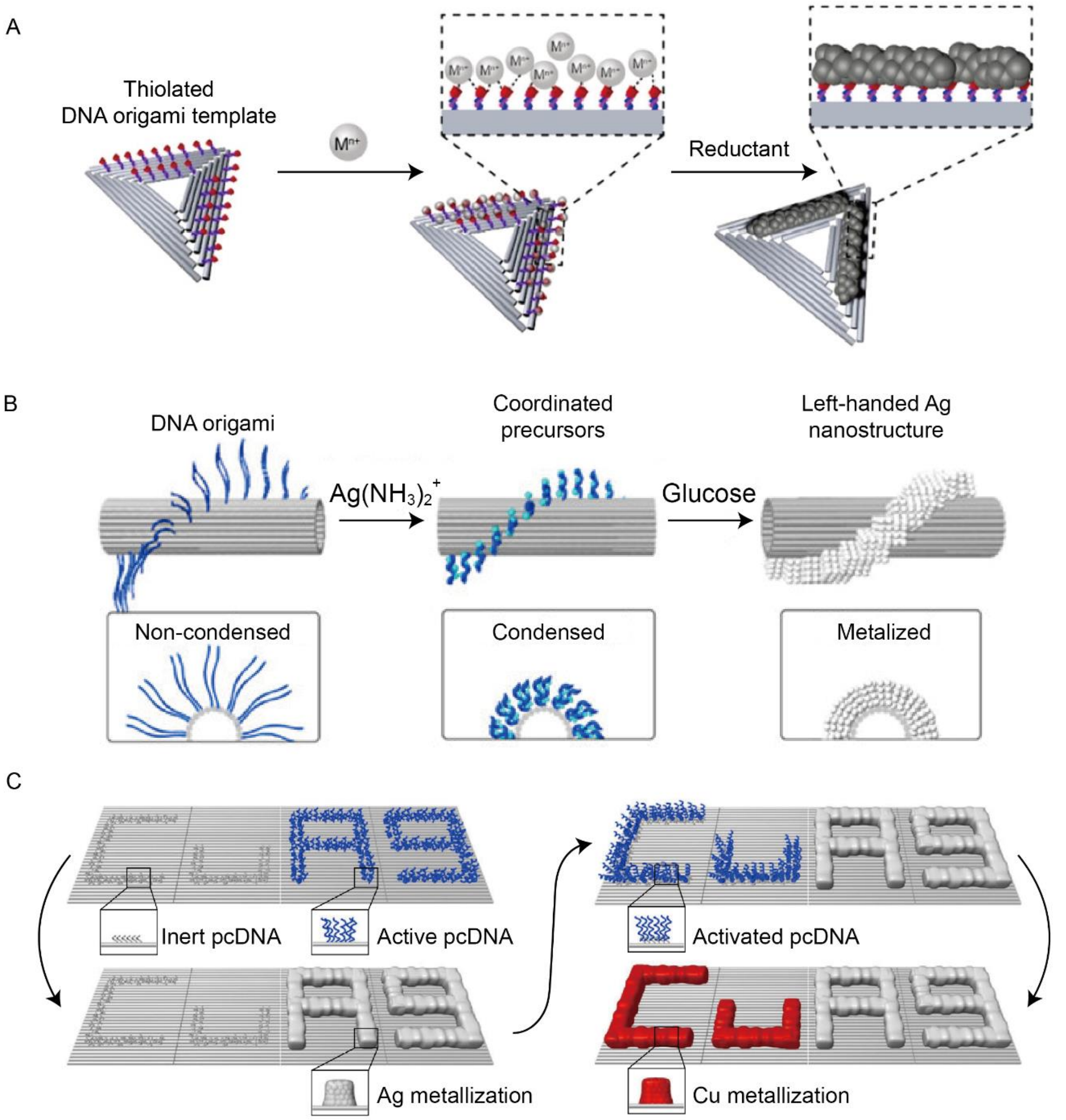

**Figure 15.** Metal/semiconductor nanoarchitectures based on DNA origami. (A) A general fabricating method for metal and metal oxide nanocluster (MMONs) with arbitrary patterns on DNA origami templates. Reproduced with permission from ref 277. Copyright 2019 American Chemical Society. (B) Helical sliver patterns on DNA filaments. Reproduced with permission from ref 278. Copyright 2021 American Chemical Society. (C) DNA origami-enabled copper and silver patterns with 10-nm precision. Reproduced with permission from ref 279. Copyright 2019 Macmillan Publishers Ltd.

Magnetic NP assemblies engineered by DNA origami have been recently reported. For instance, Meyer *et al*. assembled magnetic iron oxide nanoparticles (IONPs) using DNA origami.[282] Hydrophobic IONPs (~ 15 nm) were firstly functionalized with an azide-modified amphiphilic polymer poly(maleic anhydride-alt-octadecene) (PMAO). ssDNA sequences modified with strained alkyne dibenzocyclooctyne (DBCO) were utilized to conjugate the azide-IONPs. 16 helix bundle (16HB) DNA origami square lattices were employed as addressable templates with ssDNA extensions to bind the IONPs coated with complementary ssDNA strands. The assembly of DNA-IONPs and DNA origami lattices was evaluated using TEM and agarose electrophoresis. The resulting IONP-origami architectures showed tunable magnetic resonance imaging (MRI) contrast generation efficiency by controlling the number and spacing of the IONPs on the DNA origami.

Shape-controlled metal/metal oxide nanostructures have interesting properties and can be widely utilized in nanophotonics, light harvesting, and biosensing. In the work of Sun *et al*., an ssDNA-functionalized AuNP was assembled as a seed in the user-defined cavity of a DNA mold to cast the growth of the particle.[283] A variety of inorganic NPs, including Au cuboids, Ag cuboids, Ag triangles, Ag Y-shapes, and composite quantum dot-Ag-quantum dot sandwiched structures, were synthesized with a 3-nm resolution. In the same year, Helmi *et al*. reported a similar concept to synthesize Au nanostructures with programmable shapes using DNA origami molds.[284] Following their previous work, Bayrak *et al*. reported the realization of Au nanowires of micrometers long.[285] DNA origami structures were synthesized and used as mold bricks, which contained the prescribed cavities for the site-specific attachment of Au seeds and the interacting ends for the DNA mold adhesion. Specific staple strands at the helix ends of the mold monomers were designed with 2-nt extensions or deletions to form the corresponding interfaces for sequential docking of the long linear DNA mold superstructures. After the Au deposition was initiated by the seeds, the Au nanowires grew in the cavities of the DNA mold chains. These mold-templated metallic nanowires were conductive, although the conductance was limited by the remaining gaps between the individually grown NPs. In 2019, Ye *et al*. improved this approach to create tubular DNA mold superstructures with programmable lengths and patterns.[286] The authors designed a series of mold monomers with specific interfaces to ensure the formation of the mold polymers with specificity and affinity. After the modular assembly, the metallic wires were synthesized with mold-defined patterns. Later, Ye *et al*. continued to optimize the seeded-growth procedure for casting Au nanostructures with higher aspect ratios.[287] Inside the DNA origami mold, the number of AuNPs for deposition to form wire structures was greatly reduced, which supported the Au growth with high continuity. In a recent work of the same group, the authors reported more complex metallic structures fabricated using DNA origami building blocks as programmable mold elements.[288] A series of mold superstructures with desired geometries were constructed by assembling four different structural elements. After the Au deposition within the cavities of the mold platform, versatile Au structures, such as T-shaped, branched, dumbbell-shaped, and loop particles, were synthesized, and they showed efficient charge transport.

## 5.2 Biologically relevant nanomaterials

### 5.2.1 Polymers

DNA origami can serve as template for the arrangement of polymer chains by accurate polymer attachment *via* DNA hybridization or by site-specific polymer growth.[289] Knudsen *et al*. constructed linearly aligned patterns of individual polymer molecules with the aid of DNA origami.[290] (2,5-dialkoxy)paraphenylene vinylene-DNA conjugate, termed poly(APPV-DNA), was prepared *via* hydroxy-functionalized polymers by immobilizing APPV on a CPG solid support and subsequent automated DNA synthesis from the hydroxyl groups of the polymer. The APPV-DNA polymer chains were then hybridized with complementary ssDNA decorated on rectangular origami (Figure 16A), and toehold-mediated strand displacement reactions tuned the directions of the aligned polymer chains.[291] The same group further employed this approach to create wire-like electronic/photonic devices with arbitrary geometries.[292] Using long polymers of hundred nanometers and donor/acceptor fluorophores colocalized at defined positions along the polymer nanowires, highly efficient energy transfer over 24 nm was achieved (Figure 16B). Also, the authors prepared a similar polyfluorene-DNA graft-type polymer (poly(F-DNA)) that possessed a conjugated polyfluorene backbone with protruding ssDNA and anchored it on addressable DNA origami.[293]

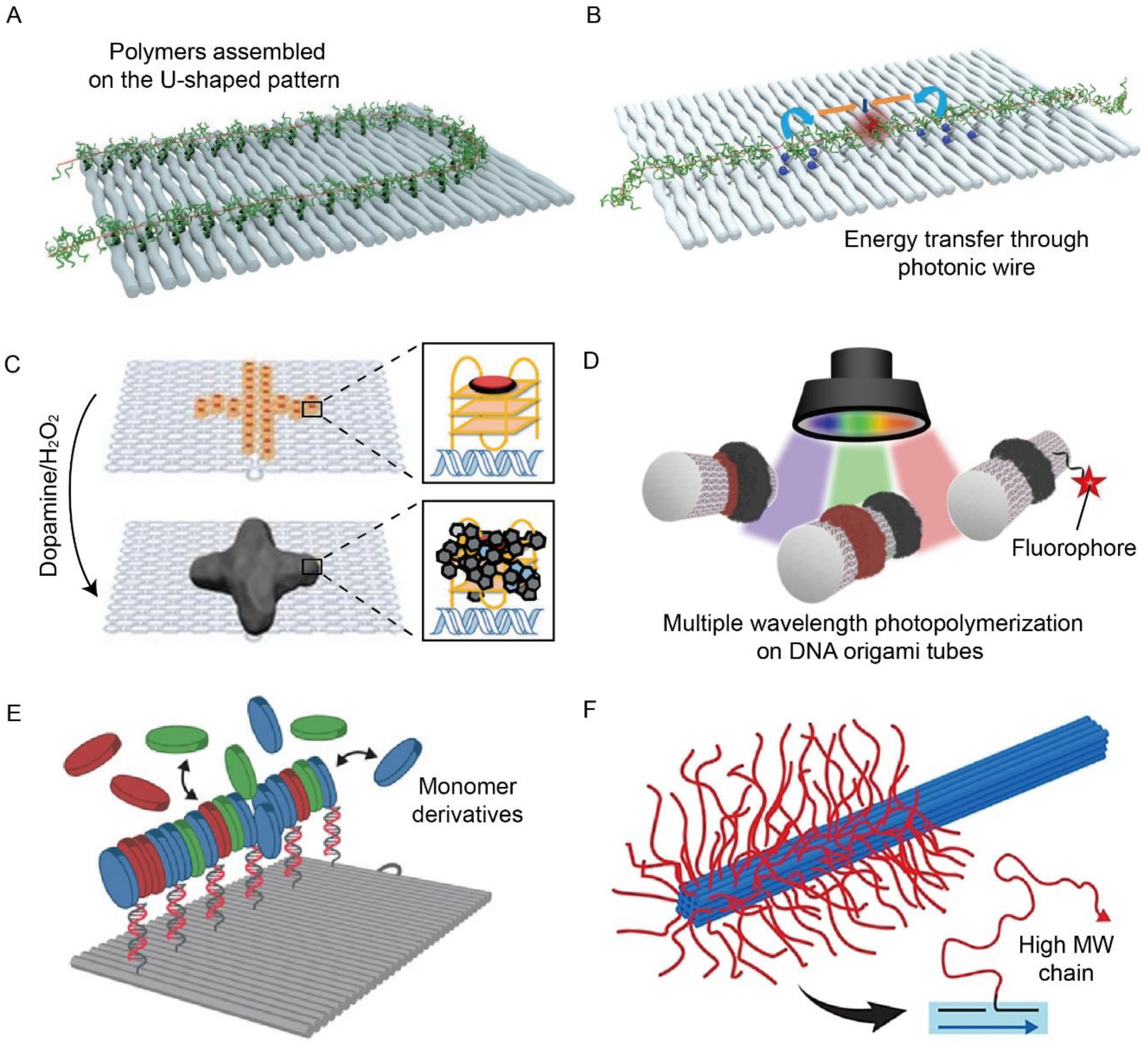

**Figure 16.** Polymer nanostructures based on addressable DNA origami template. (A) Precise linear patterns of individual polymers on rectangular DNA origami templates. Reproduced with permission from ref 290. Copyright 2015 Nature Publishing Group. (B) DNA origami-enabled polymeric nanowires for photonic investigation. Reproduced with permission from ref 292. Copyright 2021 American Chemical Society. (C) Growth of polydopamine (PD) nanoarchitectures on DNA templates. Reproduced with permission from ref 295. Copyright 2017 Wiley. (D) Photo-responsive polymer-origami hybrid nanostructures manipulated by multiple wavelength irradiation. Reproduced with permission from ref 297. Copyright 2022 Wiley. (E) A dynamic assembly of 1D supramolecules on DNA origami platforms. Reproduced with permission from ref 298. Copyright 2021 Wiley. (F) An enzyme-enabled synthesis of polynucleotide brushes on the surface of DNA nanotubes. Reproduced with permission from ref 299. Copyright 2021 Wiley.

Another approach to integrating semiconducting polymers on DNA origami was developed by Zessin *et al.*[294] A block copolymer (BCP) molecule that consisted of a water-soluble semiconducting polymer segment (polythiophene poly(3-tri(ethylene glycol)thiophene), $P3(EO)_3T$) and ssDNA was synthesized *via* click chemistry. Subsequently, the $P3(EO)_3T$-ssDNA monomers were attached *via* complementary ssDNA handles to 2D DNA origami at the desired locations, leading to densely packed BCP patterns. The optical properties of the packed BCP molecules were locally tuned by regulating the π-π stacking interactions. These interactions between the densely packed polymers could be suppressed by introducing conjugate polymers with N,N-dimethyldodecylamine N-oxide (DDAO), a zwitterionic surfactant. With the increased DDAO concentration, the fluorescence intensity of the assembled structures could be rapidly recovered.

Polymer patterns can also directly grow on DNA origami. Tokura *et al*. demonstrated DNA origami enabled-polydopamine (PD) architectures.[295] At the selected positions, G-quadruplex (G4)-containing staples were extended from the DNA origami and activated by incorporating the cofactor molecules, hemin. With horseradish peroxidase (HRP)-mimicking, $H_2O_2$-mediated oxidation, the G4/hemin-based DNA enzyme (DNAzyme) domains catalyzed the dopamine polymerization in the solution to form polymer bundles on the origami (Figure 16C). Similarly, tubular origami carrying atom transfer radical polymerization (ATRP)-initiator moieties on the outer surface was assembled to form a polymeric tubular nanoreactor. The interior space of the nanoreactor was successfully used as a mold for dopamine polymerization. In addition, the same group developed a photosensitive polymer growth approach by introducing protoporphyrin IX molecules to the origami-DNAzyme system.[296] The photosensitizers were embedded in G4 structures on the DNA origami and locally catalyzed the oxidation of dopamine to PD upon irradiating visible light, generating predefined polymeric patterns. They further demonstrated multi-wavelength photo-controlled polymer-origami hybrid structures.[297] Three types of photosensitizers, protoporphyrin IX, eosin Y (EY), and methylene blue (MB), that responded to different wavelengths were introduced to G4 on the origami tube, producing reactive oxygen species (ROS) by corresponding irradiations to initiate polymerization (Figure 16D). This approach was then employed in the patterned and layered growth of different polymers on DNA origami.

Taking one step further, Schill *et al*. reported dynamic 1D supramolecular assemblies on DNA origami.[298] Bipyridine-based C3-symmetry amphiphilic, discotic molecules were used as monomeric building blocks in the synthetic polymer system. Rectangular origami was employed as an organizing template for guiding DNA-functionalized disc-like monomers to form the discrete assembly at designated locations, where complementary ssDNA handles were extended. After recruiting additional non-DNA-tagged monomers, the continuous supramolecular assembly was realized on the origami (Figure 16E). Such templated supramolecular assemblies enabled dynamic rearrangements of monomeric building blocks and molecular cargoes by regulating their compositions. Going further, Yang *et al*. demonstrated an enzyme-aided, site-specific approach for synthesizing polynucleotide brush-decorated DNA nanotubes.[299] Staples that contained 3´ oligo-dT as the terminal deoxynucleotidyl transferase (TdT) initiator were utilized in the origami assembly, resulting in site-selectively extended initiators on the origami. In the presence of TdT and dNTP, TdT-catalyzed enzymatic polymerization enabled the growth of high-molecular-weight nucleotide chains at the designated locations (Figure 16F). The brush-like polynucleotide chains significantly improved the stability and nuclease resistance of the structures, which are essential for drug delivery and other biological applications.

### 5.2.2 Enzymes

In living organisms, biochemical functions are often achieved through highly efficient enzymatic reactions. There have been intense research activities to build artificial enzymatic systems. Nevertheless, the dynamic catalytic processes, transient interactions between enzymes and substrates, as well as complicated reaction environments all cast great challenges to mimic the key features and functions of the biological systems. Taking advantage of the inherent properties of DNA, the field has achieved remarkable progress in DNA-origami-templated enzymatic reaction systems. A series of artificial enzymatic platforms have been successfully created with precise spatial organization and dynamic controllability.

#### 5.2.2.1 Enzymatic reactions in nanocontainers

Niemeyer and coworkers reported the site-selective enzyme modification and precise assembly of enzymes on DNA origami.[300] Orthogonal tags (Snap-tag, HaloTag, or streptavidin-binding peptide) were genetically fused into recombinant enzymes (S-selective $NADP^+$/NADPH-dependent oxidoreductase Gre2 and the reductase domain BMR of the monooxygenase P450 BM3). The two tagged proteins were coupled with ssDNA in a site-specific manner and subsequently anchored on addressable DNA origami sheets at the prescribed positions. Free-flow electrophoresis (FFE) was employed to purify the enzyme-attached DNA origami. In contrast to free enzyme molecules, origami-tethered enzymes showed enhanced activities. Later, Zhao *et al*. designed a nanocaged enzyme system using DNA origami to enhance the stability and catalytic efficiency of protein molecules in a complex environment.[301] Metabolic enzymes (glucose oxidase (GOx), horseradish peroxidase (HRP) molecules) were modified through succinimidyl 3-(2-pyridyldithio) propionate (SPDP) coupling of a lysine residue on the protein surface and thiol-tagged ssDNA. Open half-cages were self-assembled based on honeycomb-lattice origami. Protein molecules with ssDNA modifications were anchored in the cavities of the half-cages *via* DNA hybridization. A complete cage containing two different enzymes was formed by linking two half-cages together through the hybridization of short DNA bridge strands extended from their edges (Figure 17A). It was demonstrated that the DNA nanocages could protect the inner enzyme payloads against protease-mediated degradation and aggregation.

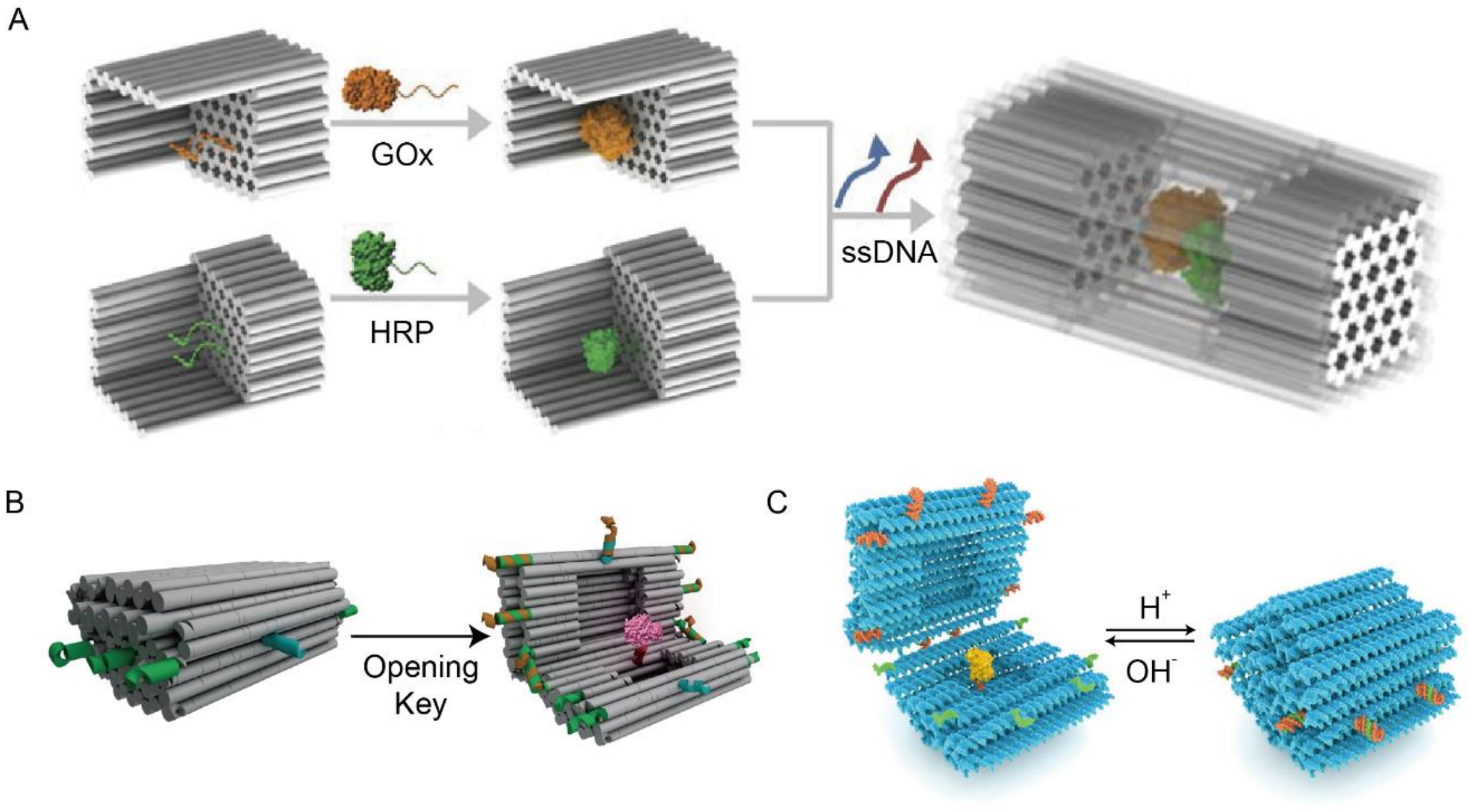

**Figure 17.** Enzyme-nanocontainer systems based on DNA origami techniques. (A) An origami nanocage incorporated enzyme for enhanced stability and catalytic efficiency. Reproduced with permission from ref 301. Copyright 2016 Nature Publishing Group. (B) A strand-triggered reconfigurable DNA nanovault for precise control enzyme reaction. Reproduced with permission from ref 302. Copyright 2017 Nature Publishing Group. (C) A pH-responsive DNA nanocapsule for precise regulation of catalytic compatibility. Reproduced with permission from ref 186. Copyright 2019 American Chemical Society.

Beyond the enzyme encapsulation using static DNA origami, Grossi *et al*. applied reconfigurable DNA nanocontainers to enclose enzyme molecules in their cavities and precisely regulate enzyme-substrate interactions.[302] An open nanovault origami structure was designed to contain a single enzyme molecule. The vault was closed by adding the corresponding locking strands, so the enzyme was shielded from its substrates in the solution. In the presence of the opening key strands, the nanovault was activated to the open state and exposed to the internal cargo, enabling the enzymatic reactions (Figure 17B). Ijas *et al*. also designed similar enzyme capsules, whose reconfigurable motions were controlled by changing the pH value of the solution.[186] Triplex containing pH-responsive DNA latches were integrated into the nanocapsules for reversible opening and closing (Figure 17C). DNA containers protected HRP molecules as inner payloads, whose catalytic compatibility was precisely regulated by the pH changes of the solutions. Furthermore, Kosinski *et al*. studied the role of the DNA container in an enzyme-loaded nanocage system.[303] Thrombin molecules were trapped in the cavities of 2D rectangular sheets or 3D boxes through the binding of enzymes with their specific aptamers. The authors showed that the catalytic reaction rates were influenced by the DNA/enzyme binding affinity and DNA/substrate electrostatic interactions. Based on the kinetic analysis of the catalytic properties, the DNA frameworks not only served as non-functional scaffolds, but also actively participated in and affected the reactions as modifiers by providing alternative kinetic routes for the substrate hydrolysis.

**5.2.2.2 Enzyme cascades**

Fu *et al*. used spatially addressable DNA origami tiles to precisely organize discrete enzyme molecules (e.g., GOx and HRP) and studied the inter-enzyme distance-dependent kinetics of the assembled enzyme pairs.[304] Also, they reported a similar enzyme cascade system that could be regulated by light.[305] The glucose6-phosphate dehydrogenase (G6pDH) and lactate dehydrogenase (LDH) molecules worked as the enzyme cascade, and nicotinamide adenine dinucleotide ($NAD^+$) worked as the cofactor. The G6pDH-LDH enzyme pairs were anchored onto the DNA origami at a controlled distance, while $NAD^+$ was conjugated to a DNA Holliday junction (HJ) that served as a swingarm located between the two enzyme molecules. The photo-responsive azobenzene (AZO) molecules were introduced into the HJ using AZO-conjugated DNA strands, which enabled light-driven hybridization by visible light and de-hybridization by ultraviolet light for the reversible control of the $NAD^+$ molecule's position. The light-driven switching of $NAD^+$ regulated the enzyme cascade activity between on and off states (Figure 18A). These results showed that addressable DNA origami-based systems allowed the rational assembly of DNA-enzyme heterostructures. The finely tuned inter-molecular distances between enzymes and substrates efficiently regulated the activity of enzyme cascades. In 2017, Sun *et al*. reported a DNA origami-based artificial raft for real-time imaging of the enzyme cascade on a supported lipid bilayer (SLB) at the single-molecule level.[306] To anchor the artificial raft on the SLB, cholesterol (CHOL)-tagged dsDNA strands were extended from the bottom surface of the 2D triangular origami. The origami raft was also functionalized with a Cy3-labelled ssDNA strand and an Atto488-modified GOx molecule on the top surface. As a downstream enzyme in the cascade reactions, Alexa647-tagged catalytic molecules were modified with ssDNA and tethered to the SLB *via* hybridization with CHOL-ssDNA (Figure 18B). The movements of dye-labeled enzyme molecules on the SLB were recorded by total internal reflection fluorescence microscopy (TIRFM). Furthermore, this real-time single-molecule characterization method was used to elucidate the correlation between the catalytic performance and the docking sites by Xu *et al*.[307]

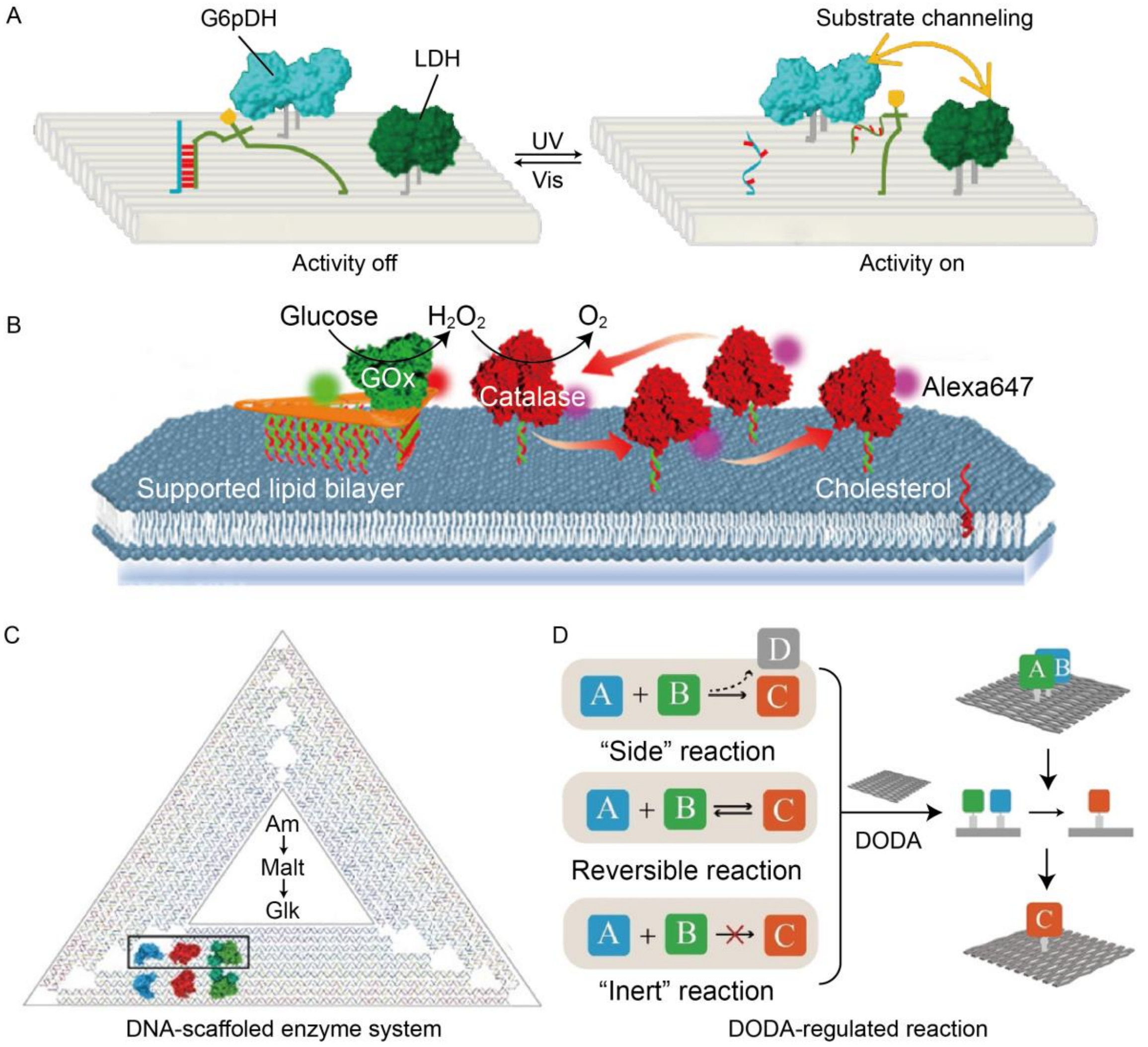

**Figure 18.** Enzyme cascade systems on DNA origami platforms. (A) Light-driven enzyme pairs assembled on DNA origami. Reproduced with permission from ref 305. Copyright 2018 American Chemical Society. (B) An artificial origami raft for real-time imaging of the enzyme cascade. Reproduced with permission from ref 306. Copyright 2017 American Chemical Society. (C) Sequential enzymatic cascade and substrate channeling based on DNA origami. Reproduced with permission from ref 308. Copyright 2019 American Chemical Society. (D) Dynamic control of molecular cascade reactions on a reconfigurable DNA origami domino array (DODA). Reproduced with permission from ref 310. Copyright 2021 Wiley.

In 2019, Klein *et al*. demonstrated a multi-enzyme cascade system with enhanced catalytic efficacy assembled on DNA origami.[308] A 2D triangular origami structure was employed to attach three ssDNA-tagged enzyme molecules (sugar metabolizing enzyme amylase, maltase, and glucokinase) in close proximity, allowing for a sequential enzymatic cascade and substrate channeling (Figure 18C). The results revealed a 30-fold enhancement of the catalytic activity, which could probably be attributed to increased enzyme stability, a localized surface affinity, or the hydration layer effect. Kahn *et al.* presented another enzyme cascade platform using a 3D DNA wireframe octahedron.[309] Similar to previous reports, GOx and HRP molecules were anchored onto the origami superlattices with precise positioning. To explore the 3D enzyme organization, the team created an enzyme-origami library with different 3D arrangements of the protein location, orientation, and intermolecular spacing. The experimental data revealed that the enzymatic cascade activity was enhanced by decreasing the spacing over the 3D origami structures. In addition, the structural continuity of the origami between the enzyme linkages affected the catalytic activity. Furthermore, Fan *et al*. designed a reconfigurable DNA origami domino array (DODA) for the dynamic control of molecular cascade reactions.[310] The DODA structure was used to anchor the reactants with nanometer spatial precision (Figure 18D). In the presence of distinct trigger strands, the DODAs transformed from narrow to wide conformations, influencing the proximity of the pre-organized reactants and self-assembly reactions. This work presented a feasible approach to regulating molecular cascade reactions based on dynamic origami platforms.

**5.2.2.3 Enzyme-based nanodevices**

Enzyme-integrated DNA origami can be used to achieve particular functions. For instance, Belcher and coworkers showed a G4/hemin DNAzyme-origami hybrid platform to create size-controlled SWNTs.[311] The DNAzyme moieties were attached to the origami with precise spatial control. Long SWNTs were wrapped with biotin-tagged ssDNA to enhance their dispersity in an aqueous solution. The ssDNA-SWNTs were then guided to assemble on the DNA origami surface with predefined biotin-streptavidin decorations. G4/hemin DNAzyme localized on the origami then catalyzed the production of free radicals and induced chemical oxidation at designated positions, resulting in size-controlled cutting of SWNTs. Masubuchi *et al*. reported a DNA origami-based logic chip to control gene expression.[312] Rectangular DNA origami containing SNAP-binding ligands was fabricated to precisely anchor SNAPf protein-fused T7 RNA polymerase (RNAP) molecules (Figure 19A). After covalent attachment of enzymes and purification, the T7-origami chip structures exhibited low affinity and transcription activity for the externally diffusing genes. The target gene was then integrated into the T7-origami chip *via* avidin-biotin interaction, and the enzyme-gene distances were precisely regulated by the origami nanosheet. Several factors, including the intermolecular spacing between the enzyme and the substrate, the tethered direction of the gene, as well as the length and rigidity of the linker, were proved to influence the transcription activity of the T7-gene chips. Multiple target-gene expression output could be rationally designed by precisely controlling the gene substrates attached to the system through varying the intermolecular distances of the enzymes. A logic-gated transcriptional chip was also constructed. It could respond to water-in-oil droplets and facilitate the expression of defined genes.

Furthermore, Hahn *et al*. demonstrated a DNA origami-based platform for RNA production and processing.[313] The functional components, including the RNA polymerases, RNA endonucleases, and DNA templates for rolling circle transcription and processing of RNA, were anchored in the cavity of a barrel-like DNA origami container (Figure 19B). The RNA-producing parts consisting of dsDNA templates and T7 RNA polymerases could generate the target RNA copies encoded by the corresponding templates. The multi-copies of premature RNA transcripts were then cleaved by the integrated RNA endonucleases in the nanofactory that served as RNA-processing units, resulting in the production of mature RNA molecules. The origami-based nanofactory could be further developed, allowing the generation of therapeutic RNA (such as shRNA and mRNA) and other types of macromolecules for biological applications. Rosier and coworkers used DNA origami to assemble caspase-9 monomers and regulate the inter-enzyme spacing. They demonstrated that DNA origami-protein assemblies could work as synthetic apoptosomes.[314] For protein-DNA modification, caspase-9 expressed with unnatural amino acid (*p*-Azidophenylalanine) was conjugated to bicyclononyne (BCN)-tagged ssDNA through azide-alkyne cycloaddition. Complementary ssDNA handles protruded from the rectangular DNA origami and were used to precisely anchor the individual caspase monomers *via* DNA hybridization (Figure 19C). After testing the distance-dependent enzyme activity of the origami-enzyme assemblies with different inter-protein spacings, the results showed that the catalytic capability was induced by proximity-driven dimerization of caspase-9. Enzymatic activity of the assemblies could be further enhanced by clustering three and four caspase-9 monomers, revealing the multivalency of the platform. This work provided a unique tool based on DNA origami to study the functions of multi-enzyme complexes in biological processes, including inflammation, innate immunity, and cell death.

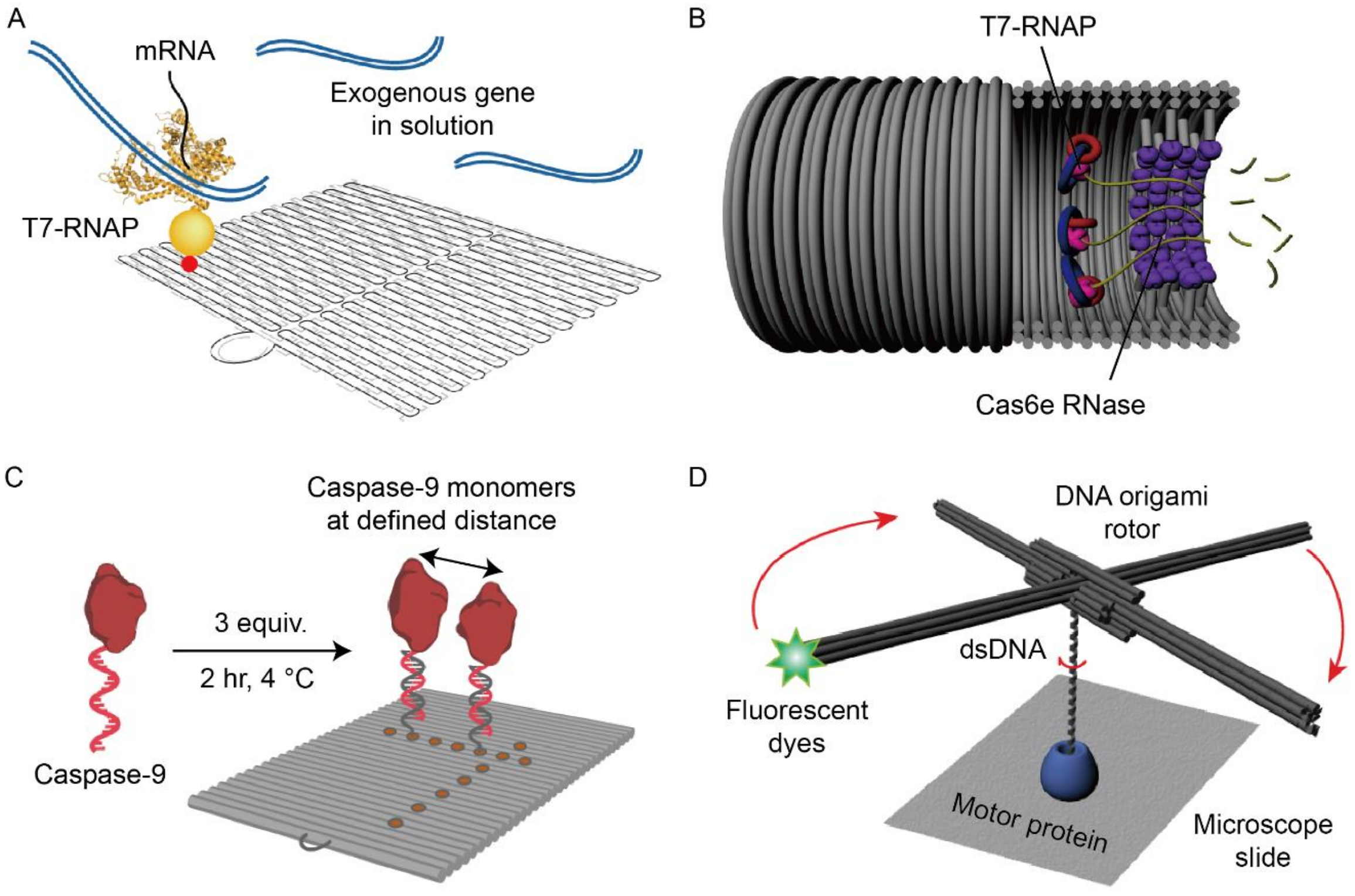

**Figure 19.** Biofunctional regulation by enzyme-origami systems on molecular levels. (A) Gene expression controlled by a DNA origami-based logic-chip. Reproduced with permission from ref 312. Copyright 2018 Nature Publishing Group. (B) RNA production and processing facilitated by a DNA origami-based nanofactory. Reproduced with permission from ref 313. Copyright 2020 American Chemical Society. (C) A synthetic apoptosome constructed by caspase-9-integrated DNA origami platform. Reproduced with permission from ref 314. Copyright 2020 Nature Publishing Group. (D) High-resolution imaging and tracking biological events by DNA origami rotors. Reproduced with permission from ref 315. Copyright 2019 Nature Publishing Group.

DNA origami can also be utilized in high-resolution imaging and biological event tracking. Kosuri *et al*. reported a fluorescence-tagged DNA origami rotor to facilitate the measurement of DNA rotation, which was usually associated with genome-processing reactions at the single-molecule level with a time resolution of milliseconds.[315] A cross-shaped honeycomb lattice origami rotor consisting of four blades (each with a length of 80 nm) was designed to amplify the DNA motions and stiffness to minimize the obscuring effect of Brownian fluctuations. One tip of the rotor blades was modified with a fluorescent dye for imaging and tracking. The center of the DNA rotor was connected with motor protein by a dsDNA segment along the axis of rotation, which served as the substrate for DNA-interacting enzymes (Figure 19D). Using the origami-rotor-based imaging and tracking (ORBIT) method, the authors tracked DNA rotations by helicase-involved DNA repair (RecBCD complex), and characterized a series of events during RecBCD-initiated DNA unwinding. For another application of tracking DNA rotations during transcription by RNAP, the rotational steps of a single base pair were detected by ORBIT.

### 5.2.3 Proteins and other biomolecules

#### 5.2.3.1 Spatial organization

Biological systems reveal that multiple types of biomolecules can co-organize into complex assemblies with diverse functions. For instance, lipids and proteins assemble into cell membranes. One of the important goals of DNA nanotechnology is to emulate biological systems using synthetic hybrid nanostructures. The scope of structural DNA nanotechnology can be further expanded by adopting different types of biomolecules as building blocks. Proteins play crucial roles in cells, including molecular recognition, catalytic activities, and transportation. In addition, proteins are more structurally diverse than DNA. Integrating proteins with DNA nanotechnology can provide functional protein nanoassemblies with prescribed geometries. Although early studies demonstrated the attachment of proteins and enzymes to DNA nanostructures for various applications, the utilization of proteins as building blocks for protein-DNA hybrid nanostructures has just begun in recent years.

Significant progress has been achieved toward creating hybrid biomaterials containing peptides/proteins and nucleic acids. Depending on the roles of proteins in constructing hybrid nanostructures, they may get involved in the hybrid structures in several ways. The most straightforward example is a protein or peptide that connects different DNA nanostructures as a linker. The Ke group demonstrated the co-assembly of DNA origami and collagen-mimetic peptides into hybrid peptide-DNA nanostructures (Figure 20A).[316] The authors found two peptides $CP^{++}$ and $sCP^{++}$ (consisting of a central neutral block and two positively charged domains at C-and N-termini), and two-layer DNA origami could assemble into 1D nanowires. The positively charged peptides on both ends could connect negatively charged DNA nanostructures into ordered 1D nanowires. The peptides and DNA nanostructures were linked through face-to-face stacking or edge-to-edge stacking by adjusting the dimension of the DNA nanostructures. Similarly, the Turberfield group used peptide-oligonucleotide as a bridge to connect two DNA origami into hybrid structures (Figure 20B).[317]

Quyang *et al*. reported 2D and 3D DNA origami with cavities to immobilize antibodies.[318] NTA-

modified ssDNA strands attached to the cavities of the origami were used to form the NTA-$Ni^{2+}$ complex with histidine clusters containing polyclonal mouse immunoglobulin G (IgG) antibodies. To enhance the stability of the antibody-origami constructs, the authors achieved a covalent linkage between origami cavities and lysine residues of antibodies by using amine-containing origami and bifunctional bis(sulfosuccinimidyl) suberate (BS3) cross-linking molecules. This work provided a method for precisely docking antibodies into rationally designed origami cavities, allowing for the fine-tuning of the distances among multiple macromolecules. Knappe *et al*. constructed virus-like origami particles using an icosahedral DNA framework that provided predefined sites for *in situ* covalent functionalization.[319] DNA origami structures were folded with scaffold strands and dibenzocyclooctyne amine (DBCO)-containing staple strands, displaying specific locations for the subsequent covalent conjugation *via* click chemistry. After the reaction with azide-modified moieties (including small molecules, carbohydrates, peptides, polymers, and proteins), the functionalized DNA origami particles were then purified by HPLC to remove the redundant nanomaterials. This facile approach for the covalent modification of DNA nanostructures post-assembly can be used as a tool for versatile biomedical applications, in which robust functionalization of DNA nanostructures is needed.

DNA origami also provides a unique platform to investigate macromolecular interactions and special nanopatterns that are crucial for biophysical and biochemical processes. Shaw *et al*. introduced two types of DNA nanostructures, an 18-helix rod and a 44-helix brick, as drawboards for the arrangement of the staple-ssDNA tagged-antigen molecules (digoxigenin, DIG; 4-hydroxy-3-iodo-5-Nitrophenylacetate, NIP; or 4-hydroxy-3-nitrophenyl, NP).[320] Through oligonucleotide hybridization, these origami-aided, precisely controlled antigen patterns were immobilized on a surface plasmon resonance (SPR) chip, followed by interactions with corresponding antibodies (Figure 20C). The antibody binding affinities varied with the antigen distances, ranging from approximately 3 nm to 17 nm. To investigate antibody-antigen interactions at a biomimetic interface, Zhang *et al*. demonstrated triangular DNA origami structures with precisely anchored artificial epitopes to capture IgG molecules.[321] Six pairs of digoxin molecules as model epitopes were anchored on the triangular origami with nanometer precision for antibody binding (Figure 20D). The dynamic process of IgG binding to epitopes-displaying origami was studied by high-speed atomic force microscopy (HS-AFM) at the single-molecule level.

Importantly, DNA nanostructures can work as templates for protein self-assembly. Inspired by the tobacco mosaic virus, where coat proteins self-assemble on an RNA template, the Wang group demonstrated the *in-situ* assembly of tobacco mosaic virus (TMV) coat proteins on DNA origami structures (Figure 20E).[322] The coat proteins were selectively assembled on different locations of the DNA origami using the TMV genome-mimicking RNA strands. Both the length of the coat protein and the number of anchored locations could be stoichiometrically controlled by the RNA strand. They further showed that the anchoring of the coat proteins on the DNA origami could be dynamically controlled.[323] The RNA was arranged with a series of path points on the DNA origami by hybridization with ssDNA protruding from the origami. The RNA sequences, which hybridized with the captured DNA, could not direct the assembly of the TMV proteins on the origami. By selectively releasing the RNA from the captured DNA *via* toehold-mediated strand displacement reactions, the assembly pathway of the coat proteins could be dynamically controlled on the triangular DNA origami. The *in-situ* assembly of the TMV protein inside a confined DNA origami barrel was challenging because the RNA strand was likely to stay out of the barrel during the protein assembly. Nevertheless, the dynamic assembly of the TMV protein into the barrel space was guided by arranging the RNA path onto the inner surface of the DNA barrel. Furthermore, DNA nanostructures can direct the assembly of proteins into ordered geometries. The Gang group reported that functional proteins could be assembled into ordered 2D and 3D protein arrays with the guidance of DNA nanostructures. They used polyhedral DNA frames to encapsulate protein ferritin and apoferritin and assembled them into 2D and 3D lattices (Figure 20F).[324]

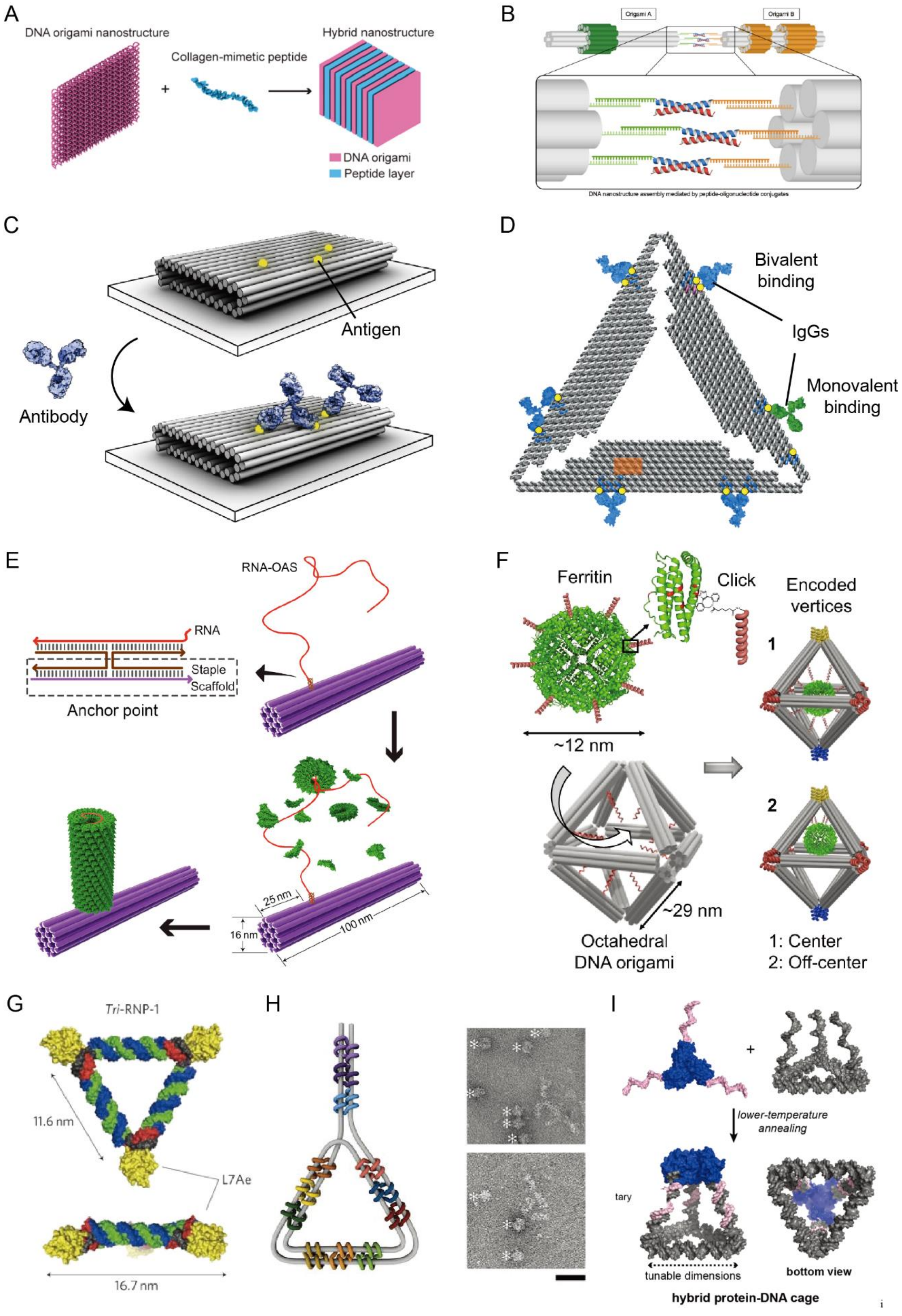
A
DNA origami nanostructure
Collagen-mimetic peptide
Hybrid nanostructure
DNA origami
Peptide layer
B
Origami A
Origami B
DNA nanostructure assembly mediated by peptide-oligonucleotide conjugates
C
Antigen
Antibody
D
Bivalent binding
IgGs
Monovalent binding
E
RNA-OAS
RNA
Staple
Scaffold
Anchor point
25 nm
16 nm
100 nm
F
Ferritin
Click
Encoded vertices
1
~12 nm
2
~29 nm
Octahedral DNA origami
1: Center
2: Off-center
G
Tri-RNP-1
11.6 nm
L7Ae
16.7 nm
H
I
lower-temperature annealing
tary
tunable dimensions
bottom view
hybrid protein-DNA cage

**Figure 20.** Spatial organization of protein on DNA origami templates. (A) Co-assembly of DNA origami and collagen-mimetic peptides into the ordered nanowire. Reproduced with permission from ref 316. Copyright 2017 American Chemical Society. (B) Formation of DNA origami dimer using peptide as the linker. Reproduced with permission from ref 317. Copyright 2019 American Chemical Society. (C) Antigen patterns guided by DNA origami for the study of antigen-antibody interactions. Reproduced with permission from ref 320. Copyright 2019 Nature Publishing Group. (D) Artificial epitopes precisely anchored on triangular DNA origami templates to capture IgG molecules. Reproduced with permission from ref 321. Copyright 2020 Nature Publishing Group. (E) DNA origami as a foundation for RNA-directed protein assembly. Reproduced with permission from ref 322. Copyright 2018 American Chemical Society. (F) DNA origami octahedron-guided protein lattice assembly. Reproduced with permission from ref 324. Copyright 2021 Macmillan Publishers Ltd. (G) Protein L7Ae induced bending of an RNA double strand into a triangle. Reproduced with permission from ref 325. Copyright 2011 Macmillan Publishers Ltd. (H) Self-assembly of genetically expressed DNA-protein hybrid nanostructures. Scale bars: 50 nm. Reproduced with permission from ref 327. Copyright 2017 AAAS. (I) A DNA cage from self-assembly of DNA and protein. Reproduced with permission from ref 328. Copyright 2019 American Chemical Society.

Proteins can also direct the assembly of hybrid nanostructures. Some DNA/RNA binding proteins can induce the bending of DNA/RNA. These proteins can thus be utilized to fold nucleic acids into desired shapes. For example, Saito and co-workers used the ribosomal protein L7Ae to fold the RNA duplex into a triangular shape (Figure 20G).[325-326] This was achieved by the fact that L7Ae could cause the K-turn RNA motif to bend ~60˚. This work established the foundation for using RNA binding proteins to fold hybrid structures. The Dietz group further generalized this strategy and invented the DNA-protein origami method.[327] Inspired by the DNA origami design, where long ssDNA acts as a scaffold and short DNA strands act as staples, the authors used dsDNA as a scaffold and transcription activator-like (TAL) effectors as staples. Each staple protein consisted of two DNA binding domains that bound to dsDNA with specific sequences. A linker connected the two DNA binding parts and the staple protein formed an antiparallel crossing after binding two parallel double DNA helices, mimicking the antiparallel crossovers used in DNA nanotechnology. They used 12 TAL proteins to fold into 18 distinct nanostructures, demonstrating the generality of this strategy for the self-assembly of DNA-protein hybrid structures into arbitrary shapes. The assembly of the DNA-protein origami was also demonstrated in a cell-free environment with genetic components (Figure 20H), suggesting that the protein-DNA origami could be produced and self-assembled *in vivo*. The hybrid DNA-protein origami method is a significant achievement, as it offers an important method for creating functional nanostructures compatible with biological systems. In a different study, Stephanopoulos and co-workers discovered that proteins could direct the hybrid nanostructures through the oriented valency.[328] They co-assembled a DNA-modified homotrimeric protein and a triangular DNA base with complementary handles into a hybrid protein-DNA tetrahedron cage (Figure 20I). The spatially oriented modified DNA on the homotrimeric protein directed the binding orientation and valency with the triangular DNA base.

#### 5.2.3.2 Regulation of cell functions

Comberlato *et al*. constructed DNA origami-based nanomaterials for cellular-level immune stimulation, decorated with CpG ODNs, a type of ligand for Toll-like receptor 9 (TLR 9).[329] On the addressable surface of the disk-like DNA origami, CpG motifs with different lengths of linkers were attached to form ligand patterns. The authors created a series of DNA structures with a controlled inter-ligand spacing of immunostimulatory CpG sequences and studied their effects in marine macrophages, Raw 264.7. In their results, the lengths of the linkers could influence the spatial binding tolerance between ligands and receptors. The increased activation of immune signaling was observed when CpG ligands were pinned at a distance of 7nm, which matched the dimer structure of TLR 9.

Based on antigen organization in a controlled manner, Veneziano *et al*. designed an origami platform to study B-cell activation.[330] They first created two types of DNA origami structures, a rigid DNA nanorod with a length of 80 nm and an icosahedral DNA framework with a diameter of 40 nm, to organize discrete antigen molecules. To trigger the B-cell response, they used the HIV-1 envelope glycoprotein antigen gp120 (termed eOD-GT8, ssDNA-modified) as a model immunogen, which was site-specifically attached to the DNA origami with ssDNA overhangs *via* hybridization (Figure 21A). The impact of the structural parameters, including the antigen copy number, inter-antigen separation, geometry and rigidity of the DNA origami, on B-cell activation was investigated.

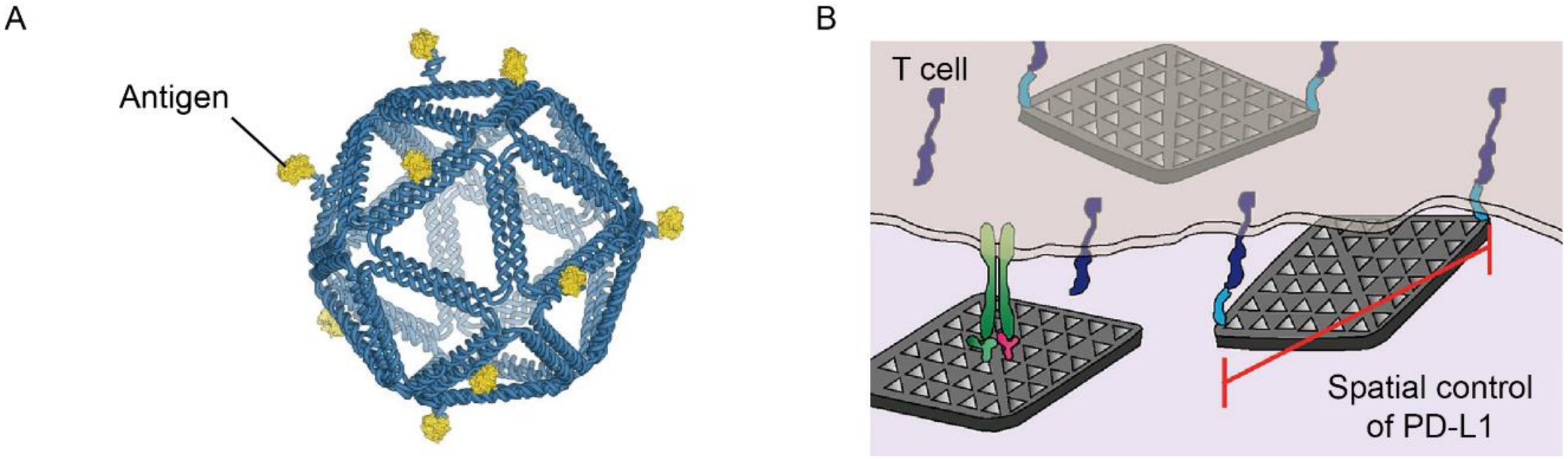

**Figure 21.** Protein decoration on DNA origami templates and regulation of biological functions. (A) Antigen displaying origami nanoplatform for B-cell activation. Reproduced with permission from ref 330. Copyright 2020 Nature Publishing Group. (B) PD-L1 decorated DNA origami sheet structures for T-cell signaling. Reproduced with permission from ref 331. Copyright 2021 American Chemical Society.

Only five copies of eOD-GT8 presented on the surface of the DNA framework triggered maximum B-cell response, while increasing the valency of antigen copies did not further enhance the cellular response. The inter-antigen spacing and rigidity of the origami were proved to be essential for the B-cell receptor activation. 25~30 nm of antigen spacing and rigid templates for antigen presentation elicited robust B-cell triggering. Fang *et al*. reported a similar design to using DNA origami wireframe sheets for facilitating the spatial control of programmed death-ligand 1 (PD-L1) molecules and studying their effects on T-cell signaling.[331] In their work, the origami-modified CD3/CD28 antibodies could trigger T-cell activation, while co-treatment with PD-L1-origami, on which protein ligands were separated by 200 nm, suppressed the T-cell signaling (Figure 21B). The authors also found that a single PD-L1 protein on the DNA origami or two ligands with an inter-distance of 13 nm or 40 nm could not inhibit the T-cell signaling mediated by CD3/CD28 antibodies. Their findings provided insights into the molecular interactions of significant immune responses and the rational design of therapeutic platforms using DNA origami. Furthermore, Hellmeier *et al*. utilized DNA origami to precisely organize macromolecules and studied the intracellular signaling triggered by the engineered extracellular ligand-origami constructs.[332] Using T cell antigen receptor (TCR)-peptide/MHC (pMHC) as a model ligand, the authors designed and implemented an origami interface that allowed spatial protein arrangement. The control over ligand distances with nanometer precision was achieved without interfering with the subsequent biological functions. By using pMHC-origami nanostructures, their work proved that the minimum unit consisting of two TCRs within a distance of 20 nm was sufficient for the T cell activation. In addition, Sun *et al*. presented several pMHC assemblies based on 2D DNA origami to regulate T cell functions.[333] The 2D triangular DNA origami with biotin sites was used to recruit streptavidin (SA) molecules and then biotinylated pMHC ligands for generating 2D pMHC multimers. The binding avidity of these assemblies with antigen-specific TCR displayed on the surface of T cells was regulated by different inter-pMHC spacing and ligand stoichiometries. Their results suggested that decreasing the nanospacing between two ligands and increasing the number of ligands on the origami could enhance the interaction of the pMHC multimers with the TCRs. The improvement of the binding avidity was then applied to animal models. The pMHC multimers exhibited greater capability to detect antigen-specific T cells with lower expression of TCRs, which were difficult to detect by equivalent tetramers.

## 6. DNA origami-based drug delivery and therapy

Delivery of functional macromolecules, such as proteins, peptides, and nucleic acids, to their targeted working environments, is crucial for treating diseases. DNA origami itself can be directly used as a functional agent for disease therapy. It can also work as intelligent carriers that integrate multiple biofunctional components, including small molecular drugs, NPs, peptides, proteins, nucleic acids, and many others, for efficient drug delivery, diagnosis, and therapy. For instance, Jiang *et al*. employed DNA origami as a therapeutic agent to alleviate acute kidney injury (AKI).[334] The authors prepared isotope-labeled DNA origami nanostructures (DONs) with Cu-64 to investigate their biodistribution and metabolism in mice. The positron emission tomography (PET) imaging results revealed that DONs accumulated in the kidneys of healthy mice or rhabdomyolysis-induced AKI mice. Due to their reactive oxygen species (ROS) scavenging effects, the DON treatment showed preferable renal-protective properties in AKI animals with cell damage caused by local ROS. Furthermore, Ma *et al*. developed a folic acid (FA)-decorated DNA origami as nanomedicine for rheumatoid arthritis (RA) therapy.[335] Taking advantage of the efficient ROS- and nitric oxide (NO)-scavenging capability of DNA, the DNA nanostructures worked as antioxidant agents for inflammations. FA molecules in the assemblies were used as targeting ligands for pro-inflammatory M1 macrophages, which predominantly promoted RA progression, including monocyte recruitment, fibroblast proliferation, and pro-inflammatory cytokine secretion. After treating FA-decorated DNA origami structures in mice with RA joints, ROS and NO in M1 macrophages were efficiently scavenged, facilitating the M1-to-M2 transition of macrophages. The effective alleviation of inflammatory damages and attenuating progression of RA were observed after the antioxidant therapy.

Also, Sigl *et al*. designed virus-sized DNA icosahedral origami shell structures to trap entire viral particles and inhibited the viral infection of host cells.[336] Based on the geometric principles of viral capsids, the authors constructed various DNA shells decorated with DNA-tagged antibodies to enhance specific virus covering. By using different binding moieties within the DNA shells, hepatitis B virus (HBV) core particles and adeno-associated virus serotype 2 (AAV2) were able to be engulfed and captured. It was demonstrated that the DNA shells could effectively inhibit HBV interaction in an *in vitro* blocking assay and neutralize the AAV2 virus in the cell cultural environment. In addition to treating viral infections, this study provided an economical and effective DNA-based nanoplatforms to deliver drugs, genes, or immunostimulatory moieties for future clinical applications. Apart from the static structures, the same group also constructed switchable DNA shells that were stabilized by IgG molecular staples and triggered to disassemble in the presence of corresponding antigens (Figure 22A).[337] The multi-layer icosahedral origami shell was assembled using 20 identical triangular origami units *via* shape-complementary stacking, which was stabilized by high magnesium concentrations (25 mM $MgCl_2$). Neighboring triangular units with antigen pairs on their triangle-triangle edges were used as bivalent binding sites to form

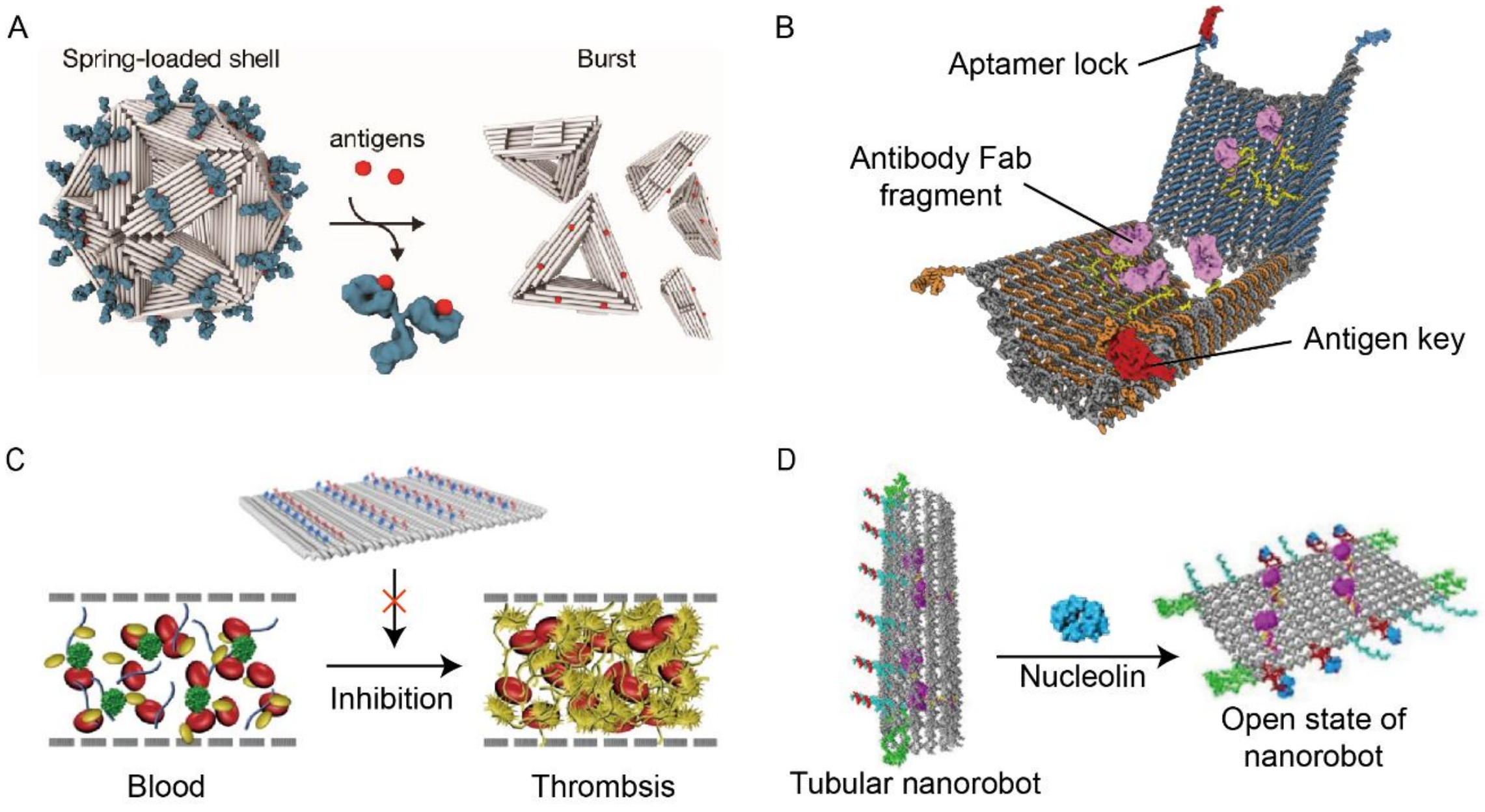

**Figure 22.** Drug delivery and therapeutic application based on DNA origami. (A) IgG-mediated stabilization and antigen-triggered disassembly of icosahedral DNA origami shells. Reproduced with permission from ref 337. Copyright 2021 American Chemical Society. (B) A barrel-shaped DNA origami nanorobot for precise delivery of antibody fragments to cell surface. Reproduced with permission from ref 183. Copyright 2012 AAAS. (C) A nanosheet displaying thrombin-binding aptamers for controllable anticoagulation. Reproduced with permission from ref 344. Copyright 2021 Nature Publishing Group. (D) A thrombin-containing DNA nanorobot for selective occlusion of tumor vessels. Reproduced with permission from ref 341. Copyright 2018 Nature Publishing Group.

IgG antibody-antigens bridges. After incubation with soluble antigen molecules, the IgG staples were replaced from the icosahedral shell, and the structural reconfiguration was triggered. This work elucidated a sound antigen-sensing strategy to use DNA origami as a novel nanocarrier for stimuli-responsive cargo release. Furthermore, Douglas *et al*. demonstrated a barrel-shaped DNA nanorobot that facilitated the payload delivery to the surfaces of target cells.[183] The origami-based nanorobots were utilized to load ssDNA-tagged NPs or antibody fragments *via* hybridization with the complementary strands extended from their inner cavities. Aptamer-containing logic gates were applied to seal the robotic origami structures, which could respond to corresponding receptors expressed on the target cells, allowing the specific origami reconfiguration for functional payload exposure (Figure 22B). By controlling the interactions between the payloads and specific receptors on the cell surface, the nanorobots were employed to manipulate the phosphorylation signaling of target cells. Arnon *et al*. utilized DNA origami robots in living insects (*Blaberus discoidalis*).[338] Their work revealed that artificial DNA nanodevices could perform complicated tasks in biological environments, even in living organisms.

The Ding group reported a series of origami-based drug delivery vehicles for chemotherapeutic agents and biofunctional macromolecules and tested their therapeutic effects *in vitro* and *in vivo*.[339-344] In one representative work, DNA origami was used as a cytotoxic protein vehicle for cancer therapy.[344] Rectangular DNA origami sheets were constructed to decorate ssDNA-tagged ribonuclease (RNase) A molecules *via* hybridization. MUC-1 aptamer strands that targeted mucin 1 (glycoproteins overexpressed on the surface of MCF 7 cells) were attached to the edges of the origami sheets to enhance the breast cancer cell internalization of the nanocarriers. In contrast to free cytotoxic proteins, origami-loaded RNase A elicited more efficient cancer cell-killing functions, which could be adapted to other protein drug delivery. They also presented a DNA origami-based, thrombin-binding aptamer nanoarray for a controllable anticoagulation study.[344] Two types of thrombin-recognizable aptamers that could bind different exosites of the protein were decorated on the surface of a rectangular DNA origami sheet with precisely controlled inter-distances. The potent protein binding affinity of the bivalent aptamer array was achieved when the inter-aptamer spacing was controlled to be 5.4 nm, which matched the dimension of the thrombin molecules (~ 4 nm). The aptamer arrays were then used in human plasma/whole blood and mice *in vivo*, which induced effective and reversible anticoagulation that could be neutralized by aptamer-complementary strands (Figure 22C). Even further, Li *et al*. developed a DNA origami-based, robotic nanocarrier to precisely deliver thrombin molecules to tumor vessels for selective occlusion therapy.[341] Thrombin molecules were modified with ssDNA strands *via* SMCC coupling and then attached to rectangular origami sheets with the complementary ssDNA strands through hybridization. DNA locking strands that responded to nucleolin proteins were integrated into the anisotropic thrombin-containing DNA origami templates, forming tubular structures to shield the inner protein cargos. After the intravenous injection to tumor-bearing mice, the robotic origami carriers recognized their targets selectively expressed on tumor-associated endothelial cells and were activated to an opening state to expose thrombin molecules (Figure 22D). The delivered thrombins induced thrombosis specifically in tumor-associated blood vessels, resulting in tumor necrosis and growth inhibition without observable toxicity and immunological stimulation.

In addition, Wang *et al*. developed a co-delivery origami vehicle for carrying small interfering RNA strands (siRNAs) and chemotherapeutic molecules to facilitate combined cancer therapy.[342] The siRNA sequences that targeted Bcl-2 and P-glycoprotein for disruption of tumor progression and multidrug resistance were anchored in the inner cavity of the tubular origami structure, while the doxorubicin molecules were intercalated into duplexes of the DNA carrier. Disulfide bond-containing strands were used as responsive locks for controlling the mechanical opening of the tubular carriers and the siRNA release in response to glutathione (GSH) inside the drug-resistant breast tumor cells. At cellular and animal levels, the multifunctional nanocarriers elicited potent RNA interference and breast tumor inhibition without observable systematic toxicity. The same group also presented a design of nanovaccine for cancer therapy based on DNA origami.[343] A reconfigurable tubular DNA origami structure was used to carry functional payloads, antigen peptides, and two types of nucleic acid adjuvants (double-stranded RNA and CpG DNA) in the inner cavity for triggering immune responses. Low pH-responsive DNA locking strands were utilized to seal the edges of the origami carriers and respond to the mild acidic environment inside the lysosomes of dendritic cells (DCs). After efficient accumulation in DCs of draining lymph nodes (dLNs), the origami nanocarriers were triggered to mechanically expose and release antigens and adjuvants, enabling T-cell activation and cancer cytotoxicity. At the animal level, the origami-based nanovaccine elicited potent tumor regression in melanoma and colon carcinoma models and generated long-term T-cell responses.

## 7. DNA origami-enabled nanoengineered membranes

Lipid membranes are essential components of biological cells. They are responsible for segregating biochemical reactions, transporting important biomolecules across compartments, as well as regulating various biochemical and biophysical signaling pathways. Membrane-associated processes are found throughout cells, including on plasma membranes, subcellular organelles, and intercellular connections.[345-346] In healthy tissues, cells maintain a delicate balance of membrane distributions and compositions to sustain homeostasis or recover after disease intrusion.[347] The importance of membrane physiology is further highlighted by diseases associated with their malfunction, including cancers, type 2 diabetes, Alzheimer's disease, and Charcot-Marie-Tooth disease.[348-349] Additionally, viruses and bacteria modify host membranes to enable infectious diseases, during which critical membrane-remodeling processes have been identified as potential therapeutic targets. In general, the remodeling of membrane bilayers is achieved by proteins that sense, stabilize, or modify membrane curvatures.[345-346, 350] These sophisticated pieces of protein machinery have thus intrigued biologists to understand their working mechanisms and inspired engineers to build artificial devices to mimic their functions. For both purposes, membrane materials with programmable and precisely controlled geometrical, biochemical, and mechanical properties are highly desirable.

DNA nanotechnology presents enormous opportunities to meet these needs by providing a synthetic framework to fabricate and manipulate membranes with nanometer-scale resolution. Several promising methods have been developed in recent years that fall under the following two general categories. The first one is to use DNA nanostructures as scaffolds or templates to guide lipid self-assembly and form bilayers of defined size and shape through membrane reconstitution. This bottom-up approach thus transduces the pre-defined geometry of the DNA templates into well-controlled membrane morphology. It is also compatible with reconstituting membrane proteins to form proteoliposomes with defined protein stoichiometry and organization. The second, top-down approach utilizes bilayer interacting DNA nanostructures to stabilize, sort, puncture, sculpt or otherwise manipulate membranes. One advantage of these techniques is that they allow the modification of specific lipid bilayer properties in a targeted manner, while preserving other existing membrane features. The two approaches complement each other, both enabling the precise engineering of membranes with customized and potentially anisotropic physicochemical properties. Furthermore, thanks to the self-assembling nature of DNA nanostructures, such DNA-engineered membrane materials can be made in large quantities and are amenable to most biochemical and microscopy analyses, paving the way to understanding the molecular mechanisms of membrane dynamics as well as to building artificial cells and drug delivery vehicles with user-defined functions.

At the foundation of these ambitious goals are the basic interactions between DNA and lipid membranes. Using model lipid systems such as micelles, supported bilayers, and liposomes, scientists found that DNA molecules electrostatically interact with lipid headgroups and that hydrophobic labels can help DNA bind to membranes more stably by inserting into the bilayer's hydrophobic core.[351-353] Many early studies that established the methods to associate DNA with membranes were motivated by the appeal of DNA-membrane materials as a promising gene delivery reagent[354-355] and by the programmability of DNA sequences to tether, aggregate and fuse membrane-bound compartments, including liposomes and cells.[356-360] In the past decade, the binding, diffusion, segregation, and self-assembly behaviors of complex DNA structures on membrane bilayers were systematically studied, with a renewed mission of creating programmable DNA-membrane hybrid materials by engineering the shape, dynamics, and chemical modification of the membrane-binding DNA nanostructures. A few milestones that showcase the unique advantages of the DNA-nanotechnology-enabled membrane engineering approach include the production of geometrically well-defined liposomes and nanodiscs on DNA templates, the controlled deformation of membranes by DNA devices, and a large variety of transmembrane DNA nanopores with customizable selectivity.

### 7.1 Guided membrane self-assembly

Monodisperse liposomes with programmable shapes and sizes are invaluable tools for studying curvature-dependent protein-membrane interactions and developing therapeutics. For instance, proteins in the endosomal sorting complex required for transport (ESCRT) family are involved in diverse cellular membrane remodeling events, including cytokinetic abscission, vesicular budding in the endosomal sorting pathway, and the release of viruses, where membranes assume complex and transient shapes.[361] To fully understand how such membrane-sculpting proteins recognize and act on various membrane structures, studying their behaviors on homogeneous liposomes with pre-defined local membrane curvatures is necessary. On the other hand, liposomal delivery systems require precise control of the vesicle size to function as safe and effective forms of medicine. Conventional physical and chemical methods to control membrane landscape can only produce homogenous liposomes restricted to a limited selection of shapes, small size ranges, and specific lipid compositions. Given that a unique advantage of DNA nanostructures is their customizable geometries, they can be utilized to template lipid bilayer assembly, producing liposomes with diverse yet controllable shapes and sizes on the nanometer scale.

A typical workflow of the DNA-templated liposome formation outlined in the work of Yang et al. is shown in Figure 23A.[362] DNA nanostructures were first designed and assembled with desired geometries. The unpaired ssDNA extensions (“handles”) were hybridized with complementary ssDNA (“anti-handles”) that was chemically conjugated with an amphiphile (e.g., lipid, amphipathic peptide, or transmembrane protein). The attachment of hydrophobic moieties occurred amongst excess free lipids and detergent to prevent aggregation. The detergent was then removed (e.g., *via* dialysis) to induce lipid bilayer formation, before gradient centrifugation was used to enrich the desired DNA-membrane complexes. TEM images taken during the membrane reconstitution revealed a series of intermediates, strongly suggesting that the DNA-conjugated amphiphiles ("seeds") served as nucleation points to recruit lipid-detergent micelles. They gradually expanded and merged into continuous bilayer membranes that finally formed a shape determined by the DNA template.

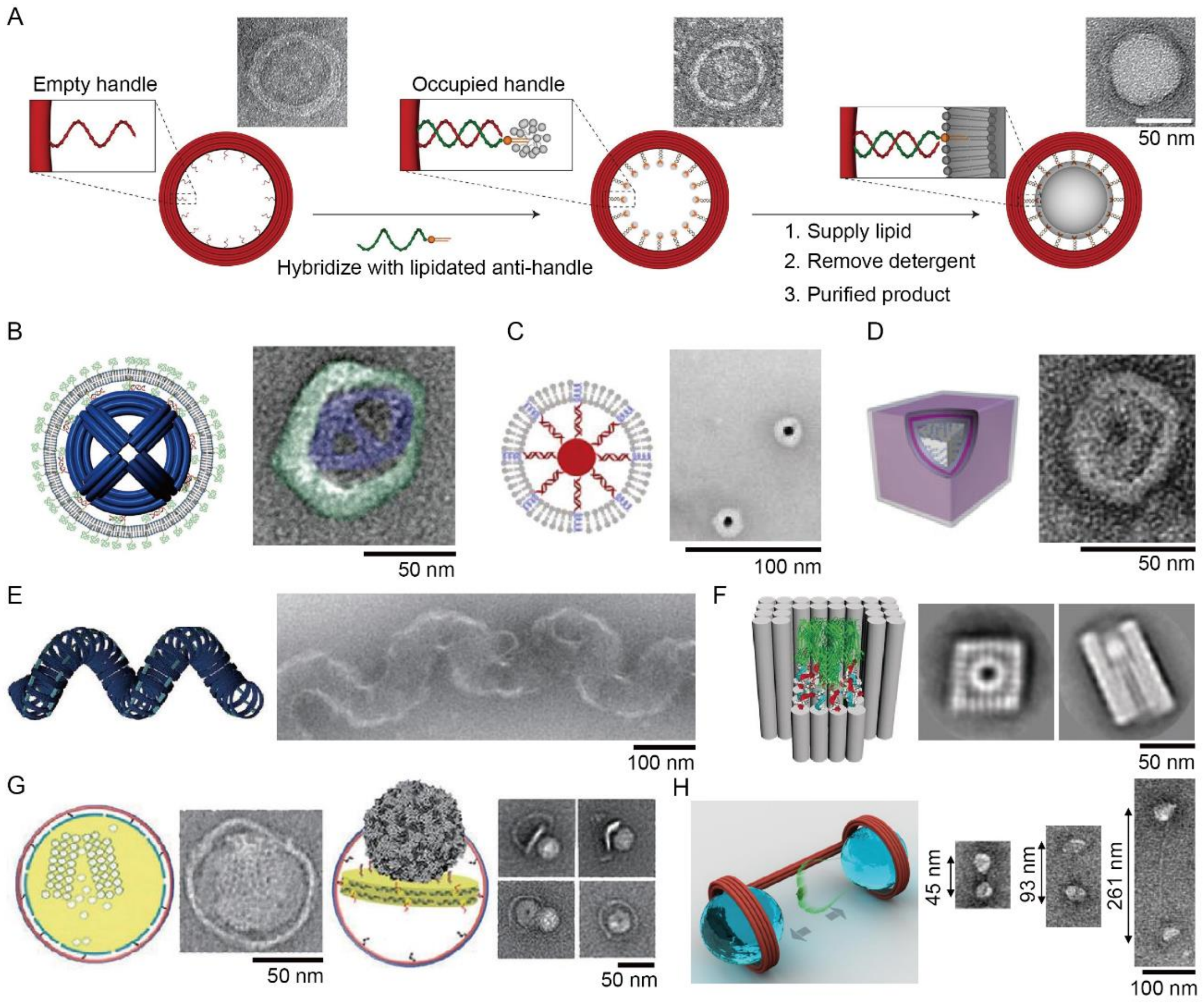

**Figure 23.** DNA-guided membrane structures. (A) Schematic of DNA-templated liposome assembly mechanism. Reproduced with permission from ref 362. Copyright 2016 Nature Publishing Group. (B) Liposome templated with a DNA origami endoskeleton. Reproduced with permission from ref 363. Copyright 2014 American Chemical Society. (C) Frame-guided assembly of liposomes around gold nanoparticles modified with pH-responsive transmembrane peptides. Reproduced with permission from ref 364. Copyright 2020 Wiley. (D) DNA origami cuboid frame decorated with hydrophobic groups guides the formation of a cuboidal vesicle. Reproduced with permission from ref 366. Copyright 2017 Wiley. (E) Helical membrane structure templated within a DNA origami cage. Reproduced with permission from ref 368. Copyright 2017 Nature Publishing Group. (F) α-hemolysin reconstituted into the lipid-modified cavity of a DNA origami barrel. Reproduced with permission from ref 372. Copyright 2022 American Chemical Society. (G) DNA origami templated nanodiscs capable of reconstituting hVDAC-1 protein clusters (left) and studying poliovirus viral particles (right). Reproduced with permission from ref 373. Copyright 2018 American Chemical Society. (H) DNA origami templated liposomes placed at various controlled distances to study the lipid transfer activity of E-Syt1. Reproduced with permission from ref 370. Copyright 2019 Nature Publishing Group.

Following the initial demonstration by Perrault et al. that a near-spherical DNA frame (53 nm in diameter) decorated by outward-facing phospholipids could guide the formation of lipid bilayers (Figure 23B),[363] there have been a number of reports in recent years to substantiate further the robustness and generalizability of the DNA-templated liposome formation approach. Examples include using DNA as an exoskeleton (outer template) or endoskeleton. Wrapping membranes around the DNA endoskeleton generates well-exposed lipid bilayers and protects the DNA core from nuclease. Wang et al. prepared AuNPs and DNA origami frames displaying hydrophobically labeled DNA molecules to guide the assembly of dendritic amphiphiles, small-molecule detergents, and phospholipids, which led to monodispersed hetero-vesicles of spherical, spheroidal, and cuboidal shapes (Figures 23C & 23D).[364-366] A study by Julin *et al*. showed that positively charged lipid membranes could assemble into multilamellar structures embedding DNA origami frames with a wide range of aspect ratios.[367] On the other hand, growing liposomes within the DNA exoskeletons seem to maximize the shape controllability of the DNA templates. It has been demonstrated that DNA origami rings carrying as few as a dozen inward-facing hydrophobic seeds could precisely define the size of spherical liposomes consisting of various lipids.[362] Non-spherical liposomes, such as membrane tubes, spirals, and tori of well-controlled dimensions, could form reliably within rationally designed DNA cages (Figure 23E).[368] Furthermore, exposed DNA templates allow further functionalization or manipulation of DNA-enclosed membranes. For example, the DNA exoskeletons were built with reconfigurable parts to change shapes dynamically in response to specific biochemical signals[369] and drove membrane remodeling, such as fusing individual liposomes or bending membrane tubes.[362] Similarly, individual liposomes or nanodiscs were brought together with defined proximity and spatial positioning *via* the oligomerization of their DNA templates.[370]

The concept of the DNA frame-guided lipid assembly can be applied to generate not only membrane-bound compartments (liposomes), but also lipid micelles and flat lipid bilayers (nanodiscs).[371] Dong et al. anchored a dozen phospholipids inside a 5-nm wide, 15-nm deep cavity on a DNA-origami barrel for the reconstitution of α-hemolysin, a pore-forming protein (Figure 23F).[372] Zhao *et al*. and Iric *et al*. independently developed similar approaches that used amphiphile-modified DNA circles to scaffold the formation of size-controlled nanodiscs up to 70 nm (Figure 23G).[373-374] Such nanodisc or nanodisc-like structures, together with the embedded proteins, were resolvable by electron microscopy. The DNA-encircled nanodiscs were also modeled computationally to understand their physical properties, such as bilayer thickness and stability.[375]

Despite being a relatively new technique, the DNA-templated lipid assembly has already shown great promise in drug delivery and understanding membrane protein structures and functions. For example, Perrault et al. showed that enclosing DNA origami structures within a lipid bilayer substantially reduced the immunogenicity of DNA structures and altered their *in vivo* distribution.[363] By controlling the copy number of vesicle-associated membrane protein-2 (VAMP2, a type of v-SNARE) on a DNA-ring-templated liposome, Xu et al. found that roughly 1–2 copies of SNARE complexes were sufficient to drive lipid mixing by vesicle fusion.[376] Bian et al. used a series of liposome pairs held at defined inter-membrane distances by stiff DNA origami rods to study the distance-dependent lipid transfer activities of extended synaptotagmin 1 (E-Syt1). This study confirmed that E-Syt1 could transport lipids well beyond the length of its lipid-transfer domain (SMP domain) and as far as its membrane tether allowed, thus supporting a model where the SMP domain of E-Syt1 repeatedly traveled between the membrane contact sites to ferry lipids across the apposing membranes (Figure 23H).[370] The potential of the DNA-templated lipid assemblies in structural biology studies was illustrated in two studies, where the DNA-lipid complexes were incorporated with transmembrane proteins like α-hemolysin[377] and VDAC-1 or a poliovirus recognizing a specific membrane receptor[373] and subjected to electron microscopy imaging.

With the help of DNA nanotechnology, one can start to recapitulate essential aspects of the cell membranes with high precision in a test tube. Progress to date has laid a solid technological foundation by generating membranes with programmable geometry and protein stoichiometry. However, current DNA-templated membrane formation methods require further developments to fully capture the complexity of biological membrane systems. For example, technologies that allow precise control over the mechanical properties of membranes (e.g., stiffness and tension), lipid bilayer asymmetry, membrane dynamics, and cross-membrane molecular transport would enable more applications in fundamental research and biotechnology. While some of these aspirations have been made possible by other DNA-nanotechnology-enabled membrane engineering techniques (reviewed in the following sections), this field will advance by developing DNA-membrane materials with increasing spatiotemporal control, possibly by integrating multiple high-precision membrane-manipulating tools.

### 7.2 Manipulation of pre-existing membranes

Functionalized DNA nanostructures (often with hydrophobic modifications) that are adhered to lipid bilayers can manipulate and modify the specific properties of membranes. Upon membrane association, such DNA nanostructures can change the membrane's rigidity, surface chemistry, buoyant density, shape, and permeability. The resulting DNA-membrane complexes may feature customized DNA surface patterns and membrane topology, opening up the opportunity to create anisotropic membrane surfaces, mimic naturally occurring membrane dynamics, and even build synthetic organelles and cells.

Just like single- or double-stranded DNA, complex DNA nanostructures can interact with membranes *via* the electrostatic interaction between the DNA phosphate backbone and the lipid headgroups, which can be modulated by ion strengths and further enhanced by labeling DNA with “membrane anchors” — hydrophobic moieties or ligands to membrane-embedded receptors.[378-382] Depending on the underlying lipid bilayer fluidity and tendency to phase separation, the membrane-attached DNA structures can diffuse in 2D and segregate into different micro/nano-sized domains,[383-385] which can promote the formation of designer DNA patterns from individual tiles.[70, 73-75, 386] The most compelling examples of surface-assisted DNA self-assembly were demonstrated by Avakyan et al. using the blunt-end mediated formation of cross-shape and hexagonal DNA lattices spanning micrometers on supported bilayers (Figure 24A).[387] It was also found that liposomes could be stabilized by DNA networks formed on their membranes. Kurokawa et al. showed that giant unilamellar vesicles (GUVs) with a thick layer of DNA network adhered to their inner surface were 50-fold more resistant to deformations by micropipette aspiration and could endure higher osmotic imbalance than liposomes without such DNA "cytoskeleton" (Figure 24B).[388] Similarly, Baumann et al. found that large unilamellar vesicles (LUVs) with an exterior coat of DNA network were less likely to burst on mica than naked liposomes and liposomes with membrane-anchored DNA duplexes, suggesting that the interconnected DNA tiles enhanced the mechanical stability of liposomes. AFM analysis revealed higher Young's moduli of liposomes coated with networks of stiffer DNA tiles, indicating that the membrane rigidity could be finetuned by changing the design of the membrane-associating DNA network.[389] Yang et al. also demonstrated the prolonged shelf-life of DNA-tile-coated small unilamellar vesicles (SUVs). Notably, the authors took advantage of the large buoyant density of the DNA to differentiate liposomes of different sizes — smaller liposomes with a higher surface-to-volume ratio gained more density from DNA coating than their larger counterparts, allowing their separation in a density gradient (Figure 24C). This method could be used to sort milligrams of liposomes into highly homogeneous populations with mean diameters from 30 to 130 nm, thereby enabling the studies of the curvature-dependent activity of ATG3/ATG7-catalyzed protein lipidation and SNARE-mediated membrane fusion with high resolution.[390]

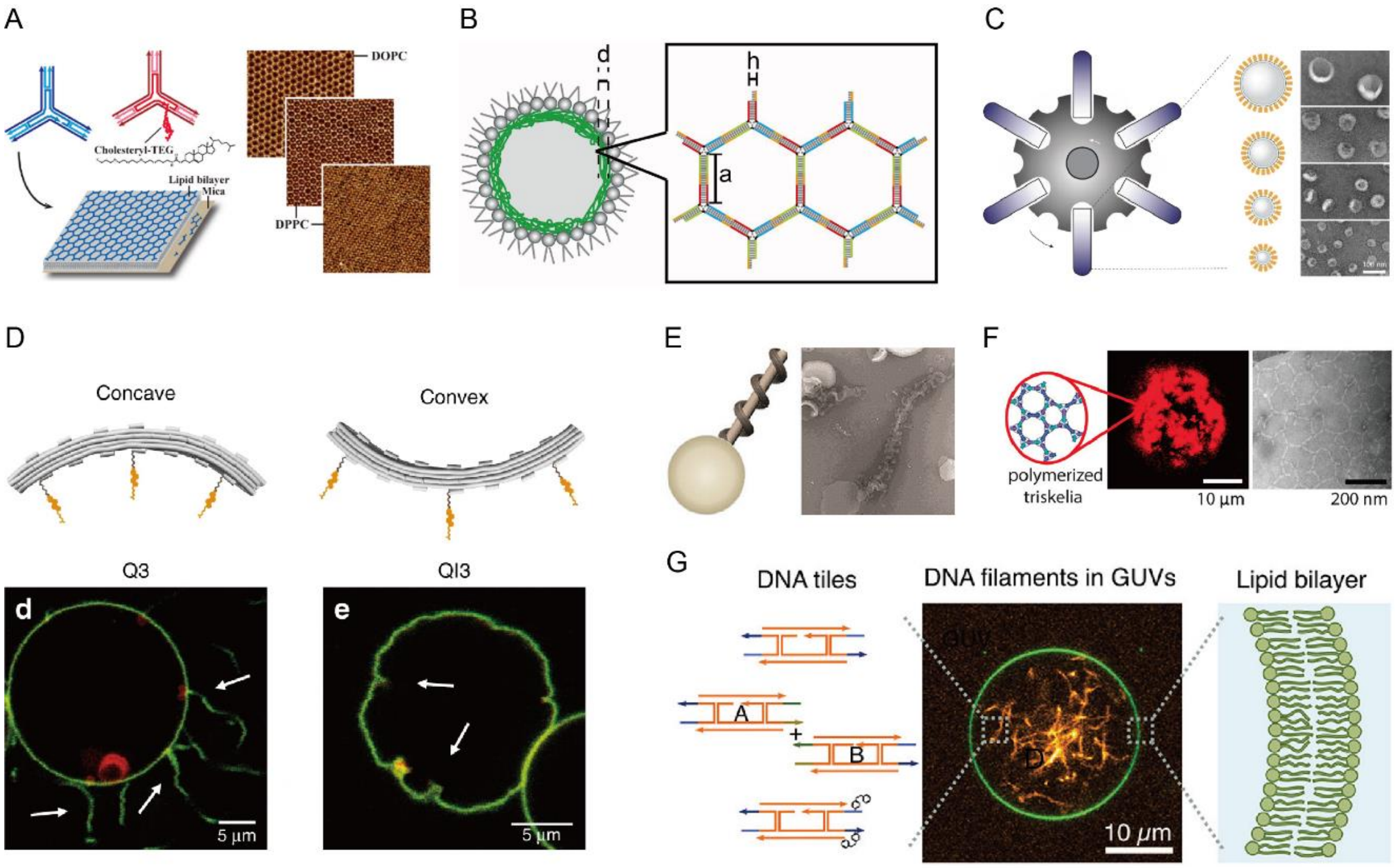

**Figure 24.** Membrane decoration and manipulation by DNA. (A) Lattice formation by DNA tiles on supported lipid bilayers. Reproduced with permission from ref 387. Copyright 2017 American Chemical Society. (B) Cytoskeleton mimic formed by a gel-like network of DNA tiles within a liposome. Reproduced with permission from ref 388. Copyright 2017 NAS. (C) DNA-brick-assisted liposome sorting generates highly homogeneous vesicles. Reproduced with permission from ref 390. Copyright 2021 Nature Publishing Group. (D) Membrane tubulations and invaginations on GUVs caused by the attachment of curved DNA origami beams. Reproduced with permission from ref 392. Copyright 2018 Nature Publishing Group. (E) Vesicle tubulation enabled by the self-assembling DNA nanosprings on membranes. Reproduced with permission from ref 393. Copyright 2018 Wiley. (F) Membrane buds generated by clathrin-like DNA triskelion dimers. Reproduced with permission from ref 395. Copyright 2019 American Chemical Society. (G) DNA tiles capable of forming cytoskeleton-like filaments (left) to deform GUVs from within after membrane attachment (right). Reproduced with permission from ref 396. Copyright 2022 American Chemical Society.

Another profound effect of covering lipid bilayers with DNA nanostructures is the dynamic deformation of membranes. In cells, peculiar membrane shapes such as buds and invaginations are primarily due to the work of proteins. Such membrane shapes are often transient and hard to reprogram. In contrast, DNA structures can be designed to mimic the membrane-sculpting proteins with programmable geometry, stiffness, membrane anchor placement, and self-assembling patterns. Analyzing how such design parameters affect their membrane-reshaping activities may thus help researchers to understand how membrane proteins have evolved to their specific functions.

While building DNA nanopores, Göpfrich et al. observed that adding large quantities of cholesterol-labeled DNA channels to GUVs generated outward-protruding tubules.[394] Birkholz et al. also observed the formation of lipid tubules parallel to the supported lipid bilayers treated with similar DNA channels.[391] A large body of work has focused on mimicking cell-membrane remodeling proteins, such as clathrin, BAR domain proteins, dynamin, ESCRT-III, and actin. For example, BAR-mimicking DNA beams with curved cholesterol-labeled surfaces induced the outward tubulation and invagination of GUVs in a manner consistent with the curvature of the cholesterol-modified DNA-origami surface, as demonstrated by Franquelim et al. (Figure 24D).[392] Similarly, Grome et al. used Snf7 (an ESCRT-III subunit) mimicking DNA curls that self-assemble into nanosprings on the membrane to draw lipid tubules from LUVs and GUVs (Figure 24E).[393] In follow-up work, the same group showed that the stiffness of the DNA nanosprings could modulate the tubulation activities of the DNA curls.[394] A common finding of the independent works carried out by the two groups was that the relatively stiff DNA structures could impose their curvatures on the membranes and stabilize the non-trivial membrane structures they generated. This discovery was also confirmed by the work from the Turberfield group, in which clathrin-like DNA triskelia formed a meshwork on vesicles and generated local membrane buds (Figure 24F).[395] Furthermore, using self-assembled DNA filaments with cholesterol tags, the Göpfrich group[396] and Schwille group[397] showed that GUVs could be deformed by cytoskeleton-like DNA structures bundling at the periphery from the inside (Figure 24G) and outside of the vesicle, respectively. Notably, the two groups controlled the assembly and disassembly of the DNA filaments by light or $Mg^{2+}$ concentration. DNA-based filaments capable of reversible assembly were also used to transport

SUVs and NPs, replicating yet another cytoskeletal function essential to the future development of synthetic cells.[398]

As promising as they are, the membrane-manipulating capabilities of the DNA-based tools are still rudimentary and sometimes cumbersome compared to their protein counterparts. The field awaits further development to build DNA devices that can generate diverse membrane topologies, respond to biochemical signals, and contain multiple coordinating components to accomplish complex tasks on the membrane. A few recent studies contributed to these goals. De Franceschi et al. developed a technique to reliably deform GUVs into stomatocyte or dumbbell-like shapes under hyperosmotic conditions using a cholesterol-labeled DNA 4-way junction.[399] Liu et al. reported prestressed DNA origami clamps, which could transform shapes in response to DNA triggers to tubulate GUV and LUV membranes.[400] Baumann *et al*. showed that the triggered contraction of a DNA network on liposome surface could lead to membrane deformation or accelerated escape of small-molecule cargos, although the mechanisms of triggered cargo release are not entirely clear.[401] Taken together, the capability of DNA nanostructures to manipulate, stabilize, and sort vesicles may find applications in therapeutics (e.g., formulating drug-delivery vehicles) and biosensing (e.g., screening for extracellular vesicles with specific biomarkers).

### 7.3 DNA nanopores

In addition to the vesicular transport pathways, transmembrane nanopores, including ion channels and transporters,[402-404] control the molecular exchange between cells and the environment or different cellular compartments. Artificial nanopores that mimic the functions of natural protein channels have shown great promise in synthesizing programmable nanoscale filters, sensing label-free biomolecules, and DNA sequencing.[405-407] As an information-rich nanomaterial, DNA can be programmed to form nanochannels of precisely controlled shapes and dimensions and be embedded into lipid bilayers through hydrophobic membrane anchors.[407-409] DNA-based nanopores also have abundant chemically addressable surfaces, which can be modified with functional and dynamic modules, such as receptor molecules and controllable switches (Figure 25A),[407-409] thus achieving customizable size and chemical selectivity. This section reviews recent progress in DNA nanopore design and application, focusing on those with conceptual novelty or advanced functionality, before offering our perspectives.

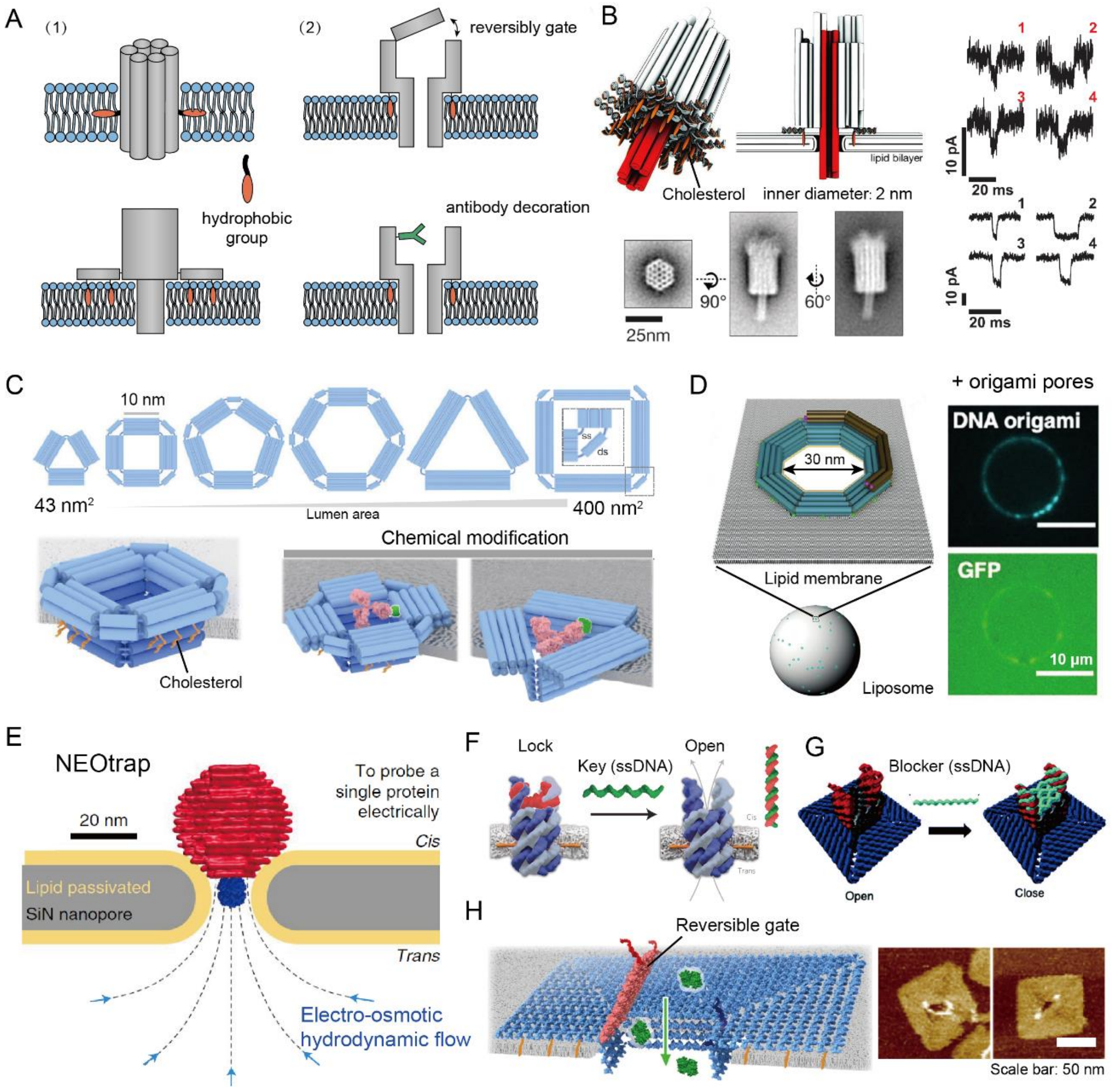

**Figure 25.** Transmembrane nanopore based on DNA origami techniques. (A) Schematics of transmembrane DNA nanopores. (1) A rod-like (top) or syringe-shaped (bottom) DNA-origami nanopore with hydrophobic anchors capable of puncturing the membrane. (2) DNA nanopores can be modified with functional modules, including controllable switches and receptor molecules. (B) A syringe-shaped DNA origami penetrates the membrane with a cholesterol-decorated DNA channel. Reproduced with permission from ref 410. Copyright 2012 AAAS. (C) A series of DNA-origami nanopores featuring a modular design and programmable shapes and sizes (up to tens of nanometers). Chemical modifications confer selectivity to DNA nanochannels for sensing applications. Reproduced with permission from ref 411. Copyright 2022 Nature Publishing Group. (D) Reconstitution of 30 nm wide DNA origami pores on liposomes for size-selective macromolecule transport. Reproduced with permission from ref 417. Copyright 2021 American Chemical Society. (E) A DNA-origami sphere-capped nanopore as an electro-osmotic trap for the label-free detection of single proteins and their conformations. Reproduced with permission from ref 416. Copyright 2021 Nature Publishing Group. (F) A DNA nanopore with a displaceable ssDNA lock controlling small molecule diffusion. Reproduced with permission from ref 426. Copyright 2016 Nature Publishing Group. (G) A 10×10 nm$^2$ DNA nanopore gated by four ssDNA blockers that control its size-selectivity. Reproduced with permission from ref 418. Copyright 2021 RSC Publishing. (H) A 20×20 nm$^2$ DNA nanopore built with a reversible gate that is dynamically controlled by external triggers. Reproduced with permission from ref 412. Copyright 2022 Nature Publishing Group.

Following the pioneering work by Langecker et al., in which a syringe-shaped DNA-origami nanopore allowed the passage of ions and nucleic acids across the membranes (Figure 25B),[410] a great variety of transmembrane DNA nanopores have been built over the years. At a minimum, a DNA nanopore consists of a channel-like architecture made of DNA and several hydrophobic membrane anchors, such as cholesterol,[410-412] porphyrin,[413-414] and alkyl chains.[415] Optionally, the DNA nanopores can be equipped with other structural components to accommodate additional membrane anchors, gating molecules, and open/close mechanisms. The insertion of DNA nanopores into membranes is mainly facilitated by the membrane anchors, although voltage pulses and mild detergents are often used to aid the membrane penetration. The key function of nanopores — the molecular translocation through them — is typically characterized by single-channel current recordings[411, 416] and fluorescence microscopy, such as total internal reflection fluorescence (TIRF) and fluorescence recovery after photobleaching (FRAP).[417-420] The effect of design and experimental conditions on the performance of nanopores have been extensively studied computationally[421-423] and experimentally.[193, 422-425]

Although early proof-of-concept studies mostly used DNA nanopores with channel widths below 3 nm,[391, 410, 413-415, 421, 426] in recent years, the field has made significant progress toward building wider nanopores with inner diameters up to 30 nm.[411-412, 416, 418, 420, 423] The main challenges to building a large nanopore are to modify the DNA channel with a sufficient number of membrane anchors (without inducing severe aggregation) to provide the energy for penetration and to build channels stiff enough to withstand lateral pressure of the membrane. The DNA origami technique offers solutions to both challenges. For example, a T-shape DNA channel can display numerous cholesterol anchors under the cap that provides steric hindrance to mitigate channel aggregation, and a multi-layer DNA origami design can provide the necessary rigidity. An illustrative example is a series of DNA origami nanopores built by the Howorka group with polygonal cross-sections and lumen areas ranging from 43 to 400 nm$^2$ (Figure 25C).[411] A similar design by the Kjems group also generated nanopores with an inner diameter of ~9 nm.[420] Different from the conventional methods that puncture preformed lipid bilayers with DNA nanopores, Fragasso et al. developed a reconstitution-based method, through which the authors embedded a 30-nm wide (the largest so far) DNA octagonal channel with dense cholesterol labels and a PEG-oligolysine coat into GUV membranes (Figure 25D).[417] Another promising method to open transmembrane nanopores with programmable sizes is to organize pore-forming peptides or proteins with a DNA scaffold.[427-429] Henning-Knechtel et al. organized a defined maximum number (7, 12, 20, or 26) of α-hemolysin with a circular DNA scaffold and thus controlled the size and conductivity of the DNA-hemolysin hybrid pores.[428] Of note, the applicability of this technique seemed to depend on the properties of the protein/peptide. For example, Spruijt et al. built a ring-shaped DNA structure that brought together up to 12 copies of Wza-derived peptide, but this construct only opened a stable octameric peptide pore.[427] In contrast, when Fennouri et al. assembled 4, 8, or 12 copies of peptide ceratotoxin A (CtxA) to a DNA scaffold, they obtained CtxA nanopores with corresponding diameters ranging from ~0.5 to 4 nm.[429] Taking advantage of pneumolysin, a potent cholesterol-dependent toxin, Shen *et al*. built DNA-origami-protein nanopores over 20 nm in diameter, which could be modified with nucleoporins to achieve programmable size-selectivity.[430] In these protein-DNA nanopores, although the DNA backbones did not have direct contact with the membrane, they played important roles to stabilize the protein pores, modulate their size, or provide attachment points to further functionalization.[428-430]

The programmable width of DNA nanopores directly dictates their permeability to molecules of different sizes. For example, Fragasso et al. monitored the diffusion of fluorescently labeled dextran molecules of various molecular weights through a 30-nm wide DNA nanopore and found that the dextran's diffusion rate was inversely correlated with their molecular weight.[417] Also, they found that within the timeframe of their measurement, the nanopore was only permeable to dextrans up to the size of its lumen. Likewise, Thomsen et al. found that a 9 nm wide DNA nanopore allowed the translocation of 40 kDa dextran but severely impeded 500 kDa dextran. However, the same nanopore with its lumen plugged by up to ten PEG molecules was impermeable even to 20 kDa dextran.[420] Therefore, the DNA nanopore's sensitivity to the transporting molecules' geometrical properties could be used for tag-free detection of biomolecules and their conformations. As an excellent example, Schmid et al. built a DNA-origami sphere-capped nanopore as an electro-osmotic trap (NEOtrap) (Figure 25E).[416] The negatively charged DNA sphere coupled to the membrane-coated nanopore induced an electro-osmotic hydrodynamic flow, which drove proteins towards the nanopore irrespective of their net charge and trapped them in the nanocavity for up to hours. The different electrical current blockades by proteins of different sizes (54–340 kDa) and by the same protein of different structural states allowed the researchers to identify the proteins and their conformations, opening up new avenues to sensing macromolecules and studying protein dynamics at the single-molecule level.

The size selectivity of the DNA nanopores can also be dynamically controlled by external triggers if they are built with an ON-OFF switch or plug. Burns et al. showed that diffusion of sulpho-rhodamine B through a ~2 nm wide DNA channel could be blocked by an ssDNA lock, which was displaced by a key DNA strand to reopen the channel (Figure 25F).[426] A similar strategy was used by Iwabuchi et al. to close a 10 nm wide DNA nanopore in a stepwise manner (Figure 25G).[418] More recently, Dey et al. built a nanopore with a 20×20 $nm^2$ lid on top of its transmembrane channel (Figure 25H).[412] Remarkably, the lid could be closed and opened for multiple rounds by sequentially adding a key and a reverse key DNA strand, which controlled the timing of molecular entry into or release from GUVs. The reversible opening and closure of nanochannel can also be achieved by light-induced DNA conformational changes, as shown by Offenbartl-Stiegert et al.[431] In another recent study, Lanphere et al. triggered the assembly of two halves of a DNA channel attached to a membrane into a complete, functional nanopore through toehold mediated strand displacement reactions,[432] providing an alternative approach for timed perforation of lipid bilayer membranes. Lastly, Mills et al. used a spring-loaded DNA device, termed "nano-winch", to mechanically unfold the plug domain of an integral membrane protein, BtuB, showing the capability of DNA devices to controllably open naturally existing protein nanopores.[433]

The negatively charged DNA nanopores are chemically selective to the translocating molecules, a property that experiments and computer simulations have supported.[426, 434] In addition, chemical modification can confer selectivity to DNA nanochannels through specific ligand-receptor interactions. For example, Xing et al. attached a biotin tag or a SARS-CoV-2 spike protein to the entrance of the 10 nm DNA nanopores, giving the nanopores the capability to detect the binding of cognate IgG antibodies (~10 nm in size) by electrical current recordings (Figure 25C).[411] In theory, a large variety of gating molecules can be grafted onto the transmembrane DNA nanopores to achieve selective permeability. This perspective is especially attractive because channel-like DNA origami structures have been modified with selective nucleoporins to reconstitute nuclear pore mimics[435-437] and with protein-binding aptamers to serve as carriers to detect protein analytes passing through a glass nanopipette.[438] Excitingly, recent studies by Shi et al. showed that DNA devices with chiral contours (like fan-blade) can harness electrochemical energy to power sustained directional rotations.[439-440] Incorporating such structures into lipid membranes would give rise to highly versatile biosensors, multi-functional drug delivery vehicles, and artificial cell systems. These goals are well within reach, considering the pressing need for programmable molecular filters in basic research and biotechnology, the availability of computational and chemical methods for DNA nanopore design and fabrication, and the advanced analytical tools generated by the fast-evolving electronics industry.

## 8. DNA-origami-enabled nanophotonics

The DNA origami technique provides a revolutionizing approach for nanofabrication based on molecular self-assembly to organize molecules and NPs in precise and well-defined 3D conformations.[441-442] Specifically, this technique offers unique platforms to build complex, hierarchical, and hybrid nanophotonic devices for the fundamental understanding of light-matter interaction processes as well as for the realization of tailored optical properties and functions.[443] Among a variety of optical elements, metallic NPs, and single quantum emitters are particularly attractive because they can be accurately organized on DNA origami with precise position and stoichiometric control on the nanometer scale.

Nobel metallic NPs, such as AuNPs and AgNPs, can support localized surface plasmon resonances (LSPRs) or particle plasmon resonances, which result from the coherent oscillations of the conduction electrons in the NPs, when interacting with light. LSPRs can be readily tuned by changing the material, size, shape, and environment of the NP.[444] When organized together by DNA origami, the LSPR coupling among the metallic NPs can lead to interesting optical phenomena, including plasmon hybridization,[445] Fano resonances,[446] magnetic resonances,[447] optical lensing and waveguiding,[448] energy transfer,[449] circular dichroism,[450] super-resolution,[451] and many others.[452]

Single quantum emitters, such as fluorescent molecules, quantum dots, nitrogen-vacancy (NV) centers, and doped NPs are at the heart of quantum optics and photonic quantum information technologies. Rational organization of single emitters on DNA origami can result in Förster resonance energy transfer (FRET), optical signal quenching, and nanoscale distance sensing, among others. Additionally, DNA origami can be used to template complex nanophotonic architectures composed of metallic NPs and single emitters. Deterministically positioning single emitters into the nanoscale hotspots of plasmonic nanoantennas can strongly modify the intensity, efficiency, spectrum, phase, polarization state, and directionality of the emission of the quantum emitters. Moreover, such hybrid systems can enhance light-matter interaction, for instance, fluorescence and Raman signals, achieve directionality engineering for spectroscopy analysis, and serve as near-field and far-field light sources in color routing.

Compared to top-down nanofabrication techniques for nanophotonics, such as electron-beam lithography and focused ion beam etching, the DNA origami technique possesses evident advantages. First, the spatial resolution between two neighboring nanoscale elements attached to DNA origami is sub-10 nm, and their spacing can be controlled in a stepwise manner within nanometer accuracy. Second, the DNA-assembled nanostructures are of high throughput and high quality. Notably, the metallic NPs from colloidal synthesis are single crystalline. This property contrasts the metals fabricated with top-down methods, which are polycrystalline and inhomogeneous with rough grains. These issues typically give rise to undesirable lossy LSPRs, thus hampering the performance of nanophotonic devices. Third, diverse nanoscale elements, such as metallic NPs of different materials,[453] single emitters of different types,[454] semiconductor nanocrystals,[455] proteins,[456] aptamers,[457] carbon nanotubes,[272, 458] among others.[443] can be hierarchically assembled on DNA origami with high scalability. For top-down methods to achieve similar goals, they need multiple alignment and processing steps with low throughput and limited chemical functionalization possibilities. Fourth, DNA-assembled nanostructures can be truly 3D, and most importantly, they can reconfigure and be dynamically controlled by various external stimuli.[459] This allows for many advanced nanophotonic applications, especially optical sensing with high molecular specificity on the single-molecule level.

## 8.1 Metallic nanoparticles

### 8.1.1 Site-specific, anisotropic functionalization

One of the significant challenges in molecular assembly is precise control over the surface functionalization of NPs with site selectivity and anisotropy, which are crucial for constructing new nanomaterials with high complexity and advanced functions. DNA nanotechnology is well suited to achieve the aims due to the sequence programmability, specific molecular recognition, and facile modification of DNA.

By taking the analogy from the atom-molecule relation in molecular chemistry, in 2015, Li *et al*. built metallic NP clusters with directional bonds and defined compositions that resembled the methane molecule (see Figure 26A).[460] Spherical AuNPs were densely decorated with two types of ssDNA. They were then encapsulated into self-assembled DNA tetrahedral frames through hybridization between one type of the ssDNA and the DNA extended on the frames. Due to the electrostatic effect and steric hindrance, the encapsulated AuNP in the frame could only be accessible and linked to four other AuNPs coated with a third type of ssDNA from the four face centers of the tetrahedron. The DNA frame thus served as a guiding agent to create tetravalent bonds for the anisotropic assembly of molecule-like metallic NP clusters.

Although the anisotropic regioselectivity of AuNPs could be enforced by the DNA frames, the AuNPs themselves did not possess geometrically controlled DNA patterns and sequences on their surfaces. One year later, Edwardson *et al*. reported an innovative approach to transfer molecular recognition information from DNA structures to AuNPs, as shown in Figure 26B.[461] Parent 3D DNA templates were created with controlled sizes, shapes, and sequence asymmetries. AuNPs were then bound to the reactive DNA arms on the templates, followed by the removal of the templates. The AuNPs subsequently obtained the molecular information from the templates after the transfer process. Remarkably, these AuNPs could be addressed site-specifically with different components, including AuNPs and fluorophores, among others. This direct printing approach to transferring DNA patterns from 3D DNA templates onto AuNPs was analogous to the concept of top-down lithography, in which electron-beam patterns are lithographically transferred from resists to specific materials on substrates. Similar to this work, Zhang *et al*. utilized 2D DNA origami to transfer oligonucleotide patterns onto the AuNP surfaces through toehold-mediated DNA displacement reactions (Figure 26C).[462] This DNA origami-based nanoimprinting excelled in controlling the valence and valence angles of AuNPs with high precision.

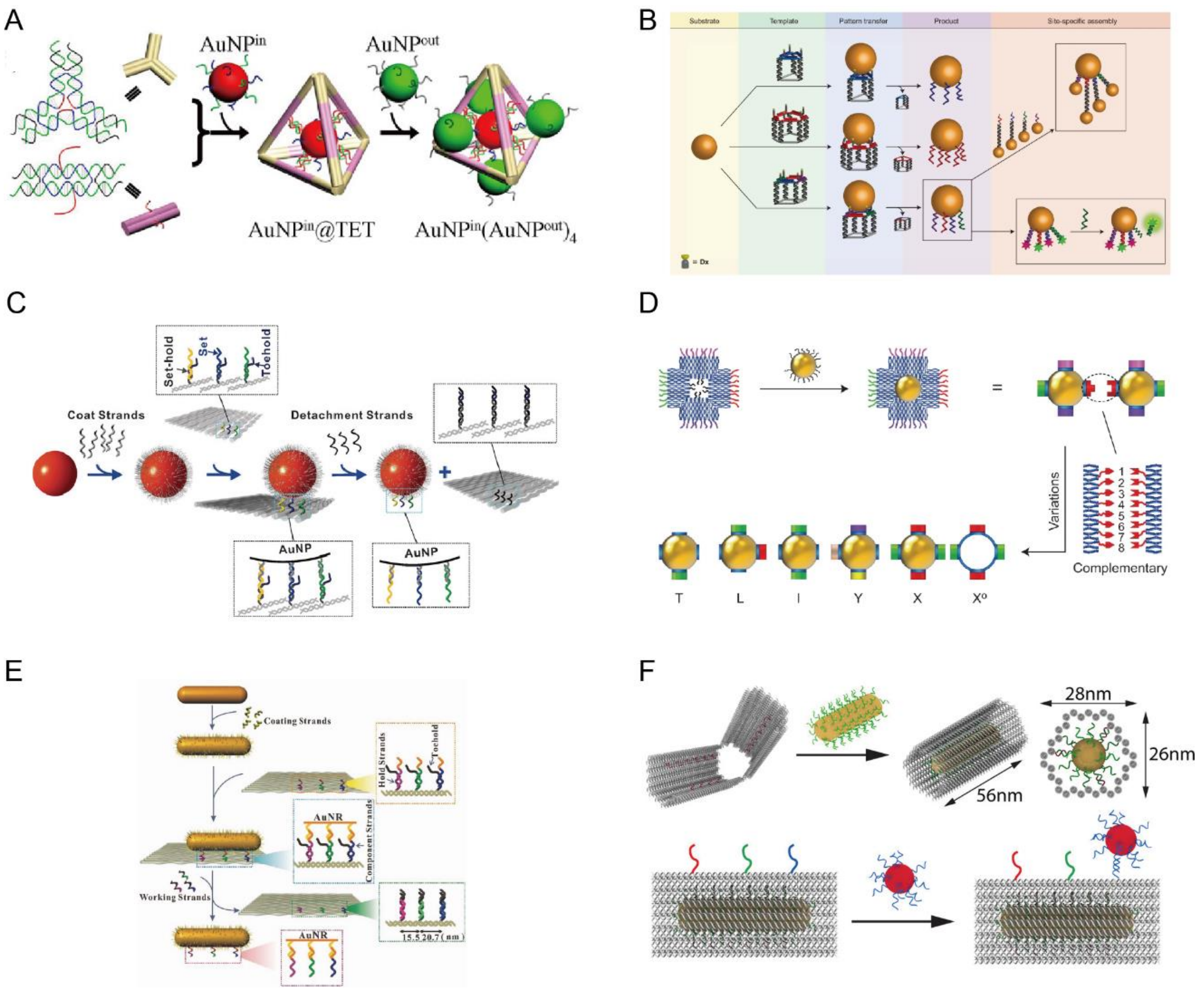

**Figure 26.** Site-specific, anisotropic functionalization of metal nanoparticles with DNA. (A) metallic NP clusters with directional bonds and defined compositions. Reproduced with permission from ref 460. Copyright 2015 American Chemical Society. (B) A molecular printing strategy to transfer molecular recognition information from DNA structures to AuNPs. Reproduced with permission from ref 461. Copyright 2016 Nature Publishing Group. (C) A DNA-origami-based nanoimprinting lithography to transfer oligonucleotide patterns onto AuNPs through toehold-mediated DNA displacement reactions. Reproduced with permission from ref 462. Copyright 2016 Wiley. (D) “chromatic bonds” for programmable assembly. Reproduced with permission from ref 463. Copyright 2016 Nature Publishing Group. (E) A DNA-origami-based nanoimprinting lithography to transfer DNA information onto AuNRs. Reproduced with permission from ref 464. Copyright 2017 Wiley. (F) AuNRs are modified with specific surface recognition sites using a DNA origami clamp. Reproduced with permission from ref 465. Copyright 2016 American Chemical Society.

Different from the site-specificity of metallic NPs encoded by coating different DNA sequences, Liu *et al*. introduced "chromatic" bonds for programmable assembly, as shown in Figure 26D.[463] Spherical AuNPs were bound inside square-shaped DNA origami frames. The outer edges of the frame were modified with different DNA strands to yield patchy NPs with selective and fully prescribed anisotropic interactions. Planar architectures with periodic and arbitrarily shaped geometries were formed, such as square-shaped clusters, cross-shaped clusters, linear chains, zigzag chains, and even non-trivial man-shaped structures, demonstrating the power of DNA origami-based patchy NPs that carried both material function through the NPs and precise binding characteristics through DNA encoding.

A more challenging task is to endow anisotropic metallic NPs, such as AuNRs, with spatial directionality and sequence asymmetry. AuNRs offer great opportunities for optical designs and applications due to their structural anisotropy, polarization dependence, and strong optical response. It was well-known that the end and side surfaces of an AuNR could be differently modified to allow end-to-end or side-by-side AuNR assemblies. Nevertheless, such domain-separated surface functionalization lacked high spatial resolution and controllability. Following their previous work, the Fan group transferred the DNA sequence configurations to the surfaces of AuNRs through specifically designed toehold-mediated displacement reactions (see Figure 26E).[464] The AuNRs were first uniformly coated with ssDNA. Three binding sites along a linear line were defined on a DNA origami template. Each site contained a pair of sequence-specific hold and component strands. After the AuNRs were mixed with the origami, they were anchored along the linear line. Upon addition of the trigger strands, the AuNRs were dissociated, carrying the component strands in the precision pattern transferred from the origami. Due to the fixed number, position, and specific sequences of the component strands, the resulting AuNRs possessed defined valence, site specificity, and sequence anisotropy. The great flexibility of this approach was demonstrated by binding AuNPs at the designated locations along the AuNRs. Using a DNA origami clamp, Shen *et al*. developed an alternative way to achieve AuNRs with specific surface recognition sites that possessed the nanometer scale addressability afforded by the origami (see Figure 26F).[465] The AuNRs were encapsulated by the origami clamps through hybridization between the DNA on the AuNRs and the complementary strands inside the clamps. The outer surface of the origami was site-specifically modified with capture strands, so that a series of well-defined heterostructures with controlled valence could be constructed. The appealing advantage of this approach lay in precise control over the site-specific functionalization of anisotropic NPs, representing a unique pathway to build higher-order metallic superstructures with well-defined components and conformations.

### 8.1.2 Plasmonic superstructures

DNA self-assembly offers a unique tool to spatially organize nanoscale objects with high accuracy. Compared to lithographic approaches, DNA-assembled plasmonic devices exhibit high crystalline quality and high position control. In particular, one of the new trends is to build plasmonic superstructures with planar, thin-layered geometries that can easily be integrated on solid substrates. This technique enables many opportunities for on-chip applications, such as building colloidal metasurfaces and combining them with top-down structures.

In 2021, Liu *et al*. demonstrated the DNA assembly of planar AuNR superstructures with complex patterns and chiroptical properties (see Figure 27A).[466] Truncated-triangular DNA origami with an edge of ~ 80 nm was used as a building block. The three edges of the triangle were specifically modified, so that six triangles could be stitched together by sequence-encoded DNA connectors, 1&2 or 2&3, to form different hexamers in one step. Capture strands were extended from the origami building blocks to assemble AuNRs functionalized with complementary DNA. Six different types of thin-layered AuNR superstructures in bi-star or pinwheel geometries were formed. The chiroptical properties, such as circular dichroism (CD), optical rotatory dispersion, and optical asymmetry factor (g-factor) of these chiral superstructures, were characterized both in solution and at the dried state on glass. Upon interaction with circularly polarized light (CPL), the plasmon hybridization between the AuNRs in a dimer assembled on an origami triangle led to bonding and antibonding plasmon modes, resulting in peak-dip or dip-peak bisignate CD features. Each bi-star or pinwheel superstructure could be viewed as a composite of six primary AuNR enantiomers with the same handedness and secondary enantiomers residing on neighboring DNA triangles with the same or opposite handedness. This feature directly influenced the net CD signals of the superstructures. For instance, the number of enantiomers of opposite handedness in the pinwheel was larger than that in the bi-stars, and thus the latter exhibited more intense CD signals. Later, this origami template was also utilized by the same group to host both spherical AuNPs and AuNRs in order to form chiral satellite-core NP superstructures (see Figure 27B).[467] In this complex system, several chiral mechanisms, including planar chirality, 3D chirality, and induced chirality transfer, were involved. Split or non-split of the characteristic CD line shape of the AuNR spiral was observed, and the induced CD responses were from the achiral AuNP.

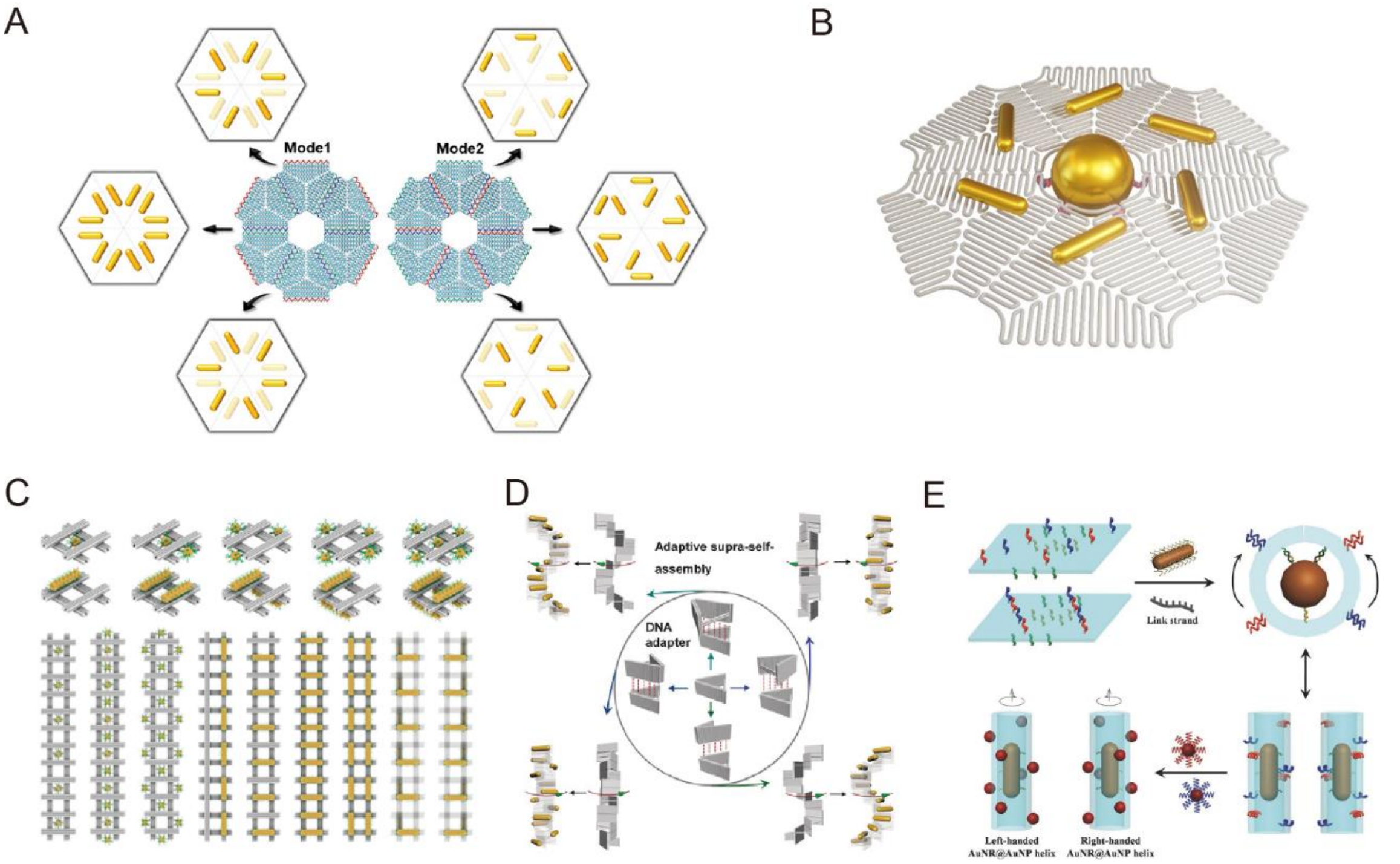

**Figure 27.** Chiral plasmonic assemblies. (A) Self-Assembly of Planar, thin-Layered chiral nanoparticle superstructures guided by DNA origami. Reproduced with permission from ref 466. Copyright 2021 American Chemical Society. (B) Chiral satellite-core nanoparticle superstructures. Reproduced with permission from ref 467. Copyright 2022 American Chemical Society. (C) A hashtag DNA origami tile polymerized into 1D rigid chains to assemble AuNPs. Reproduced with permission from ref 468. Copyright 2020 American Chemical Society. (D). A V-shaped DNA adaptor to assemble AuNRs into chiral superstructures. Reproduced with permission from ref 469. Copyright 2017 Wiley. (E). Chiral patterning of AuNPs on the site-specific functionalization of AuNRs. Reproduced with permission from ref 470. Copyright 2017 Wiley.

Although the hexamer templates based on connecting triangular tiles enabled the assembly of metallic NPs into structurally complex architectures, they lacked structural rigidity and design versatility to extend the assembly along defined directions. To this end, Wang *et al*. utilized a hashtag DNA origami tile, which could be polymerized into 1D rigid chains to assemble metallic NPs, as shown in Figure 27C.[468] This hashtag tile consisted of orthogonal stacks of two decks *via* scaffold DNA linkage. Each deck was 65 nm long, 10 nm wide, and 5 nm tall. The internal square-shaped opening formed by the four decks within a tile had a side length of 21 nm. To introduce unidirectional extension, the authors utilized connector strands to link the ends of designated decks to form hashtag chains. Specifically, the phononic band calculations revealed that the polymerized chains exhibited excellent mechanical stiffness. The hashtag chains with lengths up to several micrometers were successfully achieved and provided good platforms to organize AuNRs and spherical AuNPs with high fidelity. By assembling two AuNRs twisted by 90°, on the top and bottom decks of the hashtag monomer, respectively, chiral AuNR polymer chains with designated handedness were formed. It is worth mentioning that resonantly induced optical chirality cascaded along the chain was observed due to the coupling between the chiral dimers, which resulted in redshifted and dimmed CD spectra when compared to those of AuNR dimers.

In addition to the in-plane extension, directional self-assembly of AuNRs along the third-dimension has also been actively carried out to create plasmonic superstructures with increasing complexity. For instance, Lan *et al*. developed a V-shaped DNA adaptor, which served as a binding platform for an AuNR and allowed for the self-assembly of multiple AuNRs into superstructures (see Figure 27D).[469] The adaptor was a 3D DNA origami structure composed of two rectangular arms, forming an angle of ~45°. Four binding domains were modified along the edges of these two arms. By different combinations of linking two specific binding domains between the neighboring units, four different chiral DNA frameworks could be constructed. Assembly of AuNRs on these DNA frameworks led to stair-like LH and RH plasmonic superstructures and coil-like LH and RH plasmonic superstructures. Specifically, the authors observed that for the coil-like superstructures, the optical chirality was opposite to that predicted by the ensemble handedness of the assemblies. This resulted from the intriguing coupling effects among the AuNRs in different layers. This work suggested that the nature of the optical chirality could not be simply predicted according to the overall handedness of the DNA frameworks, but rather the interaction between the plasmonic units should be carefully considered.

Taking one step further from their previous work on site-specific functionalization of AuNRs, Shen *et al*. demonstrated highly precise chiral patterning of spherical AuNPs on AuNR motif surfaces to build plasmonic superstructures (see Figure 27E).[470] An AuNR was wrapped in a DNA origami template using linker strands. The predesigned capture strands on the outer surface of the DNA origami tube formed a helical pattern for the conjugation of AuNPs. The handedness of the AuNP helix and the AuNP size was altered to yield a series of AuNR@AuNP helices. Compared to AuNP helices assembled without the central AuNRs, higher CD intensities were acquired. The authors attributed this observation to the interaction between the plasmons from the AuNR and the surrounding AuNPs.

### 8.1.3 Dynamic plasmonic nanostructures

Spatiotemporal control over the conformations of plasmonic nanostructures by environmental cues is a unique asset that DNA nanotechnology offers. When combined with metallic NPs, the changes in environmental information can be largely amplified and revealed by the tuned optical responses of the plasmonic assemblies. In 2017, Jiang *et al*. carried out a very comprehensive study, in which a variety of external inputs were employed to control a plasmonic nanosystem based on DNA origami-templated AuNRs (see Figure 28A).[471] Two AuNRs were assembled on the opposite surfaces of a rhombus-shaped template made from two connected origami triangles. The two AuNRs formed an L-shaped plasmonic chiral structure. The key to the dynamic regulation was the nucleic acid linkages between the origami triangles, modified to respond to glutathione reduction, restriction enzyme activity, pH changes, or photon irradiation accordingly. While the former two stimuli caused irreversible changes, the pH and light irradiation triggered reversible changes in the CD signals. In particular, the glutathione-regulated plasmonic system is fascinating, because glutathione, as a ubiquitous biological tripeptide, is found to be in a high concentration in cancer cells. Therefore, glutathione-responsive plasmonic systems could have great potential in anticancer molecular sensing applications.

Due to the ease of design and fabrication, 2D single-layered origami structures have been widely used to template metallic NPs to construct plasmonic nanostructures. Nevertheless, to build dynamic plasmonic systems, rigid origami structures are much preferred. Among different designs, the 3D cross-finger template with multi-origami layers has been proven to be one of the most robust and successful platforms for creating dynamic plasmonic systems. Building on their previous achievements, the Kuzyk group reported the remote manipulation of native (non-photoresponsive) chiral plasmonic structures using light (see Figure 28B).[472] The novel aspect was the usage of a photoresponsive medium comprising a merocyanine-based photoacid, which surrounded the plasmonic structures. The two bundles of the origami cross were modified with a DNA triplex lock. Upon exposure to visible light, the pH of the medium decreased, and the DNA triple link formed, leading to the reconfiguration of the plasmonic structure. The reversed process was triggered simply by switching off the light. The authors observed that the degree of the overall CD signal changes depended on the intensity of the incident light. Interestingly, the plasmonic cross nanostructures templated by DNA origami were also encapsulated and actuated inside cell-sized microfluidic compartments (see Figure 28C).[473] These chiral assemblies were pH-sensitive and could be reversibly reconfigured inside the compartments upon adding the proton acceptor pyridine and the proton donator Krytox, respectively.

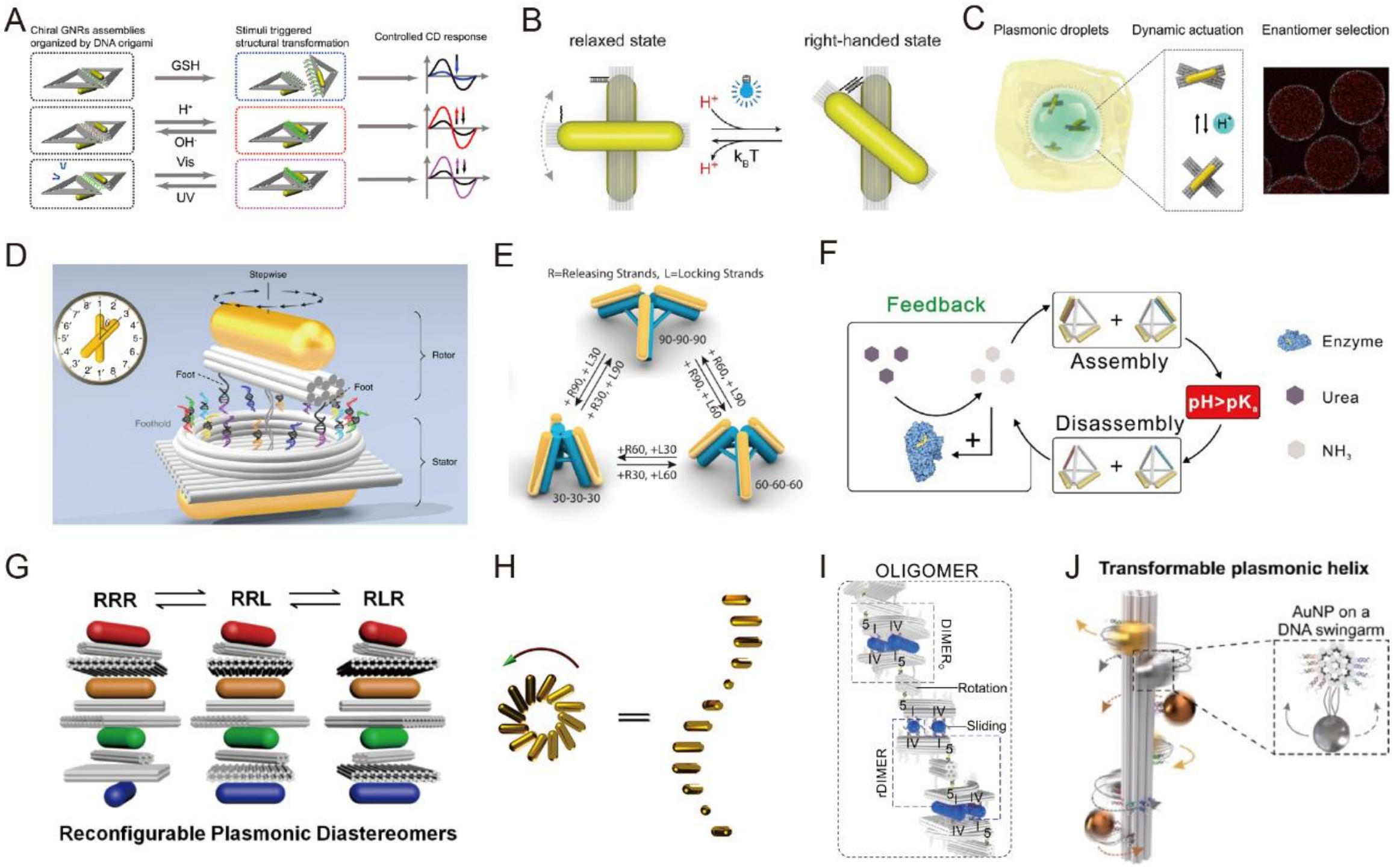

**Figure 28.** Dynamic regulation of chiral plasmonic assemblies. (A) External inputs to manipulate a plasmonic nanosystem based on DNA origami-templated AuNRs. Reproduced with permission from ref 471. Copyright 2017 American Chemical Society. (B) light-control non-photoresponsive plasmonic assemblies using a photoresponsive medium. Reproduced with permission from ref 472. Copyright 2021 Wiley. (C) DNA-assembled dynamic nanostructures encapsulated and actuated inside cell-sized microfluidic compartments. Reproduced with permission from ref 473. Copyright 2020 American Chemical Society. (D) An autonomous DNA-assembled rotary nanoclock. Reproduced with permission from ref 232. Copyright 2019 Nature Publishing Group. (E) A 3D reconfigurable tripod-shaped plasmonic nanostructure with controllable, reversible conformational transformation. Reproduced with permission from ref 207. Copyright 2017 American Chemical Society. (F) A proton-responsive plasmonic system to a positive-feedback chemical reaction network. Reproduced with permission from ref 474. Copyright 2019 American Chemical Society. (G) Reconfigurable plasmonic diastereomers organized by DNA origami. Reproduced with permission from ref 475. Copyright 2019 American Chemical Society. (H) A V-shaped DNA origami to assemble a long chiral chain. Reproduced with permission from ref 476. Copyright 2018 American Chemical Society. (I) Dimerization and oligomerization of DNA-assembled building blocks. Reproduced with permission from ref 477. Copyright 2021 Nature Publishing Group. (J) A swing arm concept to fabricate transformable plasmonic helix. Reproduced with permission from ref 478. Copyright 2022 Wiley.

Full angle tuning of the plasmonic cross nanostructures was later realized by Xin *et al*. using a DNA-assembled rotary nanoclock (see Figure 28D), in which a rotor AuNR carried out directional and reversible rotations with respect to a stator AuNR, transitioning among 16 different configurations from 0° to 360°.[232] The central component of the device was a ring-shaped track, along which the rotor could be bound and released *via* toehold-mediated strand displacement reactions in a stepwise manner. Notably, the authors also demonstrated the autonomous rotation of the plasmonic nanoclock powered by DNAzyme-RNA interactions, which could be monitored by CD spectroscopy in real time. To encode more optical information, increasing the number of AuNPs, while keeping the conformation controllability is a rigorous pathway. Zhan *et al*. designed a 3D tripod-shaped origami template, which was used to position three AuNRs, as shown in Figure 28E.[207] The interarm angle was controlled by the connecting structs, which were composed of two parallel double helices. Through toehold-mediated strand displacement reactions, the length of each strut could be altered, and thus each interarm angle could be tuned among 30°, 60°, and 90°. The transduction of the conformational changes was manifested by controlled shifts of the plasmonic resonances studied using dark-field microscopy on the single structure level.

Although great success has been achieved, switchable control over plasmonic systems by DNA, light, pH, and many others primarily operated under equilibrium conditions. Nevertheless, in nature, many cellular systems execute work out of equilibrium. For instance, the active self-assembly of microtubules and actin filaments in cells is operated by continuous energy consumption to remain far from thermodynamic equilibrium. Taking the inspiration, Man *et al*. coupled a proton-responsive plasmonic system to a positive-feedback chemical reaction network to realize nonequilibrium assembly autonomously (see Figure 28F).[474] A tetrahedron-shaped DNA origami template was used to bind AuNRs through Hoogsteen interactions. Adding chemical fuel lowered the pH below the critical assembly value $pK_a$. Plasmonic chiral structures started to form, as reflected by a CD signal increase. The hydrolysis of urea catalyzed by the enzyme produced $NH_3$, leading to the pH increase above $pK_a$ and, thus, the disassembly of the plasmonic structures.

To further increase structural complexity, Wang *et al*. reported reconfigurable plasmonic diastereomers with up to three chiral centers organized by DNA origami (see Figure 28G).[475] Three chiral centers consisting of four AuNRs were assembled vertically. Driven by programmed DNA reactions, each chiral center could be individually switched between LH and RH states *via* the rotation of the AuNR along two diagonal directions, respectively. The overall CD signals resulted from the substantial cross-talk near-field coupling among these chiral centers. The number of AuNPs in a plasmonic superstructure can be largely increased by polymerizing the DNA origami host structures and/or interconnecting AuNPs with the origami templates. In the work of Lan *et al*., AuNRs were assembled on V-shaped DNA origami monomers, which were polymerized into a long chain (see Figure 28H).[476] Here, the advance was that the interarm angle of the monomer could be controlled, so the AuNR chiral superstructures were transformed between a tightly folded state with a small interarm angle and an extended state with a larger interarm angle. This ability allowed for reversible CD signal tuning by DNA strand displacement reactions. By dynamically changing the DNA origami monomer into its mirror-image structure, inversion of the CD signals was observed accordingly.

It is also possible to implement controlled multi-motion in long-chain plasmonic superstructures. Xin *et al*. reported the dimerization and oligomerization of DNA-assembled building blocks (see Figure 28I).[477] The plasmonic chiral building block itself could exhibit walking and rotation motions. When stacked together by intercalation to form a dimer, a sliding motion could be introduced between the monomers. When oligomerized into superstructures, twisting motion among the units was subsequently introduced. This work outlined the high versatility and controllability of DNA to construct and manipulate high-order plasmonic assemblies. Furthermore, it is worth highlighting the recent work of Peil *et al*., in which the swingarm concept that had been used for substrate channeling in multi-enzyme complexes on DNA scaffolds was applied in programmable translocations of multiple spherical AuNPs with large leaps on DNA origami. Six AuNPs were tied to their respective swingarms on a DNA origami shaft to form a plasmonic helix, as shown in Figure 28J.[478] These swingarms with the extra poly-thymine segment were longer than the footholds extended from the DNA origami. Without using the swingarms, the AuNPs would have to carry out consecutive steps by rolling to bind to different footholds for helix transformations. This process involved foot-track interactions at each step by adding DNA fuels. In addition, the stepwise movements between adjacent binding sites with considerable spacing could easily lead to AuNP detachment during the translocations. In contrast, with the swingarm mechanism, the AuNPs were permanently tied to the swingarms during different translocation processes, which greatly helped to prevent AuNP detachment and facilitated foot-track interactions. Crucially, the long length and flexibility of the swingarms allowed for the direct transportation of the AuNPs to target locations over large distances, which led to reliable and distinct CD changes at different states.

### 8.1.4 Chiral plasmonic nanostructures for optical sensing

Plasmonic nanostructures have been widely used as optical sensors due to their high sensitivity, rapid responses, and simple optical readout. In particular, DNA-based dynamic plasmonic structures afford additional benefits for optical sensing because DNA structures can be designed to reconfigure upon recognition of defined nucleic acid sequences, aptamers, ions, enzymes, and pH changes. Among different geometries, the plasmonic cross that we have discussed in the previous section has proven to be one of the most efficient and successful systems, which can be chemically modified for the optical sensing of a multitude of molecules. Often, the sensing segment is included as part of the lock mechanism. For instance, in 2018, Funck *et al*. demonstrated sensitive and selective detection of a target RNA sequence from the hepatitis C virus genome (see Figure 29A).[479] The RNA could bind to a complementary sequence of the lock, which connected the two arms of the plasmonic cross. The specific RNA sequence at concentrations as low as 100 pM was detected using CD spectroscopy.

Besides nucleic acids, more universal biorecognition elements, such as aptamers, which have high affinity and specificity to target molecules, are well suited to be integrated with DNA origami for optical sensing. For instance, Huang *et al*. and Zhou *et al*. independently showcased optical sensing of adenosine (see Figure 29B), ATP, and cocaine molecules (see Figure 29C) by modifying the DNA locks of the plasmonic cross with the corresponding aptamers.[480-481] Reconfiguration of the plasmonic cross led to changes in the CD responses upon adding the specific target molecules.

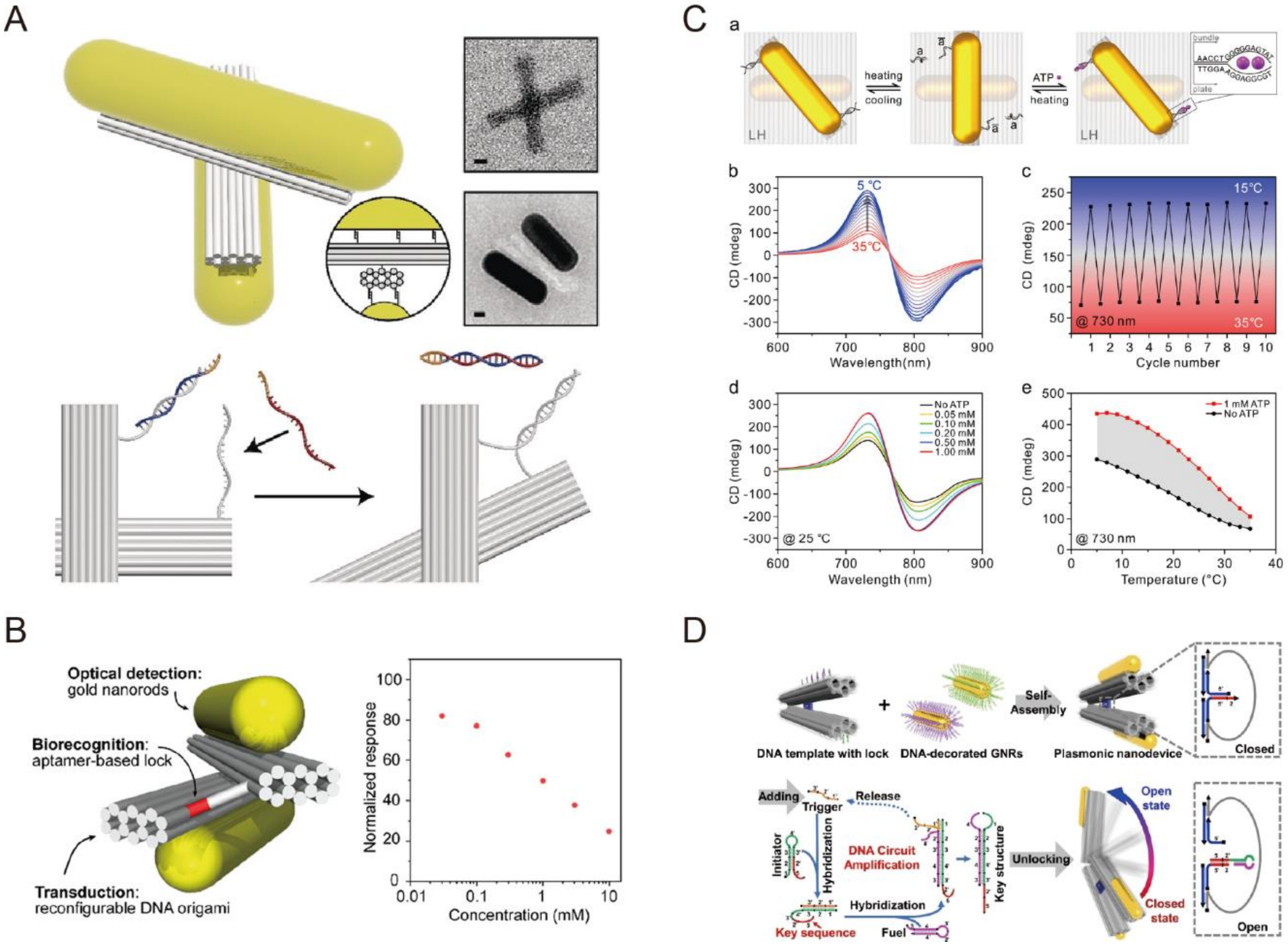

**Figure 29.** Chiral plasmonic assemblies for optical sensing. (A) Detection of a target RNA sequence from the hepatitis C virus genome using cross-shaped plasmonic nanostructure. Reproduced with permission from ref 479. Copyright 2018 Wiley. (B) Optical sensing of adenosine using plasmonic cross with aptamer. Reproduced with permission from ref 480. Copyright 2018 American Chemical Society. (C) Optical sensing of ATP and cocaine molecules, Reproduced with permission from ref 481. Copyright 2018 American Chemical Society. (D) A DNA circuit-aided plasmonic system to perform cascade amplification of weak chemical and biological signals. Reproduced with permission from ref 482. Copyright 2022 Wiley.

Profoundly, Liu *et al*. employed another type of the plasmonic cross, which could carry out DNA-regulated cascade amplification of weak chemical and biological signals (see Figure 29D).[482] Specifically designed DNA key structures were amplified by DNA circuits to dissociate DNA locks through toehold-mediated strand displacement reactions. This process drove the plasmonic devices into the open state. A variety of input signals, including nucleic acids, adenosines, chiral tyrosinamides, and specific receptors expressed by tumor cells, could be adopted, elucidating the power of the chiral plasmonic assemblies for sensitive biological signaling.

### 8.1.5 Other plasmonic nanodevices

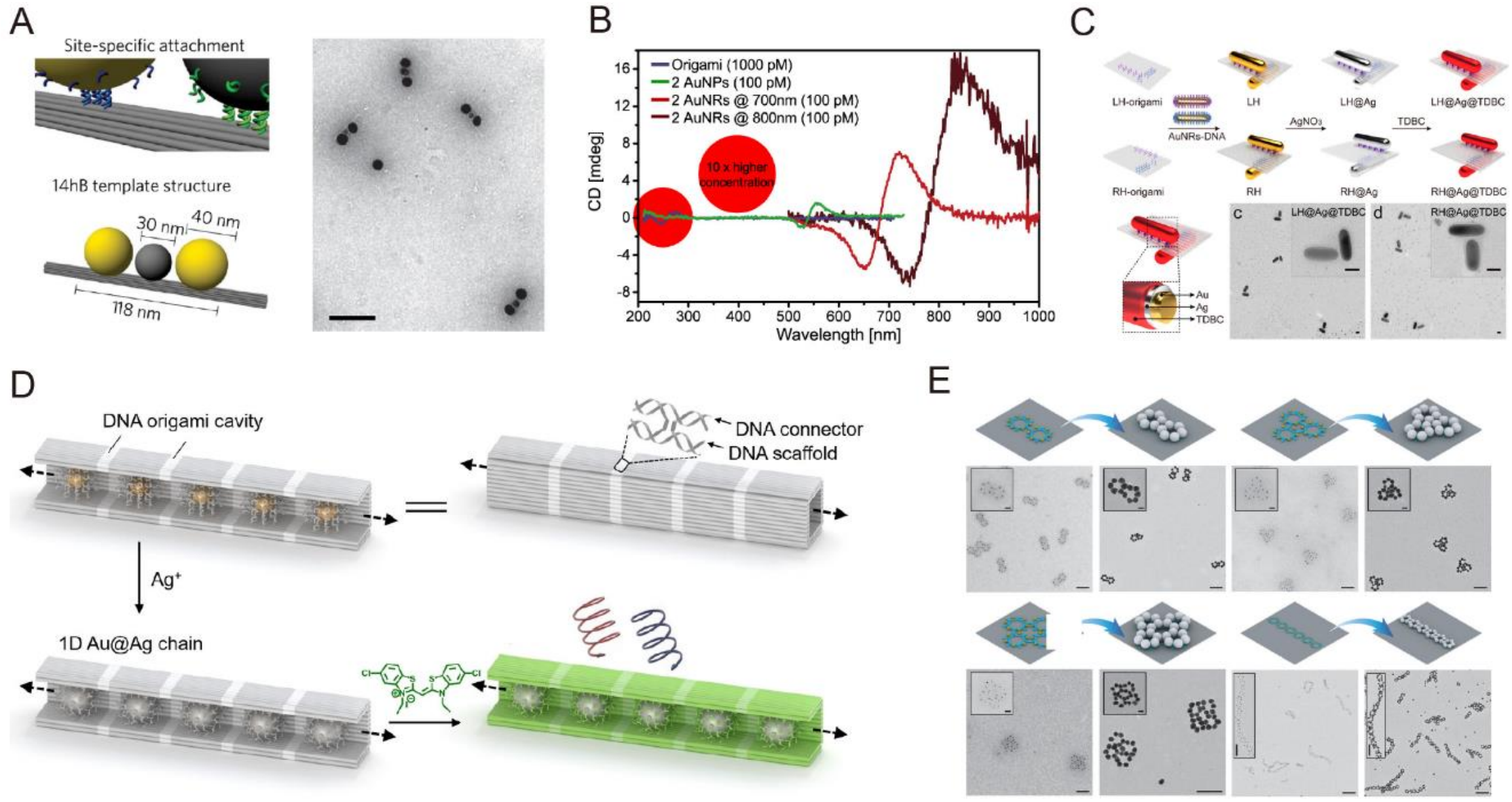

**Figure 30.** DNA-based plasmonic assemblies. (A) A plasmonic trimer composed of Au-Ag-Au NPs linearly organized along a DNA origami bundle. Reproduced with permission from ref 483. Copyright 2017 Nature Publishing Group. (B) Multiple particle types of dimer system to study this CD transfer effect. Reproduced with permission from ref 484. Copyright 2018 American Chemical Society. (C) A chiral plexcitonic hybrid system to investigate strong light-matter interactions. Reproduced with permission from ref 485. Copyright 2021 American Chemical Society. (D) 1D chain superstructures of Au@Ag core-shell nanoparticles interacted with DNA-bound chromophores (K21 J-aggregates). Reproduced with permission from ref 486. Copyright 2022 Wiley. (E) 1D chain of cyclic rings templated by a DNA origami hexagon tile. Reproduced with permission from ref 487. Copyright 2019 Wiley.

There are also other excellent examples of DNA-based plasmonic assemblies, which are advantageous platforms for investigating many interesting optical phenomena. For instance, employing the high position control over NPs by DNA origami, Roller *et al*. demonstrated a plasmonic trimer composed of Au-Ag-Au NPs linearly organized along a DNA origami bundle to study coherent ultrafast energy transfer among the NPs (see Figure 30A).[483] Without the center Ag NP, the two outer AuNPs spaced by 40 nm were not coupled. With the AgNP, the dominant resonance shifted from 549 nm to 586 nm, along with a substantial intensity increase. In this case, the coupling of the two AuNPs was mediated by the connecting AgNP with almost no energy dissipation. The authors attributed their observation to the formation of strong hotspots between these three NPs for the lossless coupling and, thus, coherent ultrafast energy transfer.

It has been shown both theoretically and experimentally that when chiral molecules are placed in close proximity, achiral metallic NPs can induce additional CD features at their plasmonic resonances, which are far from the UV CD of chiral molecules themselves. This chirality transfer is associated with the nonzero ORD of biomolecules that spectrally overlaps with the absorption of the NPs. Kneer *et al*. designed a neat system to investigate this CD transfer effect (Figure 30B).[484] Two AuNRs (aspect ratio of 3.3) that formed an achiral plasmonic dimer were placed upright on DNA origami sheets, which served as the chiral molecule medium. The strong electromagnetic fields generated in the gap of the dimer, where the DNA origami was positioned, allowed detectable CD transfer in the near-infrared which had a 300-fold enhancement compared to the UV CD of DNA.

By integrating excitonic molecules, Zhu *et al*. constructed a chiral plexcitonic hybrid system templated by DNA origami to investigate strong light-matter interactions (see Figure 30C).[485] DNA origami rectangles were used to position two crossed AuNRs coated by Ag. The excitonic medium was J aggregates. The plexcitonic hybrid structures were prepared based on electrostatic interactions. The Ag layer was used to tune the resonance of the NRs successively to the excitonic resonance of the molecules. A large Rabi splitting of ~205/199 meV for LH/RH plexcitonic systems was achieved experimentally. In a recent work from the Lan group, as shown in Figure 30D,[486] 1D chain superstructures of Au@Ag core-shell NPs that interacted with DNA-bound chromophores (K21 J-aggregates) were assembled using DNA origami. In the presence of K21, the CD spectra of the superstructures were largely modified by increasing the Ag shell thickness from a peak-dip profile to a prominent peak. Due to the complexity of the hybrid superstructure, three types of interactions were involved, including plasmon coupling, exciton coupling, and plasmon-exciton coupling. Meanwhile, the complexity of the system also gave rise to great challenges in building theoretical frameworks for describing the hybrid chiral mechanisms.

Another noteworthy example is the self-assembly of NPs into high-order networks of clusters and polymeric chains for the excitation of magnetic resonances. Wang *et al*. used a DNA origami hexagon tile to precisely assemble AuNPs (10 nm in diameter) into a monocyclic ring, and the subsequent high-order assembly led to a 1D chain of cyclic rings (see Figure 30E).[487] Critically, the Ag growth enlarged the NPs and also enhanced the coupling among the NPs. In the case of the monocyclic ring, electric dipolar resonance and magnetic dipolar resonance were both observed. The former was associated with the linear oscillations of charges in the NPs, while circulating field displacements along the rings revealed the existence of the magnetic dipolar resonance for the latter. When a long chain of cyclic rings was excited, magnetic dipoles confined in the rings along the chain could propagate in an antiparallel fashion.

Taking together, the self-assembly route for constructing plasmonic superstructures affords many superior characteristics to conventional top-down approaches, including unprecedented structural complexity, NP positioning on the nanoscale precision, dynamic 3D reconfiguration, optical sensing with molecular specificity, high-throughout and parallel fabrication and many others.

### 8.2 Quantum emitters

Single emitters can be positioned on DNA origami in a prescribed pattern with control over their relative distances and conformations on the nanoscale accuracy. This feature is handy for designing intricate experiments for understanding the fundamental photophysics of nanoscale emitters and light-matter interaction processes, such as enhanced fluorescence, single-molecule surface-enhanced Raman spectroscopy (SERS), artificial light harvesting and energy transfer, plasmon-assisted Fluorescence resonance energy transfer (FRET), and strong coupling at room temperature. It also enables new applications in optical characterizations, biomolecular imaging, and molecular diagnostics.

### 8.2.1 Nanoscale distance control

The energy transfer between a donor and acceptor fluorophore pair is highly sensitive to their proximity, enabling the determination of inter- and intramolecular distances with sub-nanometer precision. In 2019, Schröder *et al*. investigated the distance-dependent interchromophoric interactions between two dye molecules placed on a 2D DNA origami template by changing their distance in a single-nucleotide step (see Figure 31A).[488] Using single-molecule fluorescence spectroscopy, the authors observed that at smaller distances weak coupling between dyes including singlet-singlet annihilation, singlet-triplet annihilation, and singlet-radical-state annihilation played important roles. The contact quenching could be avoided when the dyes were spaced by 7 bps. The energy transfer between dyes thus could be advantageous or disadvantageous on the fluorescence, depending on the properties of fluorophores. A large number of fluorophores can also be arranged on DNA origami in finely controlled stoichiometry and configurations. As demonstrated by Nicoli *et al*., a photonic wire that consisted of a blue donor and a red acceptor with three green fluorophores in between was assembled on DNA origami (see Figure 31B).[489] The three-color cascade achieved energy transfer over a distance of 16 nm. This long-range energy transfer along fluorophore chains emulated the natural process of light harvesting, in which the photon energy absorbed by pigments in the light-harvesting complexes is transferred to the reaction center. To mimic the well-controlled multi-chromophore complexes organized on protein scaffolds for light harvesting and energy transfer, Zhou *et al*. reported a bio-inspired excitonic system templated by DNA origami to enable the long-range exciton migration (see Figure 31C).[490] A four-helix-bundle

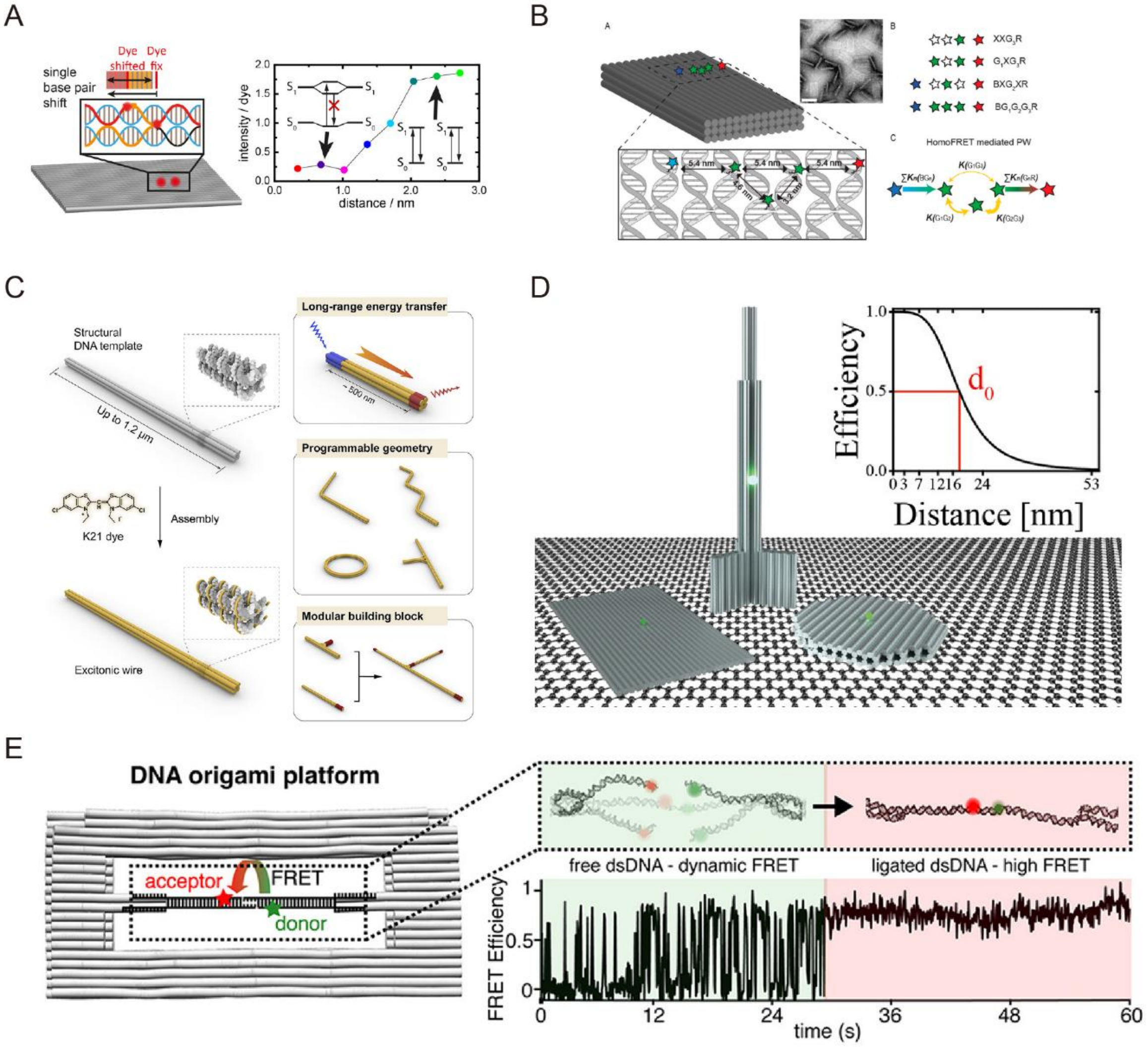

**Figure 31.** Quantum emitters with nanoscale distance control. (A) DNA origami as a rigid scaffold to arrange dye molecules in a dense pixel array for interchromophoric interactions. Reproduced with permission from ref 488. Copyright 2019 American Chemical Society. (B) DNA origami based photonic wire for dye and dye interactions. Reproduced with permission from ref 489. Copyright 2017 American Chemical Society. (C) A bio-inspired excitonic system templated by DNA origami to enable long-range exciton migration. Reproduced with permission from ref 490. Copyright 2022 Cell Press. (D) DNA origami with fluorophores at defined distances relative to the graphene layer for energy transfer. Reproduced with permission from ref 491. Copyright 2019 American Chemical Society. (E) A DNA origami platform to mimic a DNA double-strand break for the end-joining reaction optically. Reproduced with permission from ref 492. Copyright 2020 American Chemical Society.

DNA origami of 600 nm in length was used to assemble K21 dye aggregates. These excitonic wires were coupled with energy donors and acceptors to achieve directional excitation transfer over submicron distances.

Beyond the interactions between fluorophores, DNA origami can work as nanopositioners for placing single fluorophores at controlled nanoscale distances to investigate energy transfer from fluorophores to other matters, such as graphene. Graphene is an excellent energy sink to replace Au surfaces in fluorescence quenching experiments with less fluorescence background. To validate this, Kaminska *et al*. used pyrene-modified DNA strands to link DNA origami on graphene *via* π-π stacking (see Figure 31D).[491] Single, freely rotating fluorophores were positioned on the DNA origami at defined distances relative to the graphene layer to study the distance dependence of energy transfer to graphene. The $d^{-4}$ law was accurately confirmed owing to the exquisite distance tuning offered by DNA nanotechnology.

Due to the modality of DNA origami, complex multimolecular interactions such as DNA-DNA interactions, DNA-enzyme interactions, and ligation can also be studied using DNA origami. For instance, Bartnik *et al*. applied a DNA origami platform to specifically mimic a DNA double-strand break and studied the end-joining reaction optically (Figure 31E).[492] Two DNA double strands were positioned on the two sides of the DNA origami in a linear geometry. The two double strands were labeled with Atto647N as the acceptor and Cy3b as the donor, respectively. To catalyze the end-joining of the DNA duplexes, DNA ligase from bacteriophage T4 was used. The ligation processes with different lengths of complementary overhangs and sequences based on the same DNA origami framework design could be monitored using FRET on the single-molecule level.

### 8.2.2 Orientation control

The interaction efficiency between single emitters does not only depend on their relative distance but also critically on their relative orientation. For instance, the dipole-dipole interaction in FRET between two fluorophores is maximized when the two dipoles are aligned. If fluorophores are placed close to metallic NPs, especially anisotropic particles, such as AuNRs and gold triangles, both the molecular excitation rate and the radiative decay rate are closely correlated with the relative orientation between the dipole moment of the fluorophore and the AuNR. This results in different optical effects, ranging from strong enhancement to complete suppression of the photon emission.

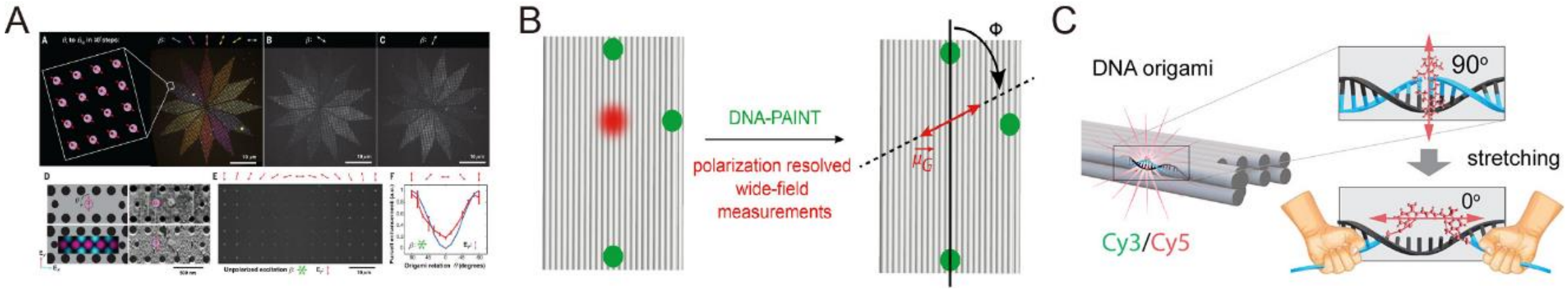

**Figure 32.** Orientation control of single emitters. (A) An asymmetric DNA origami binds to lithographically patterned sites on silica to position and orient a molecular dipole within the resonant mode of an optical cavity. Reproduced with permission from ref 493. Copyright 2021 AAAS. (B) 2D origami as a template to control the orientation of single fluorophores. Reproduced with permission from ref 494. Copyright 2021 American Chemical Society. (C) A study of the orientation of single Cy3 and Cy5 cyanine molecules relative to a DNA origami structure for different incorporation strategies. Reproduced with permission from ref 495. Copyright 2022 American Chemical Society.

There are two general strategies to assemble fluorophores on DNA origami. One is *via* noncovalent binding to dsDNA, which enables orientation control of the molecules because different molecules may bind DNA preferentially in between bases, to the minor or major groove, or along the dsDNA. Based on this, Gopinath *et al*. intercalated TOTO-3 dyes in small moon DNA origami structures (see Figure 32A).[493] The absorption transition dipole moment of the molecule formed an angle of 70°±10° with the dsDNA helix. More than 3000 small moon DNA origami structures were then bound to lithographically pattered sites on a silica substrate, yielding 12 well-controlled orientations to demonstrate a polarimeter. The other strategy is via covalent attachment to ssDNA at the 5´ or 3´ end. This enables high position and stoichiometric control of the molecules on DNA origami but gives rise to random orientations of the molecules. To address the orientation of fluorophores with respect to the DNA origami, Hübner *et al*. conducted a comprehensive study to determine the orientation of single fluorophores, including ATTO647N, ATTO643, and Cy5 covalently attached to 2D DNA origami (see Figure 32B).[494] The orientation of the absorption transition dipole of the molecule was first identified by a polarization-resolved excitation measurement and subsequently, the orientation of the DNA origami was obtained from a DNA-PAINT (points accumulation for imaging in nanoscale topography) nanoscopy measurement.

In a recent work of Adamczyk *et al*., the DNA assembly of single molecules with deterministic position and orientation was demonstrated, as shown in Figure 32C.[495] A fluorophore (Cy3 or Cy5) was doubly linked to a ssDNA staple by leaving a controlled number of unpaired bases in the scaffold, so that a high degree of stretching was applied to the fluorophore to restrain its mobility. By increasing the number of unpaired bases, there was more space for the fluorophore to accommodate and find different sites for interaction with the DNA. It was shown that 0 and 8 unpaired bases led to the molecules positioning perpendicular and parallel to the DNA double helix, respectively. This new approach is valuable for investigating orientation-dependent molecular interactions, such as energy transfer, intermolecular electron transport, exciton physics, and antenna-enhanced molecular emission with high directivity.

### 8.3 Plasmon-enhanced fluorescence of single emitters

Plasmonic nanostructures allow for enhancing spontaneous emission, altering the emission polarization, and shaping the radiation pattern of quantum emitters. A critical challenge for the experimental realizations is positioning a single emitter into the hotspot of a plasmonic antenna with nanoscale accuracy. With the help of DNA nanotechnology, in 2019, Hübner *et al*. demonstrated that a single emitter (Cy5) could be positioned in the gap of an AuNP (60 nm) dimer assembled on the two sides of a double-layered rectangular DNA origami sheet (see Figure 33A).[496] The Cy5 molecule was located near the center of the 13 nm gap between the two spherical AuNPs. The emission patterns of the Cy5 molecules showed two lobes characteristics of an in-plane dipole.

The Cy5 molecule could rotate within the gap, but the AuNP antenna enhanced the emission of the emitter, when parallel to the dimer main axis and suppressed the emission when perpendicular. As a result, the emission directionality followed the dipolar pattern according to the antenna's main resonance mode. This concept could be very useful in photon routing experiments, where the molecule (Atto647) embedded within a plasmonic nanocavity based on the NPoM geometry for the interaction of ultrafast pulses. Reproduced with permission from ref 500. Copyright 2019 Nature Publishing Group. (F) Plasmonic chain waveguides composed of eight AuNPs. Reproduced with permission from ref 501. Copyright 2018 American Chemical Society.

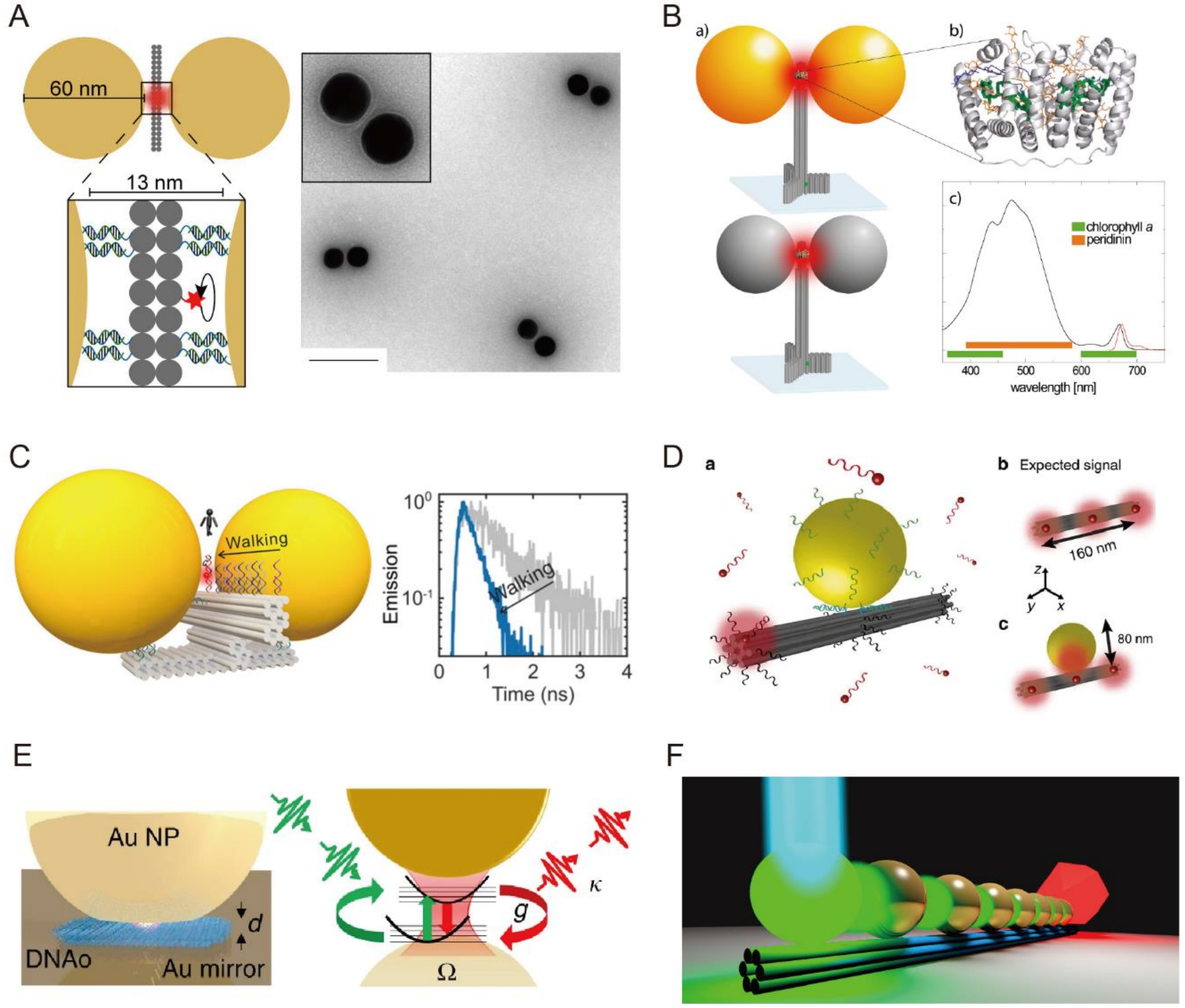

**Figure 33.** Plasmon-enhanced fluorescence of single emitters. (A) A AuNP dimer assembled on the two sides of a double-layered rectangular DNA origami for a single emitter enhancement. Reproduced with permission from ref 496. Copyright 2019 American Chemical Society. (B) An Au or Ag dimer antenna assembled on the pillar-shaped DNA origami for natural light-harvesting complexes. Reproduced with permission from ref 497. Copyright 2018 American Chemical Society. (C) A single fluorophore molecule walks autonomously and unidirectionally into the hotspot of a plasmonic dimer antenna. Reproduced with permission from ref 498. Copyright 2019 American Chemical Society. (D) A DNA origami bundle as a reference frame to acquire the relative fluorophore position for a AuNP and obtain super-resolution imaging of single molecules. Reproduced with permission from ref 499. Copyright 2017 Nature Publishing Group. (E) A single

emission directionality of single emitters is crucial. Kaminska *et al*. expanded the palette of species that could be integrated with DNA-assembled plasmonic antennas. A peridinin-chlorophyll *a*-protein (PCP) complex in the dinoflagellate *Amphidinium carterae* was localized within the hotspot of an Au or Ag dimer antenna (Figure 33B).[497] At the top part of the pillar-shaped DNA origami, a biotin molecule was incorporated to bind a PCP-streptavidin conjugate. The authors showed that the emission of the single light-harvesting complexes coupled to the dimer antenna could be enhanced 500-fold.

To exploit the reconfigurability of DNA nanotechnology, Xin *et al*. demonstrated a dynamic light-matter interaction system, in which a single fluorophore molecule could walk autonomously and unidirectionally into the hotspot of a plasmonic dimer antenna (Figure 33C).[498] Two AuNPs (60 nm) were immobilized on the two sides of a DNA origami template, which consisted of an RNA-decorated track and a bottom platform. A fluorophore molecule Cy5 was attached to a DNAzyme strand, serving as an autonomous fluorophore walker. In the presence of divalent metal ions, the DNAzyme could catalyze the cleavage of the individual RNA substrates, giving rise to unidirectional movements without any external intervention. The fluorescence decay of the fluorophore was accelerated when the walker gradually approached and eventually entered the hotspot. This design strategy allowed for optically monitoring the autonomous motion of a single DNA machine with high temporal and spatial resolution. Going further, Raab *et al*. combined the capabilities of plasmonics to control nanoscale optical fields and far-field fluorescence nanoscopy to locate single emitters and image sub-diffraction fields. To investigate the molecular localization shift caused by plasmonic coupling, the authors utilized a DNA origami bundle as a reference frame to acquire the relative fluorophore position with respect to an AuNP. Meanwhile, DNA origami served as a platform for DNA-PAINT to obtain super-resolution imaging of single molecules (Figure 33D).[499] The 12-helix bundle provided three sites for the dynamic binding of single fluorophores at three regions separated by 80 nm. Upon binding of the AuNP, the three localization spots did not appear in one line. Instead, the central localization appeared displaced from the line defined by the two extreme localizations, providing clear evidence for the single-molecule mirage.

There are other interesting platforms, such as plasmonic cavities and waveguides, for investigating the interaction processes between plasmonic structures and single emitters. Plasmonic nanocavities offer the advantage of an ultrasmall mode volume to achieve high coupling strengths. The nanoparticle-on-a-mirror (NPoM) geometry is a particularly successful platform for achieving strong cavity quantum electrodynamic effects. Ojambati *et al*. studied the interaction of ultrafast pulses with a single molecule (Atto647) embedded within a plasmonic nanocavity based on the NPoM geometry at room temperature (see Figure 33E).[500] A DNA origami sheet sandwiched between an AuNP (80 nm) and an Au mirror was used to deterministically position the Atto647 dye in the gap. The electromagnetic fields in the cavity were enhanced 170-fold and the emission of the single molecule was also significantly enhanced. The power-dependent pulsed excitation revealed Rabi oscillations due to the coupling of the oscillating electric field between the ground and excited states. The observed single-molecule fluorescent emission was split into two modes resulting from anti-crossing with the plasmonic modes. Plasmonic waveguides consisting of metallic NPs can localize and guide light well below the diffraction limit but practical applications are often hampered by high propagation losses. Gür *et al*. improved the performance of plasmonic waveguides by using chemically synthesized monocrystalline AuNPs (~ 40 nm) and shortening the interparticle spacing to about 2 nm using DNA origami as a template (see Figure 33F).[501] Such a small interparticle spacing is very challenging to be achieved by conventional lithography methods but is crucial for efficient long-distance waveguiding. The authors assembled plasmonic chain waveguides composed of eight AuNPs with a diameter roughly equal to the distance tween binding sites (42.2 nm) on a six-helix DNA origami bundle. To detect energy transport along the waveguide *via* the surface plasmon modes at a single device level, they placed a fluorescent nanodiamond at one end of the waveguide. The waveguide was excited by cathodoluminescence imaging spectroscopy, and the energy propagated toward the nanodiamond, which emitted photons into the far field. The DNA-assembled plasmonic waveguides pushed the confinement below the diffraction limit ($\lambda$/10) while retaining a propagation length of up to micrometers.

**8.4 Single-molecule surface-enhanced Raman spectroscopy**

SERS is a powerful technique based on the Raman vibrational fingerprints of molecules, which contain rich chemical and structural information for molecular analysis. The Raman signals can be enhanced by the excitation of plasmonic resonances in Au or Ag nanostructures, which form plasmonic hotspots. Due to the strong Raman enhancement, it is possible to detect single molecules using SERS, given that the number and the placement of single molecules within the plasmonic hotspots can be accurately controlled. Random adsorption of analyte molecules in the hotspots is often applied because of the significant challenges of such nanofabrication requirements. Nevertheless, this leads to the irreproducibility of the experimental results. DNA technology is well suited for single-molecule SERS, because it cannot only precisely arrange metallic NPs in sophisticated geometries, but also allows for positioning single molecules in the plasmonic hotspots with nanometer precision.

The SERS enhancement factor is strongly dependent on the geometry of the plasmonic nanostructures, the number of hotspots, as well as the gap sizes. Plasmonic dimers, consisting of two AuNPs or AgNPs, such as nanospheres, nanocubes, triangles, and nanostars, are popular platforms for SERS. Strong electromagnetic fields are created within the gap between the two NPs in a dimer. Among different anisotropic shapes, Au nanostars are particularly attractive due to their higher electromagnetic field enhancement generated at the sharp tips, when compared to spherical AuNPs or AuNRs under similar experimental conditions. In 2017, Tanwar *et al*. demonstrated the assembly of Au nanostar dimers with tunable gaps and controlled stoichiometry on DNA origami (see Figure 34A).[502] Dimerization of the rectangular DNA origami was achieved by branching staples. A single Texas red dye as the SERS reporter molecule was positioned along the connecting edge of the dimerized origami using a modified branching staple. Two Au nanostars functionalized with DNA were then assembled on the DNA origami through DNA hybridization with the dye molecule sitting in their gap. The Au nanostar dimers with various nanogaps from 7 nm to 13 nm

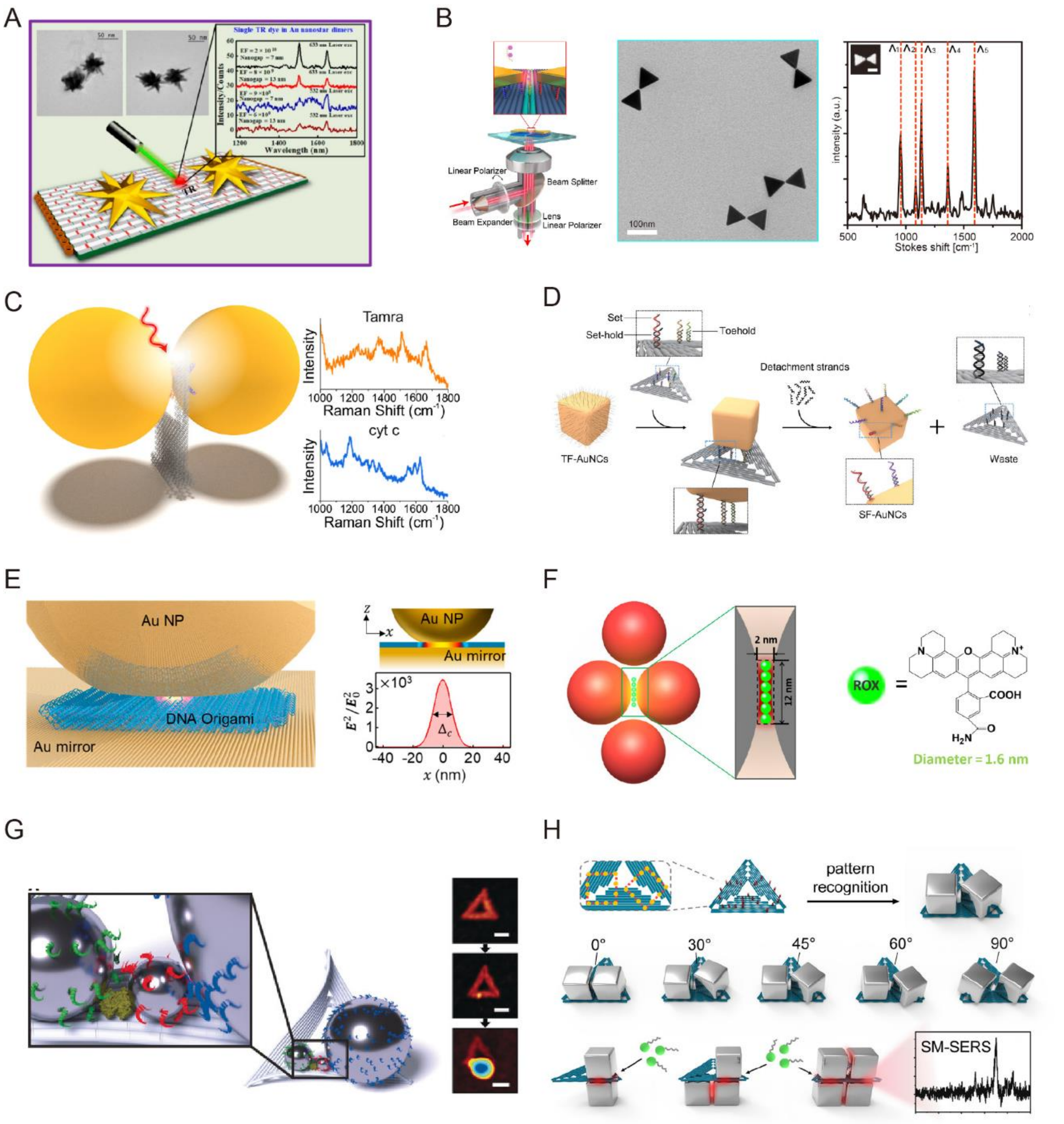

**Figure 34.** Single-molecule surface-enhanced Raman spectroscopy. (A) Au nanostar dimers with tunable gap and controlled stoichiometry for single-molecule SERS. Reproduced with permission

from ref 502. Copyright 2017 American Chemical Society. (B) Au bowtie nanostructures assembled on DNA origami for single-molecule SERS. Reproduced with permission from ref 503. Copyright 2018 Wiley. (C) Au or Ag nanoparticle dimers for single-molecule SERS. Reproduced with permission from ref 504. Copyright 2021 American Chemical Society. (D) A DNA origami-based nanoprinting strategy to create stereo-controlled plasmonic nanostructures for single-molecule SERS. Reproduced with permission from ref 505. Copyright 2021 Wiley. (E) A nanocavity with <5 nm gap achieved by a nanoparticle-on-mirror (NPoM) construct. Reproduced with permission from ref 506. Copyright 2018 American Chemical Society. (F) Four AuNPs organized by rhombus-shaped super-origami for single-molecule SERS. Reproduced with permission from ref 446. Copyright 2019 AAAS. (G) Single streptavidin molecule immobilized in the hot spot of silver nanolenses. Reproduced with permission from ref 507. Copyright 2018 Wiley. (H) A DNA origami directed pattern recognition strategy to assemble AuNCs for single-molecule SERS. Reproduced with permission from ref 508. Copyright 2022 American Chemical Society.

were examined for single-molecule SERS. AFM correlated Raman measurements revealed that with an interparticle gap of 7 nm, the SERS enhancement factor was about $2\times10^{10}$. Even with a gap size of 13 nm, the SERS enhancement factor was as high as $8\times10^{9}$, indicating that the nanostar dimers could be used for specific detection of large single biomolecules, such as proteins with sufficient SERS sensitivity.

The bowtie nanoantenna, which consists of two metallic triangles spaced by a nanometer-sized gap, is another excellent plasmonic platform that exhibits highly localized electromagnetic field enhancement in the interparticle gap. Zhan *et al*. demonstrated the assembly of Au bowtie nanostructures on DNA origami and applied them for single-molecule SERS measurements (see Figure 34B).[503] Two groups of capture strands were used to assemble the nanotriangles on two rectangular DNA origami sheets, respectively. The two origami sheets were then linked together. The Au nanotriangles (80 nm) were functionalized with thiolated ssDNA, complementary to the capture strands on the origami. A probing strand extended between the binding sites of the two nanotriangles was utilized to attach a single Raman molecule, such as Cy5, in the gap. The average gap was 5±1 nm according to the TEM analysis of the assembled structures. The SERS enhancement factor was $2.6\times10^{9}$ for the Cy5 Raman peak at 1366 $cm^{-1}$. To further push the enhancement factor for single-molecule SERS measurements, Tapio *et al*. designed a DNA origami nanofork, as shown in Figure 34C.[504] It comprised a rectangular base with two arms, on which two metallic NPs could be attached, respectively. The two arms were connected by a bridge using two DNA double helices, serving as an anchor point for target analyte molecules. A single molecule could be precisely placed into the plasmonic hotspot, where the SERS signal enhancement was the strongest for single molecule SERS measurements. The authors constructed AuNP or AgNP dimers with different gap sizes down to 1.17 nm, achieving SERS signal enhancements up to $10^{11}$.

Apart from metallic dimers, SERS hotspots can also be produced in other plasmonic geometries, such as anisotropic AuNP clusters with relative spatial directionality, NPoM, Fano clusters, self-similar chains, and many others. For instance, Niu *et al*. applied DNA nanoprinting, which transferred DNA patterns with predefined numbers, sequences, and positions to the surfaces of metallic NPs for the creation of discrete and precisely stereo-controlled nanostructures (see Figure 34D).[505] By anchoring single dye molecules in the gaps (2±1 nm) between the closely spaced NPs, the strong electromagnetic fields in the hotspots enabled the single-molecule SERS signal enhancement. The NPoM construct is another model system to investigate the coherent coupling of light and single molecules at room temperature. Chikkaraddy *et al*. created a nanocavity with <5 nm gap between an AuNP (80 nm) and an Au mirror, which could host a single molecule (Cy5) at the center of the gap by a DNA origami sheet (Figure 34E).[506] The bottom surface of the origami had thiol modifications on specific staple strands to bind to the Au mirror, while the top surface of the origami had DNA overhangs to attach the AuNP. The charge oscillations in the AuNP coupled with image charges on the Au mirror surface, resulting in enhanced electromagnetic fields in the gap by nearly 2 orders of magnitude, allowing for single-molecule SERS measurements.

Metallic NP clusters consisting of spatially coupled metallic NPs can manipulate light at the subwavelength scale and display interesting optical properties. In particular, Fano resonances can be produced, resulting from the interaction between plasmonic bright and dark modes. Due to the strong electromagnetic fields localized in hotspots at the Fano minimum, single-molecule SERS also becomes possible. Fang *et al*. precisely organized four AuNPs (80 nm) on rhombus-shaped super-origami and immobilized dye molecules site-specifically in the central gap of the AuNPs through DNA hybridization (see Figure 34F).[446] When linearly polarized light was along the shorter axis of the tetramer cluster, a Fano minimum was observed. The bright mode was associated with the in-phase charge oscillations in the AuNPs, while the dark mode showed that the charge distributions in the AuNPs were oriented in different directions. By exploiting the nanoaddressability of DNA origami, a controlled number of carboxy-X-rhodamine (ROX) molecules of up to six could be anchored in the hotspots of the tetrameric cluster. The SERS intensity quantitatively enlarged with increasing the number of the ROX molecules from one to six, characteristic of a quantized phenomenon. Notably, a single ROX molecule was detectable, and the SERS enhancement factor was on the order of $10^8$. Furthermore, with self-similar chains of differently sized spherical NPs, the so-called nanolenses can produce strong local electromagnetic fields confined in between the two smallest NPs based on a cascaded enhancement mechanism. Heck *et al*. reported the bottom-up synthesis of Ag nanolenses consisting of three AgNPs (10 nm, 20 nm, 60 nm) using DNA origami (see Figure 34G).[507] A single protein streptavidin was immobilized using a biotin-modified DNA strand and selectively placed in the gap between the 10 nm and 20 nm AgNPs, where the highest enhancement was achieved and probed by SERS spectroscopy.

Although a variety of plasmonic geometries templated by DNA origami have been reported for single-molecule SERS, the assembly of anisotropic NPs into highly hierarchical plasmonic gap nanostructures is quite challenging. Niu *et al*. reported an interesting strategy to create shape-controllable nanogaps between Au nanocubes (AuNCs) using DNA origami (see Figure 34H).[508] Due to the anisotropic nature of the AuNCs, their vertices, edges, and faces allowed for the creation of richer configurations of nanogaps than isotropic spherical AuNPs, together with significantly enhanced electromagnetic fields in these nanogaps. By tuning the position and number of capture strands on the DNA origami, pattern recognition was enabled to assemble AuNCs in different geometrical configurations with nanoscale precision and shape-controllable gaps. Followed by anchoring single Ramon probe molecules in such gaps, strong SERS signals were detected, offering a novel platform for high-sensitivity photonic devices and single-molecule biophysical studies.

## 9. Conclusions and outlook

In recent years, DNA nanotechnology has made a great leap in developing new self-assembly methodologies as well as generating DNA-based nanomaterials with diverse functions. Crucially, effective solutions have been found to remove or circumvent bottlenecks in the field. For instance, the assembly of DNA nanostructures usually requires a large number of ssDNA species of considerable quantity, and thus the high cost of DNA synthesis has been an issue for quite some time. To this end, several methods were developed to reduce the cost of DNA synthesis, including the chip-synthesized DNA followed by parallel enzymatic amplification[509] and the bacteriophage-based production of single-stranded precursor DNA followed by cleavage using DNAzymes.[510-511] Consequently, the cost of DNA-based nanomaterials is largely reduced, especially for large and complex DNA structures. This would substantially help to expand the applications of DNA nanotechnology in many research fields. However, when compared to that of conventional nanomaterials, such as lipid nanoparticles, polymers, inorganic metallic/non-metallic materials, *etc*., the cost of DNA production and structural fabrication is still higher in general. Novel approaches are needed to achieve large-scale, reliable, and good manufacturing practice-compliant production of DNA origami-based nanomaterials.

There remain many outstanding challenges and open questions in the next decade that await innovative solutions and further investigations. From the perspective of self-assembly methodologies, the development of new computational tools would be invaluable to facilitating the optimization of the assembly process, as well as the accurate prediction of the assembly pathways and outcomes. Current DNA origami design schemes still largely rely on researchers' intuition and experiences, which may differ considerably from person to person. In particular, for complicated 3D origami structures, minor design differences often lead to large variations in the assembly yield. The design optimization can thus be both time and cost-consuming. Advanced computational tools will help to standardize the design process and reduce the trial and error involved to generate desired assembly products.

Furthermore, the perspective of interfacing protein engineering with DNA nanotechnology is extremely exciting, owing to the proteins' multifaceted cellular functionalities and excellent biocompatibility. Recent work by Dietz et al. showed DNA-protein hybrid structures that may one day be produced in cells, opening exciting opportunities to encode, assemble, and operate nanodevices made of DNA-protein complexes *in vivo*.[327] Nevertheless, such DNA-protein hybrid structures mainly take advantage of the nucleic acid-binding proteins, while functional proteins are excluded from the final structures. Hence, the development of a general strategy to design genetically encoded protein assemblies that can form inside cells would be of great interest. Cell-compatible functional protein assemblies would open up a new research area, where genetically encoded proteins can autonomously assemble into target geometries and carry out designated tasks *in vivo*. One of the possible solutions would be the assembly of protein-RNA origami from co-transcription. Considering the high structural diversity of proteins, however, this ambitious goal may involve substantial challenges. The main issue is that protein folding is far more complicated than DNA self-assembly, and the correct assembly of protein structures often relies on additional molecular machineries for protein transport, modification, and quality control. Nevertheless, we remain optimistic in this direction considering the rapid development of tools for protein structure design and prediction.

While many DNA nanomachines were demonstrated, fewer RNA nanomachines were reported. The reasons might lie in the chemical instability of RNA structures *in vitro* and the limited available tools for designing complicated RNA nanostructures. However, RNA assemblies may possess advantages over their DNA counterpart because of the higher structural and functional diversity of RNA. Moreover, RNA could fold co-transcriptionally, transcriptionally *in vitro*, and *in vivo*. It would thus be appealing to develop RNA nanomachines that can form and function *in vivo* for applications such as gene regulation and signal transduction.

When considering the performance and efficacy of DNA-based drug delivery systems, the behaviors of DNA origami structures in complex biochemical environments are still not fully understood. In recent years, several groups reported PEGylated oligolysine can enhance the stability of DNA origami in physiological conditions.[512-514] For instance, polymer-covered DNA nanostructures exhibited remarkable nuclease resistance and prolonged circulating half-life *in vivo*.[515] In addition, photo-cross-linking could improve the stability of DNA nanostructures that contain acrydite modification[516] or thymines in close proximity[517]. Despite a series of careful studies,[518-523] the pharmacokinetics, biodistribution, and clearance mechanisms of DNA origami structures *in vivo* as well as their cellular uptake and trafficking, all of which are crucial for biomedical applications, need to be better characterized. In particular, the interactions between DNA origami with macromolecules in biological environments remain to be a relatively underexplored area.

As an exciting new a frontier in DNA nanotechnology, DNA-engineered membrane materials have the potential to transform basic research in membrane biology, synthetic biology, and nanomedicine. To push this frontier, researchers could devote endeavors to the following areas. First, the success of current membrane manipulation methods hinges on selecting suitable membrane anchors for DNA nanodevices. Although DNA oligonucleotides with certain hydrophobic modifications (*e.g.*, cholesterols[410]) are readily available from commercial sources, many other useful DNA modifications (*e.g.*, alkyl chains[415]) require in-house DNA synthesis or conjugation reactions. As recent studies pointed out,[400, 411] subtle differences in chemical structure could profoundly impact the membrane anchor's ability to recruit and insert into membranes. Therefore, the precision and versatility of the DNA-based membrane-engineering devices could be vastly enhanced with a library of chemically modified DNA oligonucleotides with a spectrum of size and hydrophobicity. Second, while the existing DNA-based methods focus on controlling the geometry and chemical modification of membranes (including the membrane-integrated structures, *e.g.*, nanopores), very few DNA nanodevices have been developed to directly control or respond to the mechanical and electrical properties of the membranes.[524] Because membrane tension and potential play vital roles in cell signaling and metabolism, the ability to modulate and measure these properties in biological and synthetic systems is highly desirable to fully understand the molecular determinants of cellular processes. Third, the complexity of cell membranes, such as the leaflet asymmetry, underlying cytoskeleton, as well as various peripheral and integral membrane proteins, presents both challenges and opportunities to the next generation of DNA nanodevices that act on living cells.[525] So far, DNA nanostructures have been built to deliver molecular payloads to cells,[341, 526] detect cellular contraction forces,[527] activate cell surface receptors,[4, 329, 528] puncture cell membranes,[529] and cluster cells.[530] In the future, it is conceivable that DNA nanodevices could extract patches of cell membranes, dynamically remodel cell surface landscape, as well as communicate intracellular signals among cells. Fourth, it would be possible to scale up the complexity of the membrane engineering tools by building DNA devices that work collaboratively to accomplish multi-step tasks on the membranes, such as sorting, tagging, packaging, and sending membrane-associated molecular cargos, or by organizing multiple DNA-engineered membrane compartments into interconnected chemical reactors,[531] similar to a network of neurons passing biochemical information to one another. Li *et al*. showed that transmembrane nanopores (~7 nm wide) could be coupled to micrometer-long DNA channels, which facilitated the translocation of molecules as small as 0.8 nm in radius with minimal leakage, providing a promising method for direct, long-range inter-cell communication.[532] Fifth, the DNA-enabled membrane engineering techniques may find applications in biomedicine, such as formulating therapeutic exosomes or lipid nanoparticles, as well as in DNA computing, like compartmentalizing data storage and computing units for more stable and controllable operations.

From the perspective of nanophotonics, DNA origami has solidified its role in this field in the past years, but challenges and exciting opportunities still lie ahead. First, 3D nanofabrication with arbitrary geometries is one of the extraordinary capabilities of DNA origami. However, to achieve considerable scattering or absorption cross-sections for generating pronounced plasmonic effects, large metallic NPs are needed. To date, spherical AuNPs up to 150 nm in diameter[533] and anisotropic NPs, such as AuNRs up to 50 nm in length[484], have been successfully applied in DNA origami-templated nanophotonic structures. Functionalizing larger metallic NPs and stabilizing them in salty environments that DNA origami structures prefer are tricky. New chemical functionalization strategies for stable conjugation of large NPs with DNA origami are thus desirable. Second, to build complex nanophotonic architectures, sufficiently large origami templates are required, for instance, to host big NPs or even to form lattices with long-range order. The dimensions of origami structures folded from single scaffolds are restricted by the lengths of the scaffolds. Even though longer scaffolds can be used, it does not essentially solve the problem of scaling up DNA origami. Large DNA origami can be created by the so-called superorigami approach,[2, 534] which utilizes a long scaffolding strand to link multiple origami components together. Nevertheless, the superorigami still has the issues of restricted geometry, symmetry, and size. Alternatively, large origami superstructures can be created by polymerizing origami monomers through blunt-end,[74, 76, 535-536] sticky-end,[368, 537-538] or shape complementary interactions.[59] Third, in many optical applications, the optical components in the super cell of a lattice should not be regularly reiterated. For instance, optical metasurfaces, a rapidly developing research subject in nanophotonics, comprise arrays of antennas that can modulate the amplitude, phase, and polarization of the incident light. In a super unit cell of the metasurface based on the Pancharatnam-Berry phase, the optical antennas (*e.g.*, AuNRs) are arranged with different orientations to generate a complete phase coverage from 0 to $2\pi$ for light modulation. Therefore, the individual origami components in the super cell should have high specificity to position the AuNRs with designated orientations. In addition, periodic patterning of these origami super cells in a lattice with sufficient size on the order of tens of micrometers is required for experimental characterization and good optical performance. Inspiring design strategies are highly encouraged to develop for such technical challenges.[539] The efforts along this line will be very rewarding, because DNA origami-templated optical metasurfaces could possess unprecedented properties, including addressability, programmability, and reconfigurability on the single antenna level, which cannot yet be offered by other nanotechniques at visible frequencies.

Fourth, dynamic reconfiguration is another distinct advantage of DNA origami. A multitude of external inputs to dynamically reconfigure the assembled nanostructures have been demonstrated, including temperature control,[481] pH changes,[540] ion concentration,[59] light,[541] DNA,[206] RNA,[479] aptamers,[480] enzymes, proteins,[542] small molecules,[543-544] among others.[545-547] A common limitation of the dynamic DNA structures was the reconfiguration rate, typically below 1 nm/min. A remarkable DNA-based AuNP motor powered by chemical fuels was demonstrated by the Salaita group.[548] It could processively translocate on a functionalized flat surface at an average speed of 50 nm per second. In addition, the Dietz and Simmel groups reported an electrically-driven DNA origami rotary ratchet motor with a rotational speed of up to 250 revolutions per min,[231] which approached already the speed of natural molecular rotors, ATP synthase.[549] These exciting examples open the pathway to realizing dynamic nanophotonic devices with high speed and excellent performance. Fifth, although optical elements of different materials have been successfully functionalized on DNA origami, AuNPs and AgNPs remain the major choices for optical antennas.[550] Other metals, such as palladium, platinum, magnesium (Mg), aluminum (Al), and their alloys with noble metals exhibit interesting catalytic properties, yet their good optical responses, remained vastly unexplored. Particularly, Mg and Al are also excellent candidates for ultraviolet (UV) plasmonics, as their permittivity becomes negative in the UV range. Another important aspect is that metals are very lossy due to their high absorption in the visible range. All-dielectric NPs, such as Si (silicon) and $TiO_2$ (titanium dioxide), that exhibit Mie resonances with low losses have so far been overlooked for DNA-guided assembly. Protocols for functionalizing all-dielectric NPs on DNA origami will be a promising solution to creating low-loss nanophotonic devices at visible frequencies.

Sixth, long-term high performance without degradation is the fundamental prerequisite for practical optical devices. This property is still critical for DNA-assembled nanomaterials, which generally degrade over time. Nguyen *et al*. suggested coating DNA origami with an ultrathin layer of silica to prevent the origami from degradation and aggregation.[264] However, the resulting DNA origami post silica-coating would lose most of the modification and functionalization possibilities. Strategies to enhance the rigidity and stability of DNA origami and maintain all the DNA functions are worthy of further investigation. One further aspect is the stability of the optical elements, such as metallic NPs and quantum emitters. This investigation is vital not only for DNA nanotechnology but also for self-assembly in general. Seventh, the combination of DNA self-assembly and top-down techniques is a powerful route to constructing a new generation of nanophotonic architectures with advanced optical properties. One of the significant challenges is to accurately control the positions and orientations of the DNA origami-assembled nanostructures during their immobilization on a substrate. The Gopinath, Cha and Rothemund groups are the pioneers who brought a remarkable leap forward in exploring DNA origami patterning on solid supports.[493, 551-553] To achieve nanophotonic devices with designated long-range order, for instance, the aforementioned dynamic optical metasurfaces based on the PB phase, innovative protocols have to be developed to enable the correct positioning of each antenna within a lattice and meanwhile achieve the dynamic control of each antenna orientation in a fully independent fashion.

Taking together, since the original proposal of Nadrian Seeman, the field of DNA nanotechnology has flourished on a scale that was unimaginable 40 years ago. The invention of DNA origami by Paul Rothemund has further pushed this field to a new horizon and fostered a plethora of concepts, models, methodologies, and applications that were not thought of before. In particular, the DNA origami-engineered nanomaterials have brought exciting and vastly unexplored research avenues in materials science, greatly enriching the portfolio of DNA-based applications. Undoubtedly, the synergetic efforts and collaborations among scientists with different research backgrounds will continue to bring innovations to this field in the next decade. There shall always be plenty of room at the bottom to explore and we will certainly have much to celebrate at the 50$^{th}$ anniversary of DNA nanotechnology.

## AUTHOR INFORMATION

Corresponding Authors
Baoquan Ding
*E-mail: dingbq@nanoctr.cn
National Center for Nanoscience and Technology, No 11, BeiYiTiao Zhongguancun, Beijing 100190, China
Chenxiang Lin
*E-mail: chenxiang.lin@yale.edu
Department of Cell Biology, Yale School of Medicine, 333 Cedar Street, New Haven, Connecticut 06520, United States
Nanobiology Institute, Yale University, 850 West Campus Drive, West Haven, Connecticut 06516, United States
Department of Biomedical Engineering, Yale University, 17 Hillhouse Ave, New Haven, Connecticut 06511, United States
Yonggang Ke
*E-mail: yonggang.ke@emory.edu
Wallace H. Coulter Department of Biomedical Engineering, Georgia Institute of Technology and Emory University, Atlanta, GA 30322, USA
Na Liu
*E-mail: na.liu@pi2.uni-stuttgart.de
2nd Physics Institute, University of Stuttgart, Pfaffenwaldring 57, 70569 Stuttgart, Germany
Max Planck Institute for Solid State Research, Heisenbergstr. 1, 70569 Stuttgart, Germany

### Authors

Pengfei Zhan
2nd Physics Institute, University of Stuttgart, Pfaffenwaldring 57, 70569 Stuttgart, Germany
Andreas Peil
2nd Physics Institute, University of Stuttgart, Pfaffenwaldring 57, 70569 Stuttgart, Germany
Qiao Jiang
National Center for Nanoscience and Technology, No 11, BeiYiTiao Zhongguancun, Beijing 100190, China
Dongfang Wang
School of Biomedical Engineering and Suzhou Institute for Advanced Research, University of Science and Technology of China, Suzhou 215123, China
Shikufa Mousavi
Department of Chemistry, Emory University, Atlanta, GA 30322, USA
Qiancheng Xiong
Department of Cell Biology, Yale School of Medicine, 333 Cedar Street, New Haven, Connecticut 06520, United States
Nanobiology Institute, Yale University, 850 West Campus Drive, West Haven, Connecticut 06516, United States

Qi Shen
Department of Cell Biology, Yale School of Medicine, 333 Cedar Street, New Haven, Connecticut 06520, United States
Nanobiology Institute, Yale University, 850 West Campus Drive, West Haven, Connecticut 06516, United States
Department of Molecular Biophysics and Biochemistry, Yale University, 266 Whitney Avenue, New Haven, Connecticut 06511, United States
Yingxu Shang
National Center for Nanoscience and Technology, No 11, BeiYiTiao Zhongguancun, Beijing 100190, China

## ORCID

Pengfei Zhan: 0000-0002-1929-4403
Qiao Jiang: 0000-0003-1016-9021
Dongfang Wang: 0000-0001-8112-3149
Qiancheng Xiong: 0000-0002-9897-7391
Qi Shen: 0000-0002-9062-8955
Baoquan Ding: 0000-0003-1095-8872
Chenxiang Lin: 0000-0001-7041-1946
Yonggang Ke: 0000-0003-1673-2153
Na Liu: 0000-0001-5831-3382

## Notes

The authors declare no competing financial interest.

## Biographies

Pengfei Zhan received his Ph.D. in Physical Chemistry from University of Chinese Academy of Sciences. He is currently a Postdoc at the 2. Physics Institute, University of Stuttgart.

Andreas Peil received his B.S. degree in Biochemistry and M.S. degree in Chemistry from University of Bayreuth. He is currently a Ph. D student at the 2. Physics Institute, University of Stuttgart.

Qiao Jiang received her B.S. degree from Xi'an Jiaotong University and a Ph.D. in Biology from National Center for Nanoscience and Technology. Following a period as a post-doctoral research fellow in the laboratories at Institute of Chemistry, Chinese Academy of Sciences, she joined NCNST as an assistant professor.

Dongfang Wang received his B.S. degree from Dalian Medical University and a Ph.D degree in Chemistry from Shanghai Institute of Applied Physics, University of Chinese Academy of Sciences. He is now the tenure-track professor at the School of Biomedical Engineering and Suzhou Institute for Advanced Research, University of Science and Technology of China.

Shikufa Mousavi received her B.S. degree in Chemistry from Lahore University of Management Sciences, (LUMS) Pakistan in 2021. She is currently a chemistry graduate student in the Ke lab at Emory University. Her research interests are leveraging DNA nanotechnology to create bio-inspired DNA nanostructures and studying their applications.Qiancheng Xiong received his B.A. degree in Natural Sciences from Cambridge University. He is currently a Ph.D. candidate in Cell Biology at Yale University.

Qiancheng Xiong received his B.A. degree in Natural Sciences from Cambridge University. He is currently a Ph.D. candidate in Cell Biology at Yale University.

Qi Shen obtained his Ph.D. in Physical Chemistry from Peking University in 2013. He is currently an Associate Research Scientist in the Department of Molecular Biophysics and Biochemistry and the Department of Cell Biology at Yale University.

Yingxu Shang received his B.S. degree in Material Science and Engineering from Beihang University and a Ph.D. from National Center for Nanoscience and Technology (NCNST), Chinese Academy of Sciences (CAS). He is currently working as a Special Research Assistant at NCNST.

Baoquan Ding received his B.S. degree in Chemistry from Jilin University and a Ph.D. in Chemistry from New York University. He then worked as a postdoctoral research fellow at the Molecular Foundry, Lawrence Berkeley National Laboratory. He joined the Biodesign Institute, Arizona State University as a research assistant professor. He is currently a professor at the National Center for Nanoscience and Technology.

Chenxiang Lin studied chemistry at Peking University, completed his Ph.D. thesis on DNA nanotechnology at Arizona State University, and received postdoctoral training at Dana-Farber Cancer Institute and the Wyss Institute at Harvard University. He is currently an Associate Professor of Cell Biology and Biomedical Engineering at Yale University.

Yonggang Ke received his B.S. degree in Chemistry from Peking University and Ph.D. degree in Chemistry and Biochemistry from Arizona State University. He is currently an Associate Professor at Wallace H. Coulter Department of Biomedical Engineering, Georgia Institute of Technology and Emory University.

Na Liu received her B.S. degree in Physics from Jilin University and a Ph.D. in Physics from the University of Stuttgart. She is currently the Professor and Director of the 2. Physics Institute at the University of Stuttgart.

## Acknowledgments

This work was supported by the European Union’s Horizon 2020 research and innovation program under Grant No. 964995 “DNA-FAIRYLIGHTS”, by the Deutsche Forschungsgemeinschaft (DFG, German Research Foundation)-448727036, and by the Baden-Württemberg Stiftung (Internationale Spitzenforschung, BWST-ISF2020-19).

## List of Abbreviations

| | |
|---|---|
| ssDNA | single-stranded DNA |
| 1D | one dimesion |
| 2D | two dimesion |
| 3D | three dimesion |
| DX | double-crossover |
| SSTs | single-stranded DNA tiles |
| nt | nucleotides |
| PCR | polymerase chain reaction |
| V brick | V-shaped DNA origami structure |
| DBCO | dibenzocyclooctyl |
| HDO | hexagonal prism DNA origami |
| RH | right-handed |
| LH | left-handed |
| AFM | atomic force microscopy |
| aTAM | abstract tile assembly model |
| kTAM | kinetic tile assembly model |
| ΔGSE° | the free energy change of each individual pair of sticky ends |
| ΔG° | the free energy change of single tile attachment |
| BB | boundary bricks |
| MFPT | the mean first-passage time |
| ssOrigami | ssDNA and ssRNA origami |
| mDNA | meta-DNA |
| AuNPs | gold nanoparticles |
| KL | kissing loop |
| bKL | branched kissing loops |
| HCR | hybridization chain reaction |
| TMAPS | N-trimethoxysilylpropyl-N,N,N-trimethylammonium chloride |
| TEOS | triethoxysilane |
| SAXS | small-angle X-ray scattering |
| Ag | silver |
| Nb | niobium |
| CaP | Calcium phosphate |
| ACP | amorphous calcium phosphate |
| $CaCl_2$ | calcium chloride |
| SWNTs | single-walled carbon nanotubes |
| HPLC | high-performance liquid chromatography |
| SNAs | spherical nucleic acids |
| FET | field-effect transistor |
| MMONs | metal and metal oxide nanoclusters |
| Pd | palladium |
| Fe | iron |
| Co | cobalt |
| Ni | nickel |
| Au | gold |
| AuNRs/TeNRs | gold/tellurium nanorods |
| PBI | Polybenzimidazole |
| IONPs | iron oxide nanoparticles |
| PMAO | poly(maleic anhydride-alt-octadecene) |
| HB | helix bundle |
| MRI | magnetic resonance imaging |
| APPV-DNA | (2,5-dialkoxy)paraphenylene vinylene-DNA |
| poly(F-DNA) | polyfluorene-DNA |
| BCP | block copolymer |
| P3(EO)3T | polythiophene poly(3-tri(ethylene glycol)thiophene) |
| DDAO | N,N-dimethyldodecylamine N-oxide |
| PD | polydopamine |
| G4 | G-quadruplex |
| HRP | horseradish peroxidase |
| DNAzyme | DNA enzyme |
| ATRP | atom transfer radical polymerization |
| EY | eosin Y |
| MB | methylene blue |
| ROS | reactive oxygen species |
| TdT | terminal deoxynucleotidyl transferase |
| $NADP^+$/NADPH | Nicotinamide adenine dinucleotide phosphate |
| FFE | Free-flow electrophoresis |
| GOx | glucose oxidase |
| SPDP | succinimidyl 3-(2-pyridyldithio) propionate |
| G6Pdh | glucose6-phosphate dehydrogenase |
| LDH | lactate dehydrogenase |
| $NAD^+$ | nicotinamide adenine dinucleotide |
| HJ | holliday junction |
| AZO | azobenzene |
| SLB | supported lipid bilayer |
| CHOL | cholesterol |
| TIRFM | total internal reflection fluorescence microscopy |
| DODA | DNA origami domino array |
| RNAP | RNA polymerase |
| BCN | bicyclononyne |
| ORBIT | origami-rotor-based imaging and tracking |
| NTA | nitrilotriacetate |
| IgG | immunoglobulin G |
| BS3 | bis(sulfosuccinimidyl) |
| DIG | digoxigenin |
| NIP | 4-hydroxy-3-iodo-5-Nitrophenylacetate |
| NP | 4-hydroxy-3-nitrophenyl |
| SPR | surface plasmon resonance |
| HS-AFM | high-speed atomic force microscopy |
| TMV | tobacco mosaic virus |
| TAL | transcription activator-like |
| TLR9 | Toll-like receptor 9 |
| PD-L1 | programmed death-ligand 1 |
| TCR | T cell antigen receptor |
| pMHC | peptide/major histocompatibility complex |
| AKI | acute kidney injury |
| DONs | DNA origami nanostructures |
| PET | positron emission tomography |
| ROS | reactive oxygen species |
| FA | folic acid |
| RA | rheumatoid arthritis |
| NO | nitric oxide |
| HBV | hepatitis B virus |
| AAV2 | adeno-associated virus serotype 2 |
| siRNAs | small interfering RNA strands |
| GSH | glutathione |
| DCs | dendritic cells |
| dLNs | draining lymph nodes |
| ESCRT | endosomal sorting complex required for transport |
| VAMP2 | vesicle-associated membrane protein-2 |
| E-Syt1 | extended synaptotagmin 1 |
| GUVs | giant unilamellar vesicles |

LUVs large unilamellar vesicles

SUVs small unilamellar vesicles

FRAP fluorescence recovery after photobleaching

CtxA copies of peptide ceratotoxin A

NEOtrap nanopore electro-osmotic trap

LSPRs localized surface plasmon resonances

NV nitrogen-vacancy

FRET Förster resonance energy transfer

CD circular dichroism

CPL circularly polarized light

AuNRs gold nanorods

SERS surface-enhanced Raman spectroscopy

DNA-PAINT points accumulation for imaging in nanoscale topography

PCP peridinin-chlorophyll a-protein

NPoM nanoparticle-on-a-mirror

ROX carboxy-X-rhodamine

AuNCs gold nanocubes

Mg magnesium

Al aluminium

UV ultraviolet

Si silicon

$TiO_2$ titanium dioxide

**References**

1. Seeman, N. C., Nucleic acid junctions and lattices. *Journal of theoretical biology* **1982,** *99* (2), 237-247.
2. Hong, F.; Zhang, F.; Liu, Y.; Yan, H., DNA Origami: Scaffolds for Creating Higher Order Structures. *Chem. Rev.* **2017,** *117* (20), 12584-12640.
3. Dey, S.; Fan, C.; Gothelf, K. V.; Li, J.; Lin, C.; Liu, L.; Liu, N.; Nijenhuis, M. A.; Saccà, B.; Simmel, F. C., DNA origami. *Nature Reviews Methods Primers* **2021,** *1* (1), 1-24.

4. Knappe, G. A.; Wamhoff, E.-C.; Bathe, M., Functionalizing DNA origami to investigate and interact with biological systems. *Nat Rev Mater* **2022**, 1-16.
5. Kallenbach, N. R.; Ma, R.-I.; Seeman, N. C., An immobile nucleic acid junction constructed from oligonucleotides. *Nature* **1983,** *305* (5937), 829-831.
6. Chen, J.; Seeman, N. C., Synthesis from DNA of a molecule with the connectivity of a cube. *Nature* **1991,** *350* (6319), 631-633.
7. Zhang, Y.; Seeman, N. C., Construction of a DNA-Truncated Octahedron. *J. Am. Chem. Soc.* **1994,** *116* (5), 1661-1669.
8. Mao, C.; Sun, W.; Seeman, N. C., Assembly of Borromean rings from DNA. *Nature* **1997,** *386* (6621), 137-138.
9. Fu, T. J.; Seeman, N. C., DNA double-crossover molecules. *Biochemistry* **1993,** *32* (13), 3211-3220.
10. Li, X.; Yang, X.; Qi, J.; Seeman, N. C., Antiparallel DNA Double Crossover Molecules As Components for Nanoconstruction. *J. Am. Chem. Soc.* **1996,** *118* (26), 6131-6140.
11. Winfree, E.; Liu, F.; Wenzler, L. A.; Seeman, N. C., Design and self-assembly of two-dimensional DNA crystals. *Nature* **1998,** *394* (6693), 539-544.
12. Liu, F.; Sha, R.; Seeman, N. C., Modifying the Surface Features of Two-Dimensional DNA Crystals. *J. Am. Chem. Soc.* **1999,** *121* (5), 917-922.
13. LaBean, T. H.; Yan, H.; Kopatsch, J.; Liu, F.; Winfree, E.; Reif, J. H.; Seeman, N. C., Construction, Analysis, Ligation, and Self-Assembly of DNA Triple Crossover Complexes. *J. Am. Chem. Soc.* **2000,** *122* (9), 1848-1860.
14. Reishus, D.; Shaw, B.; Brun, Y.; Chelyapov, N.; Adleman, L., Self-Assembly of DNA Double-Double Crossover Complexes into High-Density, Doubly Connected, Planar Structures. *J. Am. Chem. Soc.* **2005,** *127* (50), 17590-17591.
15. Ke, Y.; Liu, Y.; Zhang, J.; Yan, H., A Study of DNA Tube Formation Mechanisms Using 4-, 8-, and 12-Helix DNA Nanostructures. *J. Am. Chem. Soc.* **2006,** *128* (13), 4414-4421.
16. Mao, C.; Sun, W.; Seeman, N. C., Designed Two-Dimensional DNA Holliday Junction Arrays Visualized by Atomic Force Microscopy. *J. Am. Chem. Soc.* **1999,** *121* (23), 5437-5443.
17. Yan, H.; Park, S. H.; Finkelstein, G.; Reif, J. H.; LaBean, T. H., DNA-Templated Self-Assembly of Protein Arrays and Highly Conductive Nanowires. *Science* **2003,** *301* (5641), 1882-1884.
18. Chelyapov, N.; Brun, Y.; Gopalkrishnan, M.; Reishus, D.; Shaw, B.; Adleman, L., DNA Triangles and Self-Assembled Hexagonal Tilings. *J. Am. Chem. Soc.* **2004,** *126* (43), 13924-13925.
19. Ding, B.; Sha, R.; Seeman, N. C., Pseudohexagonal 2D DNA Crystals from Double Crossover Cohesion. *J. Am. Chem. Soc.* **2004,** *126* (33), 10230-10231.
20. Liu, D.; Wang, M.; Deng, Z.; Walulu, R.; Mao, C., Tensegrity: Construction of Rigid DNA Triangles with Flexible Four-Arm DNA Junctions. *J. Am. Chem. Soc.* **2004,** *126* (8), 2324-2325.
21. He, Y.; Chen, Y.; Liu, H.; Ribbe, A. E.; Mao, C., Self-Assembly of Hexagonal DNA Two-Dimensional (2D) Arrays. *J. Am. Chem. Soc.* **2005,** *127* (35), 12202-12203.
22. Lund, K.; Liu, Y.; Lindsay, S.; Yan, H., Self-Assembling a Molecular Pegboard. *J. Am. Chem. Soc.* **2005,** *127* (50), 17606-17607.
23. Park, S. H.; Pistol, C.; Ahn, S. J.; Reif, J. H.; Lebeck, A. R.; Dwyer, C.; LaBean, T. H., Finite-Size, Fully Addressable DNA Tile Lattices Formed by Hierarchical Assembly Procedures. *Angew. Chem., Int. Ed.* **2006,** *45* (5), 735-739.
24. Rothemund, P. W. K., Folding DNA to create nanoscale shapes and patterns. *Nature* **2006,** *440* (7082), 297-302.
25. Yan, H.; LaBean Thomas, H.; Feng, L.; Reif John, H., Directed nucleation assembly of DNA tile complexes for barcode-patterned lattices. *Proc. Natl. Acad. Sci. U. S. A.* **2003,** *100* (14), 8103-8108.
26. Shih, W. M.; Quispe, J. D.; Joyce, G. F., A 1.7-kilobase single-stranded DNA that folds into a nanoscale octahedron. *Nature* **2004,** *427* (6975), 618-621.

27. Douglas, S. M.; Dietz, H.; Liedl, T.; Högberg, B.; Graf, F.; Shih, W. M., Self-assembly of DNA into nanoscale three-dimensional shapes. *Nature* **2009,** *459* (7245), 414-418.
28. Douglas, S. M.; Marblestone, A. H.; Teerapittayanon, S.; Vazquez, A.; Church, G. M.; Shih, W. M., Rapid prototyping of 3D DNA-origami shapes with caDNAno. *Nucleic Acids Res.* **2009,** *37* (15), 5001-5006.
29. Ke, Y.; Douglas, S. M.; Liu, M.; Sharma, J.; Cheng, A.; Leung, A.; Liu, Y.; Shih, W. M.; Yan, H., Multilayer DNA Origami Packed on a Square Lattice. *J. Am. Chem. Soc.* **2009,** *131* (43), 15903-15908.
30. Ke, Y.; Voigt, N. V.; Gothelf, K. V.; Shih, W. M., Multilayer DNA Origami Packed on Hexagonal and Hybrid Lattices. *J. Am. Chem. Soc.* **2012,** *134* (3), 1770-1774.
31. Dietz, H.; Douglas Shawn, M.; Shih William, M., Folding DNA into Twisted and Curved Nanoscale Shapes. *Science* **2009,** *325* (5941), 725-730.
32. Han, D.; Pal, S.; Nangreave, J.; Deng, Z.; Liu, Y.; Yan, H., DNA Origami with Complex Curvatures in Three-Dimensional Space. *Science* **2011,** *332* (6027), 342-346.
33. Han, D.; Pal, S.; Yang, Y.; Jiang, S.; Nangreave, J.; Liu, Y.; Yan, H., DNA Gridiron Nanostructures Based on Four-Arm Junctions. *Science* **2013,** *339* (6126), 1412-1415.
34. Zhang, F.; Jiang, S.; Wu, S.; Li, Y.; Mao, C.; Liu, Y.; Yan, H., Complex wireframe DNA origami nanostructures with multi-arm junction vertices. *Nat. Nanotechnol.* **2015,** *10* (9), 779-784.
35. Benson, E.; Mohammed, A.; Gardell, J.; Masich, S.; Czeizler, E.; Orponen, P.; Hogberg, B., DNA rendering of polyhedral meshes at the nanoscale. *Nature* **2015,** *523* (7561), 441-444.
36. Veneziano, R.; Ratanalert, S.; Zhang, K.; Zhang, F.; Yan, H.; Chiu, W.; Bathe, M., Designer nanoscale DNA assemblies programmed from the top down. *Science* **2016,** *352* (6293), 1534-1534.
37. Iinuma, R.; Ke, Y.; Jungmann, R.; Schlichthaerle, T.; Woehrstein Johannes, B.; Yin, P., Polyhedra Self-Assembled from DNA Tripods and Characterized with 3D DNA-PAINT. *Science* **2014,** *344* (6179), 65-69.
38. Yin, P.; Hariadi Rizal, F.; Sahu, S.; Choi Harry, M. T.; Park Sung, H.; LaBean Thomas, H.; Reif John, H., Programming DNA Tube Circumferences. *Science* **2008,** *321* (5890), 824-826.
39. Wei, B.; Dai, M.; Yin, P., Complex shapes self-assembled from single-stranded DNA tiles. *Nature* **2012,** *485* (7400), 623-626.
40. Ke, Y.; Ong Luvena, L.; Shih William, M.; Yin, P., Three-Dimensional Structures Self-Assembled from DNA Bricks. *Science* **2012,** *338* (6111), 1177-1183.
41. Ke, Y.; Ong, L. L.; Sun, W.; Song, J.; Dong, M.; Shih, W. M.; Yin, P., DNA brick crystals with prescribed depths. *Nat. Chem.* **2014,** *6* (11), 994-1002.
42. Wei, X.; Nangreave, J.; Jiang, S.; Yan, H.; Liu, Y., Mapping the Thermal Behavior of DNA Origami Nanostructures. *J. Am. Chem. Soc.* **2013,** *135* (16), 6165-6176.
43. Dunn, K. E.; Dannenberg, F.; Ouldridge, T. E.; Kwiatkowska, M.; Turberfield, A. J.; Bath, J., Guiding the folding pathway of DNA origami. *Nature* **2015,** *525* (7567), 82-86.
44. Sobczak Jean-Philippe, J.; Martin Thomas, G.; Gerling, T.; Dietz, H., Rapid Folding of DNA into Nanoscale Shapes at Constant Temperature. *Science* **2012,** *338* (6113), 1458-1461.
45. Arbona, J.-M.; Aimé, J.-P.; Elezgaray, J., Cooperativity in the annealing of DNA origamis. *J. Chem. Phys.* **2013,** *138* (1), 015105.
46. Martin, T. G.; Dietz, H., Magnesium-free self-assembly of multi-layer DNA objects. *Nat. Commun.* **2012,** *3* (1), 1103.
47. Lee Tin Wah, J.; David, C.; Rudiuk, S.; Baigl, D.; Estevez-Torres, A., Observing and Controlling the Folding Pathway of DNA Origami at the Nanoscale. *ACS Nano* **2016,** *10* (2), 1978-1987.
48. Majikes, J. M.; Nash, J. A.; LaBean, T. H., Competitive annealing of multiple DNA origami: formation of chimeric origami. *New J. Phys.* **2016,** *18* (11), 115001.
49. Schneider, F.; Moritz, N.; Dietz, H., The sequence of events during folding of a DNA origami. *Sci Adv* **2019,** *5* (5), eaaw1412.

50. Zhang, H.; Chao, J.; Pan, D.; Liu, H.; Huang, Q.; Fan, C., Folding super-sized DNA origami with scaffold strands from long-range PCR. *Chem. Commun.* **2012,** *48* (51), 6405-6407.
51. Marchi, A. N.; Saaem, I.; Vogen, B. N.; Brown, S.; LaBean, T. H., Toward Larger DNA Origami. *Nano Lett.* **2014,** *14* (10), 5740-5747.
52. Jungmann, R.; Scheible, M.; Kuzyk, A.; Pardatscher, G.; Castro, C. E.; Simmel, F. C., DNA origami-based nanoribbons: assembly, length distribution, and twist. *Nanotechnology* **2011,** *22* (27), 275301.
53. Li, Z.; Liu, M.; Wang, L.; Nangreave, J.; Yan, H.; Liu, Y., Molecular Behavior of DNA Origami in Higher-Order Self-Assembly. *J. Am. Chem. Soc.* **2010,** *132* (38), 13545-13552.
54. Tigges, T.; Heuser, T.; Tiwari, R.; Walther, A., 3D DNA Origami Cuboids as Monodisperse Patchy Nanoparticles for Switchable Hierarchical Self-Assembly. *Nano Lett.* **2016,** *16* (12), 7870-7874.
55. Liu, W.; Zhong, H.; Wang, R.; Seeman, N. C., Crystalline Two-Dimensional DNA-Origami Arrays. *Angew. Chem., Int. Ed.* **2011,** *50* (1), 264-267.
56. Rajendran, A.; Endo, M.; Hidaka, K.; Sugiyama, H., Control of the two-dimensional crystallization of DNA origami with various loop arrangements. *Chem. Commun.* **2013,** *49* (7), 686-688.
57. Wang, P.; Gaitanaros, S.; Lee, S.; Bathe, M.; Shih, W. M.; Ke, Y., Programming Self-Assembly of DNA Origami Honeycomb Two-Dimensional Lattices and Plasmonic Metamaterials. *J. Am. Chem. Soc.* **2016,** *138* (24), 7733-7740.
58. Woo, S.; Rothemund, P. W. K., Programmable molecular recognition based on the geometry of DNA nanostructures. *Nat. Chem.* **2011,** *3* (8), 620-627.
59. Gerling, T.; Wagenbauer, K. F.; Neuner, A. M.; Dietz, H., Dynamic DNA devices and assemblies formed by shape-complementary, non–base pairing 3D components. *Science* **2015,** *347* (6229), 1446-1452.
60. Bruetzel, L. K.; Gerling, T.; Sedlak, S. M.; Walker, P. U.; Zheng, W.; Dietz, H.; Lipfert, J., Conformational Changes and Flexibility of DNA Devices Observed by Small-Angle X-ray Scattering. *Nano Lett.* **2016,** *16* (8), 4871-4879.
61. Tikhomirov, G.; Petersen, P.; Qian, L., Programmable disorder in random DNA tilings. *Nat. Nanotechnol.* **2017,** *12* (3), 251-259.
62. Endo, M.; Sugita, T.; Rajendran, A.; Katsuda, Y.; Emura, T.; Hidaka, K.; Sugiyama, H., Two-dimensional DNA origami assemblies using a four-way connector. *Chem. Commun.* **2011,** *47* (11), 3213-3215.
63. Endo, M.; Sugita, T.; Katsuda, Y.; Hidaka, K.; Sugiyama, H., Programmed-Assembly System Using DNA Jigsaw Pieces. *Chem. Eur. J.* **2010,** *16* (18), 5362-5368.
64. Rajendran, A.; Endo, M.; Katsuda, Y.; Hidaka, K.; Sugiyama, H., Programmed Two-Dimensional Self-Assembly of Multiple DNA Origami Jigsaw Pieces. *ACS Nano* **2011,** *5* (1), 665-671.
65. Kopielski, A.; Csaki, A.; Fritzsche, W., Surface Mobility and Ordered Rearrangement of Immobilized DNA Origami. *Langmuir* **2015,** *31* (44), 12106-12110.
66. Aghebat Rafat, A.; Pirzer, T.; Scheible, M. B.; Kostina, A.; Simmel, F. C., Surface-Assisted Large-Scale Ordering of DNA Origami Tiles. *Angew. Chem., Int. Ed.* **2014,** *53* (29), 7665-7668.
67. Ramakrishnan, S.; Subramaniam, S.; Stewart, A. F.; Grundmeier, G.; Keller, A., Regular Nanoscale Protein Patterns *via* Directed Adsorption through Self-Assembled DNA Origami Masks. *ACS Appl. Mater. Interfaces* **2016,** *8* (45), 31239-31247.
68. Woo, S.; Rothemund, P. W. K., Self-assembly of two-dimensional DNA origami lattices using cation-controlled surface diffusion. *Nat. Commun.* **2014,** *5* (1), 4889.
69. Suzuki, Y.; Endo, M.; Sugiyama, H., Mimicking Membrane-Related Biological Events by DNA Origami Nanotechnology. *ACS Nano* **2015,** *9* (4), 3418-3420.
70. Johnson-Buck, A.; Jiang, S.; Yan, H.; Walter, N. G., DNA–Cholesterol Barges as

Programmable Membrane-Exploring Agents. *ACS Nano* **2014,** *8* (6), 5641-5649.
71. Czogalla, A.; Kauert, D. J.; Seidel, R.; Schwille, P.; Petrov, E. P., DNA Origami Nanoneedles on Freestanding Lipid Membranes as a Tool To Observe Isotropic–Nematic Transition in Two Dimensions. *Nano Lett.* **2015,** *15* (1), 649-655.
72. Czogalla, A.; Kauert, D. J.; Franquelim, H. G.; Uzunova, V.; Zhang, Y.; Seidel, R.; Schwille, P., Amphipathic DNA Origami Nanoparticles to Scaffold and Deform Lipid Membrane Vesicles. *Angew. Chem., Int. Ed.* **2015,** *54* (22), 6501-6505.
73. Suzuki, Y.; Endo, M.; Yang, Y.; Sugiyama, H., Dynamic Assembly/Disassembly Processes of Photoresponsive DNA Origami Nanostructures Directly Visualized on a Lipid Membrane Surface. *J. Am. Chem. Soc.* **2014,** *136* (5), 1714-1717.
74. Suzuki, Y.; Endo, M.; Sugiyama, H., Lipid-bilayer-assisted two-dimensional self-assembly of DNA origami nanostructures. *Nat. Commun.* **2015,** *6* (1), 8052.
75. Kocabey, S.; Kempter, S.; List, J.; Xing, Y.; Bae, W.; Schiffels, D.; Shih, W. M.; Simmel, F. C.; Liedl, T., Membrane-Assisted Growth of DNA Origami Nanostructure Arrays. *ACS Nano* **2015,** *9* (4), 3530-3539.
76. Wagenbauer, K. F.; Sigl, C.; Dietz, H., Gigadalton-scale shape-programmable DNA assemblies. *Nature* **2017,** *552* (7683), 78-83.
77. Tikhomirov, G.; Petersen, P.; Qian, L., Fractal assembly of micrometre-scale DNA origami arrays with arbitrary patterns. *Nature* **2017,** *552* (7683), 67-71.
78. Tikhomirov, G.; Petersen, P.; Qian, L., Triangular DNA Origami Tilings. *J. Am. Chem. Soc.* **2018,** *140* (50), 17361-17364.
79. Lin, Z.; Xiong, Y.; Xiang, S.; Gang, O., Controllable Covalent-Bound Nanoarchitectures from DNA Frames. *J. Am. Chem. Soc.* **2019,** *141* (17), 6797-6801.
80. Lin, Z.; Emamy, H.; Minevich, B.; Xiong, Y.; Xiang, S.; Kumar, S.; Ke, Y.; Gang, O., Engineering Organization of DNA Nano-Chambers through Dimensionally Controlled and Multi-Sequence Encoded Differentiated Bonds. *J. Am. Chem. Soc.* **2020,** *142* (41), 17531-17542.
81. Zhou, Y.; Dong, J.; Zhou, C.; Wang, Q., Finite Assembly of Three-Dimensional DNA Hierarchical Nanoarchitectures through Orthogonal and Directional Bonding. *Angew. Chem., Int. Ed.* **2022,** *61* (13), e202116416.
82. Wang, X.; Jun, H.; Bathe, M., Programming 2D Supramolecular Assemblies with Wireframe DNA Origami. *J. Am. Chem. Soc.* **2022,** *144* (10), 4403-4409.
83. Li, Y.; Pei, J.; Lu, X.; Jiao, Y.; Liu, F.; Wu, X.; Liu, J.; Ding, B., Hierarchical Assembly of Super-DNA Origami Based on a Flexible and Covalent-Bound Branched DNA Structure. *J. Am. Chem. Soc.* **2021,** *143* (47), 19893-19900.
84. Liu, J.; Chen, L.; Zhai, T.; Li, W.; Liu, Y.; Gu, H., Programmable Assembly of Amphiphilic DNA through Controlled Cholesterol Stacking. *J. Am. Chem. Soc.* **2022,** *144* (36), 16598-16603.
85. Schulman, R.; Winfree, E., Synthesis of crystals with a programmable kinetic barrier to nucleation. *Proc. Natl. Acad. Sci. U. S. A.* **2007,** *104* (39), 15236-15241.
86. Li, W.; Yang, Y.; Jiang, S.; Yan, H.; Liu, Y., Controlled Nucleation and Growth of DNA Tile Arrays within Prescribed DNA Origami Frames and Their Dynamics. *J. Am. Chem. Soc.* **2014,** *136* (10), 3724-3727.
87. Sharma, J.; Ke, Y.; Lin, C.; Chhabra, R.; Wang, Q.; Nangreave, J.; Liu, Y.; Yan, H., DNA-Tile-Directed Self-Assembly of Quantum Dots into Two-Dimensional Nanopatterns. *Angew. Chem., Int. Ed.* **2008,** *47* (28), 5157-5159.
88. Rothemund, P. W. K.; Ekani-Nkodo, A.; Papadakis, N.; Kumar, A.; Fygenson, D. K.; Winfree, E., Design and Characterization of Programmable DNA Nanotubes. *J. Am. Chem. Soc.* **2004,** *126* (50), 16344-16352.
89. Zhang, D. Y.; Hariadi, R. F.; Choi, H. M. T.; Winfree, E., Integrating DNA strand-displacement circuitry with DNA tile self-assembly. *Nat. Commun.* **2013,** *4* (1), 1965.
90. Fujibayashi, K.; Hariadi, R.; Park, S. H.; Winfree, E.; Murata, S., Toward Reliable

Algorithmic Self-Assembly of DNA Tiles: A Fixed-Width Cellular Automaton Pattern. *Nano Lett.* **2008,** *8* (7), 1791-1797.
91. Rothemund, P. W. K.; Papadakis, N.; Winfree, E., Algorithmic Self-Assembly of DNA Sierpinski Triangles. *PLoS Biol.* **2004,** *2* (12), e424.
92. Wang, T.; Sha, R.; Dreyfus, R.; Leunissen, M. E.; Maass, C.; Pine, D. J.; Chaikin, P. M.; Seeman, N. C., Self-replication of information-bearing nanoscale patterns. *Nature* **2011,** *478* (7368), 225-228.
93. Schulman, R.; Yurke, B.; Winfree, E., Robust self-replication of combinatorial information *via* crystal growth and scission. *Proc. Natl. Acad. Sci. U. S. A.* **2012,** *109* (17), 6405-6410.
94. Barish, R. D.; Schulman, R.; Rothemund, P. W.; Winfree, E., An information-bearing seed for nucleating algorithmic self-assembly. *Proc. Natl. Acad. Sci. U. S. A.* **2009,** *106* (15), 6054-9.
95. Jiang, C.; Lu, B.; Zhang, W.; Ohayon, Y. P.; Feng, F.; Li, S.; Seeman, N. C.; Xiao, S.-J., Regulation of 2D DNA Nanostructures by the Coupling of Intrinsic Tile Curvature and Arm Twist. *Journal of the American Chemical Society* **2022,** *144* (15), 6759-6769.
96. Maier, A. M.; Bae, W.; Schiffels, D.; Emmerig, J. F.; Schiff, M.; Liedl, T., Self-Assembled DNA Tubes Forming Helices of Controlled Diameter and Chirality. *ACS Nano* **2017,** *11* (2), 1301-1306.
97. Zhang, Y.; Chen, X.; Kang, G.; Peng, R.; Pan, V.; Sundaresan, R.; Wang, P.; Ke, Y., Programming DNA Tube Circumference by Tile Offset Connection. *J. Am. Chem. Soc.* **2019,** *141* (50), 19529-19532.
98. Jiang, S.; Yan, H.; Liu, Y., Kinetics of DNA Tile Dimerization. *ACS Nano* **2014,** *8* (6), 5826-5832.
99. Jiang, S.; Hong, F.; Hu, H.; Yan, H.; Liu, Y., Understanding the Elementary Steps in DNA Tile-Based Self-Assembly. *ACS Nano* **2017,** *11* (9), 9370-9381.
100. Jiang, S.; Pal, N.; Hong, F.; Fahmi, N. E.; Hu, H.; Vrbanac, M.; Yan, H.; Walter, N. G.; Liu, Y., Regulating DNA Self-Assembly Dynamics with Controlled Nucleation. *ACS Nano* **2021,** *15* (3), 5384-5396.
101. Minev, D.; Wintersinger, C. M.; Ershova, A.; Shih, W. M., Robust nucleation control *via* crisscross polymerization of highly coordinated DNA slats. *Nat. Commun.* **2021,** *12* (1), 1741.
102. Ong, L. L.; Hanikel, N.; Yaghi, O. K.; Grun, C.; Strauss, M. T.; Bron, P.; Lai-Kee-Him, J.; Schueder, F.; Wang, B.; Wang, P.; Kishi, J. Y.; Myhrvold, C.; Zhu, A.; Jungmann, R.; Bellot, G.; Ke, Y.; Yin, P., Programmable self-assembly of three-dimensional nanostructures from 10,000 unique components. *Nature* **2017,** *552* (7683), 72-77.
103. Wayment-Steele, H. K.; Frenkel, D.; Reinhardt, A., Investigating the role of boundary bricks in DNA brick self-assembly. *Soft Matter* **2017,** *13* (8), 1670-1680.
104. Zhang, Y.; Reinhardt, A.; Wang, P.; Song, J.; Ke, Y., Programming the Nucleation of DNA Brick Self-Assembly with a Seeding Strand. *Angew. Chem., Int. Ed.* **2020,** *59* (22), 8594-8600.
105. Jun, H.; Wang, X.; Bricker, W. P.; Bathe, M., Automated sequence design of 2D wireframe DNA origami with honeycomb edges. *Nat. Commun.* **2019,** *10* (1), 5419.
106. Jun, H.; Shepherd, T. R.; Zhang, K.; Bricker, W. P.; Li, S.; Chiu, W.; Bathe, M., Automated sequence design of 3D polyhedral wireframe DNA origami with honeycomb edges. *ACS Nano* **2019,** *13* (2), 2083-2093.
107. Jun, H.; Zhang, F.; Shepherd, T.; Ratanalert, S.; Qi, X.; Yan, H.; Bathe, M., Autonomously designed free-form 2D DNA origami. *Sci. Adv. 5* (1), eaav0655.
108. Wang, X.; Li, S.; Jun, H.; John, T.; Zhang, K.; Fowler, H.; Doye, J. P. K.; Chiu, W.; Bathe, M., Planar 2D wireframe DNA origami. *Sci. Adv. 8* (20), eabn0039.
109. Jun, H.; Wang, X.; Parsons, Molly F.; Bricker, William P.; John, T.; Li, S.; Jackson, S.; Chiu, W.; Bathe, M., Rapid prototyping of arbitrary 2D and 3D wireframe DNA origami. *Nucleic Acids Res.* **2021,** *49* (18), 10265-10274.
110. Goodman, R. P.; Schaap, I. A. T.; Tardin, C. F.; Erben, C. M.; Berry, R. M.; Schmidt, C. F.; Turberfield, A. J., Rapid Chiral Assembly of Rigid DNA Building Blocks for Molecular

Nanofabrication. *Science* **2005,** *310* (5754), 1661-1665.
111. Aldaye, F. A.; Sleiman, H. F., Modular Access to Structurally Switchable 3D Discrete DNA Assemblies. *J. Am. Chem. Soc.* **2007,** *129* (44), 13376-13377.
112. He, Y.; Ye, T.; Su, M.; Zhang, C.; Ribbe, A. E.; Jiang, W.; Mao, C., Hierarchical self-assembly of DNA into symmetric supramolecular polyhedra. *Nature* **2008,** *452* (7184), 198-201.
113. Tian, C.; Li, X.; Liu, Z.; Jiang, W.; Wang, G.; Mao, C., Directed self-assembly of DNA tiles into complex nanocages. *Angew. Chem., Int. Ed.* **2014,** *53* (31), 8041-4.
114. Wang, W.; Chen, S.; An, B.; Huang, K.; Bai, T.; Xu, M.; Bellot, G.; Ke, Y.; Xiang, Y.; Wei, B., Complex wireframe DNA nanostructures from simple building blocks. *Nat. Commun.* **2019,** *10* (1), 1067.
115. Lin, C.; Xie, M.; Chen, J. J. L.; Liu, Y.; Yan, H., Rolling-Circle Amplification of a DNA Nanojunction. *Angew. Chem., Int. Ed.* **2006,** *45* (45), 7537-7539.
116. Lin, C.; Wang, X.; Liu, Y.; Seeman, N. C.; Yan, H., Rolling Circle Enzymatic Replication of a Complex Multi-Crossover DNA Nanostructure. *J. Am. Chem. Soc.* **2007,** *129* (46), 14475-14481.
117. Li, Z.; Wei, B.; Nangreave, J.; Lin, C.; Liu, Y.; Mi, Y.; Yan, H., A Replicable Tetrahedral Nanostructure Self-Assembled from a Single DNA Strand. *J. Am. Chem. Soc.* **2009,** *131* (36), 13093-13098.
118. Geary, C.; Rothemund Paul, W. K.; Andersen Ebbe, S., A single-stranded architecture for cotranscriptional folding of RNA nanostructures. *Science* **2014,** *345* (6198), 799-804.
119. Lin, C.; Rinker, S.; Wang, X.; Liu, Y.; Seeman Nadrian, C.; Yan, H., In vivo cloning of artificial DNA nanostructures. *Proc. Natl. Acad. Sci. U. S. A.* **2008,** *105* (46), 17626-17631.
120. Han, D.; Qi, X.; Myhrvold, C.; Wang, B.; Dai, M.; Jiang, S.; Bates, M.; Liu, Y.; An, B.; Zhang, F.; Yan, H.; Yin, P., Single-stranded DNA and RNA origami. *Science* **2017,** *358* (6369), eaao2648.
121. Wang, X.; Chandrasekaran, A. R.; Shen, Z.; Ohayon, Y. P.; Wang, T.; Kizer, M. E.; Sha, R.; Mao, C.; Yan, H.; Zhang, X.; Liao, S.; Ding, B.; Chakraborty, B.; Jonoska, N.; Niu, D.; Gu, H.; Chao, J.; Gao, X.; Li, Y.; Ciengshin, T.; Seeman, N. C., Paranemic Crossover DNA: There and Back Again. *Chem. Rev.* **2019,** *119* (10), 6273-6289.
122. Qi, X.; Zhang, F.; Su, Z.; Jiang, S.; Han, D.; Ding, B.; Liu, Y.; Chiu, W.; Yin, P.; Yan, H., Programming molecular topologies from single-stranded nucleic acids. *Nat. Commun.* **2018,** *9* (1), 4579.
123. Mueller, J. E.; Du, S. M.; Seeman, N. C., Design and synthesis of a knot from single-stranded DNA. *J. Am. Chem. Soc.* **1991,** *113* (16), 6306-6308.
124. Du, S. M.; Seeman, N. C., Synthesis of a DNA knot containing both positive and negative nodes. *J. Am. Chem. Soc.* **1992,** *114* (24), 9652-9655.
125. Du, S. M.; Stollar, B. D.; Seeman, N. C., A synthetic DNA molecule in three knotted topologies. *J. Am. Chem. Soc.* **1995,** *117* (4), 1194-1200.
126. Ohayon, Y. P.; Sha, R.; Flint, O.; Chandrasekaran, A. R.; Abdallah, H. O.; Wang, T.; Wang, X.; Zhang, X.; Seeman, N. C., Topological Linkage of DNA Tiles Bonded by Paranemic Cohesion. *ACS Nano* **2015,** *9* (10), 10296-10303.
127. Ohayon, Y. P.; Sha, R.; Flint, O.; Liu, W.; Chakraborty, B.; Subramanian, H. K. K.; Zheng, J.; Chandrasekaran, A. R.; Abdallah, H. O.; Wang, X.; Zhang, X.; Seeman, N. C., Covalent Linkage of One-Dimensional DNA Arrays Bonded by Paranemic Cohesion. *ACS Nano* **2015,** *9* (10), 10304-10312.
128. Liu, D.; Chen, G.; Akhter, U.; Cronin, T. M.; Weizmann, Y., Creating complex molecular topologies by configuring DNA four-way junctions. *Nat. Chem.* **2016,** *8* (10), 907-914.
129. Zhang, D. Y.; Yurke, B., A DNA Superstructure-based Replicator without Product Inhibition. *Nat. Comput.* **2006,** *5* (2), 183-202.
130. Chandran, H.; Gopalkrishnan, N.; Yurke, B.; Reif, J., Meta-DNA: synthetic biology *via* DNA nanostructures and hybridization reactions. *J. R. Soc. Interface* **2012,** *9* (72), 1637-1653.

131. Hamada, S.; Murata, S., Substrate-Assisted Assembly of Interconnected Single-Duplex DNA Nanostructures. *Angew. Chem., Int. Ed.* **2009,** *48* (37), 6820-6823.
132. Yao, G.; Zhang, F.; Wang, F.; Peng, T.; Liu, H.; Poppleton, E.; Šulc, P.; Jiang, S.; Liu, L.; Gong, C.; Jing, X.; Liu, X.; Wang, L.; Liu, Y.; Fan, C.; Yan, H., Meta-DNA structures. *Nat. Chem.* **2020,** *12* (11), 1067-1075.
133. Zheng, J.; Birktoft, J. J.; Chen, Y.; Wang, T.; Sha, R.; Constantinou, P. E.; Ginell, S. L.; Mao, C.; Seeman, N. C., From molecular to macroscopic *via* the rational design of a self-assembled 3D DNA crystal. *Nature* **2009,** *461* (7260), 74-77.
134. Wang, T.; Sha, R.; Birktoft, J.; Zheng, J.; Mao, C.; Seeman, N. C., A DNA Crystal Designed to Contain Two Molecules per Asymmetric Unit. *J. Am. Chem. Soc.* **2010,** *132* (44), 15471-15473.
135. Nguyen, N.; Birktoft, J. J.; Sha, R.; Wang, T.; Zheng, J.; Constantinou, P. E.; Ginell, S. L.; Chen, Y.; Mao, C.; Seeman, N. C., The absence of tertiary interactions in a self-assembled DNA crystal structure. *J. Mol. Recognit.* **2012,** *25* (4), 234-237.
136. Lu, B.; Vecchioni, S.; Ohayon, Y. P.; Sha, R.; Woloszyn, K.; Yang, B.; Mao, C.; Seeman, N. C., 3D Hexagonal Arrangement of DNA Tensegrity Triangles. *ACS Nano* **2021,** *15* (10), 16788-16793.
137. Ohayon, Y. P.; Hernandez, C.; Chandrasekaran, A. R.; Wang, X.; Abdallah, H. O.; Jong, M. A.; Mohsen, M. G.; Sha, R.; Birktoft, J. J.; Lukeman, P. S.; Chaikin, P. M.; Ginell, S. L.; Mao, C.; Seeman, N. C., Designing Higher Resolution Self-Assembled 3D DNA Crystals *via* Strand Terminus Modifications. *ACS Nano* **2019,** *13* (7), 7957-7965.
138. Li, Z.; Zheng, M.; Liu, L.; Seeman, N. C.; Mao, C., 5′-Phosphorylation Strengthens Sticky-End Cohesions. *J. Am. Chem. Soc.* **2021,** *143* (37), 14987-14991.
139. Zhao, J.; Zhao, Y.; Li, Z.; Wang, Y.; Sha, R.; Seeman, N. C.; Mao, C., Modulating Self-Assembly of DNA Crystals with Rationally Designed Agents. *Angew. Chem., Int. Ed.* **2018,** *57* (50), 16529-16532.
140. Zhao, J.; Chandrasekaran, A. R.; Li, Q.; Li, X.; Sha, R.; Seeman, N. C.; Mao, C., Post-Assembly Stabilization of Rationally Designed DNA Crystals. *Angew. Chem., Int. Ed.* **2015,** *54* (34), 9936-9939.
141. Abdallah, H. O.; Ohayon, Y. P.; Chandrasekaran, A. R.; Sha, R.; Fox, K. R.; Brown, T.; Rusling, D. A.; Mao, C.; Seeman, N. C., Stabilisation of self-assembled DNA crystals by triplex-directed photo-cross-linking. *Chem. Commun.* **2016,** *52* (51), 8014-8017.
142. Zhang, D.; Paukstelis, P. J., Enhancing DNA Crystal Durability through Chemical Crosslinking. *ChemBioChem* **2016,** *17* (12), 1163-1170.
143. Li, Z.; Liu, L.; Zheng, M.; Zhao, J.; Seeman, N. C.; Mao, C., Making Engineered 3D DNA Crystals Robust. *J. Am. Chem. Soc.* **2019,** *141* (40), 15850-15855.
144. Simmons, C. R.; Zhang, F.; Birktoft, J. J.; Qi, X.; Han, D.; Liu, Y.; Sha, R.; Abdallah, H. O.; Hernandez, C.; Ohayon, Y. P.; Seeman, N. C.; Yan, H., Construction and Structure Determination of a Three-Dimensional DNA Crystal. *J. Am. Chem. Soc.* **2016,** *138* (31), 10047-10054.
145. Simmons, C. R.; Zhang, F.; MacCulloch, T.; Fahmi, N.; Stephanopoulos, N.; Liu, Y.; Seeman, N. C.; Yan, H., Tuning the Cavity Size and Chirality of Self-Assembling 3D DNA Crystals. *J. Am. Chem. Soc.* **2017,** *139* (32), 11254-11260.
146. Zhang, F.; Simmons, C. R.; Gates, J.; Liu, Y.; Yan, H., Self-Assembly of a 3D DNA Crystal Structure with Rationally Designed Six-Fold Symmetry. *Angew. Chem., Int. Ed.* **2018,** *57* (38), 12504-12507.
147. Simmons, C. R.; MacCulloch, T.; Krepl, M.; Matthies, M.; Buchberger, A.; Crawford, I.; Šponer, J.; Šulc, P.; Stephanopoulos, N.; Yan, H., The influence of Holliday junction sequence and dynamics on DNA crystal self-assembly. *Nat. Commun.* **2022,** *13* (1), 3112.
148. Zhang, T.; Hartl, C.; Frank, K.; Heuer-Jungemann, A.; Fischer, S.; Nickels, P. C.; Nickel, B.; Liedl, T., 3D DNA Origami Crystals. *Adv. Mater.* **2018,** *30* (28), 1800273.

149. Yu, L.; Cheng, J.; Wang, D.; Pan, V.; Chang, S.; Song, J.; Ke, Y., Stress in DNA Gridiron Facilitates the Formation of Two-Dimensional Crystalline Structures. *Journal of the American Chemical Society* **2022,** *144* (22), 9747-9752.
150. Jaeger, L.; Leontis, N. B., Tecto-RNA: One-Dimensional Self-Assembly through Tertiary Interactions. *Angew. Chem., Int. Ed.* **2000,** *39* (14), 2521-2524.
151. Jaeger, L.; Westhof, E.; Leontis, N. B., TectoRNA: modular assembly units for the construction of RNA nano-objects. *Nucleic Acids Res.* **2001,** *29* (2), 455-463.
152. Chworos, A.; Severcan, I.; Koyfman Alexey, Y.; Weinkam, P.; Oroudjev, E.; Hansma Helen, G.; Jaeger, L., Building Programmable Jigsaw Puzzles with RNA. *Science* **2004,** *306* (5704), 2068-2072.
153. Nasalean, L.; Baudrey, S. p.; Leontis, N. B.; Jaeger, L., Controlling RNA self-assembly to form filaments. *Nucleic Acids Res.* **2006,** *34* (5), 1381-1392.
154. Severcan, I.; Geary, C.; Verzemnieks, E.; Chworos, A.; Jaeger, L., Square-Shaped RNA Particles from Different RNA Folds. *Nano Lett.* **2009,** *9* (3), 1270-1277.
155. Grabow, W. W.; Zakrevsky, P.; Afonin, K. A.; Chworos, A.; Shapiro, B. A.; Jaeger, L., Self-Assembling RNA Nanorings Based on RNAI/II Inverse Kissing Complexes. *Nano Lett.* **2011,** *11* (2), 878-887.
156. Severcan, I.; Geary, C.; Chworos, A.; Voss, N.; Jacovetty, E.; Jaeger, L., A polyhedron made of tRNAs. *Nat. Chem.* **2010,** *2* (9), 772-779.
157. Li, M.; Zheng, M. X.; Wu, S. Y.; Tian, C.; Liu, D.; Weizmann, Y.; Jiang, W.; Wang, G. S.; Mao, C. D., In vivo production of RNA nanostructures *via* programmed folding of single-stranded RNAs. *Nature Communications* **2018,** *9*.
158. Geary, C.; Grossi, G.; McRae, E. K. S.; Rothemund, P. W. K.; Andersen, E. S., RNA origami design tools enable cotranscriptional folding of kilobase-sized nanoscaffolds. *Nat. Chem.* **2021,** *13* (6), 549-558.
159. Liu, D.; Geary, C. W.; Chen, G.; Shao, Y.; Li, M.; Mao, C.; Andersen, E. S.; Piccirilli, J. A.; Rothemund, P. W. K.; Weizmann, Y., Branched kissing loops for the construction of diverse RNA homooligomeric nanostructures. *Nat. Chem.* **2020,** *12* (3), 249-259.
160. Wang, H.; Di Gate, R. J.; Seeman, N. C., An RNA topoisomerase. *Proc. Natl. Acad. Sci. U. S. A.* **1996,** *93* (18), 9477-9482.
161. Xu, D.; Shen, W.; Guo, R.; Xue, Y.; Peng, W.; Sima, J.; Yang, J.; Sharov, A.; Srikantan, S.; Yang, J.; Fox, D.; Qian, Y.; Martindale, J. L.; Piao, Y.; Machamer, J.; Joshi, S. R.; Mohanty, S.; Shaw, A. C.; Lloyd, T. E.; Brown, G. W.; Ko, M. S. H.; Gorospe, M.; Zou, S.; Wang, W., Top3β is an RNA topoisomerase that works with fragile X syndrome protein to promote synapse formation. *Nat. Neurosci.* **2013,** *16* (9), 1238-1247.
162. Ahmad, M.; Xue, Y.; Lee, S. K.; Martindale, J. L.; Shen, W.; Li, W.; Zou, S.; Ciaramella, M.; Debat, H.; Nadal, M.; Leng, F.; Zhang, H.; Wang, Q.; Siaw, G. E.-L.; Niu, H.; Pommier, Y.; Gorospe, M.; Hsieh, T.-S.; Tse-Dinh, Y.-C.; Xu, D.; Wang, W., RNA topoisomerase is prevalent in all domains of life and associates with polyribosomes in animals. *Nucleic Acids Res.* **2016,** *44* (13), 6335-6349.
163. Liu, D.; Shao, Y.; Chen, G.; Tse-Dinh, Y.-C.; Piccirilli, J. A.; Weizmann, Y., Synthesizing topological structures containing RNA. *Nat. Commun.* **2017,** *8* (1), 14936.
164. Mao, C.; Sun, W.; Shen, Z.; Seeman, N. C., A nanomechanical device based on the B–Z transition of DNA. *Nature* **1999,** *397* (6715), 144-146.
165. Yurke, B.; Turberfield, A. J.; Mills, A. P.; Simmel, F. C.; Neumann, J. L., A DNA-fuelled molecular machine made of DNA. *Nature* **2000,** *406* (6796), 605-608.
166. Dirks Robert, M.; Pierce Niles, A., Triggered amplification by hybridization chain reaction. *Proc. Natl. Acad. Sci. U. S. A.* **2004,** *101* (43), 15275-15278.
167. Shin, J.-S.; Pierce, N. A., A Synthetic DNA Walker for Molecular Transport. *J. Am. Chem. Soc.* **2004,** *126* (35), 10834-10835.
168. Yin, P.; Yan, H.; Daniell, X. G.; Turberfield, A. J.; Reif, J. H., A Unidirectional DNA

Walker That Moves Autonomously along a Track. *Angew. Chem., Int. Ed.* **2004,** *43* (37), 4906-4911.
169. Wang, Z.-G.; Elbaz, J.; Willner, I., DNA Machines: Bipedal Walker and Stepper. *Nano Lett.* **2011,** *11* (1), 304-309.
170. Gu, H.; Chao, J.; Xiao, S.-J.; Seeman, N. C., A proximity-based programmable DNA nanoscale assembly line. *Nature* **2010,** *465* (7295), 202-205.
171. Li, J.; Johnson-Buck, A.; Yang, Y. R.; Shih, W. M.; Yan, H.; Walter, N. G., Exploring the speed limit of toehold exchange with a cartwheeling DNA acrobat. *Nat. Nanotechnol.* **2018,** *13* (8), 723-729.
172. Goodman, R. P.; Heilemann, M.; Doose, S.; Erben, C. M.; Kapanidis, A. N.; Turberfield, A. J., Reconfigurable, braced, three-dimensional DNA nanostructures. *Nat. Nanotechnol.* **2008,** *3* (2), 93-96.
173. Yan, H.; Zhang, X.; Shen, Z.; Seeman, N. C., A robust DNA mechanical device controlled by hybridization topology. *Nature* **2002,** *415* (6867), 62-65.
174. Simmel, F. C.; Yurke, B., A DNA-based molecular device switchable between three distinct mechanical states. *Appl. Phys. Lett.* **2002,** *80* (5), 883-885.
175. Chakraborty, B.; Sha, R.; Seeman, N. C., A DNA-based nanomechanical device with three robust states. *Proc. Natl. Acad. Sci. U. S. A.* **2008,** *105* (45), 17245-17249.
176. Zhong, H.; Seeman, N. C., RNA Used to Control a DNA Rotary Nanomachine. *Nano Lett.* **2006,** *6* (12), 2899-2903.
177. Tian, Y.; Mao, C., Molecular Gears: A Pair of DNA Circles Continuously Rolls against Each Other. *J. Am. Chem. Soc.* **2004,** *126* (37), 11410-11411.
178. Ding, B.; Seeman, N. C., Operation of a DNA Robot Arm Inserted into a 2D DNA Crystalline Substrate. *Science* **2006,** *314* (5805), 1583-1585.
179. Feng, L.; Park, S. H.; Reif, J. H.; Yan, H., A Two-State DNA Lattice Switched by DNA Nanoactuator. *Angew. Chem., Int. Ed.* **2003,** *42* (36), 4342-4346.
180. Ke, Y.; Meyer, T.; Shih, W. M.; Bellot, G., Regulation at a distance of biomolecular interactions using a DNA origami nanoactuator. *Nat. Commun.* **2016,** *7* (1), 10935.
181. Lubrich, D.; Lin, J.; Yan, J., A Contractile DNA Machine. *Angew. Chem., Int. Ed.* **2008,** *47* (37), 7026-7028.
182. Benson, E.; Marzo, R. C.; Bath, J.; Turberfield, A. J., A DNA molecular printer capable of programmable positioning and patterning in two dimensions. *Sci. Robot. 7* (65), eabn5459.
183. Douglas Shawn, M.; Bachelet, I.; Church George, M., A Logic-Gated Nanorobot for Targeted Transport of Molecular Payloads. *Science* **2012,** *335* (6070), 831-834.
184. Song, J.; Li, Z.; Wang, P.; Meyer, T.; Mao, C.; Ke, Y., Reconfiguration of DNA molecular arrays driven by information relay. *Science* **2017,** *357* (6349), eaan3377.
185. Mariottini, D.; Idili, A.; Vallée-Bélisle, A.; Plaxco, K. W.; Ricci, F., A DNA Nanodevice That Loads and Releases a Cargo with Hemoglobin-Like Allosteric Control and Cooperativity. *Nano Lett.* **2017,** *17* (5), 3225-3230.
186. Ijas, H.; Hakaste, I.; Shen, B.; Kostiainen, M. A.; Linko, V., Reconfigurable DNA Origami Nanocapsule for pH-Controlled Encapsulation and Display of Cargo. *ACS Nano* **2019,** *13* (5), 5959-5967.
187. Green, L. N.; Amodio, A.; Subramanian, H. K. K.; Ricci, F.; Franco, E., pH-Driven Reversible Self-Assembly of Micron-Scale DNA Scaffolds. *Nano Lett.* **2017,** *17* (12), 7283-7288.
188. Majikes, J. M.; Ferraz, L. C. C.; LaBean, T. H., pH-Driven Actuation of DNA Origami *via* Parallel I-Motif Sequences in Solution and on Surfaces. *Bioconjugate Chem.* **2017,** *28* (7), 1821-1825.
189. Wang, W.; Yang, Y.; Cheng, E.; Zhao, M.; Meng, H.; Liu, D.; Zhou, D., A pH-driven, reconfigurable DNA nanotriangle. *Chem. Commun.* **2009,** (7), 824-826.
190. Liu, D.; Bruckbauer, A.; Abell, C.; Balasubramanian, S.; Kang, D.-J.; Klenerman, D.; Zhou, D., A Reversible pH-Driven DNA Nanoswitch Array. *J. Am. Chem. Soc.* **2006,** *128* (6), 2067-2071.

191. Modi, S.; M. G, S.; Goswami, D.; Gupta, G. D.; Mayor, S.; Krishnan, Y., A DNA nanomachine that maps spatial and temporal pH changes inside living cells. *Nat. Nanotechnol.* **2009,** *4* (5), 325-330.
192. Wang, Y.; Yan, X.; Zhou, Z.; Ma, N.; Tian, Y., pH-Induced Symmetry Conversion of DNA Origami Lattices. *Angew. Chem., Int. Ed.* **2022,** *61* (40), e202208290.
193. Arnott, P. M.; Howorka, S., A Temperature-Gated Nanovalve Self-Assembled from DNA to Control Molecular Transport across Membranes. *ACS Nano* **2019,** *13* (3), 3334-3340.
194. Yang, Y.; Endo, M.; Hidaka, K.; Sugiyama, H., Photo-Controllable DNA Origami Nanostructures Assembling into Predesigned Multiorientational Patterns. *J. Am. Chem. Soc.* **2012,** *134* (51), 20645-20653.
195. Kohman, R. E.; Han, X., Light sensitization of DNA nanostructures *via* incorporation of photo-cleavable spacers. *Chem. Commun.* **2015,** *51* (26), 5747-5750.
196. Qu, X.; Zhu, D.; Yao, G.; Su, S.; Chao, J.; Liu, H.; Zuo, X.; Wang, L.; Shi, J.; Wang, L.; Huang, W.; Pei, H.; Fan, C., An Exonuclease III-Powered, On-Particle Stochastic DNA Walker. *Angew. Chem., Int. Ed.* **2017,** *56* (7), 1855-1858.
197. Bath, J.; Green, S. J.; Turberfield, A. J., A Free-Running DNA Motor Powered by a Nicking Enzyme. *Angew. Chem., Int. Ed.* **2005,** *44* (28), 4358-4361.
198. Agarwal, N. P.; Matthies, M.; Joffroy, B.; Schmidt, T. L., Structural Transformation of Wireframe DNA Origami *via* DNA Polymerase Assisted Gap-Filling. *ACS Nano* **2018,** *12* (3), 2546-2553.
199. Darcy, M.; Crocker, K.; Wang, Y.; Le, J. V.; Mohammadiroozbahani, G.; Abdelhamid, M. A. S.; Craggs, T. D.; Castro, C. E.; Bundschuh, R.; Poirier, M. G., High-Force Application by a Nanoscale DNA Force Spectrometer. *ACS Nano* **2022,** *16* (4), 5682-5695.
200. Lauback, S.; Mattioli, K. R.; Marras, A. E.; Armstrong, M.; Rudibaugh, T. P.; Sooryakumar, R.; Castro, C. E., Real-time magnetic actuation of DNA nanodevices *via* modular integration with stiff micro-levers. *Nat. Commun.* **2018,** *9* (1), 1446.
201. Marras, A. E.; Zhou, L.; Su, H.-J.; Castro, C. E., Programmable motion of DNA origami mechanisms. *Proc. Natl. Acad. Sci. U. S. A.* **2015,** *112* (3), 713-718.
202. Wang, D.; Yu, L.; Ji, B.; Chang, S.; Song, J.; Ke, Y., Programming the Curvatures in Reconfigurable DNA Domino Origami by Using Asymmetric Units. *Nano Lett.* **2020,** *20* (11), 8236-8241.
203. Wang, D.; Yu, L.; Huang, C.-M.; Arya, G.; Chang, S.; Ke, Y., Programmable Transformations of DNA Origami Made of Small Modular Dynamic Units. *J. Am. Chem. Soc.* **2021,** *143* (5), 2256-2263.
204. Suzuki, Y.; Kawamata, I.; Mizuno, K.; Murata, S., Large Deformation of a DNA-Origami Nanoarm Induced by the Cumulative Actuation of Tension-Adjustable Modules. *Angew. Chem., Int. Ed.* **2020,** *59* (15), 6230-6234.
205. Choi, Y.; Choi, H.; Lee, A. C.; Lee, H.; Kwon, S., A Reconfigurable DNA Accordion Rack. *Angew. Chem., Int. Ed.* **2018,** *57* (11), 2811-2815.
206. Kuzyk, A.; Schreiber, R.; Zhang, H.; Govorov, A. O.; Liedl, T.; Liu, N., Reconfigurable 3D plasmonic metamolecules. *Nat. Mater.* **2014,** *13* (9), 862-866.
207. Zhan, P.; Dutta, P. K.; Wang, P.; Song, G.; Dai, M.; Zhao, S.-X.; Wang, Z.-G.; Yin, P.; Zhang, W.; Ding, B.; Ke, Y., Reconfigurable Three-Dimensional Gold Nanorod Plasmonic Nanostructures Organized on DNA Origami Tripod. *ACS Nano* **2017,** *11* (2), 1172-1179.
208. Hao, Y.; Kristiansen, M.; Sha, R.; Birktoft, J. J.; Hernandez, C.; Mao, C.; Seeman, N. C., A device that operates within a self-assembled 3D DNA crystal. *Nat. Chem.* **2017,** *9* (8), 824-827.
209. Zheng, M.; Li, Z.; Zhang, C.; Seeman, N. C.; Mao, C., Powering $\approx$50 μm Motion by a Molecular Event in DNA Crystals. *Adv. Mater.* **2022,** *34* (26), 2200441.
210. Cangialosi, A.; Yoon, C.; Liu, J.; Huang, Q.; Guo, J.; Nguyen, T. D.; Gracias, D. H.; Schulman, R., DNA sequence–directed shape change of photopatterned hydrogels *via* high-degree swelling. *Science* **2017,** *357* (6356), 1126-1130.

211. Sherman, W. B.; Seeman, N. C., A Precisely Controlled DNA Biped Walking Device. *Nano Lett.* **2004,** *4* (7), 1203-1207.
212. Yin, P.; Choi, H. M. T.; Calvert, C. R.; Pierce, N. A., Programming biomolecular self-assembly pathways. *Nature* **2008,** *451* (7176), 318-322.
213. Omabegho, T.; Sha, R.; Seeman Nadrian, C., A Bipedal DNA Brownian Motor with Coordinated Legs. *Science* **2009,** *324* (5923), 67-71.
214. Thubagere Anupama, J.; Li, W.; Johnson Robert, F.; Chen, Z.; Doroudi, S.; Lee Yae, L.; Izatt, G.; Wittman, S.; Srinivas, N.; Woods, D.; Winfree, E.; Qian, L., A cargo-sorting DNA robot. *Science* **2017,** *357* (6356), eaan6558.
215. Chao, J.; Wang, J.; Wang, F.; Ouyang, X.; Kopperger, E.; Liu, H.; Li, Q.; Shi, J.; Wang, L.; Hu, J.; Wang, L.; Huang, W.; Simmel, F. C.; Fan, C., Solving mazes with single-molecule DNA navigators. *Nat. Mater.* **2019,** *18* (3), 273-279.
216. Jung, C.; Allen, P. B.; Ellington, A. D., A stochastic DNA walker that traverses a microparticle surface. *Nat. Nanotechnol.* **2016,** *11* (2), 157-163.
217. Zhou, C.; Duan, X.; Liu, N., A plasmonic nanorod that walks on DNA origami. *Nat. Commun.* **2015,** *6* (1), 8102.
218. Urban, M. J.; Zhou, C.; Duan, X.; Liu, N., Optically Resolving the Dynamic Walking of a Plasmonic Walker Couple. *Nano Lett.* **2015,** *15* (12), 8392-8396.
219. Tian, Y.; He, Y.; Chen, Y.; Yin, P.; Mao, C., A DNAzyme That Walks Processively and Autonomously along a One-Dimensional Track. *Angew. Chem., Int. Ed.* **2005,** *44* (28), 4355-4358.
220. Wickham, S. F. J.; Endo, M.; Katsuda, Y.; Hidaka, K.; Bath, J.; Sugiyama, H.; Turberfield, A. J., Direct observation of stepwise movement of a synthetic molecular transporter. *Nat. Nanotechnol.* **2011,** *6* (3), 166-169.
221. Wickham, S. F. J.; Bath, J.; Katsuda, Y.; Endo, M.; Hidaka, K.; Sugiyama, H.; Turberfield, A. J., A DNA-based molecular motor that can navigate a network of tracks. *Nat. Nanotechnol.* **2012,** *7* (3), 169-173.
222. Yehl, K.; Mugler, A.; Vivek, S.; Liu, Y.; Zhang, Y.; Fan, M.; Weeks, E. R.; Salaita, K., High-speed DNA-based rolling motors powered by RNase H. *Nat. Nanotechnol.* **2016,** *11* (2), 184-190.
223. Bazrafshan, A.; Meyer, T. A.; Su, H.; Brockman, J. M.; Blanchard, A. T.; Piranej, S.; Duan, Y.; Ke, Y.; Salaita, K., Tunable DNA Origami Motors Translocate Ballistically Over µm Distances at nm/s Speeds. *Angew. Chem., Int. Ed.* **2020,** *59* (24), 9514-9521.
224. Yang, Y.; Goetzfried, M. A.; Hidaka, K.; You, M.; Tan, W.; Sugiyama, H.; Endo, M., Direct Visualization of Walking Motions of Photocontrolled Nanomachine on the DNA Nanostructure. *Nano Lett.* **2015,** *15* (10), 6672-6676.
225. Kopperger, E.; List, J.; Madhira, S.; Rothfischer, F.; Lamb Don, C.; Simmel Friedrich, C., A self-assembled nanoscale robotic arm controlled by electric fields. *Science* **2018,** *359* (6373), 296-301.
226. Klapper, Y.; Sinha, N.; Ng, T. W. S.; Lubrich, D., A Rotational DNA Nanomotor Driven by an Externally Controlled Electric Field. *Small* **2010,** *6* (1), 44-47.
227. Urban, M. J.; Both, S.; Zhou, C.; Kuzyk, A.; Lindfors, K.; Weiss, T.; Liu, N., Gold nanocrystal-mediated sliding of doublet DNA origami filaments. *Nat. Commun.* **2018,** *9* (1), 1454.
228. Zhan, P.; Both, S.; Weiss, T.; Liu, N., DNA-Assembled Multilayer Sliding Nanosystems. *Nano Lett.* **2019,** *19* (9), 6385-6390.
229. Zhan, P.; Urban, M. J.; Both, S.; Duan, X.; Kuzyk, A.; Weiss, T.; Liu, N., DNA-assembled nanoarchitectures with multiple components in regulated and coordinated motion. *Sci. Adv.* **2019,** *5* (11), eaax6023.
230. Ketterer, P.; Willner, E. M.; Dietz, H., Nanoscale rotary apparatus formed from tight-fitting 3D DNA components. *Sci. Adv.* **2016,** *2* (2), e1501209.
231. Pumm, A.-K.; Engelen, W.; Kopperger, E.; Isensee, J.; Vogt, M.; Kozina, V.; Kube, M.; Honemann, M. N.; Bertosin, E.; Langecker, M.; Golestanian, R.; Simmel, F. C.; Dietz, H., A DNA origami rotary ratchet motor. *Nature* **2022,** *607* (7919), 492-498.
232. Xin, L.; Zhou, C.; Duan, X.; Liu, N., A rotary plasmonic nanoclock. *Nat. Commun.* **2019,** *10* (1), 5394.
233. Peil, A.; Xin, L.; Both, S.; Shen, L.; Ke, Y.; Weiss, T.; Zhan, P.; Liu, N., DNA Assembly of Modular Components into a Rotary Nanodevice. *ACS Nano* **2022,** *16* (4), 5284-5291.
234. He, Y.; Liu, D. R., Autonomous multistep organic synthesis in a single isothermal solution mediated by a DNA walker. *Nat. Nanotechnol.* **2010,** *5* (11), 778-782.
235. Zhang, D. Y.; Winfree, E., Control of DNA Strand Displacement Kinetics Using Toehold Exchange. *J. Am. Chem. Soc.* **2009,** *131* (47), 17303-17314.
236. Ibusuki, R.; Morishita, T.; Furuta, A.; Nakayama, S.; Yoshio, M.; Kojima, H.; Oiwa, K.; Furuta, K. y., Programmable molecular transport achieved by engineering protein motors to move on DNA nanotubes. *Science* **2022,** *375* (6585), 1159-1164.
237. Stömmer, P.; Kiefer, H.; Kopperger, E.; Honemann, M. N.; Kube, M.; Simmel, F. C.; Netz, R. R.; Dietz, H., A synthetic tubular molecular transport system. *Nature communications* **2021,** *12* (1), 1-10.
238. Soong Ricky, K.; Bachand George, D.; Neves Hercules, P.; Olkhovets Anatoli, G.; Craighead Harold, G.; Montemagno Carlo, D., Powering an Inorganic Nanodevice with a Biomolecular Motor. *Science* **2000,** *290* (5496), 1555-1558.
239. Lund, K.; Manzo, A. J.; Dabby, N.; Michelotti, N.; Johnson-Buck, A.; Nangreave, J.; Taylor, S.; Pei, R.; Stojanovic, M. N.; Walter, N. G.; Winfree, E.; Yan, H., Molecular robots guided by prescriptive landscapes. *Nature* **2010,** *465* (7295), 206-210.
240. Cha, T.-G.; Pan, J.; Chen, H.; Salgado, J.; Li, X.; Mao, C.; Choi, J. H., A synthetic DNA motor that transports nanoparticles along carbon nanotubes. *Nat. Nanotechnol.* **2014,** *9* (1), 39-43.
241. Li, Q.; Liu, L.; Mao, D.; Yu, Y.; Li, W.; Zhao, X.; Mao, C., ATP-Triggered, Allosteric Self-Assembly of DNA Nanostructures. *J. Am. Chem. Soc.* **2020,** *142* (2), 665-668.
242. Amodio, A.; Adedeji, A. F.; Castronovo, M.; Franco, E.; Ricci, F., pH-Controlled Assembly of DNA Tiles. *J. Am. Chem. Soc.* **2016,** *138* (39), 12735-12738.
243. Green, L. N.; Subramanian, H. K. K.; Mardanlou, V.; Kim, J.; Hariadi, R. F.; Franco, E., Autonomous dynamic control of DNA nanostructure self-assembly. *Nat. Chem.* **2019,** *11* (6), 510-520.
244. Brown, J. W. P.; Alford, R. G.; Walsh, J. C.; Spinney, R. E.; Xu, S. Y.; Hertel, S.; Berengut, J. F.; Spenkelink, L. M.; van Oijen, A. M.; Böcking, T.; Morris, R. G.; Lee, L. K., Rapid Exchange of Stably Bound Protein and DNA Cargo on a DNA Origami Receptor. *ACS Nano* **2022,** *16* (4), 6455-6467.
245. Gentile, S.; Del Grosso, E.; Pungchai, P. E.; Franco, E.; Prins, L. J.; Ricci, F., Spontaneous Reorganization of DNA-Based Polymers in Higher Ordered Structures Fueled by RNA. *J. Am. Chem. Soc.* **2021,** *143* (48), 20296-20301.
246. Petersen, P.; Tikhomirov, G.; Qian, L., Information-based autonomous reconfiguration in systems of interacting DNA nanostructures. *Nat. Commun.* **2018,** *9* (1), 5362.
247. Woods, D.; Doty, D.; Myhrvold, C.; Hui, J.; Zhou, F.; Yin, P.; Winfree, E., Diverse and robust molecular algorithms using reprogrammable DNA self-assembly. *Nature* **2019,** *567* (7748), 366-372.
248. Mohammed, A. M.; Schulman, R., Directing Self-Assembly of DNA Nanotubes Using Programmable Seeds. *Nano Lett.* **2013,** *13* (9), 4006-4013.
249. Agrawal, D. K.; Jiang, R.; Reinhart, S.; Mohammed, A. M.; Jorgenson, T. D.; Schulman, R., Terminating DNA Tile Assembly with Nanostructured Caps. *ACS Nano* **2017,** *11* (10), 9770-9779.
250. Mohammed, A. M.; Šulc, P.; Zenk, J.; Schulman, R., Self-assembling DNA nanotubes to connect molecular landmarks. *Nat. Nanotechnol.* **2017,** *12* (4), 312-316.
251. Li, Y.; Schulman, R., DNA Nanostructures that Self-Heal in Serum. *Nano Lett.* **2019,** *19* (6), 3751-3760.

252. Jorgenson, T. D.; Mohammed, A. M.; Agrawal, D. K.; Schulman, R., Self-Assembly of Hierarchical DNA Nanotube Architectures with Well-Defined Geometries. *ACS Nano* **2017,** *11* (2), 1927-1936.

253. Lincoln Tracey, A.; Joyce Gerald, F., Self-Sustained Replication of an RNA Enzyme. *Science* **2009,** *323* (5918), 1229-1232.

254. He, X.; Sha, R.; Zhuo, R.; Mi, Y.; Chaikin, P. M.; Seeman, N. C., Exponential growth and selection in self-replicating materials from DNA origami rafts. *Nat. Mater.* **2017,** *16* (10), 993-997.

255. Zhuo, R.; Zhou, F.; He, X.; Sha, R.; Seeman Nadrian, C.; Chaikin Paul, M., Litters of self-replicating origami cross-tiles. *Proc. Natl. Acad. Sci. U. S. A.* **2019,** *116* (6), 1952-1957.

256. Zhou, F.; Sha, R.; Ni, H.; Seeman, N.; Chaikin, P., Mutations in artificial self-replicating tiles: A step toward Darwinian evolution. *Proc. Natl. Acad. Sci. U. S. A.* **2021,** *118* (50), e2111193118.

257. Jin, C.; Han, L.; Che, S., Synthesis of a DNA-silica complex with rare two-dimensional square p4mm symmetry. *Angew Chem Int Ed Engl* **2009,** *48* (49), 9268-72.

258. Liu, B.; Han, L.; Che, S., Formation of enantiomeric impeller-like helical architectures by DNA self-assembly and silica mineralization. *Angew Chem Int Ed Engl* **2012,** *51* (4), 923-7.

259. Liu, B.; Yao, Y.; Che, S., Template-assisted self-assembly: alignment, placement, and arrangement of two-dimensional mesostructured DNA-silica platelets. *Angew Chem Int Ed Engl* **2013,** *52* (52), 14186-90.

260. Cao, Y.; Kao, K.; Mou, C.; Han, L.; Che, S., Oriented Chiral DNA-Silica Film Guided by a Natural Mica Substrate. *Angew Chem Int Ed Engl* **2016,** *55* (6), 2037-41.

261. Auyeung, E.; Macfarlane, R. J.; Choi, C. H.; Cutler, J. I.; Mirkin, C. A., Transitioning DNA-engineered nanoparticle superlattices from solution to the solid state. *Advanced materials (Deerfield Beach, Fla.)* **2012,** *24* (38), 5181-6.

262. Liu, X.; Zhang, F.; Jing, X.; Pan, M.; Liu, P.; Li, W.; Zhu, B.; Li, J.; Chen, H.; Wang, L.; Lin, J.; Liu, Y.; Zhao, D.; Yan, H.; Fan, C., Complex silica composite nanomaterials templated with DNA origami. *Nature* **2018,** *559* (7715), 593-598.

263. Nguyen, L.; Doblinger, M.; Liedl, T.; Heuer-Jungemann, A., DNA-Origami-Templated Silica Growth by Sol-Gel Chemistry. *Angew Chem Int Ed Engl* **2019,** *58* (3), 912-916.

264. Nguyen, M.-K.; Nguyen, V. H.; Natarajan, A. K.; Huang, Y.; Ryssy, J.; Shen, B.; Kuzyk, A., Ultrathin Silica Coating of DNA Origami Nanostructures. *Chem. Mater.* **2020,** *32* (15), 6657-6665.

265. Shang, Y.; Li, N.; Liu, S.; Wang, L.; Wang, Z. G.; Zhang, Z.; Ding, B., Site-Specific Synthesis of Silica Nanostructures on DNA Origami Templates. *Advanced materials (Deerfield Beach, Fla.)* **2020,** *32* (21), e2000294.

266. Ober, M. F.; Baptist, A.; Wassermann, L.; Heuer-Jungemann, A.; Nickel, B., In situ small-angle X-ray scattering reveals strong condensation of DNA origami during silicification. *arXiv preprint arXiv:2204.07385* **2022**.

267. Zhao, Y.; Zhang, C.; Yang, L.; Xu, X.; Xu, R.; Ma, Q.; Tang, Q.; Yang, Y.; Han, D., Programmable and Site-Specific Patterning on DNA Origami Templates with Heterogeneous Condensation of Silver and Silica. *Small* **2021,** *17* (47), e2103877.

268. Dai, X.; Chen, X.; Jing, X.; Zhang, Y.; Pan, M.; Li, M.; Li, Q.; Liu, P.; Fan, C.; Liu, X., DNA Origami-Encoded Integration of Heterostructures. *Angew Chem Int Ed Engl* **2022,** *61* (11), e202114190.

269. Shani, L.; Michelson, A. N.; Minevich, B.; Fleger, Y.; Stern, M.; Shaulov, A.; Yeshurun, Y.; Gang, O., DNA-assembled superconducting 3D nanoscale architectures. *Nat Commun* **2020,** *11* (1), 5697.

270. Liu, X.; Jing, X.; Liu, P.; Pan, M.; Liu, Z.; Dai, X.; Lin, J.; Li, Q.; Wang, F.; Yang, S.; Wang, L.; Fan, C., DNA Framework-Encoded Mineralization of Calcium Phosphate. *Chem* **2020,** *6* (2), 472-485.

271. Wu, S.; Zhang, M.; Song, J.; Weber, S.; Liu, X.; Fan, C.; Wu, Y., Fine Customization of

Calcium Phosphate Nanostructures with Site-Specific Modification by DNA Templated Mineralization. *ACS Nano* **2021,** *15* (1), 1555-1565.
272. Maune, H. T.; Han, S.-p.; Barish, R. D.; Bockrath, M.; Rothemund, P. W.; Winfree, E., Self-assembly of carbon nanotubes into two-dimensional geometries using DNA origami templates. *Nature nanotechnology* **2010,** *5* (1), 61-66.
273. Eskelinen, A. P.; Kuzyk, A.; Kaltiaisenaho, T. K.; Timmermans, M. Y.; Nasibulin, A. G.; Kauppinen, E. I.; Törmä, P., Assembly of single‐walled carbon nanotubes on DNA‐origami templates through streptavidin–biotin interaction. *Small* **2011,** *7* (6), 746-750.
274. Zhao, Z.; Liu, Y.; Yan, H., DNA origami templated self-assembly of discrete length single wall carbon nanotubes. *Organic & biomolecular chemistry* **2013,** *11* (4), 596-598.
275. Zhang, Y.; Mao, X.; Li, F.; Li, M.; Jing, X.; Ge, Z.; Wang, L.; Liu, K.; Zhang, H.; Fan, C., Nanoparticle‐Assisted Alignment of Carbon Nanotubes on DNA Origami. *Angewandte Chemie* **2020,** *132* (12), 4922-4926.
276. Zhao, M.; Chen, Y.; Wang, K.; Zhang, Z.; Streit, J. K.; Fagan, J. A.; Tang, J.; Zheng, M.; Yang, C.; Zhu, Z., DNA-directed nanofabrication of high-performance carbon nanotube field-effect transistors. *Science* **2020,** *368* (6493), 878-881.
277. Li, N.; Shang, Y.; Xu, R.; Jiang, Q.; Liu, J.; Wang, L.; Cheng, Z.; Ding, B., Precise Organization of Metal and Metal Oxide Nanoclusters into Arbitrary Patterns on DNA Origami. *J Am Chem Soc* **2019,** *141* (45), 17968-17972.
278. Zhang, Y.; Qu, Z. B.; Jiang, C.; Liu, Y.; Pradeep Narayanan, R.; Williams, D.; Zuo, X.; Wang, L.; Yan, H.; Liu, H.; Fan, C., Prescribing Silver Chirality with DNA Origami. *J Am Chem Soc* **2021,** *143* (23), 8639-8646.
279. Jia, S.; Wang, J.; Xie, M.; Sun, J.; Liu, H.; Zhang, Y.; Chao, J.; Li, J.; Wang, L.; Lin, J.; Gothelf, K. V.; Fan, C., Programming DNA origami patterning with non-canonical DNA-based metallization reactions. *Nat Commun* **2019,** *10* (1), 5597.
280. Aryal, B. R.; Ranasinghe, D. R.; Pang, C.; Ehlert, A. E. F.; Westover, T. R.; Harb, J. N.; Davis, R. C.; Woolley, A. T., Annealing of Polymer-Encased Nanorods on DNA Origami Forming Metal–Semiconductor Nanowires: Implications for Nanoelectronics. *ACS Appl. Nano Mater.* **2021,** *4* (9), 9094-9103.
281. Aryal, B. R.; Ranasinghe, D. R.; Westover, T. R.; Calvopiña, D. G.; Davis, R. C.; Harb, J. N.; Woolley, A. T., DNA origami mediated electrically connected metal—semiconductor junctions. *Nano Res.* **2020,** *13* (5), 1419-1426.
282. Meyer, T. A.; Zhang, C.; Bao, G.; Ke, Y., Programmable assembly of iron oxide nanoparticles using DNA origami. *Nano Letters* **2020,** *20* (4), 2799-2805.
283. Sun, W.; Boulais, E.; Hakobyan, Y.; Wang, W. L.; Guan, A.; Bathe, M.; Yin, P., Casting inorganic structures with DNA molds. *Science* **2014,** *346* (6210), 1258361.
284. Helmi, S.; Ziegler, C.; Kauert, D. J.; Seidel, R., Shape-controlled synthesis of gold nanostructures using DNA origami molds. *Nano letters* **2014,** *14* (11), 6693-6698.
285. Bayrak, T. r.; Helmi, S.; Ye, J.; Kauert, D.; Kelling, J.; Schönherr, T.; Weichelt, R.; Erbe, A.; Seidel, R., DNA-mold templated assembly of conductive gold nanowires. *Nano Letters* **2018,** *18* (3), 2116-2123.
286. Ye, J.; Helmi, S.; Teske, J.; Seidel, R., Fabrication of metal nanostructures with programmable length and patterns using a modular DNA platform. *Nano Letters* **2019,** *19* (4), 2707-2714.
287. Ye, J.; Weichelt, R.; Kemper, U.; Gupta, V.; König, T. A.; Eychmüller, A.; Seidel, R., Casting of Gold Nanoparticles with High Aspect Ratios inside DNA Molds. *Small* **2020,** *16* (39), 2003662.
288. Ye, J.; Aftenieva, O.; Bayrak, T.; Jain, A.; König, T. A.; Erbe, A.; Seidel, R., Complex Metal Nanostructures with Programmable Shapes from Simple DNA Building Blocks. *Advanced Materials* **2021,** *33* (29), 2100381.

289. Hannewald, N.; Winterwerber, P.; Zechel, S.; Ng, D. Y. W.; Hager, M. D.; Weil, T.; Schubert, U. S., DNA Origami Meets Polymers: A Powerful Tool for the Design of Defined Nanostructures. *Angew Chem Int Ed Engl* **2021,** *60* (12), 6218-6229.
290. Knudsen, J. B.; Liu, L.; Bank Kodal, A. L.; Madsen, M.; Li, Q.; Song, J.; Woehrstein, J. B.; Wickham, S. F.; Strauss, M. T.; Schueder, F.; Vinther, J.; Krissanaprasit, A.; Gudnason, D.; Smith, A. A.; Ogaki, R.; Zelikin, A. N.; Besenbacher, F.; Birkedal, V.; Yin, P.; Shih, W. M.; Jungmann, R.; Dong, M.; Gothelf, K. V., Routing of individual polymers in designed patterns. *Nat Nanotechnol* **2015,** *10* (10), 892-8.
291. Krissanaprasit, A.; Madsen, M.; Knudsen, J. B.; Gudnason, D.; Surareungchai, W.; Birkedal, V.; Gothelf, K. V., Programmed switching of single polymer conformation on DNA origami. *ACS nano* **2016,** *10* (2), 2243-2250.
292. Madsen, M.; Bakke, M. R.; Gudnason, D. A.; Sandahl, A. F.; Hansen, R. A.; Knudsen, J. B.; Kodal, A. L. B.; Birkedal, V.; Gothelf, K. V., A Single Molecule Polyphenylene-Vinylene Photonic Wire. *ACS Nano* **2021,** *15* (6), 9404-9411.
293. Madsen, M.; Christensen, R. S.; Krissanaprasit, A.; Bakke, M. R.; Riber, C. F.; Nielsen, K. S.; Zelikin, A. N.; Gothelf, K. V., Preparation, Single-Molecule Manipulation, and Energy Transfer Investigation of a Polyfluorene-graft-DNA polymer. *Chem. Eur. J.* **2017,** *23* (44), 10511-10515.
294. Zessin, J.; Fischer, F.; Heerwig, A.; Kick, A.; Boye, S.; Stamm, M.; Kiriy, A.; Mertig, M., Tunable Fluorescence of a Semiconducting Polythiophene Positioned on DNA Origami. *Nano Lett* **2017,** *17* (8), 5163-5170.
295. Tokura, Y.; Harvey, S.; Chen, C.; Wu, Y.; Ng, D. Y. W.; Weil, T., Fabrication of Defined Polydopamine Nanostructures by DNA Origami-Templated Polymerization. *Angew Chem Int Ed Engl* **2018,** *57* (6), 1587-1591.
296. Winterwerber, P.; Harvey, S.; Ng, D. Y. W.; Weil, T., Photocontrolled Dopamine Polymerization on DNA Origami with Nanometer Resolution. *Angew Chem Int Ed Engl* **2020,** *59* (15), 6144-6149.
297. Winterwerber, P.; Whitfield, C. J.; Ng, D. Y. W.; Weil, T., Multiple Wavelength Photopolymerization of Stable Poly(Catecholamines)-DNA Origami Nanostructures. *Angew Chem Int Ed Engl* **2022,** *61* (8), e202111226.
298. Schill, J.; Rosier, B.; Gumi Audenis, B.; Magdalena Estirado, E.; de Greef, T. F. A.; Brunsveld, L., Assembly of Dynamic Supramolecular Polymers on a DNA Origami Platform. *Angew Chem Int Ed Engl* **2021,** *60* (14), 7612-7616.
299. Yang, Y.; Lu, Q.; Huang, C. M.; Qian, H.; Zhang, Y.; Deshpande, S.; Arya, G.; Ke, Y.; Zauscher, S., Programmable Site-Specific Functionalization of DNA Origami with Polynucleotide Brushes. *Angew Chem Int Ed Engl* **2021,** *60* (43), 23241-23247.
300. Timm, C.; Niemeyer, C. M., Assembly and purification of enzyme-functionalized DNA origami structures. *Angew Chem Int Ed Engl* **2015,** *54* (23), 6745-50.
301. Zhao, Z.; Fu, J.; Dhakal, S.; Johnson-Buck, A.; Liu, M.; Zhang, T.; Woodbury, N. W.; Liu, Y.; Walter, N. G.; Yan, H., Nanocaged enzymes with enhanced catalytic activity and increased stability against protease digestion. *Nat Commun* **2016,** *7* (1), 10619.
302. Grossi, G.; Dalgaard Ebbesen Jepsen, M.; Kjems, J.; Andersen, E. S., Control of enzyme reactions by a reconfigurable DNA nanovault. *Nat. Commun.* **2017,** *8* (1), 992.
303. Kosinski, R.; Perez, J. M.; Schöneweiß, E.-C.; Ruiz-Blanco, Y. B.; Ponzo, I.; Bravo-Rodriguez, K.; Erkelenz, M.; Schlücker, S.; Uhlenbrock, G.; Sanchez-Garcia, E., The role of DNA nanostructures in the catalytic properties of an allosterically regulated protease. *Science advances* **2022,** *8* (1), eabk0425.
304. Fu, J.; Liu, M.; Liu, Y.; Woodbury, N. W.; Yan, H., Interenzyme substrate diffusion for an enzyme cascade organized on spatially addressable DNA nanostructures. *J Am Chem Soc* **2012,** *134* (12), 5516-9.
305. Chen, Y.; Ke, G.; Ma, Y.; Zhu, Z.; Liu, M.; Liu, Y.; Yan, H.; Yang, C. J., A Synthetic Light-Driven Substrate Channeling System for Precise Regulation of Enzyme Cascade Activity

Based on DNA Origami. *J Am Chem Soc* **2018,** *140* (28), 8990-8996.
306. Sun, L.; Gao, Y.; Xu, Y.; Chao, J.; Liu, H.; Wang, L.; Li, D.; Fan, C., Real-Time Imaging of Single-Molecule Enzyme Cascade Using a DNA Origami Raft. *J Am Chem Soc* **2017,** *139* (48), 17525-17532.
307. Xu, Y.; Gao, Y.; Su, Y.; Sun, L.; Xing, F.; Fan, C.; Li, D., Single-Molecule Studies of Allosteric Inhibition of Individual Enzyme on a DNA Origami Reactor. *The journal of physical chemistry letters* **2018,** *9* (23), 6786-6794.
308. Klein, W. P.; Thomsen, R. P.; Turner, K. B.; Walper, S. A.; Vranish, J.; Kjems, J.; Ancona, M. G.; Medintz, I. L., Enhanced Catalysis from Multienzyme Cascades Assembled on a DNA Origami Triangle. *ACS Nano* **2019,** *13* (12), 13677-13689.
309. Kahn, J. S.; Xiong, Y.; Huang, J.; Gang, O., Cascaded Enzyme Reactions over a Three-Dimensional, Wireframe DNA Origami Scaffold. *Jacs Au* **2022,** *2* (2), 357-366.
310. Fan, S.; Ji, B.; Liu, Y.; Zou, K.; Tian, Z.; Dai, B.; Cui, D.; Zhang, P.; Ke, Y.; Song, J., Spatiotemporal Control of Molecular Cascade Reactions by a Reconfigurable DNA Origami Domino Array. *Angew Chem Int Ed Engl* **2022,** *61* (9), e202116324.
311. Atsumi, H.; Belcher, A. M., DNA Origami and G-Quadruplex Hybrid Complexes Induce Size Control of Single-Walled Carbon Nanotubes *via* Biological Activation. *Acs Nano* **2018,** *12* (8), 7986-7995.
312. Masubuchi, T.; Endo, M.; Iizuka, R.; Iguchi, A.; Yoon, D. H.; Sekiguchi, T.; Qi, H.; Iinuma, R.; Miyazono, Y.; Shoji, S.; Funatsu, T.; Sugiyama, H.; Harada, Y.; Ueda, T.; Tadakuma, H., Construction of integrated gene logic-chip. *Nat Nanotechnol* **2018,** *13* (10), 933-940.
313. Hahn, J.; Chou, L. Y. T.; Sorensen, R. S.; Guerra, R. M.; Shih, W. M., Extrusion of RNA from a DNA-Origami-Based Nanofactory. *ACS Nano* **2020,** *14* (2), 1550-1559.
314. Rosier, B.; Markvoort, A. J.; Gumi Audenis, B.; Roodhuizen, J. A. L.; den Hamer, A.; Brunsveld, L.; de Greef, T. F. A., Proximity-induced caspase-9 activation on a DNA origami-based synthetic apoptosome. *Nature catalysis* **2020,** *3* (3), 295-306.
315. Kosuri, P.; Altheimer, B. D.; Dai, M.; Yin, P.; Zhuang, X., Rotation tracking of genome-processing enzymes using DNA origami rotors. *Nature* **2019,** *572* (7767), 136-140.
316. Jiang, T.; Meyer, T. A.; Modlin, C.; Zuo, X.; Conticello, V. P.; Ke, Y., Structurally Ordered Nanowire Formation from Co-Assembly of DNA Origami and Collagen-Mimetic Peptides. *J. Am. Chem. Soc.* **2017,** *139* (40), 14025-14028.
317. Jin, J.; Baker, E. G.; Wood, C. W.; Bath, J.; Woolfson, D. N.; Turberfield, A. J., Peptide Assembly Directed and Quantified Using Megadalton DNA Nanostructures. *ACS Nano* **2019,** *13* (9), 9927-9935.
318. Ouyang, X.; De Stefano, M.; Krissanaprasit, A.; Bank Kodal, A. L.; Bech Rosen, C.; Liu, T.; Helmig, S.; Fan, C.; Gothelf, K. V., Docking of Antibodies into the Cavities of DNA Origami Structures. *Angew Chem Int Ed Engl* **2017,** *56* (46), 14423-14427.
319. Knappe, G. A.; Wamhoff, E. C.; Read, B. J.; Irvine, D. J.; Bathe, M., In Situ Covalent Functionalization of DNA Origami Virus-like Particles. *ACS Nano* **2021,** *15* (9), 14316-14322.
320. Shaw, A.; Hoffecker, I. T.; Smyrlaki, I.; Rosa, J.; Grevys, A.; Bratlie, D.; Sandlie, I.; Michaelsen, T. E.; Andersen, J. T.; Hogberg, B., Binding to nanopatterned antigens is dominated by the spatial tolerance of antibodies. *Nat Nanotechnol* **2019,** *14* (2), 184-190.
321. Zhang, P.; Liu, X.; Liu, P.; Wang, F.; Ariyama, H.; Ando, T.; Lin, J.; Wang, L.; Hu, J.; Li, B.; Fan, C., Capturing transient antibody conformations with DNA origami epitopes. *Nat Commun* **2020,** *11* (1), 3114.
322. Zhou, K.; Ke, Y.; Wang, Q., Selective in Situ Assembly of Viral Protein onto DNA Origami. *J. Am. Chem. Soc.* **2018,** *140* (26), 8074-8077.
323. Zhou, K.; Zhou, Y.; Pan, V.; Wang, Q.; Ke, Y., Programming Dynamic Assembly of Viral Proteins with DNA Origami. *J. Am. Chem. Soc.* **2020,** *142* (13), 5929-5932.
324. Wang, S.-T.; Minevich, B.; Liu, J.; Zhang, H.; Nykypanchuk, D.; Byrnes, J.; Liu, W.; Bershadsky, L.; Liu, Q.; Wang, T.; Ren, G.; Gang, O., Designed and biologically active protein lattices. *Nat. Commun.* **2021,** *12* (1), 3702.
325. Ohno, H.; Kobayashi, T.; Kabata, R.; Endo, K.; Iwasa, T.; Yoshimura, S. H.; Takeyasu, K.; Inoue, T.; Saito, H., Synthetic RNA–protein complex shaped like an equilateral triangle. *Nat. Nanotechnol.* **2011,** *6* (2), 116-120.
326. Osada, E.; Suzuki, Y.; Hidaka, K.; Ohno, H.; Sugiyama, H.; Endo, M.; Saito, H., Engineering RNA–Protein Complexes with Different Shapes for Imaging and Therapeutic Applications. *ACS Nano* **2014,** *8* (8), 8130-8140.
327. Praetorius, F.; Dietz, H., Self-assembly of genetically encoded DNA-protein hybrid nanoscale shapes. *Science* **2017,** *355* (6331), eaam5488.
328. Xu, Y.; Jiang, S.; Simmons, C. R.; Narayanan, R. P.; Zhang, F.; Aziz, A.-M.; Yan, H.; Stephanopoulos, N., Tunable Nanoscale Cages from Self-Assembling DNA and Protein Building Blocks. *ACS Nano* **2019,** *13* (3), 3545-3554.
329. Comberlato, A.; Koga, M. M.; Nüssing, S.; Parish, I. A.; Bastings, M. M., Spatially Controlled Activation of Toll-like Receptor 9 with DNA-Based Nanomaterials. *Nano letters* **2022,** *22* (6), 2506-2513.
330. Veneziano, R.; Moyer, T. J.; Stone, M. B.; Wamhoff, E. C.; Read, B. J.; Mukherjee, S.; Shepherd, T. R.; Das, J.; Schief, W. R.; Irvine, D. J.; Bathe, M., Role of nanoscale antigen organization on B-cell activation probed using DNA origami. *Nat Nanotechnol* **2020,** *15* (8), 716-723.
331. Fang, T.; Alvelid, J.; Spratt, J.; Ambrosetti, E.; Testa, I.; Teixeira, A. I., Spatial Regulation of T-Cell Signaling by Programmed Death-Ligand 1 on Wireframe DNA Origami Flat Sheets. *ACS Nano* **2021,** *15* (2), 3441-3452.
332. Hellmeier, J.; Platzer, R.; Eklund, A. S.; Schlichthaerle, T.; Karner, A.; Motsch, V.; Schneider, M. C.; Kurz, E.; Bamieh, V.; Brameshuber, M.; Preiner, J.; Jungmann, R.; Stockinger, H.; Schutz, G. J.; Huppa, J. B.; Sevcsik, E., DNA origami demonstrate the unique stimulatory power of single pMHCs as T cell antigens. *Proceedings of the National Academy of Sciences of the United States of America* **2021,** *118* (4).
333. Sun, Y.; Yan, L.; Sun, J.; Xiao, M.; Lai, W.; Song, G.; Li, L.; Fan, C.; Pei, H., Nanoscale organization of two-dimensional multimeric pMHC reagents with DNA origami for CD8+ T cell detection. *Nature communications* **2022,** *13* (1), 1-11.
334. Jiang, D.; Ge, Z.; Im, H. J.; England, C. G.; Ni, D.; Hou, J.; Zhang, L.; Kutyreff, C. J.; Yan, Y.; Liu, Y.; Cho, S. Y.; Engle, J. W.; Shi, J.; Huang, P.; Fan, C.; Yan, H.; Cai, W., DNA origami nanostructures can exhibit preferential renal uptake and alleviate acute kidney injury. *Nature biomedical engineering* **2018,** *2* (11), 865-877.
335. Ma, Y.; Lu, Z.; Jia, B.; Shi, Y.; Dong, J.; Jiang, S.; Li, Z., DNA origami as a nanomedicine for targeted rheumatoid arthritis therapy through reactive oxygen species and nitric oxide scavenging. *ACS nano* **2022,** *16* (8), 12520-12531.
336. Sigl, C.; Willner, E. M.; Engelen, W.; Kretzmann, J. A.; Sachenbacher, K.; Liedl, A.; Kolbe, F.; Wilsch, F.; Aghvami, S. A.; Protzer, U., Programmable icosahedral shell system for virus trapping. *Nature materials* **2021,** *20* (9), 1281-1289.
337. Engelen, W.; Sigl, C.; Kadletz, K.; Willner, E. M.; Dietz, H., Antigen-Triggered Logic-Gating of DNA Nanodevices. *J Am Chem Soc* **2021,** *143* (51), 21630-21636.
338. Arnon, S.; Dahan, N.; Koren, A.; Radiano, O.; Ronen, M.; Yannay, T.; Giron, J.; Ben-Ami, L.; Amir, Y.; Hel-Or, Y.; Friedman, D.; Bachelet, I., Thought-Controlled Nanoscale Robots in a Living Host. *PloS one* **2016,** *11* (8), e0161227.
339. Jiang, Q.; Song, C.; Nangreave, J.; Liu, X.; Lin, L.; Qiu, D.; Wang, Z. G.; Zou, G.; Liang, X.; Yan, H.; Ding, B., DNA origami as a carrier for circumvention of drug resistance. *J Am Chem Soc* **2012,** *134* (32), 13396-403.
340. Zhang, Q.; Jiang, Q.; Li, N.; Dai, L.; Liu, Q.; Song, L.; Wang, J.; Li, Y.; Tian, J.; Ding, B.; Du, Y., DNA origami as an in vivo drug delivery vehicle for cancer therapy. *ACS Nano* **2014,** *8* (7), 6633-43.

341. Li, S.; Jiang, Q.; Liu, S.; Zhang, Y.; Tian, Y.; Song, C.; Wang, J.; Zou, Y.; Anderson, G. J.; Han, J. Y.; Chang, Y.; Liu, Y.; Zhang, C.; Chen, L.; Zhou, G.; Nie, G.; Yan, H.; Ding, B.; Zhao, Y., A DNA nanorobot functions as a cancer therapeutic in response to a molecular trigger in vivo. *Nat Biotechnol* **2018,** *36* (3), 258-264.
342. Wang, Z.; Song, L.; Liu, Q.; Tian, R.; Shang, Y.; Liu, F.; Liu, S.; Zhao, S.; Han, Z.; Sun, J.; Jiang, Q.; Ding, B., A Tubular DNA Nanodevice as a siRNA/Chemo-Drug Co-delivery Vehicle for Combined Cancer Therapy. *Angew Chem Int Ed Engl* **2021,** *60* (5), 2594-2598.
343. Liu, S.; Jiang, Q.; Zhao, X.; Zhao, R.; Wang, Y.; Wang, Y.; Liu, J.; Shang, Y.; Zhao, S.; Wu, T.; Zhang, Y.; Nie, G.; Ding, B., A DNA nanodevice-based vaccine for cancer immunotherapy. *Nat Mater* **2021,** *20* (3), 421-430.
344. Zhao, S.; Tian, R.; Wu, J.; Liu, S.; Wang, Y.; Wen, M.; Shang, Y.; Liu, Q.; Li, Y.; Guo, Y.; Wang, Z.; Wang, T.; Zhao, Y.; Zhao, H.; Cao, H.; Su, Y.; Sun, J.; Jiang, Q.; Ding, B., A DNA origami-based aptamer nanoarray for potent and reversible anticoagulation in hemodialysis. *Nat Commun* **2021,** *12* (1), 358.
345. Jarsch, I. K.; Daste, F.; Gallop, J. L., Membrane curvature in cell biology: An integration of molecular mechanisms. *J Cell Biol* **2016,** *214* (4), 375-87.
346. McMahon, H. T.; Boucrot, E., Membrane curvature at a glance. *J Cell Sci* **2015,** *128* (6), 1065-70.
347. Dias, C.; Nylandsted, J., Plasma membrane integrity in health and disease: significance and therapeutic potential. *Cell Discov* **2021,** *7* (1), 4.
348. Harayama, T.; Riezman, H., Understanding the diversity of membrane lipid composition. *Nat Rev Mol Cell Biol* **2018,** *19* (5), 281-296.
349. De Matteis, M. A.; Luini, A., Mendelian disorders of membrane trafficking. *N Engl J Med* **2011,** *365* (10), 927-38.
350. Zimmerberg, J.; Kozlov, M. M., How proteins produce cellular membrane curvature. *Nat Rev Mol Cell Biol* **2006,** *7* (1), 9-19.
351. Ainalem, M. L.; Kristen, N.; Edler, K. J.; Hook, F.; Sparr, E.; Nylander, T., DNA Binding to Zwitterionic Model Membranes. *Langmuir* **2010,** *26* (7), 4965-4976.
352. Gromelski, S.; Brezesinski, G., DNA condensation and interaction with Zwitterionic phospholipids mediated by divalent cations. *Langmuir* **2006,** *22* (14), 6293-6301.
353. Malghani, M. S.; Yang, J., Stable binding of DNA to zwitterionic lipid bilayers in aqueous solutions. *J Phys Chem B* **1998,** *102* (44), 8930-8933.
354. Audouy, S.; Hoekstra, D., Cationic lipid-mediated transfection in vitro and in vivo (review). *Mol Membr Biol* **2001,** *18* (2), 129-43.
355. Allen, T. M.; Cullis, P. R., Liposomal drug delivery systems: from concept to clinical applications. *Adv Drug Deliv Rev* **2013,** *65* (1), 36-48.
356. Stengel, G.; Zahn, R.; Hook, F., DNA-induced programmable fusion of phospholipid vesicles. *J Am Chem Soc* **2007,** *129* (31), 9584-5.
357. Chan, Y. H.; van Lengerich, B.; Boxer, S. G., Effects of linker sequences on vesicle fusion mediated by lipid-anchored DNA oligonucleotides. *Proceedings of the National Academy of Sciences of the United States of America* **2009,** *106* (4), 979-84.
358. Beales, P. A.; Vanderlick, T. K., Specific binding of different vesicle populations by the hybridization of membrane-anchored DNA. *J Phys Chem A* **2007,** *111* (49), 12372-80.
359. Beales, P. A.; Nam, J.; Vanderlick, T. K., Specific adhesion between DNA-functionalized "Janus" vesicles: size-limited clusters. *Soft Matter* **2011,** *7* (5), 1747-1755.
360. Selden, N. S.; Todhunter, M. E.; Jee, N. Y.; Liu, J. S.; Broaders, K. E.; Gartner, Z. J., Chemically programmed cell adhesion with membrane-anchored oligonucleotides. *J Am Chem Soc* **2012,** *134* (2), 765-8.
361. Vietri, M.; Radulovic, M.; Stenmark, H., The many functions of ESCRTs. *Nat Rev Mol Cell Biol* **2020,** *21* (1), 25-42.
362. Yang, Y.; Wang, J.; Shigematsu, H.; Xu, W.; Shih, W. M.; Rothman, J. E.; Lin, C., Self-

assembly of size-controlled liposomes on DNA nanotemplates. *Nature chemistry* **2016,** *8* (5), 476-483.
363. Perrault, S. D.; Shih, W. M., Virus-inspired membrane encapsulation of DNA nanostructures to achieve in vivo stability. *ACS Nano* **2014,** *8* (5), 5132-40.
364. Wang, C.; Piao, J.; Li, Y.; Tian, X.; Dong, Y.; Liu, D., Construction of Liposomes Mimicking Cell Membrane Structure through Frame-Guided Assembly. *Angew Chem Int Ed Engl* **2020,** *59* (35), 15176-15180.
365. Wang, C.; Zhang, Y. Y.; Shao, Y.; Tian, X. C.; Piao, J. F.; Dong, Y. C.; Liu, D. S., pH-responsive Frame-Guided Assembly with hydrophobicity controllable peptide as leading hydrophobic groups. *Giant* **2020,** *1*, 100006.
366. Dong, Y.; Yang, Y. R.; Zhang, Y.; Wang, D.; Wei, X.; Banerjee, S.; Liu, Y.; Yang, Z.; Yan, H.; Liu, D., Cuboid Vesicles Formed by Frame-Guided Assembly on DNA Origami Scaffolds. *Angew Chem Int Ed Engl* **2017,** *56* (6), 1586-1589.
367. Julin, S.; Nonappa; Shen, B.; Linko, V.; Kostiainen, M. A., DNA-Origami-Templated Growth of Multilamellar Lipid Assemblies. *Angew Chem Int Ed Engl* **2021,** *60* (2), 827-833.
368. Zhang, Z.; Yang, Y.; Pincet, F.; Llaguno, M. C.; Lin, C., Placing and shaping liposomes with reconfigurable DNA nanocages. *Nature chemistry* **2017,** *9* (7), 653-659.
369. Zhang, Y.; Pan, V.; Li, X.; Yang, X.; Li, H.; Wang, P.; Ke, Y., Dynamic DNA Structures. *Small* **2019,** *15* (26), e1900228.
370. Bian, X.; Zhang, Z.; Xiong, Q.; De Camilli, P.; Lin, C., A programmable DNA-origami platform for studying lipid transfer between bilayers. *Nat Chem Biol* **2019,** *15* (8), 830-837.
371. Zhang, Z.; Chapman, E. R., Programmable Nanodisc Patterning by DNA Origami. *Nano Lett* **2020,** *20* (8), 6032-6037.
372. Dong, Y.; Chen, S.; Zhang, S.; Sodroski, J.; Yang, Z.; Liu, D.; Mao, Y., Folding DNA into a Lipid‐Conjugated Nanobarrel for Controlled Reconstitution of Membrane Proteins. *Angewandte Chemie* **2018,** *130* (8), 2094-2098.
373. Zhao, Z.; Zhang, M.; Hogle, J. M.; Shih, W. M.; Wagner, G.; Nasr, M. L., DNA-Corralled Nanodiscs for the Structural and Functional Characterization of Membrane Proteins and Viral Entry. *J Am Chem Soc* **2018,** *140* (34), 10639-10643.
374. Iric, K.; Subramanian, M.; Oertel, J.; Agarwal, N. P.; Matthies, M.; Periole, X.; Sakmar, T. P.; Huber, T.; Fahmy, K.; Schmidt, T. L., DNA-encircled lipid bilayers. *Nanoscale* **2018,** *10* (39), 18463-18467.
375. Maingi, V.; Rothemund, P. W. K., Properties of DNA- and Protein-Scaffolded Lipid Nanodiscs. *ACS Nano* **2021,** *15* (1), 751-764.
376. Xu, W.; Nathwani, B.; Lin, C.; Wang, J.; Karatekin, E.; Pincet, F.; Shih, W.; Rothman, J. E., A Programmable DNA Origami Platform to Organize SNAREs for Membrane Fusion. *J Am Chem Soc* **2016,** *138* (13), 4439-47.
377. Dong, Y.; Chen, S.; Zhang, S.; Sodroski, J.; Yang, Z.; Liu, D.; Mao, Y., Folding DNA into a Lipid-Conjugated Nanobarrel for Controlled Reconstitution of Membrane Proteins. *Angew Chem Int Ed Engl* **2018,** *57* (8), 2072-2076.
378. Morzy, D.; Rubio-Sanchez, R.; Joshi, H.; Aksimentiev, A.; Di Michele, L.; Keyser, U. F., Cations Regulate Membrane Attachment and Functionality of DNA Nanostructures. *J Am Chem Soc* **2021,** *143* (19), 7358-7367.
379. Hernandez-Ainsa, S.; Ricci, M.; Hilton, L.; Avino, A.; Eritja, R.; Keyser, U. F., Controlling the Reversible Assembly of Liposomes through a Multistimuli Responsive Anchored DNA. *Nano Lett* **2016,** *16* (7), 4462-6.
380. Ohmann, A.; Gopfrich, K.; Joshi, H.; Thompson, R. F.; Sobota, D.; Ranson, N. A.; Aksimentiev, A.; Keyser, U. F., Controlling aggregation of cholesterol-modified DNA nanostructures. *Nucleic Acids Res* **2019,** *47* (21), 11441-11451.
381. Arnott, P. M.; Joshi, H.; Aksimentiev, A.; Howorka, S., Dynamic Interactions between Lipid-Tethered DNA and Phospholipid Membranes. *Langmuir* **2018,** *34* (49), 15084-15092.

382. Khmelinskaia, A.; Mucksch, J.; Petrov, E. P.; Franquelim, H. G.; Schwille, P., Control of Membrane Binding and Diffusion of Cholesteryl-Modified DNA Origami Nanostructures by DNA Spacers. *Langmuir* **2018,** *34* (49), 14921-14931.
383. Czogalla, A.; Petrov, E. P.; Kauert, D. J.; Uzunova, V.; Zhang, Y.; Seidel, R.; Schwille, P., Switchable domain partitioning and diffusion of DNA origami rods on membranes. *Faraday Discuss* **2013,** *161*, 31-43; discussion 113-50.
384. Sato, Y.; Endo, M.; Morita, M.; Takinoue, M.; Sugiyama, H.; Murata, S.; Nomura, S. M.; Suzuki, Y., Environment-Dependent Self-Assembly of DNA Origami Lattices on Phase-Separated Lipid Membranes. *Adv Mater Interfaces* **2018,** *5* (14).
385. Bogawat, Y.; Krishnan, S.; Simmel, F. C.; Santiago, I., Tunable 2D diffusion of DNA nanostructures on lipid membranes. *Biophys J* **2022,** *121*, 1–9.
386. Suzuki, Y.; Kawamata, I.; Watanabe, K.; Mano, E., Lipid bilayer-assisted dynamic self-assembly of hexagonal DNA origami blocks into monolayer crystalline structures with designed geometries. *iScience* **2022,** *25* (5), 104292.
387. Avakyan, N.; Conway, J. W.; Sleiman, H. F., Long-Range Ordering of Blunt-Ended DNA Tiles on Supported Lipid Bilayers. *J Am Chem Soc* **2017,** *139* (34), 12027-12034.
388. Kurokawa, C.; Fujiwara, K.; Morita, M.; Kawamata, I.; Kawagishi, Y.; Sakai, A.; Murayama, Y.; Nomura, S. M.; Murata, S.; Takinoue, M.; Yanagisawa, M., DNA cytoskeleton for stabilizing artificial cells. *Proceedings of the National Academy of Sciences of the United States of America* **2017,** *114* (28), 7228-7233.
389. Baumann, K. N.; Piantanida, L.; Garcia-Nafria, J.; Sobota, D.; Voitchovsky, K.; Knowles, T. P. J.; Hernandez-Ainsa, S., Coating and Stabilization of Liposomes by Clathrin-Inspired DNA Self-Assembly. *ACS Nano* **2020,** *14* (2), 2316-2323.
390. Yang, Y.; Wu, Z.; Wang, L.; Zhou, K.; Xia, K.; Xiong, Q.; Liu, L.; Zhang, Z.; Chapman, E. R.; Xiong, Y.; Melia, T. J.; Karatekin, E.; Gu, H.; Lin, C., Sorting sub-150-nm liposomes of distinct sizes by DNA-brick-assisted centrifugation. *Nature chemistry* **2021,** *13* (4), 335-342.
391. Gopfrich, K.; Zettl, T.; Meijering, A. E.; Hernandez-Ainsa, S.; Kocabey, S.; Liedl, T.; Keyser, U. F., DNA-Tile Structures Induce Ionic Currents through Lipid Membranes. *Nano Lett* **2015,** *15* (5), 3134-8.
392. Franquelim, H. G.; Khmelinskaia, A.; Sobczak, J. P.; Dietz, H.; Schwille, P., Membrane sculpting by curved DNA origami scaffolds. *Nat Commun* **2018,** *9* (1), 811.
393. Grome, M. W.; Zhang, Z.; Pincet, F.; Lin, C., Vesicle Tubulation with Self-Assembling DNA Nanosprings. *Angew Chem Int Ed Engl* **2018,** *57* (19), 5330-5334.
394. Grome, M. W.; Zhang, Z.; Lin, C., Stiffness and Membrane Anchor Density Modulate DNA-Nanospring-Induced Vesicle Tubulation. *ACS applied materials & interfaces* **2019,** *11* (26), 22987-22992.
395. Journot, C. M. A.; Ramakrishna, V.; Wallace, M. I.; Turberfield, A. J., Modifying Membrane Morphology and Interactions with DNA Origami Clathrin-Mimic Networks. *ACS Nano* **2019,** *13* (9), 9973-9979.
396. Jahnke, K.; Huth, V.; Mersdorf, U.; Liu, N.; Gopfrich, K., Bottom-Up Assembly of Synthetic Cells with a DNA Cytoskeleton. *ACS Nano* **2022,** *6* (5), 7233–7241.
397. Franquelim, H. G.; Dietz, H.; Schwille, P., Reversible membrane deformations by straight DNA origami filaments. *Soft Matter* **2021,** *17* (2), 276-287.
398. Zhan, P.; Jahnke, K.; Liu, N.; Gopfrich, K., Functional DNA-based cytoskeletons for synthetic cells. *Nature chemistry* **2022,** *14* (8), 958-963.
399. De Franceschi, N.; Pezeshkian, W.; Fragasso, A.; Bruininks, B. M. H.; Tsai, S.; Marrink, S. J.; Dekker, C., Synthetic Membrane Shaper for Controlled Liposome Deformation. *ACS Nano* **2022**.
400. Liu, L.; Xiong, Q.; Xie, C.; Pincet, F.; Lin, C., Actuating tension-loaded DNA clamps drives membrane tubulation. *Sci Adv* **2022,** *8* (41), eadd1830.
401. Baumann, K. N.; Schroder, T.; Ciryam, P. S.; Morzy, D.; Tinnefeld, P.; Knowles, T. P. J.;

Hernandez-Ainsa, S., DNA-Liposome Hybrid Carriers for Triggered Cargo Release. *ACS Appl Bio Mater* **2022,** *5* (8), 3713-3721.
402. Song, L.; Hobaugh, M. R.; Shustak, C.; Cheley, S.; Bayley, H.; Gouaux, J. E., Structure of staphylococcal alpha-hemolysin, a heptameric transmembrane pore. *Science* **1996,** *274* (5294), 1859-66.
403. Mueller, M.; Grauschopf, U.; Maier, T.; Glockshuber, R.; Ban, N., The structure of a cytolytic alpha-helical toxin pore reveals its assembly mechanism. *Nature* **2009,** *459* (7247), 726-U135.
404. Beck, M.; Hurt, E., The nuclear pore complex: understanding its function through structural insight. *Nat Rev Mol Cell Biol* **2017,** *18* (2), 73-89.
405. Catterall, W. A., Structure and Function of Voltage-Sensitive Ion Channels. *Science* **1988,** *242* (4875), 50-61.
406. Deamer, D.; Akeson, M.; Branton, D., Three decades of nanopore sequencing. *Nat Biotechnol* **2016,** *34* (5), 518-24.
407. Howorka, S., Building membrane nanopores. *Nat Nanotechnol* **2017,** *12* (7), 619-630.
408. Ding, T.; Yang, J.; Pan, V.; Zhao, N.; Lu, Z.; Ke, Y.; Zhang, C., DNA nanotechnology assisted nanopore-based analysis. *Nucleic Acids Res* **2020,** *48* (6), 2791-2806.
409. Shen, B.; Piskunen, P.; Nummelin, S.; Liu, Q.; Kostiainen, M. A.; Linko, V., Advanced DNA Nanopore Technologies. *ACS Appl Bio Mater* **2020,** *3* (9), 5606-5619.
410. Langecker, M.; Arnaut, V.; Martin, T. G.; List, J.; Renner, S.; Mayer, M.; Dietz, H.; Simmel, F. C., Synthetic lipid membrane channels formed by designed DNA nanostructures. *Science* **2012,** *338* (6109), 932-6.
411. Xing, Y.; Dorey, A.; Jayasinghe, L.; Howorka, S., Highly shape- and size-tunable membrane nanopores made with DNA. *Nat Nanotechnol* **2022,** *17* (7), 708-713.
412. Dey, S.; Dorey, A.; Abraham, L.; Xing, Y.; Zhang, I.; Zhang, F.; Howorka, S.; Yan, H., A reversibly gated protein-transporting membrane channel made of DNA. *Nat Commun* **2022,** *13* (1), 2271.
413. Seifert, A.; Gopfrich, K.; Burns, J. R.; Fertig, N.; Keyser, U. F.; Howorka, S., Bilayer-Spanning DNA Nanopores with Voltage-Switching between Open and Closed State. *Acs Nano* **2015,** *9* (2), 1117-1126.
414. Burns, J. R.; Gopfrich, K.; Wood, J. W.; Thacker, V. V.; Stulz, E.; Keyser, U. F.; Howorka, S., Lipid-bilayer-spanning DNA nanopores with a bifunctional porphyrin anchor. *Angew Chem Int Ed Engl* **2013,** *52* (46), 12069-72.
415. Burns, J. R.; Stulz, E.; Howorka, S., Self-assembled DNA nanopores that span lipid bilayers. *Nano Lett* **2013,** *13* (6), 2351-6.
416. Schmid, S.; Stommer, P.; Dietz, H.; Dekker, C., Nanopore electro-osmotic trap for the label-free study of single proteins and their conformations. *Nat Nanotechnol* **2021,** *16* (11), 1244-1250.
417. Fragasso, A.; De Franceschi, N.; Stommer, P.; van der Sluis, E. O.; Dietz, H.; Dekker, C., Reconstitution of Ultrawide DNA Origami Pores in Liposomes for Transmembrane Transport of Macromolecules. *ACS Nano* **2021**.
418. Iwabuchi, S.; Kawamata, I.; Murata, S.; Nomura, S. M., A large, square-shaped, DNA origami nanopore with sealing function on a giant vesicle membrane. *Chem Commun* **2021,** *57* (24), 2990-2993.
419. Krishnan, S.; Ziegler, D.; Arnaut, V.; Martin, T. G.; Kapsner, K.; Henneberg, K.; Bausch, A. R.; Dietz, H.; Simmel, F. C., Molecular transport through large-diameter DNA nanopores. *Nat Commun* **2016,** *7*, 12787.
420. Thomsen, R. P.; Malle, M. G.; Okholm, A. H.; Krishnan, S.; Bohr, S. S.; Sorensen, R. S.; Ries, O.; Vogel, S.; Simmel, F. C.; Hatzakis, N. S.; Kjems, J., A large size-selective DNA nanopore with sensing applications. *Nat Commun* **2019,** *10* (1), 5655.
421. Maingi, V.; Burns, J. R.; Uusitalo, J. J.; Howorka, S.; Marrink, S. J.; Sansom, M. S.,

Stability and dynamics of membrane-spanning DNA nanopores. *Nat Commun* **2017,** *8*, 14784.
422. Ohmann, A.; Li, C. Y.; Maffeo, C.; Al Nahas, K.; Baumann, K. N.; Gopfrich, K.; Yoo, J.; Keyser, U. F.; Aksimentiev, A., A synthetic enzyme built from DNA flips 10(7) lipids per second in biological membranes. *Nat Commun* **2018,** *9* (1), 2426.
423. Gopfrich, K.; Li, C. Y.; Ricci, M.; Bhamidimarri, S. P.; Yoo, J.; Gyenes, B.; Ohmann, A.; Winterhalter, M.; Aksimentiev, A.; Keyser, U. F., Large-Conductance Transmembrane Porin Made from DNA Origami. *Acs Nano* **2016,** *10* (9), 8207-8214.
424. Burns, J. R.; Howorka, S., Defined Bilayer Interactions of DNA Nanopores Revealed with a Nuclease-Based Nanoprobe Strategy. *Acs Nano* **2018,** *12* (4), 3263-3271.
425. Chidchob, P.; Offenbartl-Stiegert, D.; McCarthy, D.; Luo, X.; Li, J.; Howorka, S.; Sleiman, H. F., Spatial Presentation of Cholesterol Units on a DNA Cube as a Determinant of Membrane Protein-Mimicking Functions. *J Am Chem Soc* **2019,** *141* (2), 1100-1108.
426. Burns, J. R.; Seifert, A.; Fertig, N.; Howorka, S., A biomimetic DNA-based channel for the ligand-controlled transport of charged molecular cargo across a biological membrane. *Nat Nanotechnol* **2016,** *11* (2), 152-6.
427. Spruijt, E.; Tusk, S. E.; Bayley, H., DNA scaffolds support stable and uniform peptide nanopores. *Nat Nanotechnol* **2018,** *13* (8), 739-745.
428. Henning-Knechtel, A.; Knechtel, J.; Magzoub, M., DNA-assisted oligomerization of pore-forming toxin monomers into precisely-controlled protein channels. *Nucleic Acids Res* **2017,** *45* (21), 12057-12068.
429. Fennouri, A.; List, J.; Ducrey, J.; Dupasquier, J.; Sukyte, V.; Mayer, S. F.; Vargas, R. D.; Pascual Fernandez, L.; Bertani, F.; Rodriguez Gonzalo, S.; Yang, J.; Mayer, M., Tuning the Diameter, Stability, and Membrane Affinity of Peptide Pores by DNA-Programmed Self-Assembly. *ACS Nano* **2021,** *15* (7), 11263-11275.
430. Shen, Q.; Xiong, Q.; Zhou, K.; Feng, Q.; Liu, L.; Tian, T.; Wu, C.; Xiong, Y.; Melia, T. J.; Lusk, C. P.; Lin, C., Functionalized DNA-Origami-Protein Nanopores Generate Large Transmembrane Channels with Programmable Size-Selectivity. *Journal of the American Chemical Society* **2022**.
431. Offenbartl-Stiegert, D.; Rottensteiner, A.; Dorey, A.; Howorka, S., A Light-Triggered Synthetic Nanopore for Controlling Molecular Transport Across Biological Membranes. *Angew Chem Int Ed Engl* **2022**.
432. Lanphere, C.; Ciccone, J.; Dorey, A.; Hagleitner-Ertugrul, N.; Knyazev, D.; Haider, S.; Howorka, S., Triggered Assembly of a DNA-Based Membrane Channel. *J Am Chem Soc* **2022,** *144* (10), 4333-4344.
433. Mills, A.; Aissaoui, N.; Maurel, D.; Elezgaray, J.; Morvan, F.; Vasseur, J. J.; Margeat, E.; Quast, R. B.; Lai Kee-Him, J.; Saint, N.; Benistant, C.; Nord, A.; Pedaci, F.; Bellot, G., A modular spring-loaded actuator for mechanical activation of membrane proteins. *Nat Commun* **2022,** *13* (1), 3182.
434. Diederichs, T.; Ahmad, K.; Burns, J. R.; Nguyen, Q. H.; Siwy, Z. S.; Tornow, M.; Coveney, P. V.; Tampe, R.; Howorka, S., Principles of Small-Molecule Transport through Synthetic Nanopores. *Acs Nano* **2021,** *15* (10), 16194-16206.
435. Shen, Q.; Tian, T.; Xiong, Q.; Ellis Fisher, P. D.; Xiong, Y.; Melia, T. J.; Lusk, C. P.; Lin, C., DNA-Origami NanoTrap for Studying the Selective Barriers Formed by Phenylalanine-Glycine-Rich Nucleoporins. *J Am Chem Soc* **2021,** *143* (31), 12294-12303.
436. Fisher, P. D. E.; Shen, Q.; Akpinar, B.; Davis, L. K.; Chung, K. K. H.; Baddeley, D.; Saric, A.; Melia, T. J.; Hoogenboom, B. W.; Lin, C.; Lusk, C. P., A Programmable DNA Origami Platform for Organizing Intrinsically Disordered Nucleoporins within Nanopore Confinement. *ACS Nano* **2018,** *12* (2), 1508-1518.
437. Ketterer, P.; Ananth, A. N.; Laman Trip, D. S.; Mishra, A.; Bertosin, E.; Ganji, M.; van der Torre, J.; Onck, P.; Dietz, H.; Dekker, C., DNA origami scaffold for studying intrinsically disordered proteins of the nuclear pore complex. *Nat Commun* **2018,** *9* (1), 902.

438. Raveendran, M.; Lee, A. J.; Sharma, R.; Walti, C.; Actis, P., Rational design of DNA nanostructures for single molecule biosensing. *Nat Commun* **2020,** *11* (1), 4384.
439. Shi, X.; Pumm, A. K.; Isensee, J.; Zhao, W. X.; Verschueren, D.; Martin-Gonzalez, A.; Golestanian, R.; Dietz, H.; Dekker, C., Sustained unidirectional rotation of a self-organized DNA rotor on a nanopore. *Nat Phys* **2022,** *18* (9), 1105–1111.
440. Shi, X.; Pumm, A.-K.; Maffeo, C.; Kohler, F.; Zhao, W.; Verschueren, D.; Aksimentiev, A.; Dietz, H.; Dekker, C., A nanopore-powered DNA turbine. *arXiv preprint arXiv:2206.06612* **2022**.
441. Johnson, J. A.; Dehankar, A.; Robbins, A.; Kabtiyal, P.; Jergens, E.; Lee, K. H.; Johnston-Halperin, E.; Poirier, M.; Castro, C. E.; Winter, J. O., The path towards functional nanoparticle-DNA origami composites. *Materials Science and Engineering: R: Reports* **2019,** *138*, 153-209.
442. Kuzyk, A.; Jungmann, R.; Acuna, G. P.; Liu, N., DNA origami route for nanophotonics. *Acs Photonics* **2018,** *5* (4), 1151-1163.
443. Liu, N.; Liedl, T., DNA-assembled advanced plasmonic architectures. *Chemical reviews* **2018,** *118* (6), 3032-3053.
444. Hutter, E.; Fendler, J. H., Exploitation of localized surface plasmon resonance. *Advanced materials* **2004,** *16* (19), 1685-1706.
445. Shen, B.; Kostiainen, M. A.; Linko, V., DNA origami nanophotonics and plasmonics at interfaces. *Langmuir* **2018,** *34* (49), 14911-14920.
446. Fang, W.; Jia, S.; Chao, J.; Wang, L.; Duan, X.; Liu, H.; Li, Q.; Zuo, X.; Wang, L.; Wang, L., Quantizing single-molecule surface-enhanced Raman scattering with DNA origami metamolecules. *Science advances* **2019,** *5* (9), eaau4506.
447. Wang, P.; Huh, J. H.; Lee, J.; Kim, K.; Park, K. J.; Lee, S.; Ke, Y., Magnetic Plasmon Networks Programmed by Molecular Self‐Assembly. *Advanced Materials* **2019,** *31* (29), 1901364.
448. Dass, M.; Gür, F. N.; Kołątaj, K.; Urban, M. J.; Liedl, T., DNA origami-enabled plasmonic sensing. *The Journal of Physical Chemistry C* **2021,** *125* (11), 5969-5981.
449. Glembockyte, V.; Grabenhorst, L.; Trofymchuk, K.; Tinnefeld, P., DNA origami nanoantennas for fluorescence enhancement. *Accounts of Chemical Research* **2021,** *54* (17), 3338-3348.
450. Urban, M. J.; Shen, C.; Kong, X.-T.; Zhu, C.; Govorov, A. O.; Wang, Q.; Hentschel, M.; Liu, N., Chiral plasmonic nanostructures enabled by bottom-up approaches. *Annual Review of Physical Chemistry* **2019,** *70*, 275-299.
451. Schmied, J. J.; Gietl, A.; Holzmeister, P.; Forthmann, C.; Steinhauer, C.; Dammeyer, T.; Tinnefeld, P., Fluorescence and super-resolution standards based on DNA origami. *Nature methods* **2012,** *9* (12), 1133-1134.
452. Neubrech, F.; Hentschel, M.; Liu, N., Reconfigurable plasmonic chirality: fundamentals and applications. *Advanced Materials* **2020,** *32* (41), 1905640.
453. Schreiber, R.; Do, J.; Roller, E.-M.; Zhang, T.; Schüller, V. J.; Nickels, P. C.; Feldmann, J.; Liedl, T., Hierarchical assembly of metal nanoparticles, quantum dots and organic dyes using DNA origami scaffolds. *Nature nanotechnology* **2014,** *9* (1), 74-78.
454. Maccaferri, N.; Barbillon, G.; Koya, A. N.; Lu, G.; Acuna, G. P.; Garoli, D., Recent advances in plasmonic nanocavities for single-molecule spectroscopy. *Nanoscale Advances* **2021,** *3* (3), 633-642.
455. Wang, Y.; Dai, L.; Ding, Z.; Ji, M.; Liu, J.; Xing, H.; Liu, X.; Ke, Y.; Fan, C.; Wang, P., DNA origami single crystals with Wulff shapes. *Nature communications* **2021,** *12* (1), 1-8.
456. Xu, Z.; Huang, Y.; Yin, H.; Zhu, X.; Tian, Y.; Min, Q., DNA Origami‐Based Protein Manipulation Systems: From Function Regulation to Biological Application. *ChemBioChem* **2022,** *23* (9), e202100597.
457. Sakai, Y.; Islam, M. S.; Adamiak, M.; Shiu, S. C.-C.; Tanner, J. A.; Heddle, J. G., DNA aptamers for the functionalisation of DNA origami nanostructures. *Genes* **2018,** *9* (12), 571.

458. Sun, W.; Shen, J.; Zhao, Z.; Arellano, N.; Rettner, C.; Tang, J.; Cao, T.; Zhou, Z.; Ta, T.; Streit, J. K., Precise pitch-scaling of carbon nanotube arrays within three-dimensional DNA nanotrenches. *Science* **2020,** *368* (6493), 874-877.
459. Ijäs, H.; Nummelin, S.; Shen, B.; Kostiainen, M. A.; Linko, V., Dynamic DNA origami devices: from strand-displacement reactions to external-stimuli responsive systems. *International journal of molecular sciences* **2018,** *19* (7), 2114.
460. Li, Y.; Liu, Z.; Yu, G.; Jiang, W.; Mao, C., Self-assembly of molecule-like nanoparticle clusters directed by DNA nanocages. *Journal of the American Chemical Society* **2015,** *137* (13), 4320-4323.
461. Edwardson, T. G.; Lau, K. L.; Bousmail, D.; Serpell, C. J.; Sleiman, H. F., Transfer of molecular recognition information from DNA nanostructures to gold nanoparticles. *Nature chemistry* **2016,** *8* (2), 162-170.
462. Zhang, Y.; Chao, J.; Liu, H.; Wang, F.; Su, S.; Liu, B.; Zhang, L.; Shi, J.; Wang, L.; Huang, W., Transfer of two‐dimensional oligonucleotide patterns onto stereocontrolled plasmonic nanostructures through DNA‐origami‐based nanoimprinting lithography. *Angewandte Chemie International Edition* **2016,** *55* (28), 8036-8040.
463. Liu, W.; Halverson, J.; Tian, Y.; Tkachenko, A. V.; Gang, O., Self-organized architectures from assorted DNA-framed nanoparticles. *Nature chemistry* **2016,** *8* (9), 867-873.
464. Liu, B.; Song, C.; Zhu, D.; Wang, X.; Zhao, M.; Yang, Y.; Zhang, Y.; Su, S.; Shi, J.; Chao, J., DNA‐Origami‐based assembly of anisotropic plasmonic gold nanostructures. *Small* **2017,** *13* (23), 1603991.
465. Shen, C.; Lan, X.; Lu, X.; Meyer, T. A.; Ni, W.; Ke, Y.; Wang, Q., Site-specific surface functionalization of gold nanorods using DNA origami clamps. *Journal of the American Chemical Society* **2016,** *138* (6), 1764-1767.
466. Liu, Y.; Ma, L.; Jiang, S.; Han, C.; Tang, P.; Yang, H.; Duan, X.; Liu, N.; Yan, H.; Lan, X., DNA Programmable Self-Assembly of Planar, Thin-Layered Chiral Nanoparticle Superstructures with Complex Two-Dimensional Patterns. *ACS nano* **2021,** *15* (10), 16664-16672.
467. Ma, L.; Liu, Y.; Han, C.; Movsesyan, A.; Li, P.; Li, H.; Tang, P.; Yuan, Y.; Jiang, S.; Ni, W., DNA-Assembled Chiral Satellite-Core Nanoparticle Superstructures: Two-State Chiral Interactions from Dynamic and Static Conformations. *Nano Letters* **2022**.
468. Wang, P.; Huh, J.-H.; Park, H.; Yang, D.; Zhang, Y.; Zhang, Y.; Lee, J.; Lee, S.; Ke, Y., DNA origami guided self-assembly of plasmonic polymers with robust long-range plasmonic resonance. *Nano Letters* **2020,** *20* (12), 8926-8932.
469. Lan, X.; Su, Z.; Zhou, Y.; Meyer, T.; Ke, Y.; Wang, Q.; Chiu, W.; Liu, N.; Zou, S.; Yan, H., Programmable Supra‐Assembly of a DNA Surface Adapter for Tunable Chiral Directional Self‐Assembly of Gold Nanorods. *Angewandte Chemie* **2017,** *129* (46), 14824-14828.
470. Shen, C.; Lan, X.; Zhu, C.; Zhang, W.; Wang, L.; Wang, Q., Spiral patterning of Au nanoparticles on Au nanorod surface to form chiral AuNR@ AuNP helical superstructures templated by DNA origami. *Advanced Materials* **2017,** *29* (16), 1606533.
471. Jiang, Q.; Liu, Q.; Shi, Y.; Wang, Z.-G.; Zhan, P.; Liu, J.; Liu, C.; Wang, H.; Shi, X.; Zhang, L., Stimulus-responsive plasmonic chiral signals of gold nanorods organized on DNA origami. *Nano letters* **2017,** *17* (11), 7125-7130.
472. Ryssy, J.; Natarajan, A. K.; Wang, J.; Lehtonen, A. J.; Nguyen, M. K.; Klajn, R.; Kuzyk, A., Light‐Responsive Dynamic DNA‐Origami‐Based Plasmonic Assemblies. *Angewandte Chemie* **2021,** *133* (11), 5923-5927.
473. Göpfrich, K.; Urban, M. J.; Frey, C.; Platzman, I.; Spatz, J. P.; Liu, N., Dynamic actuation of DNA-assembled plasmonic nanostructures in microfluidic cell-sized compartments. *Nano letters* **2020,** *20* (3), 1571-1577.
474. Man, T.; Ji, W.; Liu, X.; Zhang, C.; Li, L.; Pei, H.; Fan, C., Chiral metamolecules with active plasmonic transition. *ACS nano* **2019,** *13* (4), 4826-4833.

475. Wang, M.; Dong, J.; Zhou, C.; Xie, H.; Ni, W.; Wang, S.; Jin, H.; Wang, Q., Reconfigurable plasmonic diastereomers assembled by DNA origami. *ACS nano* **2019,** *13* (12), 13702-13708.
476. Lan, X.; Liu, T. J.; Wang, Z. M.; Govorov, A. O.; Yan, H.; Liu, Y., DNA-Guided Plasmonic Helix with Switchable Chirality. *Journal of the American Chemical Society* **2018,** *140* (37), 11763-11770.
477. Xin, L.; Duan, X. Y.; Liu, N., Dimerization and oligomerization of DNA-assembled building blocks for controlled multi-motion in high-order architectures. *Nature Communications* **2021,** *12* (1).
478. Peil, A.; Zhan, P.; Duan, X.; Krahne, R.; Garoli, D.; Liz-Marzán, L. M.; Liu, N., Transformable Plasmonic Helix with Swinging Gold Nanoparticles. *Angew Chem Int Ed Engl* **2022**.
479. Funck, T.; Nicoli, F.; Kuzyk, A.; Liedl, T., Sensing Picomolar Concentrations of RNA Using Switchable Plasmonic Chirality. *Angew Chem Int Edit* **2018,** *57* (41), 13495-13498.
480. Huang, Y. K.; Nguyen, M. K.; Natarajan, A. K.; Nguyen, V. H.; Kuzyk, A., A DNA Origami-Based Chiral Plasmonic Sensing Device. *ACS applied materials & interfaces* **2018,** *10* (51), 44221-44225.
481. Zhou, C.; Xin, L.; Duan, X. Y.; Urban, M. J.; Liu, N., Dynamic Plasmonic System That Responds to Thermal and Aptamer-Target Regulations. *Nano Letters* **2018,** *18* (11), 7395-7399.
482. Liu, F. S.; Li, N.; Shang, Y. X.; Wang, Y. M.; Liu, Q.; Ma, Z. T.; Jiang, Q.; Ding, B. Q., A DNA-Based Plasmonic Nanodevice for Cascade Signal Amplification. *Angew Chem Int Edit* **2022,** *61* (22).
483. Roller, E. M.; Besteiro, L. V.; Pupp, C.; Khorashad, L. K.; Govorov, A. O.; Liedl, T., Hotspot-mediated non-dissipative and ultrafast plasmon passage. *Nat Phys* **2017,** *13* (8), 761-+.
484. Kneer, L. M.; Roller, E. M.; Besteiro, L. V.; Schreiber, R.; Govorov, A. O.; Liedl, T., Circular Dichroism of Chiral Molecules in DNA-Assembled Plasmonic Hotspots. *Acs Nano* **2018,** *12* (9), 9110-9115.
485. Zhu, J.; Wu, F.; Han, Z.; Shang, Y.; Liu, F.; Yu, H.; Yu, L.; Li, N.; Ding, B., Strong Light-Matter Interactions in Chiral Plasmonic-Excitonic Systems Assembled on DNA Origami. *Nano Lett* **2021,** *21* (8), 3573-3580.
486. Yuan, Y.; Li, H.; Yang, H.; Han, C.; Hu, H.; Govorov, A. O.; Yan, H.; Lan, X., Unraveling the Complex Chirality Evolution in DNA-Assembled High-Order, Hybrid Chiroplasmonic Superstructures from Multi-Scale Chirality Mechanisms. *Angew Chem Int Ed Engl* **2022,** *61* (44), e202210730.
487. Wang, P.; Huh, J. H.; Lee, J.; Kim, K.; Park, K. J.; Lee, S.; Ke, Y., Magnetic Plasmon Networks Programmed by Molecular Self-Assembly. *Advanced materials (Deerfield Beach, Fla.)* **2019,** *31* (29), e1901364.
488. Schröder, T.; Scheible, M. B.; Steiner, F.; Vogelsang, J.; Tinnefeld, P., Interchromophoric Interactions Determine the Maximum Brightness Density in DNA Origami Structures. *Nano Lett* **2019,** *19* (2), 1275-1281.
489. Nicoli, F.; Barth, A.; Bae, W.; Neukirchinger, F.; Crevenna, A. H.; Lamb, D. C.; Liedl, T., Directional Photonic Wire Mediated by Homo-Förster Resonance Energy Transfer on a DNA Origami Platform. *ACS Nano* **2017,** *11* (11), 11264-11272.
490. Zhou, X.; Liu, H.; Djutanta, F.; Satyabola, D.; Jiang, S. X.; Qi, X. D.; Yu, L.; Lin, S.; Hariadi, R. F.; Liu, Y.; Woodbury, N. W.; Yan, H., DNA-templated programmable excitonic wires for micron-scale exciton transport. *Chem* **2022,** *8* (9), 2442-2459.
491. Kaminska, I.; Bohlen, J.; Rocchetti, S.; Selbach, F.; Acuna, G. P.; Tinnefeld, P., Distance Dependence of Single-Molecule Energy Transfer to Graphene Measured with DNA Origami Nanopositioners. *Nano Lett* **2019,** *19* (7), 4257-4262.
492. Bartnik, K.; Barth, A.; Pilo-Pais, M.; Crevenna, A. H.; Liedl, T.; Lamb, D. C., A DNA Origami Platform for Single-Pair Förster Resonance Energy Transfer Investigation of DNA-DNA Interactions and Ligation. *J Am Chem Soc* **2020,** *142* (2), 815-825.

493. Gopinath, A.; Thachuk, C.; Mitskovets, A.; Atwater, H. A.; Kirkpatrick, D.; Rothemund, P. W. K., Absolute and arbitrary orientation of single-molecule shapes. *Science* **2021,** *371* (6531).
494. Hübner, K.; Joshi, H.; Aksimentiev, A.; Stefani, F. D.; Tinnefeld, P.; Acuna, G. P., Determining the In-Plane Orientation and Binding Mode of Single Fluorescent Dyes in DNA Origami Structures. *ACS Nano* **2021,** *15* (3), 5109-5117.
495. Adamczyk, A. K.; Huijben, T.; Sison, M.; Di Luca, A.; Chiarelli, G.; Vanni, S.; Brasselet, S.; Mortensen, K. I.; Stefani, F. D.; Pilo-Pais, M.; Acuna, G. P., DNA Self-Assembly of Single Molecules with Deterministic Position and Orientation. *ACS Nano* **2022,** *16* (10), 16924-16931.
496. Hübner, K.; Pilo-Pais, M.; Selbach, F.; Liedl, T.; Tinnefeld, P.; Stefani, F. D.; Acuna, G. P., Directing Single-Molecule Emission with DNA Origami-Assembled Optical Antennas. *Nano Lett* **2019,** *19* (9), 6629-6634.
497. Kaminska, I.; Bohlen, J.; Mackowski, S.; Tinnefeld, P.; Acuna, G. P., Strong Plasmonic Enhancement of a Single Peridinin-Chlorophyll a-Protein Complex on DNA Origami-Based Optical Antennas. *ACS Nano* **2018,** *12* (2), 1650-1655.
498. Xin, L.; Lu, M.; Both, S.; Pfeiffer, M.; Urban, M. J.; Zhou, C.; Yan, H.; Weiss, T.; Liu, N.; Lindfors, K., Watching a Single Fluorophore Molecule Walk into a Plasmonic Hotspot. *Acs Photonics* **2019,** *6* (4), 985-993.
499. Raab, M.; Vietz, C.; Stefani, F. D.; Acuna, G. P.; Tinnefeld, P., Shifting molecular localization by plasmonic coupling in a single-molecule mirage. *Nat Commun* **2017,** *8*, 13966.
500. Ojambati, O. S.; Chikkaraddy, R.; Deacon, W. D.; Horton, M.; Kos, D.; Turek, V. A.; Keyser, U. F.; Baumberg, J. J., Quantum electrodynamics at room temperature coupling a single vibrating molecule with a plasmonic nanocavity. *Nat Commun* **2019,** *10* (1), 1049.
501. Gür, F. N.; McPolin, C. P. T.; Raza, S.; Mayer, M.; Roth, D. J.; Steiner, A. M.; Löffler, M.; Fery, A.; Brongersma, M. L.; Zayats, A. V.; König, T. A. F.; Schmidt, T. L., DNA-Assembled Plasmonic Waveguides for Nanoscale Light Propagation to a Fluorescent Nanodiamond. *Nano Lett* **2018,** *18* (11), 7323-7329.
502. Tanwar, S.; Haldar, K. K.; Sen, T., DNA Origami Directed Au Nanostar Dimers for Single-Molecule Surface-Enhanced Raman Scattering. *J Am Chem Soc* **2017,** *139* (48), 17639-17648.
503. Zhan, P.; Wen, T.; Wang, Z. G.; He, Y.; Shi, J.; Wang, T.; Liu, X.; Lu, G.; Ding, B., DNA Origami Directed Assembly of Gold Bowtie Nanoantennas for Single-Molecule Surface-Enhanced Raman Scattering. *Angew Chem Int Ed Engl* **2018,** *57* (11), 2846-2850.
504. Tapio, K.; Mostafa, A.; Kanehira, Y.; Suma, A.; Dutta, A.; Bald, I., A Versatile DNA Origami-Based Plasmonic Nanoantenna for Label-Free Single-Molecule Surface-Enhanced Raman Spectroscopy. *ACS Nano* **2021,** *15* (4), 7065-7077.
505. Niu, R.; Song, C.; Gao, F.; Fang, W.; Jiang, X.; Ren, S.; Zhu, D.; Su, S.; Chao, J.; Chen, S.; Fan, C.; Wang, L., DNA Origami-Based Nanoprinting for the Assembly of Plasmonic Nanostructures with Single-Molecule Surface-Enhanced Raman Scattering. *Angew Chem Int Ed Engl* **2021,** *60* (21), 11695-11701.
506. Chikkaraddy, R.; Turek, V. A.; Kongsuwan, N.; Benz, F.; Carnegie, C.; van de Goor, T.; de Nijs, B.; Demetriadou, A.; Hess, O.; Keyser, U. F.; Baumberg, J. J., Mapping Nanoscale Hotspots with Single-Molecule Emitters Assembled into Plasmonic Nanocavities Using DNA Origami. *Nano Lett* **2018,** *18* (1), 405-411.
507. Heck, C.; Kanehira, Y.; Kneipp, J.; Bald, I., Placement of Single Proteins within the SERS Hot Spots of Self‐Assembled Silver Nanolenses. *Angewandte Chemie International Edition* **2018,** *57* (25), 7444-7447.
508. Niu, R.; Gao, F.; Wang, D.; Zhu, D.; Su, S.; Chen, S.; YuWen, L.; Fan, C.; Wang, L.; Chao, J., Pattern Recognition Directed Assembly of Plasmonic Gap Nanostructures for Single-Molecule SERS. *ACS nano* **2022,** *16* (9), 14622-14631.
509. Schmidt, T. L.; Beliveau, B. J.; Uca, Y. O.; Theilmann, M.; Da Cruz, F.; Wu, C.-T.; Shih, W. M., Scalable amplification of strand subsets from chip-synthesized oligonucleotide libraries. *Nature communications* **2015,** *6* (1), 1-7.

510. Praetorius, F.; Kick, B.; Behler, K. L.; Honemann, M. N.; Weuster-Botz, D.; Dietz, H., Biotechnological mass production of DNA origami. *Nature* **2017,** *552* (7683), 84-87.
511. Zhang, Q.; Xia, K.; Jiang, M.; Li, Q.; Chen, W.; Han, M.; Li, W.; Ke, R.; Wang, F.; Zhao, Y., Catalytic DNA‐Assisted Mass Production of Arbitrary Single‐Stranded DNA. *Angewandte Chemie International Edition* **2022**.
512. Anastassacos, F. M.; Zhao, Z.; Zeng, Y.; Shih, W. M., Glutaraldehyde Cross-Linking of Oligolysines Coating DNA Origami Greatly Reduces Susceptibility to Nuclease Degradation. *Journal of the American Chemical Society* **2020,** *142* (7), 3311-3315.
513. Agarwal, N. P.; Matthies, M.; Gür, F. N.; Osada, K.; Schmidt, T. L., Block copolymer micellization as a protection strategy for DNA origami. *Angewandte Chemie International Edition* **2017,** *56* (20), 5460-5464.
514. Bertosin, E.; Stömmer, P.; Feigl, E.; Wenig, M.; Honemann, M. N.; Dietz, H., Cryo-Electron Microscopy and Mass Analysis of Oligolysine-Coated DNA Nanostructures. *ACS Nano* **2021,** *15* (6), 9391-9403.
515. Ponnuswamy, N.; Bastings, M.; Nathwani, B.; Ryu, J. H.; Chou, L. Y.; Vinther, M.; Li, W. A.; Anastassacos, F. M.; Mooney, D. J.; Shih, W. M., Oligolysine-based coating protects DNA nanostructures from low-salt denaturation and nuclease degradation. *Nature communications* **2017,** *8* (1), 1-9.
516. Wu, C.; Han, D.; Chen, T.; Peng, L.; Zhu, G.; You, M.; Qiu, L.; Sefah, K.; Zhang, X.; Tan, W., Building a multifunctional aptamer-based DNA nanoassembly for targeted cancer therapy. *Journal of the American Chemical Society* **2013,** *135* (49), 18644-18650.
517. Gerling, T.; Kube, M.; Kick, B.; Dietz, H., Sequence-programmable covalent bonding of designed DNA assemblies. *Science advances* **2018,** *4* (8), eaau1157.
518. Perrault, S. D.; Shih, W. M., Virus-Inspired Membrane Encapsulation of DNA Nanostructures To Achieve In Vivo Stability. *ACS Nano* **2014,** *8* (5), 5132-5140.
519. Jiang, D.; Ge, Z.; Im, H.-J.; England, C. G.; Ni, D.; Hou, J.; Zhang, L.; Kutyreff, C. J.; Yan, Y.; Liu, Y., DNA origami nanostructures can exhibit preferential renal uptake and alleviate acute kidney injury. *Nature biomedical engineering* **2018,** *2* (11), 865-877.
520. Zhang, Q.; Jiang, Q.; Li, N.; Dai, L.; Liu, Q.; Song, L.; Wang, J.; Li, Y.; Tian, J.; Ding, B.; Du, Y., DNA Origami as an In Vivo Drug Delivery Vehicle for Cancer Therapy. *ACS Nano* **2014,** *8* (7), 6633-6643.
521. Wang, P.; Rahman, M. A.; Zhao, Z.; Weiss, K.; Zhang, C.; Chen, Z.; Hurwitz, S. J.; Chen, Z. G.; Shin, D. M.; Ke, Y., Visualization of the Cellular Uptake and Trafficking of DNA Origami Nanostructures in Cancer Cells. *Journal of the American Chemical Society* **2018,** *140* (7), 2478-2484.
522. Bastings, M. M. C.; Anastassacos, F. M.; Ponnuswamy, N.; Leifer, F. G.; Cuneo, G.; Lin, C.; Ingber, D. E.; Ryu, J. H.; Shih, W. M., Modulation of the Cellular Uptake of DNA Origami through Control over Mass and Shape. *Nano Letters* **2018,** *18* (6), 3557-3564.
523. Lacroix, A.; Vengut-Climent, E.; de Rochambeau, D.; Sleiman, H. F., Uptake and Fate of Fluorescently Labeled DNA Nanostructures in Cellular Environments: A Cautionary Tale. *ACS Central Science* **2019,** *5* (5), 882-891.
524. Blanchard, A. T.; Salaita, K., Emerging uses of DNA mechanical devices. *Science* **2019,** *365* (6458), 1080-1081.
525. Keller, A.; Linko, V., Challenges and perspectives of DNA nanostructures in biomedicine. *Angewandte Chemie International Edition* **2020,** *59* (37), 15818-15833.
526. Zhao, Y.-X.; Shaw, A.; Zeng, X.; Benson, E.; Nystrom, A. M.; Högberg, B. r., DNA origami delivery system for cancer therapy with tunable release properties. *ACS nano* **2012,** *6* (10), 8684-8691.
527. Dutta, P. K.; Zhang, Y.; Blanchard, A. T.; Ge, C.; Rushdi, M.; Weiss, K.; Zhu, C.; Ke, Y.; Salaita, K., Programmable Multivalent DNA-Origami Tension Probes for Reporting Cellular Traction Forces. *Nano Lett* **2018,** *18* (8), 4803-4811.

528. Sun, Y.; Sun, J.; Xiao, M.; Lai, W.; Li, L.; Fan, C.; Pei, H., DNA origami–based artificial antigen-presenting cells for adoptive T cell therapy. *Science Advances* **2022,** *8* (48), eadd1106.
529. Birkholz, O.; Burns, J. R.; Richter, C. P.; Psathaki, O. E.; Howorka, S.; Piehler, J., Multi-functional DNA nanostructures that puncture and remodel lipid membranes into hybrid materials. *Nat Commun* **2018,** *9* (1), 1521.
530. Ge, Z.; Liu, J.; Guo, L.; Yao, G.; Li, Q.; Wang, L.; Li, J.; Fan, C., Programming cell–cell communications with engineered cell origami clusters. *Journal of the American Chemical Society* **2020,** *142* (19), 8800-8808.
531. Li, X.; Wang, T.; Sun, Y.; Li, C.; Peng, T.; Qiu, L., DNA-Based Molecular Engineering of the Cell Membrane. *Membranes* **2022,** *12* (2), 111.
532. Li, Y.; Maffeo, C.; Joshi, H.; Aksimentiev, A.; Menard, B.; Schulman, R., Leakless end-to-end transport of small molecules through micron-length DNA nanochannels. *Sci Adv* **2022,** *8* (36), eabq4834.
533. Vietz, C.; Lalkens, B.; Acuna, G. P.; Tinnefeld, P., Functionalizing large nanoparticles for small gaps in dimer nanoantennas. *New Journal of Physics* **2016,** *18* (4), 045012.
534. Zhao, Z.; Liu, Y.; Yan, H., Organizing DNA origami tiles into larger structures using preformed scaffold frames. *Nano letters* **2011,** *11* (7), 2997-3002.
535. Hayakawa, D.; Videbaek, T. E.; Hall, D. M.; Fang, H.; Sigl, C.; Feigl, E.; Dietz, H.; Fraden, S.; Hagan, M. F.; Grason, G. M.; Rogers, W. B., Geometrically programmed self-limited assembly of tubules using DNA origami colloids. *Proceedings of the National Academy of Sciences* **2022,** *119* (43), e2207902119.
536. Monferrer, A.; Kretzmann, J. A.; Sigl, C.; Sapelza, P.; Liedl, A.; Wittmann, B.; Dietz, H., Broad-Spectrum Virus Trapping with Heparan Sulfate-Modified DNA Origami Shells. *ACS Nano* **2022,** *16* (12), 20002-20009.
537. Tikhomirov, G.; Petersen, P.; Qian, L., Triangular DNA origami tilings. *Journal of the American Chemical Society* **2018,** *140* (50), 17361-17364.
538. Lin, T.; Yan, J.; Ong, L. L.; Robaszewski, J.; Lu, H. D.; Mi, Y.; Yin, P.; Wei, B., Hierarchical Assembly of DNA Nanostructures Based on Four-Way Toehold-Mediated Strand Displacement. *Nano Letters* **2018,** *18* (8), 4791-4795.
539. Wintersinger, C. M.; Minev, D.; Ershova, A.; Sasaki, H. M.; Gowri, G.; Berengut, J. F.; Corea-Dilbert, F. E.; Yin, P.; Shih, W. M., Multi-micron crisscross structures grown from DNA-origami slats. *Nature Nanotechnology* **2022**.
540. Kuzyk, A.; Urban, M. J.; Idili, A.; Ricci, F.; Liu, N., Selective control of reconfigurable chiral plasmonic metamolecules. *Science Advances* **2017,** *3* (4), e1602803.
541. Kuzyk, A.; Yang, Y.; Duan, X.; Stoll, S.; Govorov, A. O.; Sugiyama, H.; Endo, M.; Liu, N., A light-driven three-dimensional plasmonic nanosystem that translates molecular motion into reversible chiroptical function. *Nature communications* **2016,** *7* (1), 1-6.
542. Funke, J. J.; Ketterer, P.; Lieleg, C.; Schunter, S.; Korber, P.; Dietz, H., Uncovering the forces between nucleosomes using DNA origami. *Science advances* **2016,** *2* (11), e1600974.
543. Kuzuya, A.; Sakai, Y.; Yamazaki, T.; Xu, Y.; Komiyama, M., Nanomechanical DNA origami'single-molecule beacons' directly imaged by atomic force microscopy. *Nature communications* **2011,** *2* (1), 1-8.
544. Marras, A. E.; Shi, Z.; Lindell III, M. G.; Patton, R. A.; Huang, C.-M.; Zhou, L.; Su, H.-J.; Arya, G.; Castro, C. E., Cation-activated avidity for rapid reconfiguration of DNA nanodevices. *ACS nano* **2018,** *12* (9), 9484-9494.
545. Turek, V. A.; Chikkaraddy, R.; Cormier, S.; Stockham, B.; Ding, T.; Keyser, U. F.; Baumberg, J. J., Thermo-Responsive Actuation of a DNA Origami Flexor. *Advanced Functional Materials* **2018,** *28* (25), 1706410.
546. Nickels, P. C.; Wünsch, B.; Holzmeister, P.; Bae, W.; Kneer, L. M.; Grohmann, D.; Tinnefeld, P.; Liedl, T., Molecular force spectroscopy with a DNA origami–based nanoscopic force clamp. *Science* **2016,** *354* (6310), 305-307.

547. Chauhan, N.; Xiong, Y.; Ren, S.; Dwivedy, A.; Magazine, N.; Zhou, L.; Jin, X.; Zhang, T.; Cunningham, B. T.; Yao, S.; Huang, W.; Wang, X., Net-Shaped DNA Nanostructures Designed for Rapid/Sensitive Detection and Potential Inhibition of the SARS-CoV-2 Virus. *Journal of the American Chemical Society* **2022**.
548. Bazrafshan, A.; Kyriazi, M.-E.; Holt, B. A.; Deng, W.; Piranej, S.; Su, H.; Hu, Y.; El-Sagheer, A. H.; Brown, T.; Kwong, G. A., DNA gold nanoparticle motors demonstrate processive motion with bursts of speed up to 50 nm per second. *ACS nano* **2021,** *15* (5), 8427-8438.
549. Senior, A. E., ATP synthase: motoring to the finish line. *Cell* **2007,** *130* (2), 220-221.
550. Liu, N., DNA Nanotechnology Meets Nanophotonics. ACS Publications: 2020; Vol. 20, pp 8430-8431.
551. Shetty, R. M.; Brady, S. R.; Rothemund, P. W.; Hariadi, R. F.; Gopinath, A., Bench-Top Fabrication of Single-Molecule Nanoarrays by DNA Origami Placement. *ACS nano* **2021,** *15* (7), 11441-11450.
552. Kershner, R. J.; Bozano, L. D.; Micheel, C. M.; Hung, A. M.; Fornof, A. R.; Cha, J. N.; Rettner, C. T.; Bersani, M.; Frommer, J.; Rothemund, P. W. K.; Wallraff, G. M., Placement and orientation of individual DNA shapes on lithographically patterned surfaces. *Nature Nanotechnology* **2009,** *4* (9), 557-561.
553. Hung, A. M.; Micheel, C. M.; Bozano, L. D.; Osterbur, L. W.; Wallraff, G. M.; Cha, J. N., Large-area spatially ordered arrays of gold nanoparticles directed by lithographically confined DNA origami. *Nature Nanotechnology* **2010,** *5* (2), 121-126.